\documentclass[12pt,openright, ] {article}
\usepackage[T1]{fontenc}
\usepackage{fourier}
\usepackage{graphicx} 
\usepackage[utf8]{inputenc}
\usepackage[french]{babel}
\usepackage{amsfonts}
\usepackage{amssymb}
\usepackage{array,nccmath,mathtools}
\usepackage{pstricks-add}
\usepackage{pst-all}
\usepackage{epsf}
\usepackage{float}
 \usepackage{textcomp}
 \usepackage{amsthm}
\usepackage[shortlabels]{enumitem}
\setlist[enumerate, 1]{label = \upshape\arabic*. }
\usepackage{nccmath} 
\usepackage{empheq}
\DeclarePairedDelimiter{\abs}{\lvert}{\rvert}
\DeclarePairedDelimiter{\norm}{\lVert}{\rVert}
\DeclarePairedDelimiterX\innerp[2]{\lparen}{\rparen}{#1∣#2}
\def\pds(#1,#2){\innerp{#1}{#2}}
\usepackage{accents} 
\DeclareMathSymbol{\varkappa}       {\mathord}{AMSb}{"7B}
\newtheorem{theorem}{Th\'eor\`eme}[section]

\newtheorem{exo}{Exercice}
\newtheorem{sol*}{Solution}
\newtheorem{lemma}[theorem]{Lemme}
\newtheorem{corollary}[theorem]{Corollaire}
\newtheorem{definition}{D\'efinition}[section]
\newtheorem{proposition}[theorem]{Proposition}
\newtheorem{principe}[theorem]{Principe}
\newenvironment{preuve}[1]{\par\noindent{\textbf{\bf
D\'emonstration}
#1} \par\noindent}%
{\unskip\nobreak\hfil\penalty50\hskip2em\null\nobreak\hfil%
$\Box$\parfillskip0pt\par\medskip}
\newenvironment{sol}[1]{\par\noindent{\textbf{Solution}
#1} \par}%
{\unskip\nobreak\hfil\penalty50\hskip2em\null\nobreak\hfil%
$\Box$\parfillskip0pt\par\medskip}
\newenvironment{preuvera}[1]{\par\noindent{\textbf{D\'emonstration rapide}
#1} \par\noindent}%
{\unskip\nobreak\hfil\penalty50\hskip2em\null\nobreak\hfil%
$\Box$\parfillskip0pt\par\medskip}
\newenvironment{exemple}{\par\noindent
\textbf{Exemple}{}\;\quad}{\unskip\nobreak\hfil\penalty50\hskip2em\null\nobreak\hfil%
$\clubsuit$\parfillskip0pt\par\medskip}
\newenvironment{cexemple}{\par\noindent
\textbf{Contre-exemple}{}\;\quad}{\unskip\nobreak\hfil\penalty50\hskip2em\null\nobreak\hfil%
$\clubsuit$\parfillskip0pt\par\medskip}
\newenvironment{exemples}{\par\noindent
\textbf{Exemples {}}\quad}{\unskip\nobreak\hfil\penalty50\hskip2em\null\nobreak\hfil%
$\clubsuit$\parfillskip0pt\par\medskip}
\newtheorem{remark}[theorem]{Remarque}

\newcommand{\cc}{\mathbb C}
\newcommand{\nn}{\mathbb N}
\newcommand{\rr}{\mathbb R}
\newcommand{\rrp}{\mathbb R\mathbf P}
\newcommand{\f}{\varphi}
\newcommand{\al}{\alpha}
\newcommand{\be}{\beta}
\newcommand{\ga}{\gamma}
\newcommand{\sy}{\sigma}
\newcommand{\lm}{\lambda}

\newcommand{\zz}{\mathbb Z}
\newcommand{\la}{\langle}
\newcommand{\ra}{\rangle}
\newcommand{\ep}{\varepsilon}
\newcommand{\sat}{\mathrm{Sat}}
\newcommand{\su}{\mathrm{Supp}}
\newcommand{\Int}{\mathrm{Int}}

\let\ker\Ker
\DeclareMathOperator{\img}{Im}
\DeclareMathOperator{\vect}{vect}
\DeclareMathOperator{\grad}{grad}

\newcommand{\vv}{\mathcal {V}}
\newcommand{\ma}{\mathcal}
\newcommand{\vol}{\mathrm{Vol}}
\usepackage{pst-node}
\newpsobject{ncemptyline}{ncline}{linestyle=none}

\newcommand{\supp}{\mathrm{supp}}
\newcommand{\mbf}{\mathbf}

\newcommand{\rot}{\mathrm{rot}}
\newcommand{\Div}{\mathrm{div}}
\newcommand{\tr}{\mathrm{Tr}}
\DeclareMathOperator{\opop}{End}

\DeclareMathOperator{\I}{I}
\DeclareMathOperator{\J}{J}
\usepackage{amssymb}
\renewcommand*\emptyset{\varnothing}

\let \le \leqslant

\let \ge \geqslant
\author{Jean-Marc Rinkel\\
}
\title{De la structure des espaces fibrés aux équations fondamentales de la physique.
\center{Une introduction}}
\date{\empty}
\usepackage{makeidx}
\makeindex
\usepackage{epsf}
\begin{document}
\maketitle

\begin{tabular}[]{cccccc}
\'{A}&&&&&\\
&Hélène&&&&Pierre\\
&&Séverin&&Maud&\\
&&&Jean&&\\
\end{tabular}
\vspace{3cm}

\noindent  \emph{Schüler}

Ich wünschte recht gelehrt zu werden

Und möchte gern, was auf der Erden

Und in dem Himmel ist, erfassen,

Die Wissenschaft und die Natur.

\noindent\emph{Mephistopheles}

Da seid Ihr auf der rechten Spur;

Doch müsst Ihr Euch nicht zerstreuen lassen.

\noindent
\hspace{5cm} {Goethe }

\hspace{5cm} {(Faust) }
Je remercie mes collègues du groupe de travail Toeplitz, Bernard Alfonsi Jean Chanzy, Philippe Rambour et Abdellatif Seghier
pour leur écoute patiente, indulgente. Leurs remarques  ont guidé la rédaction de ces pages. 

This text describes the fiber bundle structure and shows its universality for writing the laws of classical physics: newtonian, relativistic and quantum mechanics.

 \tableofcontents

\section{Introduction}\label{intro}
 
 
Le calcul différentiel est enseigné au cours des trois premières années de licence dans le cadre des espaces vectoriels qui peuvent être de dimension infinie (espaces de Banach). Nous allons étendre ce cadre  tout en restant en dimension finie. 
Les définitions basées sur le taux d'accroissement 
 ne convien\-nent plus lorsqu'il s'agit par exemple de dériver le vecteur vitesse d'une particule se mouvant sur une surface car on ne dispose pas dans ce cas d'addition \og naturelle\fg. Pour ce faire nous devons introduire des structures nouvelles.
 On a choisi de développer dans ce livre l'extension dont il est question dans deux directions. 
 
 Dans la première, on montre comment passer d'une structure réelle à une structure complexe en décrivant le lien entre espaces vectoriels réels euclidiens et espaces vectoriels complexes hermitiens appelé \emph{complexification}. Nous exposerons par la suite cette complexification dans le cadre de structures plus générales telles que les fibrés vectoriels. Ces notions sont développées  au chapitre \ref{evrouc}, ainsi qu'aux paragraphes \ref{comfiture} et \ref{ciment} et \ref{cru}. 

 La deuxième étape d'élargissement se cantonne à des structures réelles. On y  introduit les notions de \emph{variété}  et par suite de \emph{fibré vectoriel sur une variété}. La seule mécanique newtonienne nous impose cet élargissement : un système de points matériels soumis à des forces extérieures et à des contraintes géométriques gagne à être décrit par un point décrivant une \og surface\fg \; (\emph{espace de configuration du système}) dont la dimension dépend précisément des contraintes géométriques. Cette surface est un exemple de variété  et la vitesse associée sera définie rigoureusement dès lors que l'on disposera de la notion  de fibré tangent.  L'intérêt de cette démarche est essentielle dans l'étude de la dynamique d'un système mécanique : celle-ci est décrite par des équations différentielles qu'on ne sait pas ou ne peut pas généralement résoudre (penser en particulier au problème des trois corps en interactions newtoniennes). Cependant les propriétés topologiques et géométrique de l'espace de configuration donnent des indications, au moins qualitative, sur  l'orbite du point qui synthétise le système mécanique. Les chapitres \ref{warheit} et \ref{etap2} réalisent cette deuxième étape.
  
Les lois fondamentales de la physique font intervenir des opérateurs différentiels du second degré tels que les dérivations secondes, l'opérateur  laplacien. Les définir amène à munir les fibrés vectoriels d'un opérateur de dérivation covariante  sur les 
champs de vecteurs, les formes différentielles et plus généralement les champs de tenseurs. 
La construction de la dérivée covariante  dépend d'un outil construit dans le chapitre \ref{cacudeux} appelé \emph{connexion}. Dans ce chapitre on introduit également l'opérateur laplacien.

Cependant il n'y a pas de connexion canonique sur un fibré vectoriel, tout comme il n'y a pas de base canonique dans un espace vectoriel.
Par conséquent le calcul différentiel à l'ordre $2$ dépend du choix arbitraire d'une connexion. Mais dans le cas particulier du fibré tangent à une variété pseudo-riemannienne, Levi-Civita a mis en évidence l'existence d'une unique connexion compatible avec la métrique : cette connexion (appelée \emph{connexion de Levi-Civita}) est entièrement déterminée par le tenseur métrique.
Elle permet  de définir différentes notions de courbure de nature tensorielle, avec des propriétés de symétrie intéressante, ou  de nature scalaire. Ces courbures aboutissent à la construction d'un tenseur fondamental pour la théorie de la relativité générale  appelé dans ce livre \emph{tenseur de Hilbert-Einstein-Cartan}. C'est ce tenseur, lié à la géométrie de l'espace-temps  qui va permettre à partir des idées d'Einstein d'établir l'équation du champ de gravitation en relativité générale.
 Le chapitre \ref{mollusk} traite en détaille la notion de courbure en recourant aux idées de \'{E}lie Cartan.
Revenons maintenant aux liens avec la physique.

 Le chapitre \ref{conclaqua}
développe l'interprétation du calcul différentiel de Levi-Civita dans le cadre de la mécanique newtonienne et quantique.

\noindent Dans le cadre de la mécanique newtonienne, la connexion de Levi-Civita permet d'interpréter les équations de Lagrange comme des équations de Newton intrinsèques sur l'espace de configuration. De même dans le cas où la transformée de Legendre est régulière peut-on interpréter les équations de Hamilton comme des équations de Newton intrinsèques à l'aide la connexion de Levi-Civita sur le fibré cotangent. 

\noindent Les équations d'Hamilton sont à la base de l'équation fondamentale d'évolution de la fonction d'onde en mécanique quantique, à savoir l'équation de Schrödinger, construite sur une représentation du hamiltonien par un opérateur, l'opérateur hamiltonien. Celle-ci est est une équation de propagation d'onde qui met en regard une dérivée temporelle de la fonction d'onde et l'action de l'opérateur hamiltonien sur cette fonction d'onde. Le fait que cette équation doive être maintenue en toute circonstance impose à l'opérateur hamiltonien une transformation établie par H. Weyl, lors de l'apparition d'un champ électromagnétique, transformation appelé \emph{principe d'invariance de jauge}. Ce principe a une interprétation géométrique : il est équivalent  à l'existence d'une connexion sur le fibré trivial en droites complexes sur l'espace de Minkowski.  Ainsi dans le cas de la mécanique quantique la notion de connexion intervient en rapport avec \og l'équation fondamentale de la dynamique\fg \;quantique sans être liée directement à la dérivée  de l'impulsion comme c'est le cas dans les équations de Lagrange.

Dans la construction de la théorie relativiste de la gravitation, exposée au chapitre \ref{equachation}, l'équation du champ de gravitation est une équation des milieux continus mettant en regard le tenseur de Hilbert-Einstein-Cartan tenseur, lié à la géométrie de l'espace-temps
et un tenseur décrivant la distribution des énergies dans un milieu continu appelé \emph{tenseur impulsion-énergie}.
Cette équation tensorielle contient en particulier comme information une correction relativiste de l'équation classique de Poisson qui évalue le laplacien du potentiel de gravitation à partir de la densité de masse.
Dans le cas particulier d'une métrique $g_{ij}$ statique de l'espace temps  les équations du champs associées à un théorème de Levi-Civita estimant le laplacien de $\sqrt{-g_{00}}$, on retrouve une correction de la loi de Poisson du cadre newtonien.

 De même que la géométrie permet de ré-interpréter le principe d'invariance de jauge, elle permet en relativité générale de donner une interprétation de la courbure scalaire :
 on pense par exemple au corollaire \ref{relhypgeoesp} p.
\pageref{relhypgeoesp} du chapitre \ref{geoequachation} qui évalue
 la courbure scalaire d'une hypersurface de l'espace-temps en termes d'énergie.
 
 Ce texte comporte six appendices. Ils ont trois utilités.
 \begin{enumerate}
  \item Ce sont des démonstrations de théorèmes du texte dont la lecture peut être passée ou remise sans nuire à la compréhension des énoncés lors d'une première lecture (voir les appendices \ref{cosssi}, \ref{museio} et le paragraphe \ref{demlevci} de l'appendice \ref{appquatro}).

\item Ils rappellent  des notions dont la connaissance est nécessaire à la compréhension du texte principal. Mais ils les rappelle avec les mêmes conventions et notations que celles du texte. Leur lecture évite ainsi la consultation d'autres sources et la gymnastique de leur traduction mais ne l'exclut pas. Citons dans ce groupe les appendices \ref{maxwiwi} et \ref{ap6} qui exposent le socle de la relativité restreinte sur lequel s'appuie la construction de théorie relativiste de la gravitation.

 \item Quant à l'appendice \ref{appquatro}, il expose, dans le but d'illustrer son  lien avec le transport parallèle, la notion d'orientation d'un fibré vectoriel. Sa lecture est facultative.
\end{enumerate}

\noindent Ce texte est émaillé de soixante-trois exercices  qui font font partie de la lecture. La lecture de leur énoncé s'impose donc et leur résolution immédiate dépendra de l'énergie du lecteur. Je  conseille au lecteur d'en tenter la résolution le plus systématiquement possible quitte à le faire à un moment qui  convient : ils testent 
bien entendu l'assimilation du texte, les résoudre ne comporte aucun risque sinon celui de consulter leur correction, \emph{car ils sont tous corrigés}. Lire le corrigé après une réflexion solitaire est source de progrès dans la compréhension des notions.

\noindent Un guide de lecture indique une lecture linéaire possible.





\vspace{1cm}

\noindent\textbf{Notation générale}

\emph{Dans tout le document on utilise la notation d'Einstein : si on dispose de $n$ éléments $a_i$ et de $n$ éléments $b^j$ dans une structure multiplicative et additive alors $$a_ib^{i}=:\sum_{i=1}^na_ib^{i}$$
Il y a sommation quand le même indice apparaît en position \og covariante\fg\; et position \og contravariante\fg.
Par exemple au lieu décrire dans $\rr^3$ $$x=a^1e_1+a^2 e_2+a^3 e_3,$$ on écrira $x=a^{i}e_i$ s'il est clair que la sommation comporte trois termes.}

\section{Leitfaden}
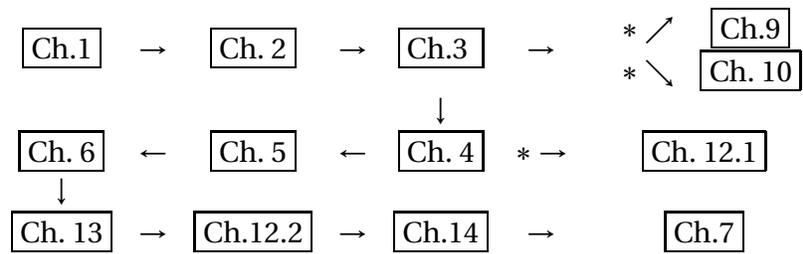
\[
\begin{matrix}
\fbox{Ch.1}&\rightarrow&\fbox{Ch. 2}&\rightarrow&\fbox{Ch.3 }&\rightarrow&\begin{matrix}&*\nearrow&\fbox{Ch.9}\\
&*\searrow&\fbox{Ch. 10}
\end{matrix}\\
&&&&\downarrow&\\
\fbox{Ch. 6}&\leftarrow&\fbox{Ch. 5}&\leftarrow&\fbox{Ch. 4}&*\rightarrow&\fbox{Ch. 12.1}\\
\downarrow&&&&&\\
\fbox{Ch. 13}&\rightarrow&\fbox{Ch.12.2}&\rightarrow&\fbox{Ch.14}&\rightarrow&\fbox{Ch.7}
\end{matrix}
\]

\noindent $\rightarrow $ et$ \leftarrow$ signifient \og lecture suivante\fg.\\
\noindent$*\rightarrow$, $*\nearrow$ et  $*\searrow$ signifient \og lecture suivante facultative\fg.
\section{Un détour par les espaces vectoriels}\label{evrouc}
\subsection{Complexifié d'un espace vectoriel euclidien.}
\subsubsection{Structure complexe sur un espace euclidien}
Le couple $(E,\langle\;\rangle)$ désigne un espace vectoriel euclidien sur $\rr$. On suppose qu'il existe sur cet espace une isométrie $\J$ telle que ${\J}^{2}=-\I$. Ceci impose une dimension paire pour $E$ car si $\dim E=n$, alors $\det(-I)=(-1)^n$. Le couple $(E,\langle\;\rangle,\J)$ est appelé structure complexe sur $E$. Si $\J^{*}$ désigne l'opérateur adjoint de $\J$, on a : $\J^{*}=\J^{{-1}}=-\J$.
Parfois on parlera de la structure complexe $\J$ sur $E$ en faisant référence à $(E,\langle\;\rangle,\J)$.\index{structure complexe sur un espace vectoriel}

\begin{proposition}\label{first}
Soit $E$ un espace vectoriel sur $\rr$ de dimension $2n$ et $(E,\langle\;\rangle,\J)$ une structure complexe sur $E$.  Il existe une base orthonormée de $E$ dans laquelle $\J$ est représentée par la matrice
\[\mathcal{J}_n=\left(
\begin{array}{cccc}
\begin{matrix*}[r]
0&-1\\
1&0
\end{matrix*}&&&\\
&\begin{matrix*}[r]
0&-1\\
1&0
\end{matrix*}&&\\ 
&&\ddots&\\
&&&\begin{matrix*}[r]
0&-1\\
1&0
\end{matrix*}
\end{array}\right)
\]
\end{proposition}
\begin{preuve}{}
 On considère un vecteur $e_{1}$ de norme $1$. Alors $\langle e_{1},\J(e_{1})\rangle =0$. On note $F_{1}$ le plan engendré par $(e_{1},\J(e_{1}))$. Si $n=2$, la base recherchée est $(e_1,J(e_1))$. Sinon,
on choisit $e_{2}$ un vecteur de norme $1$ dans $F_{1}^{\perp}$. Alors $\J(e_{2})\in F_{1}^{\perp}$ et le plan $F_{2}$ engendré par $(e_{2},\J(e_{2}))$ est inclus dans $F_{1}^{\perp}$.
Si $n=4$, $F_{1}\bigoplus F_{2}=E$ et la base recherchée est $(e_1,J(e_1),e_2,J(e_2))$. Sinon le processus continue en choisissant $e_{3 }$ de norme $1$ dans $(F_{1}\bigoplus F_{2})^{\perp}$.

 Ainsi par récurrence on construit une somme finie $\bigoplus_{k=1}^{n} F_{k}$ de plans deux à deux orthogonaux tels que $F_{k}=\vect\{e_{k},\J(e_{k})\}$ et $\vert\vert e_{k}\vert\vert=1$.
Alors le $2n$-uplet $\bigl(e_{1},\J(e_{1}), e_{2},\J(e_{2}),\cdots,e_{n},\J(e_{n})\bigr)$ est une base orthonormée de $E$ dans laquelle l'endomorphisme  $\J$ est représenté par la matrice $\mathcal{J}_n$.
\end{preuve}
\begin{corollary}{(\emph{unicité de la structure complexe})}

Soit $(E,\langle\;\rangle)$ un espace euclidien de dimension $2n$. 
Si $f$ et $g$ sont deux structures complexes sur $E$, alors $f$ et $g$ sont  conjugués : il existe un automor\-phisme $\psi$ tel que $g=\psi^{{-1}}f\psi$.
\end{corollary}
\begin{preuve}{}
Donnons-nous une base $\mathcal{B}$ de $E$. L'endomorphisme $f$ y est représenté par la matrice $A$ et $g$ par la matrice $A'$. Mais $A$ et $A'$ sont semblables à $\mathcal{J}$.
Donc elles sont semblables. Il existe une matrice inversible $P$ telle que $A'=P^{{-1}}AP$. Soit $\psi$ l'automorphisme représenté par la matrice $P$ dans la base  $\mathcal{B}$.
On a $g=\psi^{{-1}}f\psi$.
\end{preuve}
\begin{exemple}
On considère l'espace euclidien standard $\rr^{2}$. Les endomorphismes $\J_{1}$ et $\J_{2}$  respectivement représentés dans la base canonique par les matrices 
$\left(\begin{matrix*}[r]
0&-1\\
1&0
\end{matrix*}\right)$ et $\left(\begin{matrix*}[r]
0&1\\-1&0
\end{matrix*}\right)$ sont les deux  structures complexes de $\rr^{2}$ et $\J_{1}=\psi^{{-1}}J_{2}\psi$  pour n'importe quelle symétrie $\psi\in O(2)$.

\end{exemple}

\subsubsection{Complexifié d'un espace vectoriel réel}\index{complexifié d'un espace vectoriel réel}
\begin{definition}\label{azeza}
Un espace vectoriel $E_{\cc}$ sur $\cc$ est appelé complexifié de $(E,\langle\rangle,\J)$ s'il existe une bijection $\rr$-linéaire $\varphi$ de $E_{\cc}$ vers $E$ telle que le diagramme suivant 
\[
\begin{array}{ccc}
E_{\cc}&\stackrel{\varphi}{\longrightarrow}&E\\
\vcenter{\llap{i}}\Big\downarrow&&\Big\downarrow\vcenter{\rlap{$\J$}}\\
E_{\cc}&\xrightarrow[\enspace i\enspace]&E
\end{array}
\]
\noindent soit commutatif où $i$ désigne la multiplication par $i$.
\end{definition}
\begin{exemple}\label{russe}
On considère $\rr^{{2n}}$ muni du produit scalaire stantard. On considère $(e_{1},\ldots,e_{2n})$ la base canonique et $\J$ l'endomorphisme représenté dans cette base par la matrice
$\mathcal{J}$ dans cette base. 
On a donc $\J(e_{1})=e_{2},\J(e_{2})=-e_{1},\ldots,\J(e_{2n-1})=e_{2n},\J(e_{2n})=e_{2n-1}$.
On a bien  $\J^{2}=-\I$ et $\J$ transforme une base orthonormée en une base orthonormée. C'est une structure complexe sur $E$.
 
 En notant pour tout $k\in\{1,\ldots,n\}, z_{k}=x_{k}+iy_{k}$, on pose 
 \begin{equation}\label{fifi}
\varphi(z_{1},\ldots,z_{n})=(x_{1},y_{1},\ldots,x_{n},y_{n}).
\end{equation}
On vérifie que $\varphi$ est $\rr$-linéaire de $\cc^{n}$ vers $\rr^{2n}$ et que pour 
$Z=(z_{1},\ldots,z_{n}), \varphi(iZ)=\J\varphi(Z)$. L'espace $\cc^{n}$ est donc sans surprise un complexifié de $\rr^{2n}$.
\end{exemple}
\begin{proposition}
Deux complexifiés d'un espace vectoriel réel sont des espaces vectoriels complexes canoniquement isomorphes
\end{proposition}
\begin{preuve}{}
Soit $E^{1}_{\cc}$ et $E^{2}_{\cc}$ deux complexifiés de $E$. On a le diagramme commutatif
\[
\begin{array}{ccccc}
E^{1}_{\cc}&\stackrel{\varphi_{1}}{\longrightarrow}&E&\stackrel{\varphi_{2}^{{-1}}}{\longrightarrow}&E^{2}_{\cc}\\
\Big\downarrow\vcenter{\rlap{i}}&&\Big\downarrow\vcenter{\rlap{J}}&&\Big\downarrow\vcenter{\rlap{i}}\\
E^{1}_{\cc}&\stackrel{\varphi_1}{\longrightarrow}&E&\stackrel{\varphi_{2}^{{-1}}}{\longrightarrow}&E^{2}_{\cc}
\end{array}
\]
où $\varphi_{1}$  et $\varphi_{2}$ sont des bijections $\rr$-linéaires. On a :
\[\varphi_{2}^{{-1}}\varphi_{1}(i\ u)=\varphi_{2}^{-1}\J\varphi_{1}(u)=i\varphi_{2}^{{-1}}\varphi_{1}(u).\]
Ainsi $\varphi_{2}^{{-1}}\varphi_{1}$ est un isomorphisme d'espace vectoriel complexe.
\end{preuve}
\begin{proposition}\label{repu}
Si $(E,\J)$ est une structure complexe sur $E$, cet espace acquiert une structure d'espace vectoriel sur $\cc$ si on le munit de la multiplication externe définie par 
\[ \forall u\in E,\forall \lambda,\mu\in \rr,(\lambda+i\mu)\ast u=\lambda u+\mu \J(u).\]
\end{proposition}
\begin{preuve}{}
Il faut vérifier les axiomes. Par exemple, montrons que $$(a+ib)\ast\big((c+id)\ast u\big)=\big((a+ib)(c+id)\big)\ast u.$$
On a :

  $(a+ib)\ast\big((c+id)\ast u\big)=(a+ib)\ast\big(c u+d\J(u)\big)=ac u+ad\J(u)bc\J(u)-bdu=(ac-bd)u+(ad+bc)\J(u)=\big((a+ib)(c+id)\big)\ast u.$

  Les autres axiomes se démontrent identiquement.
\end{preuve}
\begin{corollary}
Tout complexifié de $E$ est canoniquement isomorphe à $(E,+,\ast)$.
\end{corollary}
\begin{preuve}{}
En effet $(E,+,\ast)$ est un complexifié de $E$ en prenant $\varphi=\I$ dans la définition \ref{aze}.
\end{preuve}
\begin{proposition}\label{hermit}
Soit $(E,\langle\;\rangle)$ un espace euclidien de dimension paire muni d'une structure complexe \;$\J$. Alors tout complexifié $E_\cc$ de $E$ hérite d'une structure hermitienne $\langle\;\rangle_{\cc}$ définie par 
\[\forall u,v\in E_{\cc}, \langle u,v\rangle_{\cc}=\langle\varphi(u),\varphi(v)\rangle +i\langle\varphi(u),\J\varphi(v)\rangle,\]
où $\varphi$ est la bijection linéaire de la définition \ref{aze}.
\end{proposition}
\begin{preuve}{}
Clairement,$\langle\; \;\rangle_{\cc}$ est additif à gauche. Pour tout complexe $\lambda=a+ib$, pour tous vecteurs $ u,v$, on a 
$\langle\lambda u,v\rangle_{\cc}=\lambda\langle u,v\rangle_{\cc}$. En effet,

$\langle\lambda u,v\rangle_{\cc}=\langle(a+ib) u,v\rangle_{\cc}=\langle a\varphi(u)+b\varphi(iu),\varphi(v)\rangle + i\langle a\varphi(u)+b\varphi(iu),\J\varphi(v)\rangle=$

$\noindent a\langle \varphi(u),\varphi(v)\rangle+ib\langle \varphi(iu),\J\varphi(v)\rangle+b\langle \varphi(iu),\varphi(v)\rangle +ia\langle \varphi(u),\J\varphi(v)\rangle=$

$\noindent a\langle \varphi(u),\varphi(v)\rangle+ib\langle\J\varphi(u),\J\varphi(v)\rangle+b\langle \J\varphi(u),\varphi(v)\rangle +ia\langle \varphi(u),\J\varphi(v)\rangle=$

$\noindent a\langle \varphi(u),\varphi(v)\rangle+ib\langle\varphi(u),\varphi(v)\rangle+b\langle \J\varphi(u),\varphi(v)\rangle +ia\langle \varphi(u),\J\varphi(v)\rangle=$

$\noindent (a+ib)\bigl(\langle \varphi(u),\varphi(v)\rangle+i\langle \varphi(u),\J\varphi(v)\rangle\bigr)=\lambda\langle u,v\rangle_{\cc}.$

\noindent De plus pour tous vecteurs $u,v$ on a $\langle u,v\rangle_{\cc}=\overline{\langle v,u\rangle_{\cc}}$ car $\langle u,v\rangle_{\cc}=\langle \varphi(u),\varphi(v)\rangle +i\langle \varphi(u),\J\varphi(v)\rangle=\langle \varphi(v),\varphi(u)\rangle -i\langle \varphi(v),\J\varphi(u)\rangle=\overline{\langle v,u\rangle_{\cc}}$.

\end{preuve}
\begin{proposition}
Si $\dim_{_{\rr}}(E)=2n$, alors $\dim_{_{\cc}}(E_{\cc})=n$.
\end{proposition}
\begin{preuve}{}
On considère la base orthonormée $(e_{1},\J(e_{1}), e_{2},\J(e_{2}),\cdots,e_{n},\J(e_{n}))$ de la démonstration de la proposition\ref{first}.
Alors $(e_{1}, e_{2},,\cdots,e_{n})$ est une base orthonormée de $E_{\cc}$, en prenant comme représentant de $(E_{\cc},\varphi)$ l'espace $(E,+,\ast)$ de la proposition \ref{repu} pour lequel $\varphi$ est l'identité.
\end{preuve}
\subsubsection{Complexifié d'un endomorphisme}\index{complexifié d'un endomorphisme}
\begin{proposition}\label{youyou}
Soit $E_{\cc}$ un complexifié de $E$ avec les notations de la définition \ref{aze} et $f$ un endomorphisme de $E$.
Alors $f_{\cc}\equiv \varphi^{-1}f\varphi$ est un endomorphisme de $E_{\cc}$ si et seulement si $\J f=f\J $. 
De plus, si $f$ est orthogonal, alors $f_{\cc}$ est unitaire.
\end{proposition}
\begin{preuve}{}
Supposons que $\J f=f\J $. L'application $f_{\cc}$ est clairement additive.  La seule chose à monter est la $\cc$-linéarité de $f_{\cc}$ c'est à dire $f_{\cc}(iu)=if_{\cc}(u)$. Or $\varphi^{-1}f\varphi(iu)=\varphi^{-1}f\J\varphi(u)=\varphi^{-1}\J f\varphi(u)=i(\varphi^{-1}f\varphi)(u)$. Inversement si l'on suppose que $\varphi^{-1}f\varphi$ est un endomorphisme complexe, le calcul précédent montre que $\J f=f\J $.
Supposons $f$ orthogonal. Pour des vecteurs $u,v$ de $E_{\cc}$, on a :
\begin{fleqn}[0.5em]
\begin{align*}
 \langle \varphi^{{-1}}f\varphi(u),\varphi^{{-1}}f\varphi(v)\rangle_{\cc}&= \langle f\varphi(u),f\varphi(v)\rangle
+i \langle f\varphi(u),\J f\varphi(v)\rangle\\
&= \langle \varphi(u),\varphi(v)\rangle+i\langle \varphi(u),\J\varphi(v)\rangle &&(\mathrm{puisque} \J f=f\J)\\
&=\langle u,v\rangle_{\cc}
\end{align*}
\end{fleqn}
\end{preuve}
\begin{definition}\index{endomorphisme complexe}
Un endomorphisme  $f$ de $(E,\langle\rangle,\J)$ tel que $\J f=f\J $ est appelé endomorphisme complexe de $E$ et $f_{\cc}$ est le complexifié de $f$.
\end{definition}
\begin{remark}\label{lavache}
\begin{enumerate}
\item Si $E_{\cc}=(E,+,\ast)$ alors $f_{\cc}=f $ et $f_{\cc}((a+ib)\ast u)=a f(u)+bi\ast f(u)=af(u)+b\J f(u).$
\item Considérons l'espace $\rr^{{2n}}$ muni de sa structure euclidienne standard et $J$ la structure complexe sur $\rr^{{2n}}$ donné dans l'exemple \ref{russe}, p.\pageref{russe}. 
 Le sous-ensemble $G$ constitué des endomorphismes complexes  appartenant à  $SO(2n)$ pour la structure $J$ constitue un sous-groupe \emph{propre} de $SO(2n)$ isomorphe à $U(n)$ d'après la proposition \ref{youyou}..
 
 En effet, c'est clairement un sous-groupe et il est propre car pour $n=2$, par exemple si $f\in SO(4)$ est représenté dans la base canonique de $\rr^{4}$ par la matrice\[ \left(\begin{matrix}1&0&0&0\\0&-1&0&0\\0&0&1&0\\0&0&0&-1
 \end{matrix}\right),\] alors $fJ\not=Jf$.
\end{enumerate}
\end{remark}
\begin{exemples}
\begin{enumerate}
\item Soit dans l'espace vectoriel $\rr^{2}$ la rotation d'angle $\theta$, notée $r_{\theta}$ et $\J=r_{\pi/2}$. Pour tout $\theta$, $r_{\theta}\J=\J r_{\theta}$. Ainsi $r_{\theta}$ est un endomorphisme complexe orthogonal de complexifié l'endomorphisme unitaire $z\to e^{i\theta}  z$.
\item Soit dans l'espace vectoriel $\rr^{2}$, l'endomorphisme $\Phi$ défini par $\Phi((1,0))=(a,0)$ et $ \Phi((0,1))=(0,b)$. Alors $\Phi\J=\J\Phi$ si et seulement si $a=b$. Si $a\not= b$, $\Phi$ n'est pas un endomorphisme complexe. Cela peut être vu également comme un corollaire de la proposition suivante.
\end{enumerate}
\end{exemples}

\begin{proposition}
Soit $f$ un endomorphisme complexe de $(E,\langle\rangle,\J)$. Les sous-espaces propres réels de $f$ sont de dimension paire.
\end{proposition}
\begin{preuve}{}
Du fait de l'égalité $f\J=\J f$, les espaces propres réels  sont stables par $\J$. Ils sont donc nécessairement de dimension paire.
\end{preuve}
\begin{remark}\label{dingue}
Avec les notations de la proposition précédente, \emph{si le complexifié de $f$ est diagonalisable sur $\cc$}, alors il existe une base de $E$ dans laquelle $f$ est représenté par une matrice réelle de la forme
\[
\left(\begin{array}{cccccc}
\begin{matrix}
a_{1}&-b_{1}\\
b_{1}&a_{1}
\end{matrix}&&&&&\\
&\ddots&&&&\\
&&\begin{matrix}
a_{p}&-b_{p}\\
b_{p}&a_{p}
\end{matrix}&&&\\
&&&\lambda_{1}&&\\
&&&&\ddots &\\
&&&&&\lambda_{\sigma}
\end{array}\right)
\]
où chaque $\lambda_{i}$ appara\^{i}t $2s_{i}$ fois de sorte que \[\sum_{i=1}^{\sigma}s_{i}+p=\dfrac{\dim E}{2}.\]
En effet, parmi les valeurs propres de $f_{\cc}$ il y a d'une part les valeurs propres complexes non réelles. Si $\lambda$ en est une si $u$ est un vecteur propre associé de $E_{\cc}=(E,+,\ast)$, $f_{\cc}(u)=f(u)=\lambda u=(a+ib)\ast u=a u+biu= au +b\J(u)$. Donc $f(\J(u))=a\J(u)-bu$.
 Le plan $\vect\{u,\J(u)\}$ est stable par $f$ et la restriction de $f$ à ce plan représentée dans la base $(u,\J(u))$ par la matrice \[\left(\begin{matrix}
a&-b\\b&a
\end{matrix}\right)
\]
Si $\lambda$ est de multiplicité $k$, on peut trouver une famille libre $\{e_{1},\cdots,e_{k}\}$ de vecteurs propres associés à $\lambda$, puisque $f_{\cc}$ est diagonalisable, ce qui produira $k$ blocs $(2,2)$ de la forme précédente. En considérant l'ensemble des valeurs pro\-pres complexes non réelles on forme une famille libre de vecteurs, base d'un sous-espace 
$F$ de $E_{\cc}$ stable par $f_{\cc}$ dans laquelle la restriction de $f_{\cc}$ est représentée par une matrices de blocs $(2,2)$ du type précédent. 

Il y a d'autre part les valeurs propres réelles $\{\lambda_{i}\}_{i=1,\cdots ,\sigma}$.
Les vecteurs propres associés engendrent un sous-espace $G$ de $E_{\cc}$ supplémentaire de $F$, puisque $f_{\cc}$ est diagonalisable. Si $\epsilon_j$ est un vecteur propre de $f_{\cc}$ associé à la valeur propre réelle $\lambda_j$, alors $i*\epsilon_j$ est également un vecteur propre de $f_{\cc}$ associé à la valeur propre réelle $\lambda_j$. Donc si $s_j$ est la multiplicité de $\lambda_j$, cette valeur propre apparaît $2 s_j$ fois dans la diagonale de la matrice précédente. 
En complétant la base de $F$ par une base de $G$, on obtient une base de $E$ dans laquelle $f$ est représenté par la matrice annoncée.
\end{remark}
\begin{exo}\label{matcom}
Soit $f$ un endomorphisme antisymétrique de l'espace euclidien $\rr^{4}$. Alors $f$ est un endomorphisme complexe si et seulement si $f$ est représenté dans la base canonique de 
$\rr^{4}$ par une matrice de la forme \begin{equation}\label{labonne}M=\left(\begin{matrix}A_{1}&A_{2}\\A_{4}&A_{3}\end{matrix}\right)
\end{equation}
avec $ A_{1}=-\mu \mathcal{J}_1 ,A_{3}=-\nu\mathcal{J}_1, A_{2}=-\alpha I -\beta \ma{J}_1, A_{4}=\alpha I -\beta \ma{J}_1$.

\noindent Alors l'endomorphisme complexifié de $f$ est représenté dans la base canonique de l'espace complexe $\cc^{2}$ par la matrice 
$M_{\cc}=\left(\begin{matrix}-i\mu&-\alpha-i\beta\\\alpha-i\beta&-i\nu\end{matrix}\right).$
\end{exo}
\subsection{Produit tensoriel de deux espaces vectoriels  de dimension finie.}
Les espaces vectoriels concernés ici sont réels ou complexes. Si $E$ et $F$ sont de tels espaces, on note $\mathcal{L}^2(E\times F)$ l'espace vectoriels des formes bilinéaires sur $E\times F$.
\begin{definition}\label{potence}
Soient $E$ et $F$ deux espaces vectoriels de dimension finie. Le produit tensoriel $E\times F$ est l'espace $\mathcal{L}^2(E^*\times F^*)$ où $E^*$ et $F^*$ sont respectivement les espaces vectoriels duaux de $E$ et $F$ (c.à.d. les espaces des formes linéaires sur $E$ et $F$).
\end{definition}
\begin{remark}
Dans les cours d'algèbre le produit tensoriel de deux espaces est défini par une propriété universelle et il est unique à un isomorphisme près. On pourra consulter l'appendice de [\ref{rinrin}]. Ici la définition propose directement une réalisation du produit tensoriel qui vérifie la propriété universelle.
\end{remark}
\begin{exemple}
Soit $(x,y)\in E\times F$ et  $x\otimes y : (x^*,y^*)\in E^*\times F^*\mapsto x^*(x) y^*(y)$. 
Alors $x\otimes y\in E\otimes F$. La vérification est directe.
\end{exemple}
\begin{proposition}\label{badpotence}
Soient $\{e_i\}$ une base de $E$ et $\{\varepsilon _j\}$ une base de $F$. Alors $\{e_i\otimes \varepsilon_j\}$ est une base de $E\otimes F$.
\end{proposition}
\begin{preuve}{}
On note $\{e^*_i\}$ et $\{\varepsilon^* _j\}$ les bases duales de $\{e_i\}$ et $\{\varepsilon _j\}$ :
$e^*_i(e_j)=\delta_{ij}$ et $\varepsilon _i(\varepsilon_j)=\delta_{ij}$.
On note $a_{ij}=\varphi(e^*_i,\varepsilon_j^*)$. On vérifie alors que $\varphi=\sum_{i,j}a_{ij}e_i\otimes \varepsilon_j$. En effet  si $x^*=\sum_ix_ie^*_i$ et $y^*=\sum_jy_j\varepsilon^*_j$, on a : $\varphi(x^*,y^*)
=\sum_{i,j}x_iy_ja_{ij}=\sum_{i,j}a_{ij}e_i\otimes \epsilon_j(x^*,y^*).$
La famille $\{e_i\otimes \varepsilon_j\}$ est donc génératrice. Elle est libre car si $\varphi=\sum_{i,j}a_{ij}e_i\otimes \varepsilon_j=0$, alors pour tout couple $(i,j)$, on a : $\varphi(e^*_i,\varepsilon^*_j)=a_{ij}=0$ ce qui permet de conclure.
\end{preuve}
\begin{proposition}\label{endopotence}
L'objet de cette proposition est l'obtention d'endomorphismes de $E\otimes F$ à partir d'endomorphismes de $E$ et de $F$.
\begin{enumerate}
\item Soit $f$ un endomorphisme de $E$, $g$ un endomorphisme de $F$ et $\varphi\in\mathcal{L}^2(E^*\times F^*)$. Alors l'application $\Phi$ qui à $\varphi$ associe $(x^*,y^*)\mapsto \varphi(x^*\circ f,y^*\circ g)$ réalise un endomorphisme de $\mathcal{L}^2(E^*\times F^*)$.
\item Si $\varphi=x\otimes y$ alors $\Phi(x\otimes y)=f(x)\otimes g(y)$. On notera $\Phi=(f,g)$.
\item Si $f$ et $g$ sont des automorphimes de $E$ et $F$respectivement, alors $(f,g)$ est un automorphisme de $E\otimes F$.
\end{enumerate}
\end{proposition}
\begin{preuve}{}
\begin{enumerate}
\item On vérifie que $(x^*,y^*)\mapsto \varphi(x^*\circ f,y^*\circ g)$  est bilinéaire. C'est immédiat. De même la linéarité par rapport à $\varphi$ est immédiate. Voilà pour le premier item.
\item On a :
$\left((f,g)(x\otimes y)\right)(x^*,y^*)=(x\otimes y)(x^*\circ f,y^*\circ g)=(x^*\circ f)(x)(y^*\circ g)(y)=$

\noindent $x^*\left(f(x)\right)y^*\left(g(y))\right)=(f(x)\otimes g(y))(x^*,y^*)$
\item Un élément de $E\otimes F$ s'écrit $\sum_{i,j}\lambda_{ij}e_i\otimes \varepsilon_j=
\sum_{i,j}\lambda_{ij}f\left(f^{-1}(e_i)\right)\otimes f\left(f^{-1}(\varepsilon_j)\right)=$

\noindent $(f,g)\left(\sum_{i,j}\lambda_{ij}f^{-1}(e_i)\otimes f^{-1}(\varepsilon_j)\right)$. Ainsi $(f,g)$ étant surjective est un automorphisme de l'espace de dimension finie $E\otimes F$.
\end{enumerate}
\end{preuve}
\begin{exo}\label{exotens}
Soit $E$ et $F$ deux espaces vectoriels, $E^*$ le dual de $E$.
Montrer qu'il existe un isomorphisme canonique entre $E^*\otimes F$ et $\mathcal{L}(E,F)$.

\noindent\textbf{NB} Canonique signifie que cet isomorphisme se construit sans passer par des bases sur $E$ et $F$.
\end{exo}
\begin{remark}Dans le vocabulaire de l'algèbre tensoriel un vecteur de $E^*\otimes F$ est un \emph{tenseur $1$ fois covariant et $1$ fois contravariant}. L'exercice nous enseigne que les applications linéaires peuvent être vues comme de tels tenseurs.
\end{remark}
\section{ Première étape du calcul différentiel à l'ordre un : structure de variété.}\label{warheit}
\subsection{Cadre des espaces vectoriels}\label{zob}
\subsubsection{Révision}
\begin{definition}\label{dedifa}
Soit $f$ une application d'un ouvert $I$de $\rr$ à valeurs dans $\rr$ et $x\in I$. On dit que $f$ est dérivable en $x$ s'il existe un réel $a$ tel que $$f(x+h)=f(x) +ah+h\epsilon(h)$$ où 
$\lim_{h\to 0}\epsilon(h)=0$.
Le réel $a$ se note habituellement $f'(x) $ ou encore $df(x)$ ou encore $Df(x)$.
\end{definition}
Rappelons encore que l'application $\varphi : \rr\longrightarrow\cal{L}(\rr,\rr)$ définie par $$\varphi(x) : h\mapsto xh$$
( où $h$ est un vecteur de $\rr$ et $x$ un scalaire de $\rr$) est un isomorphisme qui permet d'identifier $\rr$ avec $\cal{L}(\rr,\rr)$.
D'où la généralisation naturelle :
\begin{definition}\label{didif}
Soit $f$ une application d'un ouvert $U$ de $\rr^n$ à valeurs dans $\rr^p$ et $x=(x^1,\ldots,x^n)\in U$. On dit que $f$ est dérivable (ou différentiable) en $x$ s'il existe une application linéaire notée $f'(x)$ (ou $df(x)$) telle que pour $h\in \rr^n$ tel que $x+h\in U$ on ait :
\begin{equation}\label{diff}f(x+h)=f(x)+f'(x).h+h\epsilon(h),\end{equation}avec $\lim_{h\to 0}\epsilon(h)=0$.
De fa\c con équivalente si on note $\Vert.\Vert_{n}$ et $\Vert.\Vert_{p}$ les normes euclidiennes de $\rr^n$ et de $\rr^p$, l'égalité (\ref{diff}) équivaut à 
\begin{equation}\label{aze}
\lim_{h\to 0}\frac{1}{\Vert h\Vert_{n}}\Vert f(x+h)-f(x)+f'(x).h\Vert_{p}=0
\end{equation}
L'application $f'$ (ou $df$) de $\rr^n$ vers $\mathcal{L}(\rr^n,\rr^p)$ est la dérivée ( ou différentielle ) de $f$ sur $U$.\index{différentielle d'une application}\index{dérivée d'une application}

\end{definition}
\subsubsection{\'{E}criture en coordonnées et exemples}
L'application $f$ est celle de la définition \ref{didif}. On choisit dans $\rr^n$ et $\rr^p$ les bases canoniques, ce qui permet d'associer à $f'(x)$ une matrice $(a_{{ij}})$ à $p$ lignes et $n$ colonnes. Si $h=(h^1,\ldots h^n)$, alors 
\[f'(x).h=\left(\begin{matrix}
a_{11}h^1+\ldots+a_{n}h^n\\
..\\
a_{p1}h^1+\ldots+a_{pn}h^n
\end{matrix}\right).
\]
En particulier si $h^{i}=0$ sauf pour $i=j$, on a : \[f'(x).h=h^j\left(\begin{matrix}
a_{1j}\\a_{2j}\\
.\\
.\\
a_{pj}
\end{matrix}\right).
\]
Posons $f(x)=(f^1(x)\ldots,f^p(x))$. L'équation (\ref{aze}) équivaut à 
\[\forall i \in \{1,\ldots,p\}\;\;\lim_{h^j\to 0} \frac{f^{i}(x+h)-f^{i}(x)}{h^j}-a_{ij}=0\]ou encore \[a_{ij}=\frac{\partial f^{i}}{\partial x^j}(x).\]
La différentielle de $f$ en $x$ est représentée par sa matrice jacobienne $J(f)(x)$.
\begin{remark}\label{classik}
Notons immédiatement que $f$peut admettre en un point une matrice jacobienne sans être différentiable en ce point. Par exemple la fonction définie sur $\rr^2$ par :
\[
f(x)=\begin{cases}\frac{xy}{x^2+y^2}\;\;\mathrm{si}(x,y)\not=(0,0)\\
0\;\;\;\;\;\;\;\;\;\;\;\;\mathrm{si}\; \;\;x=y=0
\end{cases}
\]
a deux dérivées partielles à l'ordre $1$  nulles en $(0,0)$ et elle n'est pas différentiable en $(0,0)$.
Cependant l'existence et la continuité des dérivées partielles sur un voisinage de $x$  équivaut à la différentiabilité de $f$  et  à la continuité de $f'$ sur ce voisinage. On dit alors que $f$ est continûment différentiable (voir tout cours de calcul différentiel).
\end{remark}

\begin{exemples}
$1$)
Soit $f:\rr^n\longrightarrow \rr$ différentiable en tout point$x$ d'un ouvert $U$ de $\rr^n$.

\noindent Pour tout $h=(h^1,\ldots,h^n)\in \rr^n$ et tout $x\in\rr^n$, on a : $df(x).h=\sum_{i=1}^n\frac{\partial f}{\partial x^{i}}(x)h^{i}$.

\noindent Notons $dx^{i}$ la forme linéaire définie par $dx^{i}(h)=h^{i}$. 
On peut écrire sur $U$ l'égalité : \[df=\sum_{i=1}^n\frac{\partial f}{\partial x^{i}}dx^{i}.\]
$2$)
Soit $f=(f^1,\ldots,f^n) :\rr\longrightarrow \rr^n$ une courbe paramétrée différentiable de $\rr^n$.
Pour tout $t$ et tout $h\in\rr$, on a : $f'(t).h=h \big({(f^1)}'(t),\ldots,{(f^n)}'(t)\big)$. 

\noindent La différentielle en $t$ s'identifie au vecteur vitesse à l'instant $t$. 

\noindent$3$) Soit $f\in \mathcal{L}(\rr^n,\rr^p)$. Alors $f$ est différentiable partout et pour tout $x\in \rr^n$ on a 
\[f'(x)=f.\]
\end{exemples}
\subsection{Généralisation du cadre vectoriel}
Les deux concepts à définir dans une généralisation du calcul différentiel sur les espaces vectoriels sont :\begin{enumerate}
\item La différentiabilité d'une fonction qui nécessite l'introduction de la structure de \emph{variété différentiable},
\item la différentielle (ou la dérivée) d'une fonction différentiable, qui  nécessite l'introduction de la structure de \emph{fibré vectoriel}.
\end{enumerate}
\subsubsection{ Variété et différentiabilité}
Le concept de différentiabilité  des applications exige de mesurer des taux d'accroissement. Il est donc nécessaire d'avoir des coordonnées sur les ensem\-bles sources de ces  applications. La structure la plus générale 
pour obtenir des coordonnées est la structure de variété que nous allons définir. Le prix à payer pour cette généralisation est 
\emph{qu'il n'y a plus de choix canonique des coordonnées au voisinage d'un point sur une variété}
comme c'est le cas pour $\rr^n$.
Il faudra donc vérifier que notre définition de la différentiabilité  d'une fonction définie sur une variété ne dépend pas du choix d'un système de coordonnées.
Pour qu'il en soit ainsi, nous donnons d'une variété la définition suivante.
\begin{definition}\label{varan}\index{variété topologique}
Une \emph{variété topologique} $M$, de dimension $n$ est un espace topologique séparé\footnote{ un espace topologique est séparé si deux points distincts appartiennent à deux ouverts disjoints} tel que tout point $x$ possède un voisinage ouvert homéomorphe à un ouvert de $\rr^n$. Notons $U$ ce voisinage et $\varphi$ l'homéomporphisme de $U$ sur l'ouvert $\varphi(U)$ de $\rr^n$. On dit que le couple $(U,\varphi)$ est \emph{une carte locale en $x$}
et $U$ est \emph{domaine de la carte} $(U,\varphi)$.\index{carte locale}\index{domaine de carte}
\end{definition}
\noindent Cette définition répond à la nécessité d'avoir des coordonnées pour les points de $M$ : si $y\in U$, on a $\varphi(y)=(y^1,\ldots,y^n)$. Les $y^{i}$ sont les coordonnées de $y$, dans la carte $(U,\varphi)$.
Soit donc $M$ une variété (l'adjectif topologique sera omis dans la suite). Un point point $x$ de $M$ peut appartenir à deux domaines de cartes. Supposons que $x\in U\cap V$ où $(U,\varphi)$ et $(V,\psi)$ sont deux cartes locales. Alors l'application $\psi\circ\varphi ^{-1}: \varphi(U\cap V)\subset\rr^n\longrightarrow\psi
(U\cap V)\subset\rr^n$ est appelée \emph{changement de coordonnées }: elle exprime les coordonnées d'un point de $(V,\psi)$ en fonction de ses coordonnées dans $(U,\varphi)$.\index{changement de coordonnées (variétés)}
Dans une variété topologiques les  changements de coordonnées sont des applications continues de $\rr^n$ dans lui-même. Plus généralement la régularité des changements de coordonnées définit la régularité de la variété $M$. Ces changements de coordonnées étant des applications définies  $\rr^n$ vers $\rr^n$ elles peuvent être différentiables : voir section \ref{zob}.

\noindent Si $\{U_i\}_{i\in I}$ est un recouvrement ouvert de $M$ tel que les couples $(U_i,\varphi_i)$ soient des cartes locales, on dit que la famille $\{(U_i,\varphi_i)\}_{i\in I}$ est un atlas de $M$.\index{atlas sur une variété}

\begin{exo}\label{topinembour}
Soit $M$ une variété topologique et $(U,\varphi)$ une carte locale en $x\in M$.
Soit $V$ un ouvert de $M$ contenant $x$.
Alors $\left(U\cap V,\varphi_{\vert U\cap V}\right)$ est une carte locale en $x$ de la variété $M$.
 Par conséquent si $\mathcal{A}=\{(U_i,\varphi_i)\}_{i\in I}$ est un atlas sur $M$, alors la famille  $\mathcal{A}\cup \{(U_i\cap U, {\varphi_i}_{\vert U_i\cap U})\}_{i\in I}$ est un atlas sur $M$.
\end{exo}
\begin{definition}\label{vavaran}
Si les changements de cartes d'une variété topologique $M$
 sont différentiables, on dit que $M$ est une variété différentiable.\index{variété différentiable}
\end{definition}
Notons encore la définition d'un \emph{atlas défini sur une variété} $M$ :  c'est une famille de cartes locales $\{(U_i,\varphi_i)\}_{i\in \nn}$ telle que $\{U_i\}_{i\in\nn}$ soit un recouvrement ouvert de $M$.
\begin{exemple}
On considère $S^1$, le cercle unité  de $\rr^2$ de centre $(0,0)$. On pose $U=S^1\setminus \{(0,1)\}$
et $V=S^1\setminus \{(0,-1)\}$. On définit $\varphi : U\longrightarrow \rr$ et  $\psi : V \longrightarrow \rr$ par $\varphi(x,y)=\frac{x}{1-y}$ et $\psi(x,y)=\frac{x}{1+y}$. Alors $\{(U,\varphi),(V,\psi)\}$ est un atlas sur $S^1$ pour lequel il a une structure de variété différentiable. Plus précisément, si on note $X$ la coordonnée locale dans la carte $(U,\varphi)$ et  $Y$ la coordonnée locale dans la carte $(V,\psi)$ alors le changement de coordonnées $ \psi\circ\varphi ^{-1}:\varphi(U\cap V)\to\psi(U\cap V)$ s'écrit $X\mapsto Y=\frac{1}{X}$.
\end{exemple}
\begin{exo}\label{vaca}
Détailler les preuves dans  l'exemple précédent.
\end {exo}
Nous pouvons maintenant introduire la notion de différentiabilité d'une fonction $f:M\longrightarrow N$
où $M$ et $N$ sont deux variétés différentiables de dimensions respectives $n$ et $p$.\index{fonction différentiable (variétés)}
\begin{definition}\label{divan}
On dit que $f$ est différentiable en $x$ s'il existe en $x$ une carte $(U_1,\varphi_1)$ et en $f(x)$ une carte 
$(U_2,\varphi_2)$ telles que $\varphi_2\circ f\circ\varphi_1^{-1}$ soit une fonction de $\rr^n$ vers $\rr^p$ 
différentiables en $\varphi_1(x)$.

\noindent On dit que $\varphi_2\circ f\circ\varphi_1^{-1}$ est l'expression locale de $f$ dans les coordonnées liées aux cartes $(U_1,\varphi_1)$ et $(U_2,\varphi_2)$
\end{definition}
Et voici la cohérence des définitions précédentes :
\begin{proposition}
La définition précédente ne dépend pas du choix des cartes locales $(U_1,\varphi_1)$ et $(U_2,\varphi_2)$.
\end{proposition}
\begin{preuve}{}
Soit $(V_1,\psi_1)$ et $(V_2,\psi_2)$ deux autres cartes locales respectivement en $x$ et en $f(x)$.
Alors $\psi_2\circ f\circ \psi_1^{-1}=(\psi_2\circ\varphi_2^{-1})\circ(\varphi_2\circ f\circ\varphi_1^{-1})\circ (\varphi_1\circ\psi_1^{-1})$ est différentiable puisque $M$ et $N$ étant des variétés différentiables, les applications $\varphi_1\circ\psi_1^{-1}$ et $\psi_2\circ\varphi_2^{-1}$ sont différentiables.
\end{preuve}
\begin{exo}\label{didi}
Soit $f$ une fonction différentiable sur $\rr^2$. On note $g$ sa restriction à $S^1$. Avec les notations de l'exemple lié à l'exercice \ref{vaca}, montrer que $g$ s'écrit dans la coordonnée $X$ :
\[g(X)=f\Big(\frac{2X}{1+X^2},\frac{X^2-1}{1+X^2}\Big). \]
En déduire que $g$ est différentiable sur $S^1$.
\end{exo}
Il convient maintenant de définir les sous-structures induites par les structures  de variété topologique et différentiable.

\begin{definition}\label{sousstruc}\index{sous-variété}
Soit $M$ une variété de dimension $n$, $p$ un entier naturel inférieur ou égal à $n$ et $N$ un sous-ensemble de $M$. On le munit de la topologie induite par celle de $M$. Alors $N$ est une sous-variété de dimension $p$ s'il existe une sous- famille  $\{(U_i,\varphi_i)_{i\in I}\}$ d'un atlas de M tel que pour tout $i\in I$ on ait :
\begin{equation}\label{souvar}\varphi_i(N\cap U_i)=\rr^p\cap\varphi_i(U_i)\end{equation}
où $\rr^p=\{(x^1,\ldots,x^p,0,\ldots0)\in\rr^n\}$.
\end{definition}
 Les $p$-uplets $(x^1,\ldots,x^p)$ de la définition précédentes apparaissent comme des coordonnées locales sur une variété, ce que justifie la proposition suivante.
 \begin{proposition}\label{sousstrucva!}
Avec les notations de la définition \ref{sousstruc}, $N$ est une variété de dimension $p$.
\end{proposition}
\begin{preuve}{}
Si $\{(U_i,\varphi_i)\}_i$ est un atlas de $M$, alors $\{N\cap U_i\}_i$ est un recouvrement ouvert de $N$. On note $\psi_i$ la restriction de $\varphi_i$ à $N\cap U_i$. Soit $x\in N\cap U_i \cap U_j=( N\cap U_i)\cap (N\cap U_j)$. On note $(x^1,\ldots,x^n)$ les coordonnées locales dans $U_i$ et $(y^1,\ldots,y^n)$ les coordonnées locales dans $U_j$.
Supposons que $M$ une variété topologique (respec. différentiable). Alors $\varphi_j\circ\varphi_i^{-1} : \varphi_i(U_i\cap U_j)\longrightarrow \varphi_j(U_i\cap U_j)$ est un homéomorphisme (respec.un difféomorphisme) tel que  
$\varphi_j\circ\varphi_i^{-1} (x^1,\ldots,x^n)=(y^1,\dots,,y^n)$. On a $\psi_j\circ\psi_i^{-1}=
{\varphi_j\circ\varphi_i^{-1}}_{\left\vert\varphi_i \left((N\cap U_i)\cap (N\cap U_j)\right)\right.} :  (x^1,\ldots,x^p,0\ldots,0)\mapsto(y^1,\dots,,y^p,0\ldots,0)$ est un homéomorphisme (respec.un difféomorphisme) entre deux ouverts de $\rr^p$. Ainsi l'atlas $\{(N\cap U_i,\psi_i)\}$ munit $N$ d'une structure de variété topologique (respec.différentiable).
\end{preuve}
\begin{exemples}
$i)$ Si $M$ est une variété, tout ouvert de $M$ en est une sous-variété de même dimension. Ceci est une conséquence directe de l'exercice \ref{topinembour}.

$ii)$ Une sous-variété de $M$ de dimension $(\dim M)-1$ est appelée une hypersurface de $M$.
\end{exemples}

\subsubsection{Compléments : variété à bord}\label{varabobo}\index{variété à bord}
\emph{On peut passer ici cette section dont la lecture peut s'avérer utile pour le lecteur de la section \ref{crocs}.}

Le disque unité ouvert de $\rr^2$ est une variété topologique de dimension $2$. Ce n'est pas le cas du disque unité fermé. En effet les points de son bord (topologique), $S^1$, n'ont pas dans la topologie induite de voisinage ouvert dans $\rr^2$. Par contre il admettent un voisinage homéomorphe à l'intersection d'un ouvert de $\rr^2$ avec $H^2=\{(x,y)\in \rr^2\;;\;y\ge 0\}$.\footnote{nous attendons qu'à la lecture de cette phrase, le lecteur fasse un dessin}

\noindent Nous allons  généraliser cet exemple.

\noindent On pose $H^n=\{x=(x^1,\ldots, x^{n-1},x^n)\in\rr^n\;;\;x^n\ge 0\}$. Dans $H^n$ tous les points $x$ tels que $x^n>0$ admettent un voisinage ouvert dans $\rr^n$. Les points $x$ pour lesquels $x^n=0$ n'admettent pas de voisinage inclus dans $H^n$, qui soit un ouvert de $\rr^n$. On dira qu'ils constituent le bord  de $H^n$.
\begin{definition}\label{bordeau}\index{variété à bord}
Une variété topologique à bord de dimension $n$ est un espace topologique dont tous les points admettent un ouvert homéomorphe à un ouvert de $H^n$ muni de la topologie induite de $\rr^n$ et dont les changements de cartes sont continus.
\end{definition}
\noindent Une variété à bord $M$ est recouverte par une famille $\{U_i\}_{i\in I}$ d'ouverts chaque $U_i$ étant homéomorphe par $\varphi_i$   à un ouvert  de $H^n$.
\emph{La famille $\{(U_i,\varphi_i)\}_{i\in I}$ est encore appelée un \emph{atlas} de $M$.}

\noindent Si on veut définir \emph{une variété différentiable à bord} il faut pouvoir dire que les changements de cartes locales sont différentiables. Pour cela il suffit de définir une application différentiable d'un ouvert de $H^n$ à valeurs dans $\rr^n$. Ceci est l'objet de la définition suivante.
\begin{definition}
Soit $U$ un ouvert de $H^n$. Une application $f$ de $U$ vers $\rr^n$ est différentiable s'il existe un ouvert $V$ de $\rr^n$ et une application différentiable $g$ de $V$ vers $\rr^n$ tels que $U\subset V$ et $f=g_{\left\vert U\right.}$.
\end{definition}
Soit $M$ est une variété topologique à bord.  
Notons $\Int(M)$ les points de $M$ qui admettent un voisinage ouvert $U$ homéomorphe à un ouvert de $\rr^n$. Si $\varphi$ est cet homéomorphisme, on a nécessairement pour tout $x$ de $U$ : $\varphi(x)=(x^1,\ldots,x^n)$ avec $x^n>0$. Remarquons que : \begin{enumerate}
\item
 $\Int M$ est un ouvert de $M$ et par conséquent $\Int M$ est une variété de dimension $n$.
\item Si $M$ est une variété topologique ou différentiable (au sens des définitions \ref{varan} et \ref{vavaran}) alors $\Int M=M$
\end{enumerate}

La structure de variété à bord induit également la sous-structure de sous-variété (éventuellement à bord). 
La définition  d'une sous-variété dans la catégorie des variétés à bord s'énonce naturellement comme suit.
\begin{definition}\label{sousstruccu}
Un sous-ensemble $N$ d'une variété à bord $M$ de dimension $n$ en est une sous-variété (éventuellement à bord) de dimension $p$ s'il existe sur $M$ un atlas $\{U_i,\varphi_I\}_I$ tel que $\{U_i\cap N,{\varphi_i}_{\vert N}\}_I$ soit un atlas sur $N$ lui conférant un structure de variété (éventuellement à bord) de dimension $p$.
\end{definition}
De façon analogue à l'exemple de $H^n$, on appelle \emph{bord de $M$},\index{bord d'une variété} noté $\partial M$ le fermé  $M\setminus \Int(M)$ de $M$. 
Cette définition implique que la variété $\Int M$ a un bord vide.

 Un point $x$ de $\partial M$ admet donc, de par la définition de $\partial M$, un voisinage $U$ homéomorphe à l'ouvert $\varphi(U)$ de $H^n$ de sorte que $\varphi(x)=(x^1,\ldots,x^{n-1},0)$. Il existe donc une boule ouverte  $B_{n-1}$ de $\rr^{n-1}$ et et réel $\varepsilon>0$ tel que $B_{n-1}\times [0,\varepsilon[\subset \varphi(U)$. Quitte à rétrécir $U$, on peut le choisir égal à 
$\varphi^{-1}(B_{n-1}\times [0,\varepsilon[)$. On a alors 
\begin{equation}\label{bour}\partial M\cap U=\{x\in U\;;\;\varphi(x)=(x^1,\ldots,x^{n-1},0)\}\end{equation}
L'équation (\ref{bour})  permet  de dire, à partir de la définition \ref{sousstruccu} que \emph{$\partial M$ est une sous-variété de $M$ de dimension $n-1$}. De plus, en utilisant les mêmes arguments que dans la proposition \ref{sousstrucva!} on établit que $\partial M$ est une variété de dimension $n-1$ (topologique ou différentiable selon la structure de $M$). Par conséquent $\partial(\partial M)=\emptyset$. On peut écrire symboliquement
$\partial\circ\partial=\emptyset$. 

\noindent On dira que $\partial M$ est une hypersurface de $M$.
\begin{exo}\label{sousous}
 On suppose que $M_1$ est une sous-variété différentiable de $M_2$ et que $M_2$ est une sous-variété (éventuellement  à bord) de la variété différentiable $M_3$ vérifiant $\dim M_1<\dim M_2\le \dim M_3$. Montrer que $M_1$ est une sous-variété  de $M_3$. En déduire que si $N$ est une sous-variété à bord de $M$ alors $\partial N$ en est une sous-variété de $M$  de dimension $(\dim  N )-1$.
\end{exo}

Nous allons montrer que la structure naturelle pour définir la différentielle est celle de fibré vectoriel. 
Le chapitre \ref{etap2}  développe succintement le concept de fibré vectoriel et définit la notion de différentielle dans les exemples \ref{true}.

\section{Deuxième étape du calcul  différentiel à l'ordre un : structure de fibré vectoriel}\label{etap2}
\subsection{Quelques définitions.}
\begin{definition}\label{fifibre}\index{fibré vectoriel}
On considère un espace topologique $E$, une variété $M$ (qui sera ici différentiable), $p$ une surjection continue de $E$ vers $M$, $F$ un espace vectoriel  (ici $F=\rr^{n} $ ou $\cc^{n}$) et enfin $G$ un sous-groupe (topologique) du groupe des automorphismes de $F$. Le triplé $\tau=(E,p,M)$  est
\emph{un fibré vectoriel de fibre} $F$ et de \emph{groupe} $G$, d'espace total $E$  si on dispose d'un recouvrement ouvert de $M$, $\{U_{i} \}_{i}$ de sorte que :
\begin{enumerate}
\item Il existe un homéomorphisme $\Phi_i$ de $p^{{-1}}(U_{i})$ sur $U_{i}\times F$, de sorte que 
si  $p_{1}$ désigne la projection de $U_{i}\times F$ sur $U_{i}$ alors 
\begin{equation}\label{resto}p_1\circ\Phi_i=p.\end{equation}

\item Si $U_{i}$ et $U_{j}$ sont des domaines de cartes non disjoints les changements de cartes $\Phi_{i}\Phi_{j}^{{-1}}$ sont de la forme :
\begin{equation}\label{chang}
 \begin{array}{cccc}
\Phi_{i}\Phi_{j}^{{-1}} :& U_{j}\cap U_{i}\times F&\longrightarrow&U_{i}\cap U_{j}\times F\\
&(x,f)&\longmapsto&\bigl(x, c_{ij}(x)f\bigr)
\end{array}
\end{equation}

\noindent  où $c_{ij}$ est une application continue de $M$ vers $G$.
\end{enumerate}
 La variété $M$ est \emph{la base du fibré} $\tau$.\index{base d'un fibré} Le couple $(U_{i},\Phi_{i})$ est appelé \emph{une carte locale} du fibré ou aussi une trivialisation de $\tau$. \index{carte locale d'un fibré}\index{trivialisation d'un fibré}L'ensemble des cartes $\{(U_{i},\Phi_{i})\}_{i}$ est appelé \emph{un atlas} du fibré.\index{atlas sur un fibré}
\end{definition}
\noindent En notant $p_{2}$ la projection de $U_{j}\cap U_{i}\times F$ sur $F$  on déduit de (\ref{chang}) :
\begin{equation}\label{changg}
\forall e\in p^{-1}(U_{i}\cap U_{j}),\;\; c_{ij}\Big(p_{1}\big(\Phi_j(e)\big)\Big)p_{2}\Phi_{j}(e)=p_{2}\Phi_{i}(e).
\end{equation}
\begin{remark}\label{elleestbonne}
Si $\mathcal{A}=\{(U_{i},\Phi_{i})\}_{i}$ est un atlas de $\tau$ et si $U$ est un ouvert de $M$, alors 
$\mathcal{A}\cup\left\{(U_{i}\cap U, {\Phi_{i}}_{\vert U_i\cap U})\right\}_{i}$ est encore un atlas de $\tau$. 
Cette remarque est à rapprocher de l'exercice \ref{topinembour}.
\end{remark}
\begin{exo}\label{fivar}
Soit $\tau=(E,p,M)$ un fibré vectoriel de fibre un espace vectoriel de dimension $d$ sur une variété différentiable de dimension $n$. Montrer que $E$ hérite de $M$ d'une structure de variété différentiable de dimension $n+d$.

\end{exo}

 \begin{definition}\index{section d'un fibré}
 Une section de $\tau=(E,p,M)$ est une application $s$ de $M$ vers $E$ telle que $p\circ s=I_{M}$. Si $M$ est une variété différentiable et $s$ est une application différentiable de $M$ vers $E$ (voir exercice \ref{fivar}), on dit que $s$ est une section différentiable.
 \end{definition}

 La définition \ref{fifibre} montre que les changements de cartes sont définis par des applications   $c_{ij}$  de $M$ dans le groupe linéaire de $F$ vérifiant la relation de Chasles $c_{ki}(x)c_{ij}(x)=c_{kj}(x)$ pour tout $x\in U_{j}\cap U_{i}\cap U_{k}$.  

 \begin{definition}\label{cyclotouriste}\index{cocycle sur un fibré}
 Une famille d'applications continues $c_{ij}$ définies à partir d'un recouvrement ouvert de $M$, $\{U_{i} \}_{i}$ à valeurs dans  le groupe linéaire d'un espace vectoriel et vérifiant
 $c_{ki}(x)c_{ij}(x)=c_{kj}(x)$ pour tout $x\in U_{j}\cap U_{i}\cap U_{k}$ s'appellent un \emph{cocycle} sur $M$.
 \end{definition}
 Les changements de cartes définissent donc un cocycle sur $M$. L'importance de cette notion appara\^{\i}t dans le troisième item de la liste suivante des quatre propriétés et définitions qu'il faut avoir en tête lorsqu'on s'occupe de fibrés vectoriels.
 \subsection{Liste de propriétés et définitions}\label{licite}

\subsubsection {Morphisme de fibrés vectoriel}\label{morpion} 
\begin{definition}\label{morfic}\index{morphisme de fibré vectoriel}
Soient $\tau=(E,p,M)$ et $\tau'=(E',p',M')$ deux fibrés localement triviaux de fibres respectives $F$ et $F'$ et d'atlas respectifs  $\{(U_{i},\Phi_{i})\}_{i}$ et  $\{(U'_{j},\Phi'_{j})\}_{j}.$  Soit $h$ une application différentiable de $M$ vers $M'$. Un morphisme $H$ de $\tau$ vers $\tau'$ au-dessus de $h$ est une application continue de $E$ vers $E'$ telle que :
\begin{enumerate}
\item si $e\in p^{-1}(b)$, alors $H(e)\in p'^{-1}\big(h(b)\big)$,
\item si $e$ s'écrit dans la carte locale $(U_{i},\Phi_{i})$ : $e=\Phi_{i}^{-1}(b,f)$, alors $H(e)$ s'écrit dans une carte locale  $(U'_{j},\Phi'_{j})$ telle que $h(b)\in U'_{j}$ :
$H(e)=\Phi'^{-1}_{j}\big(h(b),h_{ji}(b)f\big)$
où $h_{ji}$ est une application continue de $U_{i}\cap h^{{-1}}(U'_{j})$ dans $\mathcal{L}(F,F')$.
\end{enumerate}
\end{definition}
On a donc avec ces notations l'égalité :
\begin{equation}\label{morfer}
\forall (b,f)\in U_{i}\cap h^{-1}(U'_{j})\times F,\;\;\Phi'_{j}H\Phi^{-1}_{i}(b,f)=\big(h(b), h_{ji}(b)f\big).
\end{equation}
L'équation (\ref{morfer}) est l'écriture de $H$ dans des cartes locales.

\subsubsection{Caractérisation locale d'un morphisme de fibré vectoriel }

On vérifie immédiatement les relations de compatibilité suivantes : si $\{c_{ij}\} $ est la famille des cocycles associés aux changements de cartes de $\tau$, $\{c'_{ij}\} $  la famille des cocycles associés aux changements de cartes de $\tau'$, alors 
 \begin{equation}\label{concon}
h_{ij}c_{jk}=h_{ik}, \;\;c'_{ij}h_{jk}=h_{ik}
\end{equation}
Nous attirons l'attention que l'écriture utilisée dans les relations (\ref{concon}) est un abus mnémotechnique. Voici une écriture exhaustive de la première relation de (\ref{concon}).
 
\noindent Soit $(U_j,\Phi_j)$, $(U_k,\Phi_k)$ deux cartes de $\tau$ de domaines non disjoints, $(U'_i,\Phi_i)$ une carte de $\tau'$ telle que $W_{ijk}=h^{-1}(U'_i)\cap U_j\cap U_k\not=\emptyset $. Alors, 

 \[\forall x\in W_{ijk}\;\;h_{ij}(x)\circ c_{jk}(x)=h_{ik}(x).\]
 Voici une écriture exhaustive de la deuxième relation de (\ref{concon}).
 
 \noindent Soit $(U'_i,\Phi'_i)$ et $(U'_j,\Phi'_j)$ deux cartes de $\tau'$ de domaines non disjoints, $(U_k,\Phi_k)$ une carte locale de $\tau$ telle que $h^{-1}(U'_i\cap U'_j)\cap U_k\not=\emptyset$. Alors,
 \[\forall x \in h^{-1}(U'_i\cap U'_j)\cap U_k\;\;c'_{ij}(h(x))\circ h_{jk}(x)=h_{ik}(x).\]
 \begin{exo}\label{ladeuz}
 \'{E}crire explicitement la deuxième relation de (\ref{concon}) et vérifier effectivement une des deux relations.
 \end{exo}
Ce qui est remarquable  c'est la réciproque : 
\begin{proposition}\label{crac}
Si $h$ est une application de $M$ vers $M'$ et si une famille d'applications continues $\{h_{ij}\}$ vérifie les équations (\ref{concon}), où $h_{ij}$  est une application continue de  $h^{-1}(U'_{i})\cap U_{j}$ vers $\mathcal{L}(F,F')$, alors elles définissent par l'équation (\ref{morfer}) un morphisme de fibré $\tau$ vers le fibré $\tau'$ au dessus de $h$.
\end{proposition}
On trouvera au besoin la démonstration, généralisée au cas des fibrés localement triviaux dans l'appendice $2$.
\begin{definition} Avec les notations de la définition \ref{morfic},
 si un morphisme $H$ de fibré vectoriel au-dessus de $h$ est un homéomorphisme de $E$ sur $E'$, on dit que c'est un isomorphisme de fibré vectoriel.	
\end{definition}
\begin{exo}\label{job}\index{isomorphisme de fibré vectoriel}
Les notations sont celles de la définition \ref{morfic}.

\noindent Si $H$ est un isomorphisme de fibré vectoriel au-dessus de $h$, alors $h$ est un homéomorphisme de $M$ sur $M'$.

\noindent Indication : on montrera que la projection d'un fibré vectoriel est une application ouverte.
\end{exo}

\subsubsection{Caractérisation d'un fibré par ses cocycles}\label{ratoi}
\begin{definition}\label{fivecteq}
Soient deux fibrés $\tau$ et $\tau'$ de même base, une variété topologique $M$ de même fibre, un espace vectoriel $F$. Un morphisme $(H,Id_M)$ de $\tau$ vers $\tau'$ au dessus de l'identité de $M$ est une équivalence des fibrés $\tau$ et $\tau'$ s'il existe un atlas $\{(U_i,\Phi_i)\}$ de $\tau$ et un atlas 
$(U'_j,\Phi'_j)\}$ de $\tau'$ tels que les trivialisations de $H$, $h_{ji}$, définies sur $U_i\cap U'_j$ par l'équation (\ref{morfer}) soient à valeurs dans le groupe linéaire de $F$, $\mathcal{G}l(F)$.
\end{definition}
Une équivalence est donc un isomorphisme de fibré vectoriel au-dessus de l'identité entre deux fibrés vectoriels de même base et de même fibre.
\begin{theorem}\label{cossi}
Soit $M$ une variété topologique. Un cocycle sur $M$ à valeurs dans un sous-groupe $G$ d'automorphismes d'un espace vectoriel $F$, associée à un recouvrement ouvert $\{U_{i}\}$ de $M$  définit  un unique, à un isomorphisme (de fibré) près,   fibré vectoriel de fibre $(E,p,M)$ de fibre $F$ pour lequel ce cocycle définit les changements de cartes. 
\end{theorem}

Cette proposition est fondamentale. Elle permet de définir  un fibré sur une variété à partir d'un cocycle défini sur elle. Elle est utilisée pour introduire efficacement tous les fibrés liés aux propriétés topologiques des variétés, comme le fibré tangent et au calcul différentiel sur les variétés différentiables, comme le fibré cotangent. Ceci sera détaillé dans les exemples de la section \ref{true}.
Du fait de l'importance particulière de cette proposition, une démonstration détaillée en est donnée dans l'appendice $1$. On peut l'éviter en première lecture sans que cela nuise à la compréhension de ce chapitre.

Ce théorème permet d'introduire la notion de  fibré \emph{associé} à un fibré vectoriel
$(E,p,M))$ de fibre $F$ et de groupe $G$: si ce dernier est défini par un cocycle $\{c_{ij}\}$ à valeurs dans $G$ et si $\mu$ est un morphisme de groupe de $G$ vers $G'$, groupe d'automorphismes d'un espace vectoriel $F'$, alors $\{\mu\circ c_{ij}\}$ est un cocycle sur $M$ définissant ainsi un fibré 
$(E',p',M)$ de fibre $F'$ et de groupe $G'$. Nous avons en fait avec ces fibrés associés la voie royale pour définir  les fibrés tangent et cotangent qu'on vient d'évoquer. 
Donnons immédiatement une conséquence du théorème \ref{cossi}.
\begin{corollary}[ produit tensoriel de deux fibrés vectoriels de même base]\index{produit tensoriel de deux fibrés}
\hspace{1cm}\\
On reprend les notations de la proposition \ref{endopotence}.
 Soit $\tau=(E,p,M)$ et $\tau'=(E',p',M)$ deux fibrés vectoriels de même base $M$ de fibres respectives les espaces vectoriels  $F$ et $F'$, définis respectivement par les cocycles $\{c_{ij}\}$ à valeurs dans un sous-groupe $G$ de  $\mathcal{G}l(F)$ et $\{c'_{ij}\}$ à valeurs dans un sous-groupe $G'$ de  $\mathcal{G}l(F')$ 
 (compte tenu de la remarque \ref{elleestbonne}, on peut faire en sorte (quitte à rajouter des intersections de domaine des cartes)  que les atlas de $\tau$ et $\tau'$ aient la même famille de domaines).
 Alors,
 \begin{enumerate}
 \item $\{t_{ij}\}=\{(c_{ij}, c'_{ij})\}$ est un cocycle sur $M$ à valeurs dans $G\times G'$.
 \item Il existe un unique fibré vectoriel, à un isomorphisme près, noté $\tau\otimes \tau'$ de fibre $F\otimes  F'$ de groupe $G\times G'$ dont les changements de cartes soient la famille $\{t_{ij}\}$.
  \end{enumerate}
  \end{corollary}
  \begin{preuve}{}
  Soit $x\in M$. Les vecteurs de $F\otimes F'$ sont des combinaisons linéaires de vecteurs de la forme $f\otimes f',f\in F,f'\in F'$. L'application linéaire $(c_{ij},c'_{ij})(x)=(c_{ij}(x),c'_{ij}(x))$ vérifie l'égalité $(c_{ij},c'_{ij})(x)(f\otimes f')=c_{ij}(x)(f)\otimes c'_{ij}(x)(f')$, ce qui la définit entièrement sur $F\otimes F'$.
   On a ainsi :
   
   \noindent $(t_{ij} t_{jk})(x)(f,f')= \left(c_{ij}(x), c'_{ij}(x)\right)\left((c_{jk}(x)(f), c'_{jk}(x)(f')\right)=$
   
   \noindent $(c_{ij}(x), c_{jk}(x))(f)\otimes (c'_{ij}(x),c'_{jk}(x))(f')=c_{ik}(f)\otimes c'_{ik}(x)(f')=t_{ik}(f\otimes f')$.
   D'où on déduit l'égalité $t_{ij}\circ t_{jk}=t_{ik}$, ce qui montre le premier item.
   Le second est une conséquence du théorème \ref{cossi}.
   \end{preuve}
\subsubsection{Structure d'espace vectoriel des fibres}

\begin{proposition}\label{ev} Pour tout $b\in M, p^{{-1}}(\{b\})$ hérite de $F$ d'une structure d'espace vectoriel de même dimension que $F$. On notera au besoin $0_{p^{-1}(\{b\})}$ son vecteur nul. De plus 
Si $b$ est dans le domaine d'une carte $(U,\Phi_{U})$ alors la restriction de $\Phi_{U}$ à $p^{{-1}}(b)$ est un isomorphisme de  $p^{{-1}}(b)$ sur $F$.
\end{proposition}
\begin{preuvera}{}
On suppose que $F$ est un espace vectoriel sur $K$. Soit $e,e' \in p^{-1}(b)$ où $b\in U$ et $(U,\Phi_{U})$ est une carte locale. On note $p_{2}$ la deuxième projection sur $U\times F$. On pose $e+e'=\Phi_{U}^{{-1}}(b, p_{2}\Phi_{U}(e)+p_{2}\Phi_{U}(e'))$. En utilisant l'égalité (\ref{changg}), on montre que cette égalité est indépendante de la carte locale choisie et définit ainsi la somme sur $p^{{-1}}(b)$.

De même pour $\lambda\in K$, on définit le produit $\lambda e$ à l'aide de l'égalité  $\lambda e=\Phi_{U}^{{-1}}(b, \lambda p_{2}\Phi_{U}(e))$. Cette égalité ne dépend pas du choix de la carte locale. 
Le couple $\{b\}\times F$ a une structure naturelle d'espace vectoriel isomorphe à $F$, dont le vecteur nul est $(b,0)$. Pour cette structure $\Phi_{\vert p^{{-1}}(\{b\})}$ est un isomorphisme de $p^{{-1}}(\{b\})$ sur $\{b\}\times F$.

\end{preuvera}
\begin{corollary}\label{modsec}
L'ensemble des sections d'un  fibré vectoriel $\tau=(E,p,M)$ a une structure naturelle de module sur l'anneau des fonctions différentiables sur $M$.
\end{corollary}
\begin{preuve}{}
Si $s_{1} $ et $s_{2}$ sont deux sections de $\tau$ et $f$ une fonction différentiable sur $\tau$, alors $s_{1}+fs_{2}$ est la section de $\tau$ définie par $x\mapsto s_{1}(x)+f(x)s_{2}(x)$. La vérification des axiomes est évidente.
\end{preuve}
On note $\mathcal{D}(M)$ l'anneau des fonctions réelles différentiables sur $M$. Par la suite le module des sections sur $\tau$ désignera la structure de $\mathcal{D}(M)$-module.

\begin{exo}\label{jobi}
 Les notations sont celles de la définition \ref{morfic}.
 Soit $H$ un morphisme de fibré au-dessus de $h$, déterminé par les applications $h_{ji}$  de l'équation (\ref{morfer}). 
 
$1$) Montrer que pour tout $b\in M$, la restriction de $H$ à $p^{-1}(b)$ est une application linéaire 
de $p^{-1}(b)$ sur $p'^{-1}(h(b))$ avec les structures d'espace vectoriel décrites par la proposition \ref{ev}.

$2$) Le morphisme $H$ est un isomorphisme de fibré si et seulement si $h$ est un homéomorphisme et pour tout $i$, $j$ et $b\in U_{i}\cap h^{-1}(U'_{j})$, l'application linéaire $h_{ji}(b)$ est un isomorphisme d'espace vectoriel.
 \end{exo}
\begin{remark}\label{chant}
Soit $\tau=(E,p,M)$ un fibré vectoriel de fibre $\rr^{n}$. Si l'on considère un élément $e$ de $E$, on peut le représenter par ses coordonnées $(\lambda^{1}_{U},\cdots,\lambda^{n}_{U})$
 dans une carte locale $(U,\Phi_{U})$. Cela signifie que si $e\in p^{-1}(U),$ alors
 $\Phi_{U}(e)=(b,(\lambda^{1}_{U},\cdots,\lambda^{n}_{U}))$.
  Soit $\Lambda_U$ la matrice colonne définie par  $^{t}\Lambda_{U}=(\lambda^{1}_{U},\cdots ,\lambda^{n}_{U}).$
  
   \noindent On peut également, après avoir choisi une base de sections $(s_{1,U},\cdots,s_{n,U})$ au-dessus de $U$ (qu'on appelle un \emph{référentiel local}) représenter $e$ par ses composantes dans la base $(s_{1,U}(b),\cdots,s_{n,U}(b))$. Avec la convention d'Einstein on écrira :
    \begin{equation}\label{eins}
    e=l^{i}_{U}s_{i,U}(b).
    \end{equation}
 Notons $l_U$ la matrice colonne définie par $^{t}l_{U}=(l^{1}_{U},\cdots ,l^{n}_{U})$. 
 Comme le choix des référentiels et des bases n'est pas canonique se posent deux problèmes distincts : celui changement de coordonnées (c'est à dire un problème de changement de cartes locales) et celui  de changement de composantes (c'est à dire un problème de changement de bases). Détaillons.
Pour le changement de coordonnées on suppose que $b$ est dans l'intersection des domaines des cartes $(U,\Phi_{U})$ et $(V,\Phi_{V})$. Alors  $\Phi_{U}\Phi_{V}^{-1}(b,^{{t}}\Lambda_{V})=(b,c_{UV}(b)^{{t}}\Lambda_{V})=(b,^{{t}}\Lambda_{U})$, en se référant à l'égalité (\ref{chang}). Ainsi, $^{{t}}\Lambda_{U}=c_{UV}(b)^{{t}}\Lambda_{V}$ . Soit $C_{UV}(b)$ la matrice qui représente $c_{UV}(b)$ dans la base canonique de $\rr^{n}$ (\emph{matrice de changements de coordonnées}). Alors 
\begin{equation}\label{chacord}
\Lambda_{U}=C_{UV}(b)\Lambda_{V}.
\end{equation}
 Pour le changement de composantes, on se donne une deuxième base de sections $(s_{1,V},\cdots,s_{n,V})$ au dessus du domaine $V$ de la carte $(V,\Phi_{V})$.

 \noindent Posons  $\Phi_{U}(s_{i,U}(b))=(b,\ep_{i,U})$ et $\Phi_{V}(s_{i,V}(b))=(b,\ep_{i,V})$. On a compte tenu de l'équation (\ref{eins}) et de la structure vectorielle des fibres, $\Phi_{U}(e)=\big(b,l^{i}_{U}\ep_{i,U}(b)\big)$. Cette équation représente sur $\rr^{n}$, l'équation (\ref{eins}) sur $E$.
 On a de même $\Phi_{V}(e)=\big(b,l^{i}_{V}\ep_{i,V}(b)\big)$.
 On note respective\-ment  $P_{U}$  et $P_{V}$ les matrices de passage de la base canonique de $\rr^{n}$ aux bases $\{\ep_{i,U}\}$ et $\{\ep_{i,V}\}$.  À l'équation (\ref{chacord}) se rajoutent les équations de changements de composantes de l'algèbre linéaire :
 \begin{equation}\label{chacon}
\Lambda_{U}=P_{U}l_{U}\;\mathrm{et}\;\Lambda_{V}=P_{V}l_{V}
 \end{equation}
 D'où on déduit la matrice \emph{de changement de composantes} $\tilde{C}_{UV}(b)$ définie par :
 \begin{align}\label{chapo}
 l_{U}&= \tilde{C}_{UV}(b) l_{V}\\
 \tilde{C}_{UV}(b)&=P_{U}^{-1}C_{UV}(b)P_{V} \end{align}
 On peut toujours définir les changements de cartes $(U,\Phi_{U})$ et $(V,\Phi_{V})$ de sorte que $P_{U}$ et $P_{V}$ soient la matrice identité (quitte à composer $\Phi_{U}$ à gauche par $
 \varphi_{U } : (b,\varepsilon_{i,U})\in U\times \rr^{n}\mapsto(b,e_{i})\in U\times \rr^{n}$ où $e_{i}$ est le $i$-ème vecteur de la base canonique de $\rr^{n}$. Idem avec $\Phi_{V}$) auquel cas la matrice de changement de coordonnées est égale à celle de changement de bases, c'est à dire $ \tilde{C}_{UV}(b)=C_{UV}(b)$ 
 Pour finir considérons les matrices de vecteurs \begin{equation}\label{chicmol}
 e_{U}=\big(s_{1,U}(b),\cdots,s_{n,U}(b)\big) \;\;\mathrm{et} \;\;e_{V}=\big(s_{1,V}(b),\cdots,s_{n,V}(b)\big).
 \end{equation} L'algèbre linéaire nous enseigne qu'à l'équation (\ref{chapo}), on peut associer, dans le cas où on a fait le choix  $ \tilde{C}_{UV}(b)=C_{UV}(b)$, l'équation
 (\ref{superchapo}) :
 \begin{equation}\label{superchapo}
 e_{V}=e_{U\;} {C}_{UV}(b).
 \end{equation}
\end{remark} 
\begin{exo}\label{bofbof}
Il s'agit d'établir la version duale de la relation(\ref{superchapo}) dans un espace vectoriel $E$.
Soit $e=(e_1,\ldots,e_n)$ et $e'=(e'_1,\ldots,e'_n)$ deux bases de $E$ et $A$ la matrice de passage de $e$ vers $e'$. La relation (\ref{superchapo}) s'écrit $e'=eA$. Notons $\epsilon=(\epsilon_1,\ldots,\epsilon_n)$ la base duale de $e$ et $\epsilon'=(\epsilon'_1,\ldots,\epsilon'_n)$ la base duale de $e'$. 
Montrer que $\epsilon'=\epsilon.\;\;{^tA^{-1}}$.
\end{exo}
\subsection{Fibrés associés et calcul différentiel.}\label{illicite}
Nous allons exploiter dans cette section la proposition \ref{cossi} et  énoncer  la proposition \ref{dual} suivie de deux exemples fondamentaux. Commençons préalablement par un petit coup de projecteur sur la \emph{dualité dans les espaces vectoriels euclidiens}.
\subsubsection{Espaces euclidiens et dualité}\label{eude}\index{dualité}
On considère un espace vectoriel euclidien $(\mathcal{E},\la,\ra)$. On sait que pour tout endomorphisme $f$ de $\mathcal{E}$, il existe un unique endomorphisme $^tf$ de $\mathcal{E}$ tel que 
\begin{equation}\label{trala} 
\forall x,y\in \mathcal{E},\;\la x,f(y)\ra=\la^tf(x),y\ra.
\end{equation}
Dans une base orthonormée de $\mathcal{E}$, si $f$ est représenté par la matrice $A$, alors $^tf$ est représenté par la matrice $^tA$.  

On considère le dual de $\mathcal{E}$, noté habituellement $\mathcal{E}^*$ : il s'agit de l'espace vectoriel des applications linéaires de $\mathcal{E}$ vers $\rr$, qu'on appelle également les \emph{formes linaires} sur $\mathcal{E}$ ou en vue d'une extension à l'algèbre multilinéaire les \emph{$1$-formes} sur $\mathcal{E}$. Retenons ici les deux énoncés suivants concernant $\mathcal{E}^*$ :
\begin{enumerate}
\item L'espace $\mathcal{E}^*$ est canoniquement isomorphe à $\mathcal{E}$, l'isomorphisme canonique $j$ (lié à la structure euclidienne) étant défini par :
\begin{equation}\label{croup}\forall x\in \mathcal{E},\;j(x)=\la x,.\ra.\end{equation}
\item Tout endomorphisme  $f$de $\mathcal{E}$ définit l'endomorphisme $f^*$ de $\mathcal{E}^*$ suivant : 
\begin{equation}\label{croupette}
\forall \alpha\in \mathcal{E}^*,\;f^*\alpha=\alpha\circ f.\end{equation}
\end{enumerate}
D'où la proposition qui suit.

  \noindent L'espace vectoriel des endomorphismes de $\mathcal{E}$ est noté $\opop(\mathcal{E})$.
\begin{proposition}\label{ident}
L'application $\Phi$ de $\opop(\mathcal{E})$ vers $\opop(\mathcal{E}^*)$ définie par l'égalité $\Phi(f)=(^tf)^*$ est un isomorphisme d'espace vectoriel vérifiant pour tous endomorphimes $f$et $g$ de $\mathcal{E}$ : $\Phi(f\circ g)=\Phi(f)\circ\Phi(g).$

De plus, pour tout $f\in\opop(\mathcal{E})$, on a le diagramme commutatif

\[
\begin{array}{ccc}
\mathcal{E}&\stackrel{j}{\longrightarrow}&\mathcal{E}^*\\
\vcenter{\llap{$^tf$}}\Big\downarrow&&\Big\downarrow\vcenter{\rlap{$f^*=\Phi(^tf)$}}\\
\mathcal{E}&\xrightarrow[\enspace j\enspace]&\mathcal{E}^*
\end{array}
\]
qui montre que si on identifie $\mathcal{E}$ à $\mathcal{E}^*$ alors $^tf$ s'identifie à $f^*$.
\end{proposition}
\begin{preuve}{}
La linéarité de $\Phi$ est immédiate. Cherchons le noyau de $\Phi$.
L'égalité (\ref{trala}) peut se réécrire à partir des égalités (\ref{croup}) et (\ref{croupette}):
\begin{equation}\label{dicotif}f^*\circ j=j\circ ^tf,\end{equation}
qui montre que le diagramme est commutatif.
En particulier on obtient en remplaçant dans la dernière équation $f$ par $^tf$ :
\[\Phi(f)\circ j=j\circ f.\]
Ainsi $\Phi(f)=0$ équivaut à : $\forall x\in \mathcal{E},\;j\big(f(x)\big)=0$ ou encore
$\forall x,y\in \mathcal{E},\;\la f(x),y\ra=0$, ce qui signifie que $f(x)$ est orthogonal à $\mathcal{E}$, donc nul et ceci pour tout $x\in\mathcal{E}$. D'où $f=0$. On en conclut que $\Phi$ est un isomorphisme.
\end{preuve}
Pour terminer notons que par l'identification de $\mathcal{E}$ et $\mathcal{E}^*$ par $j$, $^tf$ se prolonge en un endomorphisme $f^*$ sur l'espace vectoriel $\Lambda^p(\mathcal{E})$ des $p$-formes sur $\mathcal{E}$ de la façon suivante :
soient $\alpha_1,\ldots,\alpha_p$ des $1$-formes sur $\mathcal{E}$.
On considère la $p$-forme $\omega_p=\alpha_1\wedge\ldots\wedge\alpha_p$. Alors par définition
$$f^*(\alpha_1\wedge\ldots\wedge\alpha_p)=:f^*\alpha_1\wedge\ldots\wedge f^*\alpha_p$$
Les $p$-formes telles que $\omega_p$ constituant une partie génératrice de $\Lambda^p(\mathcal{E})$, l'endomorphisme $f^*$ est définie sur $\Lambda^p(\mathcal{E})$ si on l'étend par linéarité.
On a en particulier la propriété  :
$$
\forall x^1,\ldots,x^p\in \mathcal{E},\;f^*\big(\alpha_1\wedge\ldots\wedge\alpha_p\big)(x^1,\ldots, x^p)
=(\alpha_1\wedge\ldots\wedge\alpha_p)(f(x^1),\ldots,f(x^p)).
$$
Pour toute révision sur l'algèbre tensoriel et extérieur on pourra consulter l'appendice de [ \ref{rinrin}].

Une dernière remarque : tout se qui a été exposé ici dans le cadre des espaces euclidiens peut se 
reprendre sur des espaces vectoriels hermitiens à des modifications mineures près (par exemple, avec les  mêmes notations,  dans une base orthonormée la matrice associée à $^tf$ est $^t\bar{A}$ si $A$ est la matrice associée à $f$).

Dans la suite, faisant implicitement référence à la proposition \ref{ident} on écrira systématiquement $f^*$ pour l'application adjointe à $f$ dans une structure euclidienne.
\subsubsection{Un énoncé productif}

On considère un fibré vectoriel $\tau$ 
de fibre $\rr^n$ ou $\cc^{n}$ elle est munie canoniquement d'une structure  euclidienne dans le cas de $\rr^{n}$, hermitienne dans le cas de $\cc^{n}$ et ainsi on définit pour chaque automorphisme $f$de $F$ un automorphisme adjoint $f^{*}$. Supposons que le groupe $G$ du fibré vérifie $G=G^{*}$ où $G^{*}=\{f^{*}\;;\;f\in G\}$. C'est le cas par exemple si $G=SO(n)$ ou $G=U(n)$.
On considère l'application $\mu$ qui à tout cocycle $\{c_{ij}\}_{ij}$ définissant le fibré $\tau$ associe la famille $\{(c_{ij}^*)^{{-1}}\}_{ij}$.
\begin{proposition}\label{dual}
La famille $\{\mu(c_{ij})\}_{i,j}$ est un cocyle sur $M$ à valeurs dans $G$, associé au même recouvrement ouvert de $M$ que les $\{c_{ij}\}.$
\end{proposition}
\begin{preuve}{}
$\mu(c_{ij})(b)\mu(c_{jk})(b)=(c_{ij}^*)^{{-1}}(b)(c_{jk}^*)^{{-1}}(b)=\big((c_{jk}^*(b)(c_{ij}^*(b)\big)^{{-1}}=\Big(\big(c_{ij}(b)c_{jk}(b)\big)^{*}\Big)^{-1}$

\noindent$=\big((c_{ik}(b))^{*}\big)^{{-1}}
=\mu(c_{ik})(b)$.
\end{preuve}
\begin{definition}\label{cofiassoc}\index{fibré associé}
Le cocycle $\{\mu(c_{ij})\}_{ij}$ est le cocycle associé au cocycle $\{c_{ij}\}_{ij}$ par $\mu$. Il définit un unique fibré, à un isomorphisme près, appelé fibré associé au fibré défini par $\{c_{ij}\}_{ij}$.
\end{definition}
\subsubsection{Deux exemples fondamentaux}\label{true}
\noindent \textbf{ a) Le fibré tangent}\index{fibré tangent}

 Si $\{(U_{i},\varphi_{i})\}$ est un atlas de la variété $M$ et conformément aux notations de la définition \ref{dedifa}, la famille $\{c_{ij}\}=\{D(\varphi_{i}\varphi_{j}^{{-1}})\}$ est un  cocycle sur $M$ à valeurs dans $\ma{G}l(\rr^{n})$ définissant le fibré tangent de $M$, noté $\tau(M)$. Plus précisément, les $c_{ij}$ étant définies sur $U_i\cap U_j$ on a pour tout $x\in U_i\cap U_j$  : 
$c_{ij}(x)=D(\varphi_i\varphi_j^{-1})(\varphi_j(x))$.
Les sections du fibré tangent sont appelés \emph{ champs de vecteurs sur $M$.}\index{champ de vecteurs}

Cette vision des champs de vecteurs a un mérite : elle leur donne rigoureusement et  à moindre co\^{u}t toutes les propriétés des sections d'un fibré vectoriel. Par exemple, par le corollaire \ref{modsec}, les champs de vecteurs sur une variété différentiable  ont une structure de module sur l'anneau des fonctions différentiables définies sur cette variété et par conséquent ils ont également une structure d'espace vectoriel sur $\rr$. Il semble que le prix à payer pour une telle définition est  qu'elle est loin de l'intuition qu'on a d'un champ de vecteurs en particulier d'un champ de vecteurs  dans $\rr^n$. Il n'en est rien (après la petite réflexion que voici). En effet un champ de vecteurs sur $\rr^n$ est élémentairement une application (différentiable) $X : x=(x^1,\ldots,x^n)\mapsto (X^1(x),\ldots,X^n(x))$. Un tel champ définit une dérivation sur les fonctions différentiables de la façon suivante : si $f$ est une telle fonction sa dérivée (de Lie) dans la direction $X$ est la fonction \begin{equation}\label{derive}
x\mapsto X.f=X^{i}\frac{\partial f}{\partial x^{i}}.
\end{equation}
Il est en effet immédiat que pour toutes fonctions différentiables $f$et $g$ et tout réel $\lambda$, on a  $X.(f+\lambda g)=X.f+\lambda X.g$ et $X.(fg)=(X.f)g+f(X.g)$.
 
\noindent \textbf{Notation :} L'écriture $X.f=L_Xf$ est très utlisée, la lettre L faisant référence à Sophus Lie et on parle également pour cette expression de \emph{dérivée de Lie dans la direction $X$.}\index{dérivée de Lie d'une fonction}

\noindent L'égalité (\ref{derive}) induit l'égalité suivante entre dérivations à savoir :
\begin{equation}\label{derision}
X=X^{i}\frac{\partial }{\partial x^{i}}.
\end{equation}
Faisant appara\^{i}tre la famille $\{\frac{\partial }{\partial x^{i}}\}$, comme une base de l'espace vectoriel des dérivations sur les fonctions dérivables sur $\rr^n$.

\noindent Plaçons-nous maintenant  sur une variété $M$, au voisinage d'un point $x$. En ce point, on peut avoir le choix de diverses coordonnées locales. Supposons que $x$ appartienne à l'intersection de deux 
cartes locales $(U_j,\varphi_j)$ et $(U_i,\varphi_i)$ avec des coordonnées locales $$\varphi_i(x)=(x^1,\ldots,x^n)\;\mathrm{et}\;
\varphi_j(x)=(y^1,\ldots,y^n).$$
Toute fonction différentiable $f$ sur $M$  s'écrit respectivement $f\circ{\varphi_i}^{-1}$ et $f\circ{\varphi_{j}}^{-1}$ dans les cartes $(U_i,\varphi_i)$  et $(U_j,\varphi_j)$ conformément à la définition \ref{divan}. Dans les habitudes du calcul différentiel on écrit $f(x^1,\ldots,x^n)$ pour $f\circ\varphi_i^{-1}$. Le calcul différentiel nous enseigne que 
\begin{equation}\label{reglo}
\frac{\partial( f\circ \varphi_i^{-1})}{\partial x^{i}}=\frac{\partial (f\circ\varphi_j^{-1})}{\partial y^{j}}\circ(\varphi_j\circ\varphi_i^{-1})\;\;\frac{\partial (\varphi_j\circ\varphi_i^{-1})}{\partial x^{i}}.
\end{equation}
que l'on écrit avec plus de légèreté (et plus d'imprécision)
\begin{equation}\label{reg}
\frac{\partial f}{\partial x^{i}}=\frac{\partial f}{\partial y^{j}}\;\frac{\partial y^j}{\partial x^{i}}.
\end{equation}
Notons  $C(x)$ la matrice de  $\left(D(\varphi_{i}\varphi_{j}^{{-1}})(x)\right)$. Alors on peut écrire l'égalité suivante entre dérivations :
\begin{equation}\label{chaele}
\left(\frac{\partial}{\partial x^{1}},\ldots,\frac{\partial}{\partial x^{n}}\right)=\left(\frac{\partial}{\partial y^1},\ldots,\frac{\partial}{\partial y^1}\right)C(x)
\end{equation}
qui est conforme à l'équation (\ref{superchapo}) qui décrit un changement de bases locales de sections  dans un fibré vectoriel défini par le cocycle $\{D(\varphi_{i}\varphi_{j}^{{-1}})\}$, c'est à dire un changement de bases locales de sections du fibré tangent (i.e de champs de vecteurs).
Ceci montre que   la  dérivation est un objet intrinsèque du fibré $\tau(M)$ de même nature qu'un champ de vecteurs. Il y a donc équivalence à parler de section de $\tau(M)$, de champ de vecteurs sur $M$ ou encore de dérivation sur les fonctions de $\mathcal{C}^1(M)$. Le sens algébrique de cette équivalence est précisé dans 
[\ref{rinrin}] : les champs de vecteurs et les dérivations sur les fonctions différentiables ont des structures d'espace vectoriel réel, isomorphes.
 \`{A} la lumière de cette vision des champs de vecteurs nous pouvons énoncer  : 

\noindent\emph{Les cordonnées locales $(x^1,\ldots,x^n)$ définissent une base locale de champs de vecteurs  $\left(\frac{\partial}{\partial x^{1}},\ldots,\frac{\partial}{\partial x^{n}}\right)$. L'équation (\ref{derision}) est l'écriture du champ $X$ dans ces coordonnées locales.}

\noindent Nous pouvons maintenant introduire la notion de différentielle annoncée p.\pageref{didi}.

\noindent Considérons deux variétés différentiables $M$ de dimension $n$, $N$ de dimension $p$ et $h:M\longrightarrow N$ une application différentiable. Considérons un atlas $\{(U_i,\varphi_i)\}$ de $M$
 et $\{(V_j,\psi_j)\}$ un atlas de $N$.
Soit $x\in M$, $(U_i,\varphi_i)$ une carte locale en $x$ et $(V_j,\psi_j)$ une carte locale en $h(x)$. On pose $h_{ji}(x)=d(\psi_j h\varphi_i^{-1})(\varphi_i(x)$. De par la définition \ref{didif}, $h_{ji}(x)\in \mathcal{L}(\rr^n,\rr^p)$.
Si $(U_k,\varphi_k)$ est une autre carte en $x$ et $(V_l,\psi_l)$ une autre carte en $f(x)$, on a :

$h_{ij}(x)\circ c_{jk}((x)=d(\psi_ih\varphi_j^{-1})(\varphi_j(x)\circ d(\varphi_j\varphi_k^{-1})(x)=
d(\psi_i h\varphi_k^{-1})(x)=h_{ik}(x).$ De même si on note $\tilde{c}_{lj}(h(x))=
d(\psi_l\psi_j^{-1})(h(x)$, on a :

$\tilde{c}_{lj}(h(x))\circ h_{ji}(x)=d(\psi_l\psi_j^{-1})(h(x))\circ d(\psi_j h \varphi_i^{-1})(\varphi_i(x))$

$=d(\psi_l h\varphi_i^{-1})(\varphi_i(x))=h_{li}(x)$.

\noindent Les égalités précédentes ayant lieu pour tout $x$ sur un ouvert $U$ de $M$,
 d'après la proposition \ref{crac}, les applications linéaires $\{h_{ij}\}$ définissent  un morphisme 
 $(h,dh)$ de $\tau(M)$ vers $\tau(N)$  au dessus de $U$, appelé  \emph{l'application tangente de $h$.}\index{application tangente} Elle est localement représentée par la différentielle de $h$ exprimée dans les cartes locales. On l'appellera également la \emph{différentielle de $h$}.\index{différentielle}
  La proposition suivante étend ce que l'on sait sur la différentielle de la composée de deux d'applications différentiables sur des espaces vectoriels normés. 
 \begin{proposition}\label{normi}
 Soient $M,N,L$ trois variétés différentiables, $x$ un point de $M$.
 Si $ h : M\longrightarrow N$ et $k :N\longrightarrow L$ sont deux applications différentiables
 alors $k\circ h$ est différentiable et 
  $d(k\circ h)=dk\circ dh$. On note $dh(x)$ la restriction de $dh$ à la fibre $T_xM$. Alors $dh(x)\in \mathcal{L}( T_xM,T_{h(x)}N)$,
 $dk(h(x))\in\mathcal{L}( T_{h(x))}M,T_{k\circ h(x)}L)$ et on a : $d(k\circ h)(x)=dk(h(x))\circ dh(x)$.
 \end{proposition}
 \begin{preuve}{}
$1$)Une application entre deux variétés différentiables est différentiable si et seulement si son expression dans un couple de cartes locales l'est. Alors puisque les changements de cartes sont différentiables l'expression de cette application dans tout couple de cartes locales est différentiable. On en déduit que la composée $k\circ h$ est différentiable.

\noindent $2$)
On considère $(U,\varphi), (V,\psi), (W,\theta)$ trois cartes locales respectivement en $x,h(x)$ et $(k\circ h)(x)$.
Les différentielles $d(k\circ h)(x), dh(x)$ et $dk)(h(x))$ sont respectivement représentées par 
\begin{itemize}
\item[i)] $h_{WU}(x)=d(\theta\circ(k\circ h)\circ \varphi^{-1}) (\varphi(x))$ dans le couple de cartes $\left((U,\varphi),(W,\theta)\right)$,
\item[ii)] $h_{VU}(x)=d(\psi\circ h\circ \varphi^{-1})(\varphi(x))$ dans le couple de cartes $\left((U,\varphi), (V\psi))\right)$,
\item[iii)] $h_{WV}(h(x))=d(\theta \circ k\circ \psi^{-1})(\psi\circ h(()x))$ dans le couple de cartes $\left((V,\psi),(W,\theta))\right)$.
\end{itemize} 
On a : $h_{WU}(x)=d\left((\theta\circ k\circ\psi^{-1})\circ(\psi\circ h\circ\varphi^{-1})\right)(\varphi(x))=
d(\theta\circ k\circ \psi^{-1})(\psi\circ h(x))\circ d(\psi\circ h\circ \varphi^{-1})(\varphi(x))= h_{WU}(x)\circ h_{VU}(x)$. On conclut à l'aide de la proposition \ref{crac}, constatant que $d(k\circ h)(x)$ et $dk(h(x))\circ dh(x)$ ont les mêmes représentations locales. Les deux égalités de la proposition sont ainsi démontrées.
\end{preuve}

La proposition \ref{normi} montre que la structure de fibré vectoriel est la bonne structure pour le  calcul différentiel d'ordre $1$. 
 Revenons maintenant sur la représentation locales des champs de vecteurs pour préciser l'équation (\ref{derision}).
 \begin{proposition}\label{replo}
 Soit $X$ un champ de vecteurs sur $M$. Considérons sa restriction au domaine de la carte locale $(U,\varphi)$ définissant les coordonnées $(x^1,\ldots,x^n)$. Notons $d\varphi(x)(X)=(\varphi(x),X^1(x),\ldots,X^n(x))\in \rr^n\times \rr^n$. Alors pour toute fonction différentiable sur $M$, on a 
 \begin{equation}\label{broc}
 df(x)\left(X(x)\right)=X^{i}(x)\frac{\partial f}{\partial x^{i}}(x)=:(X.f)(x)=:L_Xf(x).
 \end{equation}
 \end{proposition}
 \begin{preuve}{}
 D'après la proposition \ref{normi}, on a : $df(x)=d(f\circ\varphi^{-1})(\varphi(x))\circ d\varphi(x)$. 
 
 Ainsi,
 
 \noindent $ df(x)\left(X(x)\right)=d(f\circ\varphi^{-1})(\varphi(x))\circ d\varphi(x)(X(x))=$
 
 \noindent$=d(f\circ\varphi^{-1})(\varphi(x))(X^1(x),\ldots,X^n(x))
 =\frac{\partial f}{\partial x^{i}}(x^1,\ldots,x^n)dx^{i}(X^1(x),\ldots,X^n(x))=$
 
 \noindent$X^{i}(x)\frac{\partial f}{\partial x^{i}}(x)=X.f(x).$
 \end{preuve}
 
 \noindent Pour terminer sur le fibré tangent, notons que l'équivalence entre champs de vecteurs, sections du fibré tangent, et dérivations permet d'introduire naturellement une loi interne sur les champs de vecteurs 
 : le \emph{crochet de Lie }. En effet,
supposons que l'on ait deux champs de vecteurs $X$ et $Y$ sur M. Conformément à l'équation (\ref{derive}), on a 
$Y(Xf)=Y^{j}\left(\dfrac{\partial X^{i}}{\partial x^{j}}\dfrac{\partial f}{\partial x^{i}}+X^{i} \dfrac{\partial^{{2}} f}{\partial x^{i}\partial x^{j}}\right)$. Cette expression montre que $ Y\circ X$ pas une dérivation (donc pas un champ de vecteurs) sur $M$. Par contre si on symétrise l'égalité précédente en évaluant $(XY-YX)f$, on obtient :
\begin{equation}
[X,Y]f=:(XY-YX)f=\left(X^{j}\frac{\partial Y^{i}}{\partial x^{j}}-Y^{j}\frac{\partial X^{i}}{\partial x^{j}}\right)\frac{\partial f}{\partial x^{i}}
\end{equation}
 $[X,Y]$ est bien  une dérivations sur les fonctions différentiables donc un champ de vecteurs. En coordonnées locales ce champ s'écrit
\begin{equation}\label{crobar}
 [X,Y]=\left(X^{j}\frac{\partial Y^{i}}{\partial x^{j}}-Y^{j}\frac{\partial X^{i}}{\partial x^{j}}\right)\frac{\partial }{\partial x^{i}}=[X,Y]^{i}\frac{\partial }{\partial x^{i}}.
\end{equation}
   Il définit \emph{le crochet de Lie} des champs de vecteurs $X$ et $Y$. \index{crochet de Lie de deux champs de vecteurs}
  Notons encore, à la lumière de l'équation (\ref{crobar}) que $\left[\frac{\partial}{\partial x^{i}},\frac{\partial}{\partial x^{j}}\right]=0$ pour tout $(i,j)$.
  
\noindent Notons enfin qu'avec le crochet de Lie, les champs de vecteurs acquièrent une structure d'algèbre.
 \begin{exo}\label{nok}
Montrer que pour toute fonction différentiable $f$ et tous champs de vecteurs $X,Y$ sur $M$, on a :
\begin{enumerate}
\item  \[[fX,Y]=f[X,Y]-(L_Yf)X\]
\item \[[X,Y]=-[Y,X]\]
\end{enumerate}
 \end{exo}
 \begin{exo}\label{kon}
 On considère l'espace euclidien standard $(\rr^n,\la,\ra)$. On note $(e_1,\ldots,e_n)$ la base canonique de $\rr^n$. Soit $X$ et $Y$ deux champs de vecteurs sur $\rr^n$. On peut écrire pour tout $x\in\rr^n$ :
$ Y(x)=Y^{i}(x)e_i$ ou encore $Y=Y^{i}e_i$. On pose $$L_XY=(L_XY^{i})e_i.$$ (voir notation p. \pageref{derive}).
 \begin{enumerate}
 \item Montrer que $(X,Y)\mapsto L_XY$ est additive par rapport à $X$ et par rapport à $Y$.
 \item Montrer que pour toute fonction réelle différentiable $f$, on a : 
 
 $$L_{fX}Y=fL_XY\;\mathrm{et}\;
 L_X(fY)=fL_XY+(X.f)Y.$$
 \item Montrer que pour tous champs de vecteurs $X,Y,Z$, on a : \[X.\la Y,Z \ra=\la L_XY,Z \ra+\la X,L_XZ \ra\].
 \item Montrer que $L_XY-L_YX=[X,Y].$
 \item Soit $S$ une hypersurface de $\rr^n$ munie de la métrique induite de celle de $\rr^n$ et pour laquelle on suppose l'existence d'un champ unitaire diiférentiable de $\rr^n$ le long de $S$, orthogonal en tout point $x$ de $S$ à $T_xS$. 
 Soient $X$ et $Y$ deux champs de vecteurs du fibré tangent $\tau_S=(TS,p,S)$ de $S$ (i.e. deux champs sur $\rr^n$ le long de $S$ et tangents à $S$). A priori $L_XY$ n'est pas une section de $\tau_S$ et on peut l'écrire sous la forme $L_XY=D_XY+w(X,Y)N$ où $D_XY$ est la composante de $L_XY$ sur l'espace tangent à $S$ ($D_XY$ est donc un champ de vecteurs tangent à $S$).
 \begin{enumerate}
 \item  Montrer que $(X,Y)\mapsto D_XY$ est additive par rapport à $X$ et par rapport à $Y$, de même que 
 $(X,Y)\mapsto w(X,Y)$.
 \item Montrer que pour toute fonction réelle différentiable $f$, on a : 
 
 $$D_{fX}Y=fL_XY\;\mathrm{et}\;
 D_X(fY)=fD_XY+(X.f)Y.$$
 $$w(fX,Y)=fw(X,Y)\;\mathrm{et}\;w(X,fY)=fw(X,Y).$$

 \item Montrer que pour tous champs de vecteurs $X,Y,Z$ tangents à $S$, on a : \[X.\la Y,Z \ra=\la D_XY,Z \ra+\la X,D_XZ \ra.\]
 \item Montrer que pour tous champs de vecteurs $X,Y$ tangents à $S$, on a : $w(X,Y)=w(Y,X).$
 En déduire que $w$ est bilinéaire symétrique pour la structure de modules des champs de vecteurs sur les applications différentiables.
 \item Montrer que pour tous champs de vecteurs $X,Y$ tangents à $S$, on a : $D_XY-D_YX=[X,Y]$
 \end{enumerate}
 \end{enumerate}
 
 \end{exo}
\noindent \textbf{b) Le fibré cotangent.}

Si $p\in\nn$, $\Lambda^p(\rr^n)$ désigne l'espace vectoriel des $p$-formes sur $\rr^n$.
Donnons-nous une base $(e_1,\ldots,e_n)$  de $\rr^n$ et $(\varepsilon_1,\ldots,\varepsilon_n)$ sa base duale 
toute $p$-forme sur $\rr^n$ peut se mettre sous la forme $\alpha=\lambda_{i_1\cdots i_p}\varepsilon^{i_1}\wedge\ldots\wedge\varepsilon ^{i_p}$. Si dans l'écriture de $\al$ il existe deux indices $i_k$ et $i_l$ où $k,l\in\{1,\ldots,p\}$ tels que $\varepsilon ^{i_k}=\varepsilon^ {i_l}$ alors $\al= 0$. Ainsi $\Lambda^p(\rr^n)=\{0\}$ dès que $p>n$.

\noindent Si $f$ est un endomorphisme de $\rr^n$, nous avons vu dans la section \ref{eude} que pour tout entier natuel positif $p$, l'endomorphisme $f^*$ de $\Lambda^p(\rr^n)$ est défini par :

\[f^*(\alpha)(u_1,\ldots,u_p)=\alpha(f(u_1),\ldots,f(u_p)),\]
 En utilisant la proposition  \ref{dual} et les notations de l'exemple précédent, nous avons le cocycle $\{c_{ij}\}=\{D(\phi_i\phi_j^{-1}\}$ et par suite 
le cocycle $\{(c_{ij}^*)^{-1}\}$. Pour $p=1$, ce cocycle  définit   le fibré dual de $\tau(M)$, noté $\tau^*(M)$
de fibre  l'espace $\Lambda^1(\rr^n)$ des formes linéaires sur $\rr^n$. Une section de $\tau^*(M)$ est une $1$-forme 
différentielle sur $M$.\index{forme différentielle de degré $p$}

\noindent Pour un entier naturel $p$ quelconque, le cocycle $\{(c_{ij}^*)^{-1}\}$ définit le fibré des $p$-formes sur $M$ de fibre $\Lambda^p(\rr^n)$. Toute section est une $p$-forme différentielle sur $M$. En particulier si $p=n$ on a le fibré des formes volume sur $M$ dont la fibre $\Lambda^n(\rr^n)$ est de dimension $1$.

\emph{On notera $\Lambda^p(M)$ le module des $p$-formes différentielles sur $M$. Notons que $\Lambda^p(M)=\{0\}$ dès que $p>\dim M$.}
\begin{remark}\label{forvarb}
Nous venons de définir ce qu'est une $p$-forme différentielle sur une variété différentiable. Cette définition ne concerne par pas
les variétés à bord. Nous allons adopter une définition utile dans le cas particulier des sous-variétés :

\noindent soit $M$ une variété différentiable et $N$ une sous-variété à bord de $M$ (voir définition \ref{sousstruccu}).
Une forme différentielle sur $N$ est la restriction à $N$ d'une forme différentielle définie sur un voisinage ouvert de $N$ dans $M$.
\end{remark}
\begin{exo}\label{cotan}
Pour justifier élémentairement le sens du fibré dual :

\begin{enumerate}\item \'{E}tablir, comme on l'a fait avec le fibré $\tau(M)$, avec les notations de l'équation (\ref{chaele}), l'égalité suivante :
\[\left(\begin{matrix}dy^1\\\vdots\\dy^n\end{matrix}\right)=\left(\begin{matrix}
\frac{\partial y^1}{\partial x^1}&\ldots&&\frac{\partial y^1}{\partial x^n}\\
\vdots&&\vdots\\
\frac{\partial y^n}{\partial x^1}&\ldots&&\frac{\partial y^n}{\partial x^n}
\end{matrix}\right)
\left(\begin{matrix}dx^1\\\vdots \\dx^n\end{matrix}\right)\]
c'est à dire 
\[\left(\begin{matrix}dx^1,\ldots, dx^n\end{matrix}\right)=\left(\begin{matrix}dy^1,\ldots ,dy^n\end{matrix}\right).^tC_{ij}^{-1}(x)\]
\item Conclure avec l'exercice \ref{bofbof}.\end{enumerate}
\end{exo}
On peut introduire la définition d'une\emph{ forme exacte}.\index{forme exacte}
\begin{definition}
Une $1$-forme différentielle sur $M$ est exacte s'il existe une fonction à valeurs réelles, différentiable sur $M$ telle que $\al= df$.
\end{definition}
Suite à l'exercice (\ref{cotan}), on aura dans une carte locale $(U,\varphi)$ définissant les coordonnées locales $(x^1,\ldots,x^n)$ l'écriture  suivante d'une $p$-forme $\al^p$ :
\begin{equation}\label{redrerision}
\al^p=\sum_{i_1<i_2<\ldots<i_p} a_{i_1\ldots i_p}dx^{i_1}\wedge\ldots\wedge dx^{i_p}.
\end{equation}

 \noindent\textbf{Notation :} Soit $\tau=(E,p,M)$ et $\tau'=(E',p';M')$ deux fibrés vectoriels et $h$ une application différentiable de $M$ vers $M'$.
On notera $\mathcal{L}_{h}(\tau,\tau')$ l'ensemble des morphismes  de $\tau$ vers  $\tau'$ au-dessus de $h$. 
 \begin{exo}\label{exau2}
 Montrer que $\mathcal{L}_{h}(\tau,\tau')$ est muni canoniquement d'une structure d'espace vectoriel.
\end{exo}

Comme autre application de la proposition \ref{cossi}, nous avons la définition suivante.
\begin{definition}\label{protens}
Soient $\tau=(E,p,M)$ et $\tau'=(E',p',M)$ deux fibrés vectoriels de même base $M$. On note $F_{1}$ et $F_{2}$ leurs fibres respectives. On suppose que $\{c_{UV}\}$ et $\{c'_{U'V'}\}$ sont les cocycles associés respectivement à deux atlas de $\tau$ et $\tau'$.  Le fibré $\tau_{1}\otimes \tau_{2}$ est l'unique (à un isomorphisme près) fibré de fibre $F_{1}\otimes  F_{2}$ et de base $M$ déterminé par la famille de cocycles ${c_{UV}}\otimes {c'}_{U'V'}$.
\end{definition}
Cette définition est justifiée par l'exercice qui suit.
\begin{exo}\label{secsec}
Soient $\tau$ et $\tau'$ deux fibrés vectoriels de fibres $F$ et $F'$ respectivement. 
Si $\{c_{ij}\}$ est un cocycle à valeurs dans $\mathcal{G}l(F)$ et $\{c'_{i'j'}\}$ est un cocycle à valeurs dans $\mathcal{G}l(F')$, alors $\{c_{ij}\otimes c'_{i'j'}
\}$ est un cocycle à valeurs dans $\mathcal{G}l(F\otimes F')$.
\end{exo}
\begin{proposition}\label{moresek}
Soient $\tau$,$\tau'$ deux fibrés vectoriels sur $M$ et $\Phi\in\mathcal{L}_{I}(\tau,\tau')$.
Alors $\Phi$ définit canoniquement un morphisme $\varphi\in \mathcal{L}(\mathcal{S}_{\tau}, \mathcal{S}_{\tau'})$ par :
\[\forall s\in\mathcal{S}_{\tau},\;\;\varphi(s)=\Phi\circ s.\]
\end{proposition}
\begin{preuve}{}
si $\tau=(E,p,M)$, $\tau'=(E',p',M)$ et $s\in\mathcal{S}_\tau$, alors $p'(\varphi(s))=p'\circ(\Phi\circ s)=(p'\circ \Phi)\circ s=p\circ s=Id(M).$ Ceci montre que $\varphi(s)\in\mathcal{S}_{\tau'}$. Par ailleurs $\varphi$ est clairement linéaire.
\end{preuve}
\noindent \textbf{c}) \textbf{ Définition fleuve. }
 
 Profitons de l'introduction des fibrés tangents et cotangents pour énoncer la définition suivante.
\begin{definition}\label{assocpul}
Soit $M$ et $N$ deux variétés différentiables. Soit $\alpha\in \Lambda^p(N)$, $X$ un champ de vecteurs sur $M$ et $Y$ un champ de vecteurs sur $N$. On considère une application différentiable $f$ de $M$ vers $N$, une famille $\{X_i\}$ de champs de vecteurs sur $M$ et $\{Y_j\}$ une famille de champs de vecteurs sur $N$.
\begin{enumerate}
\item
On dit que $X$ et $Y$ sont associés par $f$ si pour tout $x\in M$, on a : \begin{equation}\label{association}df(x)\left(X(x)\right)
=Y\left(f(x)\right).\end{equation}\index{champs de vecteurs associés par une application différentiables}
Si $p>0$, on définit l'image réciproque de $\alpha$ par $f$,\index{image réciproque d'une forme différentielle par une application différentiable} comme la $p$-forme $f^*\alpha$ sur $M$ définie par :
\begin{equation}\label{imarec}
f^*\alpha(X_1,\ldots,X_p)=\alpha\circ f\;\left(df(X_1),\ldots,df(X_p)\right).
\end{equation}

Si $p=0$, la forme $\al$ est une fonction (notée habituellement par une lettre latine, $h$ par exemple) et on pose $f^*(h)=f\circ h.$
\item 
On définit le produit intérieur de $\alpha$ par $Y$ comme la $(p-1)$-forme sur $N$ notée $i(Y)\alpha$ définie par : $$i(Y)\alpha (Y_1,\ldots,Y_{p-1})=\alpha(Y,Y_1,\ldots,Y_{p-1}).$$\index{produit intérieur}
\item
Si $M=N$ et si $Y=X$,  on dit que $X$ est invariant par $f$ si la relation (\ref{association}) est vérifiée.\index{champ de vecteurs invariant par une application différentiable}

\noindent Si $M=N$ et si $f^*\alpha=\alpha$, on dit que $\alpha$ est invariante par $f$.\index{forme différentielle invariante par une application différentiable}

\noindent Si on a invariance pour toutes les transformations différentiables constituant un groupe $G$ 
on dit que $X$ ( resp. $\alpha$) est invariant (resp. invariante) par $G$.
\end{enumerate}
\end{definition}
\begin{proposition}\label{loloc}{Expression de l'image réciproque d'une forme différentielle en coordonnées locales.}
\hspace{1cm}\\
\noindent On reprend les notations du premier item de la définition. Soit $(U,\varphi)$ une carte locale en $x$ de la variété $M$ de dimension $n$ et $(V,\psi)$ une carte locale en $f(x)$ de la variété $N$ de dimension $m$. L'expression de $f$ dans ces cartes se note $x=(x^1,\ldots,x^n)\mapsto f(x)=(f^1(x),\ldots,f^m(x))$. 
Notons $(y^1,\ldots,y^m)$ les coordonnées locales sur $V$ de sorte que si  $p_i$ est la $i$-ème projection de $\rr^m$ (c.à d.$p_i(\lambda^1,\ldots,\lambda^m)=\lambda^{i}$), on a : $y^{i}=p_i\circ\psi$. Alors on peut écrire 
$f^{i}=y^{i}\circ f$. Soit $p$ une entier naturel au plus égal à $\min(m,n)$. On suppose que l'écriture d'une  $p$-forme $\alpha $ en coordonnées locales est 
\[\alpha=\alpha_{i_1,\ldots,i_p}dy^{i_1}\wedge\ldots\wedge dy^{i_p}.\]
Alors 
\begin{equation}\label{eqloimre}
f^*\alpha=\alpha_{i_1,\ldots,i_p}df^{i_1}\wedge\ldots\wedge df^{i_p}.
\end{equation}
\end{proposition}
\begin{preuve}{}
Rappelons que $dy^{i_1}\wedge\dots\wedge dy^{i_p}=\sum_{\sigma\in\mathcal{S}_p}dy^{i_{\sigma(1)}}\otimes\ldots\otimes dy^{i_{\sigma(p)}}$ où $\mathcal{S}_p$ est le groupe des permutations de $\{1,2,\ldots,p\}$
(voir au besoin [\ref{rinrin}]).

\noindent Soit $\{X_1,\ldots,x_p\}$ $p$ champs de vecteurs  sur $M$. \'{E}valuons $f^*\alpha(X_1,\ldots,X_p)$ en utlisant l'égalité (\ref{imarec}). On a :

\noindent $f^*\alpha(X_1,\ldots,X_p)=\alpha_{i_1,\ldots,i_p}dy^{i_1}\wedge\ldots\wedge dy^{i_p})\left(df(X_1),\ldots,df(X_p)\right)=$

\noindent$=\alpha_{i_1,\ldots,i_p}\sum_{\sigma\in\mathcal{S}_p}dy^{i_{\sigma(1)}}\otimes\ldots\otimes dy^{i_{\sigma(p)}}\left(df(X_1),\ldots,df(X_p)\right)=$\

\noindent$=\alpha_{i_1,\ldots,i_p}\sum_{\sigma\in\mathcal{S}_p}
dy^{i_{\sigma(1)}}(df(X_1))\ldots dy^{i_{\sigma(p)}}(df(X_p))=$

\noindent$=\alpha_{i_1,\ldots,i_p}\sum_{\sigma\in\mathcal{S}_p}
d(y^{i_{\sigma(1)}}\circ f)(X_1)\ldots d(y^{i_{\sigma(p)}}\circ f)(X_p)=$

\noindent$=\alpha_{i_1,\ldots,i_p}df^{i_1}\wedge\ldots\wedge df^{i_p}(X_1,\ldots,X_p).$
Et on conclut.
\end{preuve}
\noindent\textbf{Notation}

La matrice jacobienne à $p$ lignes et $q$ colonnes, de terme général $\frac{\partial f^{i}}{\partial x^j}$ où $i$ est l'indice de ligne et $j$ l'indice de colonne
se notera $\left\vert\frac{\partial(f^1\ldots f^p)}{\partial(x^1\ldots x^q)}\right\vert$
\begin{corollary}\label{edc}{de la proposition \ref{loloc}}

\noindent 
Avec les notations de la proposition \ref{loloc}, l'écriture de $f^*\al $ dans les coordonnées $x$ est:
\[f^*\al=\left\vert\frac{\partial(f^{k_1}\ldots f^{k_p})}{\partial(x^{i_1}\ldots x^{i_p})}\right\vert a_{k_1\ldots k_p}dx^{i_1}\wedge\ldots \wedge d x^{i_p}.\]
\end{corollary}
\begin{exo}\label{pupulette}
Démontrer le corollaire \ref{edc}.
\end{exo}

\noindent Afin de digérer la définition \ref{assocpul}, on pourra faire les utiles exercices \ref{exercice} et \ref{cucu}.
\begin{exo}\label{exercice}
Avec les notations de la définition \ref{assocpul}, on suppose $X$ et $Y$ associés par $f$. Vérifier  l'égalité :\begin{equation}\label{paliezno}
i(X)f^*\alpha=f^*\left(i(Y)\alpha\right)
\end{equation}
\end{exo}
\begin{exo}\label{cucu}
Soit $M$ une variété différentiable et $N$ une sous-variété à bord de $M$ telle que $\dim N=\dim M=n$. Soit $\omega$ une 
$n$-forme sur $N$. On note $j$ l'injection canonique de $\partial N$ dans $N$. Si $T$ est un champ de vecteurs  tangent à la variété $\partial N$ et si $\tilde{T}$ est n'importe quel prolongement de $T$ sur un voisinage ouvert de $\partial N$ dans $N$, alors $j^*\left(i(\tilde{T})\omega\right)=0$.
\end{exo}
\begin{remark}\label{colopul} Revenons sur la représentation d'une forme différentielle en coordonnées locales.
Supposons que  $\al_\varphi $ soit la $p$-forme différentielle sur $\rr^n$ qui représente $\alpha$ dans une carte locale $(U,\varphi)$.
Alors si une famille de $p$ champs de vecteurs $\{X_1,\ldots, X_p\}$ admet comme représentation locale dans la carte $(U,\varphi)$ sont représentés par la famille de champs sur $\rr^n$, $\{X_1^\varphi,\ldots,X_p^\varphi\}$ le réel 
$\al_\varphi(X_1^\varphi,\ldots,X_P^\varphi)$ doit être indépendant du choix de la carte $(U,\varphi)$. Plaçons-nous sur une carte locale $(U,\varphi)$. On considère les restrictions  à $U$ des champs  de vecteurs $X_1,\ldots,X_p$ et de la  $p$-forme $\al$. Nous noterons avec la même lettre ces objets et leurs restrictions. On a, en utilisant la définition (\ref{imarec}) l'égalité : $$(\varphi^{-1})^*\al\left(d\varphi (X_1),\ldots,d\varphi (X_p)\right)=\al(X_1,\ldots,X_p).$$
\emph{Ceci montre que la représentation de $\alpha$ dans la carte locale $(U,\varphi)$ est $(\varphi^{-1})^*\al$.}
\end{remark} 
\subsection{\'{E}léments de calcul différentiel sur les formes.}
\subsubsection{Dérivation sur l'algèbre graduée des formes différentielles}
L'ensemble des $p$-formes (différentielles) a une structure de module sur l'anneau des fonctions différentiables : cela signifie que cet ensemble est muni d'une addition (interne) et d'une multiplication (externe) par les fonctions différentiables
et que les règles de calcul vérifient les axiomes que l'on attend sur un espace vectoriel. On note cette structure $(\Lambda^p(M),+,.)$.
L'ensemble des formes différentielles, noté $\Lambda(M)$  se définit comme la somme directe des modules $\Lambda^p(M)$ : $$\Lambda(M)=\bigoplus_p\Lambda^p(M).$$
Ainsi toute forme s'écrit  formellement (de façon unique) comme somme de formes de différents degrés.
Mais alors  $\Lambda(M)$ acquiert \og naturellement\fg \;une deuxième loi interne (une multiplication) : le produit extérieur.
Si $\al\in\Lambda^p(M)$ et $\beta\in \Lambda^q(M)$ alors $\al\wedge \beta\in\Lambda^{p+q}(M)$. La structure $
(\Lambda(M),+,\wedge,.)$, appelée \emph{algèbre extérieure sur $M$}, est construite en détail lorsque $M$ est un espace vectoriel dans [\ref{rinrin}] et plus généralement dans le cadre des variétés dans [\ref{FW}]. On peut ici faire l'économie de cette construction car si localement on a :
 $\al=a_{i_1,\ldots,i_p}dx^{i_1}\wedge \ldots \wedge dx^{i_p}$ et $\be= b_{j_1,\ldots,j_q}dx^{j_1}\wedge \ldots \wedge dx^{j_q}$, il suffit de savoir dans ce qui suit que  $\al \wedge\be$ s'écrit localement 
 $ a_{i_1,\ldots,i_p}b_{j_1,\ldots,j_q}dx^{i_1}\wedge \ldots \wedge dx^{i_p}\wedge dx^{j_1}\wedge \ldots \wedge dx^{j_q}$.
\begin{definition}\label{dederalex}\index{antidérivation sur les formes différentielles}
Soit $d$ un entier naturel impair. Une antidérivation de degré $d$  sur l'algèbre extérieur $(\Lambda(M),+,\wedge,.)$ est une application $D$, $\rr$-linéaire de $\Lambda(M)$ dans lui-même qui à toute $p$-forme associe une $(p+d)$-forme de sorte que pour toutes formes $\al\in\Lambda^p(M)$ et $\beta\in\Lambda(M)$ on ait :
\begin{equation}\label{sedx}D(\al\wedge\beta)=D\al\wedge\beta+(-1)^p\al\wedge D\beta.\end{equation}
\end{definition}
Le résultat essentiel concernant les dérivations sur $\Lambda(M)$ est le suivant.
\begin{theorem}\label{deffetdf}
Une antidérivation $D$ sur $\Lambda(M)$ est entièrement déterminée par ses valeurs sur les fonctions différentiables et sur les différentielles de ces fonctions, c'est à dire par ses valeurs sur $\Lambda^0(M)\oplus d\Lambda^0(M)$.
\end{theorem}
La preuve passe par les deux lemmes qui suivent.
\begin{lemma}\label{gasp}Soit $M$ une variété différentiable.
Pour toute fonction différentiable $f$ définie sur un ouvert $U$ d'une carte locale $(U,\varphi)$ de $M$, pour tout point $x\in U$ il existe un voisinage $V$ de $x$ dans $U$ et une fonction différentiable  $g$ sur $M$ telle que $f_{\vert V}=g_{\vert V}$.
\end{lemma}
\begin{preuve}{du lemme \ref{gasp}}
Soit $x\in U$. Il existe un voisinage $V$ de $x$ dans $U$ et une fonction différentiable $\sigma$ sur $M$ telle que 
$\sigma =1$ sur $V$ et $\sigma =0$ sur $M\setminus U$ et $0<\sigma<1$ sur $U\setminus V$. On considère en effet les ouverts $V=\varphi^{-1}(]-1,1[)$ et $W=\varphi^{-1}(]-2,2[)$ de $U$. Soit la fonction $f$ de $\rr$ vers $\rr$ définie par $f(t)=e^{-1/t}$ si $t>0$ et $f(t)=0$ si $t<0$ prolongée par continuité en $0$. On pose $h(t)=\frac{f(2-\abs{ t})}{f(2-\abs{t})+f(\abs{t}-1)}$. La fonction $\sigma=h\circ \varphi$ vaut $1$ sur $V$ et $0$ sur $M\setminus W$, donc $0$ sur $M\setminus U$.
La fonction $g=f \sigma$ vérifie l'énoncé du lemme.
\end{preuve}
\begin{lemma}\label{gaspp}
Soit $U$ un ouvert de $M$, $\al$ et $\beta$ deux formes différentielles telles que $\al=\beta$ sur $U$. Alors $D\al=D\beta$ sur $U$
\end{lemma}
\begin{preuve}{du lemme \ref{gaspp}}

$i$) Soit $x\in U$ et $V$ un voisinage de $x$ dans $U$ tel qu'il existe une fonction différentiable $\sigma $ sur $M$ valant $1$ sur $V$ et $0$ sur $M\setminus U$ (voir démonstration du lemme \ref{gasp}).
Si $\al=\beta$ sur $U$, alors $\al-\beta=(\al-\beta)(1-\sigma)$ sur $M$. On en déduit que $D(\al-\beta)=0$ sur $U$ en utilisant l'égalité (\ref{sedx}). Et on conclut.
\end{preuve}
\noindent Le lemme \ref{gaspp} permet de définir un opérateur $D_{\vert U} $ sur $\Lambda (M)_{\vert U}$ par 
$D_{\vert U}(\al_U)=D(\al)_{\vert U}$. On vérifie immédiatement que $D_{\vert U}$ est une dérivation sur $\Lambda (M)_{\vert U}$.
\begin{preuve}{du théorème \ref{deffetdf}}
Une $p$-forme $\al$ sur une variété différentiable $M$ de dimension $n$, étant une section d'un fibré,  est déterminée par sa valeur en chaque point  de $M$. Soit donc $x$ un point de $M$ et $(U,\varphi)$ une carte locale en $x$. Pour la valeur de $D\al$ en $x$, il suffit de considérer $(D\al)_{\vert U}$, c'est à dire $D_U(\al_{\vert U})$. or (voir la remarque \ref{colopul}) $\al_U=\varphi^*\left((\varphi^{-1})^*(\al_U)\right)=\varphi^*\left(a_{i_1\ldots\i_p}\;dx^{i_1}\wedge\ldots\wedge dx^{i_p}\right)$
$=(a_{i_1\ldots i_p}\circ\varphi)\; d(x^{i_1}\circ\varphi)\wedge\ldots\wedge d(x^{i_p}\circ\varphi).$  Il existe donc des fonctions $\al_{i_1\ldots i_p}, f^{i_1},\ldots,f^{i_p}$ définies sur $U$ avec $\{i_1,\ldots, i_p\}\subset \{1,2,\ldots, n\}$, telles que 
\[\al_U=\al_{i_1\ldots i_p}\;df^{i_1}\wedge\ldots\wedge df^{i_p}.\]

\noindent Du lemme \ref{gasp}, on déduit directement que $x$ admet un voisinage $V$ tel que les restrictions à $V$ des fonctions $\al_{i_1\ldots i_p}, f^{i_1},\ldots,f^{i_p}$ se prolongent à $M$ en fonctions respectivement notées $\beta_{i_1\ldots i_p}, g^{i_1},\ldots,g^{i_p}$. 
Nous pouvons maintenant évaluer $(D\al)_{\vert V}$ en utilisant les règles de dérivation :

\noindent$(D\al)_{\vert V}=D_V\al_{\vert V}=
D_V \beta_{i_1\ldots i_p}\wedge dg^{i_1}\wedge\ldots\wedge dg^{i_p}
+ \beta_{i_1\ldots i_p}\left(D dg^{i_1}\wedge  dg^{i_2}\wedge\ldots\wedge  dg^{i_p}+\right.$

$\left.\ldots+( dg^{i_1}\wedge  dg^{i_2}\wedge\ldots\wedge D dg^{i_p}\right)$, ce qui montre que le calcul de $D\al$ au point $x$ est entièrement déterminé par les valeurs de $D$ sur les fonctions et leur différentielles.
\end{preuve}
\subsubsection{Dérivée extérieure des formes différentielles}\index{dérivée extérieure d'une forme différentielle}
Le théorème \ref{deffetdf} donne un sens à la définition suivante.
\begin{definition}\label{derex}
La dérivée extérieure sur $\Lambda(M)$ est la dérivation $D$ de degré $1$ définie pour toute fonction différentiable sur $M$ par :
\begin{enumerate}
\item[(d1)] Df=df,
\item [(d2)]D(df)=0.
\end{enumerate}
\end{definition}
\textbf{Notation} : s'inspirant de l'item (d1), on notera dans tout ce qui suit,  $d$, la dérivée extérieure. .
\begin{proposition}\label{benutile}
Soit $\al$une $p$-forme extérieure , on a pour tous champs de vecteurs $X_1, \ldots,X_{p+1}$:
\begin{align*}
d\al(X_1, \ldots,X_{p+1})&=\sum_{i=1}^{p+1}(-1)^{(i-1)}X_i.\al\left(X_1,\ldots,\widehat{X_i},\ldots,X_{p+1}\right)\\
&+\sum_{i<j}(-1)^{i+j}\al\left([X_i,X_j],X_1,\ldots \widehat{X_i},\ldots,\widehat{X_j},\ldots X_{p+1}\right).
\end{align*}
En particulier pour $p=1$ on peut écrire :
\[d\al (X_1,X_2)=X_1.\al(X_2)-X_2.\al(X_1)-\al([X_1,X_2]).\]
\end{proposition}
\begin{preuve}{}
Une fois que l'on a vérifié que $d$ est une antidérivation de degré $1$ sur $\Lambda(M)$, on applique le théorème \ref{deffetdf}. La vérification de la propriété d'antidérivation est un calcul direct en coordonnées locales.
\end{preuve}
\noindent Ceci nous donne les définitions d'une forme différentielle exacte : une $(p+1)$-forme $\al$ est exacte s'il existe une $p$-forme $\beta$ telle que $\al=d\beta$ et la définition d'une $p$-forme fermée : une $p$- forme $\al$ est fermée si $d\al=0$. L'axiome (d2) nous montre que toute forme exacte  est fermée. La réciproque est généralement fausse. Le défaut d'exactitude des $p$-formes fermées est mesuré par le module 
$H^p(M)=Z^p(M)/dZ^{p-1}(M)$ où $Zp(M)$ désigne le module sur l'anneau des fonctions différentiables  des $p$- formes fermées sur $M$.
Si $M=\rr^n$, Poincaré a montré que $H^p(M)=0$.
\subsubsection{Dérivée de Lie des formes différentielles}\index{dérivée de Lie des formes différentielles}
\begin{definition}\index{dérivation sur les formes différentielles}
Soit $d$ un entier naturel pair. Une dérivation $D$ de degré $d$ sur $\Lambda(M)$ est une application $\rr$-linaire de 
$\Lambda(M)$ dans lui-même associant à une $p$-forme différentielle $\al$ une $p+d$ forme différentielle de sorte que 
\[D(\al\wedge\be)=D\al\wedge\be+\al\wedge D\be.\]
\end{definition}
\noindent On a pour les dérivation un théorème analogue au théorème \ref{deffetdf}.
\begin{definition}\label{derlily}
Soit $X$ un champ de vecteurs sur $M$. La dérivation de Lie suivant un champ de vecteurs $X$ sur $\Lambda(M)$ est la dérivation notée $L_X$ de degré $0$ définie pour toute fonction différentiable sur $M$ par :
\begin{enumerate}
\item[(l1)] $L_Xf=X.f$,
\item [(l2)]$L_X(df)=d(X.f).$
\end{enumerate}
\end{definition}
\begin{proposition}Avec les notations précédentes on a pour toute forme différentielle $\al$ :
\[L_X\al=i(X)d\al+di(X)\al.\]
\end{proposition}
\begin{preuve}{}
Il suffit de montrer que $i(X)\circ d+d\circ i(X)$ est une $0$-dérivation qui vérifie  
 les axiomes (l1) et (l2), ce qui est immédiat.
\end{preuve}
La définition la plus intuitive de $L_X$ est donnée à partir du flot engendré par le champ $X$. Rappelons le théorème fondamental qui régit la théorie des équations différentielles(voir [\ref{rinrin}])
\begin{theorem}\label{eqdif}
Soit $X$ un champ de vecteurs sur $M$ et $(x,t_0)\in M\times \rr$. Il existe un réel $\epsilon >0$, un voisinage $U$ de $x$ et une application 
$\Phi : ]t_0-\epsilon,t_0+\epsilon[\to M$ tels que : 
\begin{enumerate}
 \item Pour tout $y\in U$, $t\mapsto \Phi(t,y)$ est une courbe intégrale de $X$ définie sur l'intervalle ouvert $]t_0-\epsilon,t_0+\epsilon[$ et vérifiant $\Phi(t_0,y)=y$
 \item Si $\epsilon',U',\Phi'$ vérifient l'item précédent , alors on a l'égalité  $\Phi=\Phi'$ sur l'intersection $\left(]t_0-\epsilon,t_0+\epsilon[\times U\right)
 \cap \left(]t_0-\epsilon',t_0+\epsilon'[\times U'\right).$
 \end{enumerate}
\end{theorem}
\noindent Pour $t_0=0$, on notera $\varphi_t $ le  \emph{flot (local)} de $X$ défini par l'égalité $\varphi_t(y)=\Phi(t,y)$ et $\varphi_0(y)=y$. On a donc $$\frac{d \varphi_t(x)}{dt}=X(\varphi_t(x))$$ et pour toute fonction différentiable $f$, \begin{equation}\label{essence}X.f(x)=\frac{d}{dt}(f\circ\varphi_t(x)_{\vert t=0}.\end{equation}
\begin{proposition}\label{grr}
Soit $x_0\in M$ et $U$ un voisinage ouvert  de $x_0$ sur lequel le flot $\varphi_t$ est défini pour $t\in ]-\varepsilon,\varepsilon[$.
Alors 
\begin{enumerate}
\item \begin{equation}\label{grouppe}\varphi_0=Id_{\left\vert U\right.}\end{equation}
\item Si $t,t',t+t'\in ]-\epsilon,\epsilon[$ et $\varphi_{t'}(U)\subset U$, alors on a sur $U$ l'égalité :
\begin{equation}\label{grappe}
\varphi_t\circ\varphi_{t'}=\varphi_{t+t'}
\end{equation}

\item $\forall t\in ]-\varepsilon,\varepsilon[,\;\varphi_t$ est\ inversible sur $U$ et \begin{equation}\label{grippe}
(\varphi_t )^{-1}=\varphi_{-t}.
\end{equation}
\end{enumerate}
\end{proposition}
\begin{preuve}{}
C'est une conséquence directe de la proposition \ref{eqdif}
\end{preuve}
Une famille de difféomorphismes locaux vérifiant les égalités (\ref{grouppe}),(\ref{grippe}) et (\ref{grappe}) est appelé un\emph{ groupe local de difféomorphismes à un paramètre}.
\begin{proposition}\label{altderli}
Soit $X$ un champ de vecteurs sur $M$, engendrant un flot $\varphi_t$ et $\al$ une $p$-forme différentielle sur $M$. Alors 
\begin{equation}\label{redeli}L_X\al =\lim_{t\to 0}\frac{\varphi_t^*\al-\al}{t}.\end{equation}
\end{proposition}
\begin{preuve}{}
Le plan de la démonstration consiste en trois points : $1$) montrer que le terme de droite de l'équation (\ref{redeli}), qu'on notera $D_X\al$  détermine une dérivation de degré zéro, $2$) montrer que cette dérivation coïncide avec $L_X$ sur les fonctions, $3$)
montrer que cette dérivation coïncide avec $L_X$ sur les différentielles des fonctions,
\begin{enumerate}
\item Posons $\delta(\al)=\varphi_t^*\al-\al$. Alors un calcul direct montre que pour deux formes $\al$ et $\beta$, on a :
$\delta (\al\wedge\beta)=\delta\al\wedge\varphi_t^*\be+\al\wedge \delta \be$.  On a $\lim_{t\to 0}\varphi_t^*\be=\be$. Ainsi

$D_X(\al\wedge\be)=\lim_{t\to 0}\frac{\varphi_t^*\al-\al}{t}\wedge\be+\al\wedge \lim_{t\to 0}\frac{\varphi_t^*\be-\be}{t}=D_X(\al)\wedge\be+\al\wedge D_X(\be)$.

 $D$ est une dérivation, de degré $0$.
\item $\lim_{t\to 0}\frac{\varphi_t^*f-f}{t}=\frac{d}{dt}(f\circ \varphi_t)_{\vert t=0}=X.f(x)=L_Xf(x)$ d'après l'égalité (\ref{redeli}).
\item $\varphi_f^*df-df=d\left(\varphi_t^*f-f\right)$. D'où,

 $\lim_{t\to 0}\frac{\varphi_f^*df-df}{t}=
d\left(\lim_{t\to 0}\frac{\varphi_t^*f-f}{t}\right)
=d(X.f)(x).$
\end{enumerate}
\end{preuve}
\begin{exo}\label{konar}{Suite de l'exercice \ref{kon} p.\pageref{kon}}
\hspace{1cm}\\
On considère les champs de vecteurs $X,Y$ sur $\rr^n$ et on note $\varphi_t$ le flot local de $X$ tel qu'il est défini dans le théorème \ref{eqdif}. Alors : \[L_XY(x)=\lim_{t\to 0}\frac{Y\circ\varphi_t(x)-Y(x)}{t}.\]
\end{exo}
 \subsection{Complexifié d'un fibré vectoriel.}\label{comfiture}
   
\begin{lemma}\label{con}
Il existe un isomorphisme $\lambda$ du sous-groupe $G$ des isométries complexes de $SO(2n)$ vers $U(n)$.
\end{lemma}
\begin{preuve}{}
En se basant sur la proposition \ref{youyou},
on définit l'isomorphisme $\lambda$ par $\lambda(f)=f_{\cc}=\varphi^{{-1}}f\varphi$ où $\varphi$ est la bijection de la définition \ref{aze}. On vérifie directement que $\lambda$ est un morphisme de groupe. \end{preuve}
La fabrication d'un fibré associé exposée dans  la section \ref{ratoi}, conduit au corollaire suivant.
\begin{corollary}\label{gloup}
On suppose que l'on dispose d'une structure de fibré vectoriel $\tau=(E,p,M)$ définie par une famille de cocycles $\{c_{ij}\}$ à valeurs dans le sous-groupe des isométries complexes de $SO(2n)$. Cette famille définit une  structure de fibre vectoriel $\tau^{c}=(E^{c},p,M)$ de fibre $\cc^{n}$ déterminée par la famille de  cocycles $\{\lambda(c_{ij})\}$  à valeurs dans $U(n)$.
\end{corollary}
\begin{definition}\label{fico}\index{complexifié d'un fibré vectoriel}
Le fibré  $\tau^{c}$ est le complexifié de $\tau$.
\end{definition}
\noindent
\textbf{Notation} :
Dans la suite, avec les notations de la démonstration  du lemme \ref{con}, si $c_{ij}$ désigne un cocycle du fibré $\tau$, alors \emph{$\tilde c_{ij}$ désignera le cocycle $\varphi^{-1}c_{ij}\varphi$ de $\tau^{c}$.}

Le fibré $\tau^{c}$ est unique à un isomorphisme près. Dans la proposition suivante, nous construisons une réalisation de $\tau^{c}$ qui représentera cette classe d'isomorphie dans la suite de ce texte.
\begin{proposition}\label{readefico} 
Soit $\tau=(E,p,M)$ un fibré de fibre $\rr^{2n}$, soit $\{(U_{i},\Phi_{i})\}$ un atlas  de $\tau$ et $\{c_{ij}\}$ la famille des cocycles associés à cet atlas.
Soit $\varphi$ l'application de $\cc^{n}$ vers $\rr^{{n}}$ définie par l'équation (\ref{fifi}). Si $\Phi_{i} :p^{{-1}}(U_{i})\longrightarrow U_{i}\times \rr^{{2n}}$ est une trivialisation  de $\tau$, alors $\Phi_{i}^{{c}} :p^{{-1}}(U_{i})\longrightarrow U_{i}\times \cc^{{n}}$ définie par 
$\Phi_{i}^{{c}}=(I\times \varphi^{{-1}})\circ\Phi_{i}$ est une carte locale d'un fibré $\zeta$ défini par le même triplet $(E,p,M)$ que $\tau$ mais de  fibre $\cc^{n}$.  Si on considère l'atlas $\{(U_{i},\Phi_{i}^{c})\}$, ses changements de cartes sont précisément les cocycles $\{\tilde c_{ij}\}$ définis dans la démonstration du corollaire \ref{gloup}. 
\end{proposition}
\begin{preuve}{}
Les applications $\Phi_{i}^{c}$ sont bien des homéomorphismes.
Il suffit donc de montrer que si $(U_{i},\Phi_{i}) $ et  $(U_{j},\Phi_{j}) $ sont deux cartes locales de $\tau$, on a avec les notations de la démonstration du corollaire \ref{gloup} : 
\[\forall Z\in \cc^{n}\forall b\in M ,\;\;\Phi_{i}^{c}({\Phi_{j}^{c}})^{{-1}}(b,Z)=\big(b,\tilde c_{ij}(b)Z\big).\]
Or

 \noindent$\Phi_{i}^{c}({\Phi_{j}^{c}})^{{-1}}(b,Z)=\Phi_{i}^{c}\circ\Phi_{j}^{{-1}}\circ (I\times \varphi)(b,Z)=\Phi_{i}^{c}\circ\Phi_{j}^{-1}(b,\varphi(Z))$
 
 $=(I\times \varphi^{-1})\Phi_{i}\Phi_{j}^{-1}(b,\varphi(Z))=
\big(b, \varphi^{-1}c_{ij}(b)\varphi(Z)\big)=(b,\tilde c_{ij}(b)Z)$.
\end{preuve}
\begin{corollary}\label{rearea}
Le fibré $\zeta$ est isomorphe à $\tau^{c}$. On peut donc l'utiliser pour démontrer des propriétés de $\tau^{c}$ invariantes par les isomorphismes de fibres vectoriels.
\end{corollary}
Compte tenu de la construction du complexifié d'un espace vectoriel réel,  la question suivante est naturelle : peut-on définir le complexifié d'un fibré  $\tau$  qui vérifie les hypothèses du corollaire \ref{gloup} de fa\c con analogue à la définition du complexifié d'un espace vectoriel par un schéma commutatif de la forme
\begin{equation}\label{coco}
\begin{array}{ccc}
\tau^{c}&\stackrel{\enspace\tilde{\varphi}\enspace}{\longrightarrow}&\tau\\
\vcenter{\llap{i}}\Big\downarrow&&\Big\downarrow\vcenter{\rlap{$\tilde{\J}$}}\\
\tau^{c}&\xrightarrow[\enspace\tilde{\varphi}\enspace]{}&\tau
\end{array}
\end{equation}
où les flèches représentent des morphismes de fibrés vectoriels \emph{au-dessus de l'identité}, le morphisme $\tilde{\J}$ vérifiant de plus l'égalité  $\tilde{\J}^{2}=-\I$? 
Il faudrait dans ce cas conformément à la définition \ref{azeza} introduire une structure riemannienne sur $E$. Cela se fait en toute dimension.
\begin{proposition}\label{ririta}
 Soit $\tau=(E,p,M)$ un fibré de fibre $\rr^{{2n}}$. Toute structure euclidienne de $\rr^{{2n}}$ induit sur $E$ une structure riemannienne.
\end{proposition}
\begin{preuve}{}
On considère un atlas localement fini $\{(U_{i},\Phi_{i})\}$ de $\tau$  et $\{\theta_{i}\}$ une partition de l'unité subordonnée au recouvrement $\{U_{i}\}$ de $M$\footnote{ Pour ces notions de recouvrement localement fini et de partition de l'unité, on pourra consulter [\ref{BG} ] ou [\ref{schw}]}.
Soit $q$ la forme quadratique qui définit une structure euclidienne de $\rr^{{2n}}$. Soit $e\in p^{{-1}}(U_{i})$; écrivons $\Phi_{i}(e)=(b,f)\in U_{i}\times \rr^{2n}$. Notons $p_{2}$ la deuxième projection de $U_{i}\times \rr^{{2n}}$ sur $\rr^{2n}$ et $q_{i}(e)=q p_{2}\Phi_{i}(e)=q(f).$ On définit maintenant la structure riemannienne par \begin{equation}\label{ririe}
Q(e)=\sum_{i}\theta_{i}(b)q_{i}(e). 
\end{equation}
Pour tout $e\in  E, Q(e)\ge 0$. Supposons que $Q(e)=0$. Il existe au moins un indice $i$ pour lequel $\theta_{i}(b)\not= 0$ et par conséquent $q_{i}(e)=0$. Mais alors $e=0_{p^{{-1}}}(\{b\})$. Et pour finir $Q(\lambda e)=\lambda^{2}Q(e)$ pour tout $\lambda\in \rr$ et tout $e\in E$, ce qui permet de conclure.
\end{preuve}
Pour illustrer la proposition \ref{ririta}, l'exercice suivant construit une structure riemannienne sur le fibré tangent à la sphère $S^2$.
\begin{exo}\label{abdel}
Avec la vision de champ de vecteurs comme dérivation, illustrer la proposition \ref{ririta}, en donnant l'expression du tenseur métrique de la sphère $S^2$ (de rayon 1) plongée dans $\rr^3$ sur l'ouvert
$U= 
\{(\theta,\varphi)\in]0,\pi[\times ]0,2\pi[\}$ où $(\theta,\varphi)$ représentent les coordonnées sphériques sur $S^2$.
\end{exo}
Par la suite on notera par des crochets $\langle \;\rangle$ la structure riemannienne sur $E$.
Indiquons comment exprimer ce crochet en utilisant les cartes locales du fibré.  
\begin{proposition}\label{ririder}
On considère $(U_{i},\Phi_{i})$ et $(U_{j},\Phi_{j})$ deux cartes locales telles que $U_{i}\cap U_{j}\not=\emptyset$.  Soit $b\in U_i\cap U_j$. On a l'égalité suivante :
\[\forall e,e'\in p^{{-1}}(b),\enspace \langle p_{2}\Phi_{i}(e),p_{2}\Phi_{i}(e')\rangle_{\rr^{2n}}=\langle p_{2}\Phi_{j}(e),p_{2}\Phi_{j}(e')\rangle_{\rr^{2n}}.\]
Ainsi $\langle p_{2}\Phi_{i}(e),p_{2}\Phi_{i}(e')\rangle_{\rr^{2n}}$ ne dépend pas de $i$ tel que $e\in p^{-1}(U_{i})$. Cette quantité intrinsèque est égale à $\la e,e'\ra$.
\end{proposition} 
\begin{preuve}{}

Soit $b\in U_{j}\cap U_{i}$. On a d'après l'équation (\ref{changg}):
 \[\la p_{2}\Phi_{i}(e),p_{2}\Phi_{i}(e')\ra=\la c_{ij}(b)p_{2}\Phi_{j}(e),c_{ij}(b)p_{2}\Phi_{j}(e')\ra=\la 
p_{2}\Phi_{j}(e),p_{2}\Phi_{j}(e')\ra\] puisque les $c_{ij}(b)$ sont des isométries.
En considérant que $e,e'\in p^{-1}(\{b\})$, on a d'après l'équation (\ref{ririe}) :

 $$\la e,e'\ra(b)=\sum_{i}\theta_{i}(b)\la p_{2}\Phi_{i}(e),p_{2}\Phi_{i}(e')\ra.$$ 
 Donc d'après ce qui précède on a pour un $j$ particulier tel $b\in U_{j}$:
 
\[\la e,e'\ra(b)=\sum_{i}\theta_{i}(b)\la p_{2}\Phi_{j}(e),p_{2}\Phi_{j}(e')\ra =\la p_{2}\Phi_{j}(e),p_{2}\Phi_{j}(e')\ra\] puisque $\sum_{i}\theta_{i}(b)=1$.
 \end{preuve}
\begin{proposition}\label{stcon}
L'espace $\rr^{2n}$ est muni de sa structure euclidienne standard et de la structure complexe définie dans l'exemple de la page \pageref{russe}. Soit $G$  le groupe des isométries complexes de $SO(2n)$.
 Soit $\tau=(E,p,M)$ un fibré vectoriel de fibre $\rr^{2n}$ défini par des cocycles à valeurs dans  $G$. On considère l'application $\tilde{J} :E\longrightarrow E$ définie dans les cartes locales $(U_{i},\Phi_{i})$ et $(U_{j},\Phi_{j})$ par l'équation
 \begin{equation}\label{snuc}
 \forall (b,f)\in U_{i}\cap U_{j}\times \rr^{2},\;\;\Phi_{i}\tilde{J}\Phi_{j}^{{-1}}(b,f)=(b,J c_{ij}(b)f).
 \end{equation}
Alors $\tilde{J}$ est un endomorphisme du fibré $\tau$ tel que $\tilde{J}^{{2}}=-\I$.
\end{proposition}
\begin{preuve}{}
On note $h_{ij}(b)=Jc_{ij}(b)$. Il suffit de vérifier conformément  à l'équation (\ref{concon}) du deuxième item de la section \ref{licite}  les relations pour tout $b\in U_{i}\cap U_{J}\cap U_{k}$ :
\begin{align}
c_{ij}(b)h_{jk}(b)&=h_{ik}(b)\label{qs}\\
h_{ij}(b)c_{jk}(b)&=h_{ik}(b)\label{qss}
\end{align}
L'égalité (\ref{qss}) résulte directement d'une relation de cocycles.  L'égalité (\ref{qs}) résulte du choix de $G$ : les éléments de $G$ commutent avec $J$ (voir proposition\ref{youyou}).

\noindent Pour finir,

$\Phi_{i}\tilde{J}^{2}\Phi_{j}^{{-1}}(b,f)=\big(\Phi_{i}\tilde{J}\Phi_{j}^{{-1}}\big)\big(\Phi_{j}\tilde{J}\Phi_{j}^{{-1}}\big)(b,f)=(\Phi_{i}\tilde{J}\Phi_{j}^{{-1}})(b,\J f)$

$=
(b,\J c_{ij}(b)\J f)=-(b,c_{ij}(b)f)=-(\Phi_{i}\Phi_{j}^{{-1}})(b,f)$, d'où on conclut  $\tilde{\J}^{{2}}=-\I$.
\end{preuve}
\begin{proposition}\label{dideux}
On reprend les notations et hypothèses de la proposition \ref{stcon}. La structure  complexe $\tilde{\J}$ introduite dans la proposition \ref{stcon}
 est une isométrie de $E$ muni de sa structure riemannienne (décrite dans la proposition \ref{ririta}).
\end{proposition}
\begin{preuve}{}
Soient $e,e'\in p^{{-1}}(b)$ avec  $b\in U_i$ et $(U_{j},\Phi_{j})$ une trivialisation de $\tau$. Le crochet $\la\;\;\ra_{E}$ désigne la métrique riemannienne sur $E$ et $\la\;\ra$ la structure euclidienne sur $\rr^{2n}$.  Les éléments $e$ et $e'$ peuvent également appartenir à un autre domaine de carte $p^{{-1}}(U_{i})$. Dans ce cas 
$\tilde{\J}(e)=\tilde{\J}\Phi_{j}^{{-1}}(b,f)=\Phi_{i}^{{-1}}(b,\J c_{ij}(b)f)$ et de même $\tilde{\J}(e')=\tilde{\J}\Phi_{j}^{{-1}}(b,f')=\Phi_{i}^{{-1}}(b,\J c_{ij}(b)f')$. D'où 

$\la\tilde{\J}(e),\tilde{\J}(e')\ra_{E}(b)=\sum_{i}\theta_{i}(b)\la p_{2}\Phi_{i}\tilde{\J}(e),p_{2}\Phi_{i}\tilde{\J}(e')\ra=\sum_{i}\theta_{i}(b)\la f,f'\ra $ (car $\J c_{ij}$ est une isométrie)

$= \la f,f'\ra=\sum_{i}\theta_{i}(b)\la p_{2}\Phi_{i}(e),p_{2}\Phi_{i}(e')\ra=\la e,e'\ra_{E}(b) $ (en utilisant la proposition \ref{ririder} et le fait que $\sum_{i}\theta_{i}(b)=1$).
\end{preuve}
\begin{proposition}\label{kaziol}
On reprend les notations et hypothèses de la proposition \ref{stcon}.  Il existe un morphisme bijectif \; $\rr$-linéaire $\tilde{\varphi}$ du fibré \;$\zeta$ (isomorphe à
 $\tau^{c}$)  vers 
$\tau$ tel que le diagramme (\ref{coco}) soit commutatif.
\end{proposition}
\begin{preuve}{}
Soit $(U_{k},\Phi_{k})$ une carte de $\tau$ et $(U_{k},\Phi_{k}^{c})$ la carte correspondante de $\zeta$ (voir le corollaire \ref{rearea}).
On considère $\tilde\varphi$ défini dans les cartes $(U_{l},\Phi_{l})$ et  $(U_{k},\Phi_{k}^{c})$ par
 \[\forall z\in \cc^{n},\;\;\Phi_{l}\tilde\varphi (\Phi_{k}^{{c}})^{{-1}}(b,z)=\big(b,c_{kl}(b)\varphi(z)\big),\] 
 où $\varphi$ est la bijection canonique de $\cc^{n}$ vers $\rr^{2n}$. On vérifie que $\tilde\varphi$ est bien un morphisme de fibré. En effet, si on pose $h_{kl}(b)=c_{kl}(b)\varphi$, alors 
pour tout $b\in U_{k}\cap U_{l}\cap U_{s}$ on a en utilisant les notations de la démonstration du premier item du corollaire \ref{gloup} :
\begin{align}
h_{kl }(b)\tilde c_{ls}(b)&=h_{ks}(b) \label{sd}\\
c_{kl}(b) h_{ls}(b)&=h_{ks}(b)\label{idd}
\end{align}
Vérifions par exemple l'égalité  (\ref{sd}). On a 

\noindent$h_{kl}(b)\tilde c_{ls}(b)=c_{kl}(b)\circ\varphi\circ(\varphi^{{-1}}\circ c_{ls}(b)\circ \varphi)=c_{kl}(b)\circ c_{ls}(b)\circ \varphi=c_{ks}(b)\circ\varphi=h_{ks}(b)$.
L'égalité (\ref{idd}) se vérifie de même. Ainsi $\tilde\varphi$ est bien un morphisme de fibré. Sa bijectivité est claire. Reste à vérifier
 la commutativité du diagramme (\ref{coco}).
 Si $b\in U_{k}\cap U_{l}$, on a : $\Phi_{k}(\tilde\varphi\circ {i})(\Phi_{l}^{{c}})^{{-1}}(b,z)=\big(b, c_{kl}(b)\circ\varphi(iz)\big)=\big(b,c_{kl}(b)\circ\J\circ \varphi(z)\big)=\big(b,\J\circ c_{kl}
(b)\circ \varphi(z)\big)=\Phi_{k}(\tilde\J\circ\tilde\varphi)(\Phi_{l}^{{c}})^{{-1}}(b,z)$.
\end{preuve}
Nous allons maintenant nous servir de cette construction de $\tau^{c}$  pour doter $E^{c}$ d'une métrique kählérienne.
\begin{proposition}\label{kael}
On reprend les notations des propositions \ref{stcon}.
  On a alors une métrique kählérienne sur $E^{c}$ définie par 
 \begin{equation*}
 \forall e,e'\in E^{c},\;\; \la  e, e'\ra^{c}=\la \tilde\varphi e,\tilde\varphi e'\ra+i\la \tilde\varphi e,\tilde\J \tilde\varphi e'\ra.
  \end{equation*}
 \end{proposition}
\begin{preuve}{}
 Elle est identique à celle de la proposition (\ref{hermit}).
  \end{preuve}
  \begin{remark}\label{bauf}
 La structure de fibré vectoriel, permet d'écrire les équations de la physique classique ou quantique qui font intervenir des dérivées d'ordre $1$ ou $2$ : on pense par exemple aux équations de Lagrange ou de Hamilton liées aux équations de Newton, ou aux équations qui font intervenir un opérateur laplacien, par exemple à l'équation de Schrödinger. Tout ceci est développé dans la suite de ce texte.
 
 Cependant la physique quantique  met en évidence dans le phénomène de quantification
 des structures de fibré qui ne sont pas des fibrés vectoriels. La structure   de ces fibrés est juste une extension de celle de fibrés vectoriels mais la fibre est en toute généralité  un espace topologique $F$ et le groupe un groupe opérant sur $F$. Par exemple $F=\cc\setminus\{0\}$.
 Vu la proximité avec l'exposé sur les fibrés vectoriels, nous donnons en appendice  \ref{ququ} un court exposé de cette structure de fibré localement trivial et un exemple qui concerne la quantification.
  \end{remark}

 \section{Vers un calcul différentiel à l'ordre $2$ : connexion sur un fibré vectoriel.}\label{cacudeux}
 \subsection{Introduction}
Revenons sur le sens de \og calcul différentiel à l'ordre $1$\fg.\; Le calcul différentiel à l'ordre $1$ introduit la notion de différentiabilité ( avec la structure de variété) et la notion de différentielle ( avec la structure de fibré vectoriel).
Dans le cadre des fibrés vectoriels on peut transporter une section d'un fibré $\tau$ en une section d'un fibré $\tau'$ si on dispose d'un morphisme de fibré de $\tau$ vers $\tau'$. La dérivée de Lie $L_Xf$ d'une fonction différentiable est un cas particulier de cette opération.

 L'objet d'un calcul différentiel à l'ordre $2$ est de définir la dérivée d'un champ $Y$
  de vecteurs dans la direction d'un autre champ de vecteurs $X$ en demandant que cet objet qu'on peut noter par analogie avec la dérivée de Lie précédent $\nabla_XY$ soit encore un champ de vecteurs: le champ $Y$ étant une dérivation cela explique le sens de calcul différentiel à l'ordre $2$.
 
 \noindent L'exercice \ref{kon}  donne une piste. On y définit pour deux champs de vecteurs $X$ et $Y$ sur $\rr^n$, le champ de vecteurs sur $\rr^n$ noté $L_XY$ ou encore le champs de vecteurs sur une hypersurface $S$ de $\rr^n$, lorsque $X$ et $Y$ sont tangents à $S$. 
 Parmi les propriétés mises en évidence pour $L_XY$ ou $D_XY$, certaines ne sont pas liées à la métrique :
  ce sont les essentiellement les propriétés $1.$ et $2.$ pour $L$ et les propriétés $(a)$ et $(b)$ pour $D$.
  Les propriétés $3.$ et $(c)$ sont liées à la métriques.
 
 Si on considère le résultat de l'exercice \ref{konar}, on sait que l'on peut obtenir $L_XY(x)$ comme limite quand $t$ tend vers $0$ du taux $\theta=\frac{Y(\varphi_t(x))-Y(x)}{t}$ où $\varphi_t(x)$ est la solution de l'équation
 $\frac{d\varphi_t(x)}{dt}=X(\varphi_t(x))$ telle que $\varphi_0(x)=x$. Cette écriture a été possible en utilisant la structure vectorielle de $\rr^n$. Mais si on analyse le sens géométrique de $\theta$ il faut considérer $Y(\varphi_t(x))$ et $Y(x)$ comme deux vecteurs appartenant à $T_x\rr^n$ (car on additionne deux vecteurs du même espace vectoriel et que les champs de vecteurs sont des sections du fibré tangent à $\rr^n$ qui est $\rr^n\times\rr^n$), et donc voir $Y(\varphi_t(x))$ comme un transporté parallèle dans l'espace affine $\rr^n$ au point $x$ du  vecteur 
 $Y(\varphi_t(x))$ qui est un vecteur de $T_{\varphi_t(x)}\rr^n$.
 
 \noindent Tout comme pour le calcul différentiel à l'ordre $1$ le cadre fondamental de l'étude est la catégorie des  fibrés vectoriels. 
 Le programme du calcul différentiel à l'ordre $2$ propose donc dans ce cadre de
 \begin{enumerate}
 \item définir  un opérateur analogue à $D$  tel que pour deux sections  $X$ et $Y$  d'un  fibré vectoriel  $D_XY$ soit une section de ce fibré, ayant les propriétés décrites par les  items rappelés ci-dessus de l'exercice \ref{kon},
 \item monter l'existence et l'unicité de tels opérateurs vérifiant les les propriétés  métriques quand il s'agit du fibré tangent à une variété  pseudo-riemanienne,
 \item montrer que ces opérateurs permettent de généraliser le naturel transport parallèle des espaces affines aux variétés et de généraliser ainsi le résultat de l'exercice \ref{konar}.
 \end{enumerate}

\subsection{Connexion sur un fibré}
\subsubsection{Définition}
Rappelons que si $M$ est une variété différentiable, $\tau(M)$ désigne son fibré tangent et $\tau^{*}(M)$ son fibré cotangent.

\begin{definition}\label{fomvalvec}
Soit $\tau=(E,p,M)$ un fibré vectoriel sur une variété différen\-tiable $M$. 
Le fibré $\tau\otimes \tau^{*}(M)$ est appelé \emph{le fibré des $1$-formes sur $M$ à valeurs dans $E$.}
\end{definition}\index{formes différentielles à valeurs vectorielles}
Justifions la définition \ref{fomvalvec}.  Soit $U$ un ouvert d'une carte locale à la fois de $\tau$ et de $\tau^{*}(M) $. On considère une base locale de sections de $\tau$ au-dessus de $U$, $(s_{1,U},s_{2,U},\cdots, s_{p,U})$ et une base locale de $1$-formes sur $U$, $(\alpha_{1,U},\alpha_{2,U},\cdots,\alpha_{n,U})$. Alors toute section $s$  de $\tau\otimes \tau^{*}(M)$ au-dessus de $U$ s'écrit sous la forme $s=\sum_{i,j}\lambda_{i,j}s_{i,U}\otimes \alpha_{j,U}$.
Ainsi pour tout champ de vecteurs $X$ sur $U$, on a 
\[s(X)=\sum_{i,j}\lambda_{i,j}\alpha_{j,U}(X)s_{i,U}=\sum_{i}\mu_{i}s_{i,U}\]
qui est une section de $\tau$, donc à valeurs dans $E$.
\begin{exo}\label{sectense}
Montrer que les $\mathcal{D}(M)$-modules $\mathcal{S}_{\tau}\otimes \mathcal{S}_{\tau^{*}(M)}$ et $\mathcal{S}_{\tau\otimes \tau^{*}(M)}$ sont isomorphes.
\end{exo}
Pour motiver la définition générale d'une connexion sur un fibré vectoriel, considérons le cas particulier du fibré trivial $\tau=(M\times  \rr,p_{1},M).$
On peut voir une fonction différentiable $f$ sur la variété $M$, comme une section de $\tau$. En effet l'application qui à toute fonction $f$ sur $M$ associe la section $s_{f}$ de $\tau$ définie par $s_{f}(x)=(x,f(x))$ est un isomorphisme d'anneaux.
 Et on sait définir une opération de dérivation sur les fonctions différentiables. On peut considérer sa dérivée $df$ comme un élément de 
  $\mathcal{S}_{\tau^{*}(M)}\subset \mathcal{S}_{\tau}\otimes \mathcal{S}_{\tau^{*}(M)}$ de sorte que si $f$ et $g$ sont deux fonctions différentiables, alors 
  \[d(fg)=fdg+gdf=f\otimes dg+g\otimes df\in \mathcal{S}_{\tau}\otimes \mathcal{S}_{\tau^{*}(M)}\]
  L'ambition de la notion de connexion est de généraliser cette dérivation aux  sections d'un fibré vectoriel. L'exemple précédent associé à l'exercice \ref{sectense}
 nous fait accepter la définition suivante   
  \begin{definition}\label{connex}\index{connexion sur un fibré vectoriel}
Soit $\tau$ le fibré de la définition \ref{fomvalvec}. Une connexion $\nabla$ sur $\tau$  est une application additive du module $\mathcal{S}_{\tau}$ vers le module  $\mathcal{S}_{\tau\otimes \tau^{*}(M)}$ vérifiant pour toute section $s$ de $\tau$ et toute fonction différentiable $f$ sur $M$ la relation de Leibnitz :
\begin{equation}\label{leibi}
\nabla(fs)=s\otimes df+f\nabla s
\end{equation}
\end{definition}

\emph{L'exposé qui suit montre que l'apparition de la notion de connexion dans les concepts de la géométrie différentielle a été un progrès essentiel non seulement pour le développement des mathématiques mais également la compréhension (géométrique) des équations fondamentales de la physique classique et quantique.}

Pour la physique en particulier, il est important d'avoir une écriture dans des coordonnées locales ou plus généralement dans des référentiels locaux et d'avoir des formules de changements de cartes pour ces représentations locales.

\subsubsection{Dérivée covariante, écritures locales}
On se place sur un ouvert $U$ de $M$ sur lequel existe une base $\{s_{1},s_{2},\cdots,s_{p}\}$ de sections de $\tau$ (c'est  à dire un repère local) et une base $\{\alpha^{1},\alpha^{2},\cdots,\alpha^{n}\}$  de sections de $\tau^{*}(M)$ si $M$ est une variété de dimension $p$. La famille $ \mathcal{B}=\{s_{i}\otimes \alpha^{j}\}_{i,j}$ est une base de sections de $\tau\otimes\tau^{*}(M)$ au dessus de $U$.  Pour toute section $s_{i}\in \mathcal{S}_\tau$ on a : $\nabla s_{i}=\sum_{k,j}\lambda_{ij}^{k}s_{k}\otimes \alpha^{{j}}$ que l'on écrit avec la convention d'Einstein et $\omega^{k}_{i}=\lambda_{ij}^{k}\alpha^{{j}}$ :
 \begin{equation}\label{loloo}
 \nabla s_{i}=s_{k}\otimes \omega^{k}_{i}
 \end{equation}
 \noindent  On pose $e=(s_{1},s_{2}, \cdots,s_{p})$ et $\omega=\left(\omega^{k}_{i}\right)$, matrice $p\times p$ où $k$ est l'indice de ligne et $i$ l'indice de colonne. L'équation (\ref{loloo}) devient \begin{equation}\label{loli}\nabla e=e\otimes \omega.\end{equation}
 La matrice $\omega$ s'appelle la \emph{matrice (locale) de connexion}.\index{matrice de connexion} Les formes  $\omega^{k}_{i}$ sont appelées les \emph{formes (locales) de connexion}. \index{formes de connexion}

 \noindent Exprimons les formes de connexion par leur composantes dans la base de $1$-formes  $A=\{\alpha^{1},\alpha^{2},\cdots,\alpha^{n}\}$. On a :  
  \begin{equation}\label{coefcon}
  \omega_{i}^{k}=\omega_{ji}^{k}\alpha^{{j}}.
  \end{equation} 
  Soit $\{\varepsilon_{1},\varepsilon_{2},\cdots,\varepsilon_{n}\}$  la base duale de $A$.  
 Alors l'équation (\ref{loloo}) aboutit à 
\begin{equation}\label{lolita}
\nabla s_{i}(\varepsilon_{j})=: \nabla_{\varepsilon_{j}}s_{i}=\omega_{ji}^{k}s_{k}
\end{equation}
On pense  $\nabla_{\varepsilon_{j}}s_{i}$ comme la \emph{dérivée (covariante) de la section $s_{i}$ suivant $\varepsilon_{j}$.}

\noindent La relation de Leibnitz (\ref{leibi}) appliquée à l'écriture locale
 $s=\lambda^{{i}}s_{i}$, donne :\begin{equation}\label{caroule}
\nabla s=s_{i}\otimes \nabla \lambda^{{i}} 
\end{equation}
où  \begin{equation}\label{dedecor}
\nabla \lambda^{{i}}=d \lambda^{{i}}+\lambda^{{k}}\omega^{{i}}_{k}.
 \end{equation}

\begin{definition}L'expression $\nabla \lambda^{{i}}$ est la \emph{dérivée covariante de $\lambda^{{i}}$ }et $\nabla s$ la \emph{dérivée covariante de} $s$.\end{definition}\index{dérivée covariante}
Transformons l'égalité précédente en une égalité matricielle. 

\noindent Posons $\Lambda={^{t}(\lambda^{1},\cdots,\lambda^{p})}$. Alors
\begin{equation}\label{lololo}
s=e\Lambda\;\;\mathrm{et}\;\;\nabla s=e\otimes (d\Lambda+\omega\Lambda)=e\otimes \nabla\Lambda.
\end{equation}
Soit $\{x^{1},x^{2},\cdots,x^{n}\}$ un système de coordonnées locales sur un ouvert $U$ de $M$. On peut écrire : 
\begin{equation}\label{der}
\nabla \lambda^{i}=\big(\mfrac{\partial \lambda^{i}}{\partial x^{j}}+\lambda^{k}\omega^{i}_{jk}\big)dx^{j}
\end{equation} 
Dans la littérature on trouve les différentes notations suivantes :
\begin{equation}\label{nonotes}
\frac{\partial \lambda^{i}}{\partial x^{j}}+\lambda^{k}\omega^{i}_{jk}=:\lambda^{i}_{/j}=:\nabla_{j}\lambda^{i}=:\frac{\nabla \lambda^{i}}{\partial x^j}
\end{equation}
Alors
\begin{equation}\label{dersec}
\nabla s=\big(\nabla_{/j}\lambda^{i}\big)s_{i}\otimes dx^{j}
\end{equation}

\subsubsection{Un cas particulier particulièrement utile : $\tau=\tau(M)$.}
Dans ce cas  $s$ est un champ de vecteurs.

 \noindent\textbf{\'{E}criture locale d'une connexion sur $\tau(M)$.}
 
 Avec la notation habituelle notons  $X$ un champ de vecteurs et $\{s_{i}\}_{i}=\left\{\dfrac{\partial}{\partial x^{i}}\right\}_{i}$ une base locale de champs sur un domaine de carte muni de coordonnées $(x^{i})_{i} : X=X^{{i}}\dfrac{\partial}{\partial x^{i}}$.
 Localement si $X$ est un champ sur $U$, 
 \begin{equation}\label{meca}
 \nabla X=\frac{\partial}{\partial x^{{i}}}\otimes \left(\frac{\partial X^{{i}}}{\partial x^{{k}}}+ \omega^{i}_{kj}X^{j}\right)dx^{k}=
 \frac{\partial}{\partial x^{{i}}}\otimes  X^{i}_{/k}dx^{k}.
 \end{equation} 
 L'équation (\ref{meca}) montre $\nabla X(x)$ ne dépend que des valeurs des $X^{i}$ et de ses dérivées partielles en $x$.
 Si on note  $\nabla_{{\partial}_{k}}X=X^{i}_{/k}\;\frac{\partial}{\partial x^{{i}}},$ alors l'équation (\ref{meca}) se réécrit sous la forme : \[\nabla X=\bigl(\nabla_{{\partial} _{k} }X\bigr)\otimes dx^{k}.\]
 En particulier \begin{equation}\label{kronek}\nabla \frac{\partial}{\partial x^{i}}=\omega^{l}_{ki} \frac{\partial}{\partial x^{l}}\otimes dx^{k}\;\mathrm{ et}\; \nabla_{ \frac{\partial}{\partial x^{j}}} \frac{\partial}{\partial x^{i}}
 =\nabla \frac{\partial}{\partial x^{i}}\left( \frac{\partial}{\partial x^{j}}\right)=
\omega^{l}_{ji} \frac{\partial}{\partial x^{l}}.\end{equation}
Soit $Y=Y^{i}\frac{\partial}{\partial x^{i}}$ un \emph{vecteur} de $T_{x}M$. L'équation (\ref{meca}) donne  la dérivée covariance de $X$ dans la direction $Y$ au point $x$:
\begin{equation}\label{esse}
\nabla_{Y}X=Y^{{j}}X^{{i}}_{/j}\frac{\partial}{\partial x^{i}}
\end{equation}
L'équation (\ref{esse}) montre en particulier que $\nabla_{Y}X$ est un champ de vecteurs si $Y$ est un champ de vecteurs.
\subsubsection{Produit tensoriel de deux connexions}\label{protenscon}\index{produit tensoriel de deux connexions}

On se donne deux fibrés $\tau=(E,p,M)$ et $\tau'=(E',p',M)$ sur la même variété différentiable $M$. On dispose sur chaque fibré d'une connexion : $\nabla$ connexion sur $\tau$ et $\nabla'$ connexion sur $\tau'$. 
\begin{proposition}\label{totau}
 L'application vérifiant
 \begin{equation}\label{protensconne}
 \begin{array}{cccc}
\nabla'' :&\ma{S}_{\tau\otimes \tau'}&\longrightarrow&\ma{S}_{\tau\otimes \tau'\otimes \tau^{{*}}(M)}\\
& s\otimes s'&\to&(\nabla s)\otimes  s'+s\otimes \nabla s'\\
\end{array}
\end{equation}
prolongée par additivité sur $\ma{S}_{\tau\otimes \tau'}$, définit une connexion sur $\tau\otimes \tau'$
\end{proposition}
\begin{preuve}{}
Les notations sont celles de la définition \ref{fomvalvec}.

\noindent La seule propriété qu'il vaut la peine d'établir est la relation de Leibnitz. Soit $f$ une fonction différentiable sur $M$. Pour deux sections $s\in \ma{S}$ et $s'\in \ma{S}'$, on a :

\noindent$\nabla''\big(f(s\otimes s')\big)=\nabla(fs)\otimes s'+(fs)\otimes\nabla'(s')
=(s\otimes df+f\nabla s)\otimes s'+ fs\otimes \nabla' s'=$

\noindent$(s\otimes s')\otimes df+f(\nabla s\otimes s'+s\otimes \nabla' s')$ car pour tout champ de vecteurs $X$ sur $M$
 on a $(s\otimes df\otimes s')(X)=(s\otimes s'\otimes df) (X)
=df(X) s\otimes s'$ (ce qui identifie $\ma{S}_{\tau\otimes \tau'\otimes \tau^{{*}}(M)}$ et $\ma{S}_{\tau\otimes  \tau^{{*}}(M)\otimes \tau'}$).
 \end{preuve}
 \begin{definition}
 La connexion $\nabla''$ est le produit tensoriel des connexions $\nabla$ et $\nabla'$. On la note $\nabla\otimes \nabla'$.\index{produit tensoriel de deux connexions}
 \end{definition}
  \noindent Si l'on choisit sur $M$ des coordonnées locales, la connexion $\nabla''$ est déterminée conformément à l'équation (\ref{caroule}) par les dérivées covariantes des composantes des sections dans une base de sections locale de $\tau\otimes \tau'$. La proposition suivante détermine ces dérivées en fonctions des formes de connexion.
 \begin{proposition}\label{calculasse}
 On considère les données du début de la section \ref{protenscon}.
 On considère un ouvert $U$ d'une carte locale de $M$ sur lequel on dispose de coordonnées locales $\{x^{i}\}$. On suppose que l'on a deux cartes locales $(U,\Phi_{U})$ et $(U,\Phi'_{U})$ de $\tau$ et $\tau'$ sur lesquelles existent des bases de sections, respectivement $\{s_{i}\}$ et $\{s'_{j}\}$. On a donc : 
 $s=\sigma^{i}s_{i},s'=\sigma'^{j}s'_{j},s\otimes s'=\lambda^{ij}s_{i}\otimes s'_{j}$ avec $\lambda^{{ij}}=\sigma^{i}\sigma'^{j}$. 
 On note pour finir $\bigl(\omega^{i}_{j}\bigr)$ et 
 $\bigl(\omega'^{i}_{j}\bigr)$ les matrices de connexion de $\nabla$ et $\nabla'$ dans ces cartes et on utilise les notations des équations (\ref{lolita}) et (\ref{der}). Alors
 \begin{equation}\label{contessa}
 (\nabla\otimes \nabla')_{j}\lambda^{kl}=\mfrac{\partial\lambda^{{kl}}}{\partial x^{{j}}}+\omega^{k}_{js}\lambda^{sl}+\omega'^{l}_{js}\lambda^{ks}
 \end{equation}
  \end{proposition}
  \begin{preuve}{}
  D'après les égalités (\ref{caroule}) et (\ref{nonotes}), $(\nabla\otimes \nabla')(s\otimes s')=(s_{k}\otimes s'_{l})\otimes (\nabla\otimes \nabla')_{j}\lambda^{kl}dx^{{j}}$.
  Par ailleurs 
 $ \nabla s=s_{k}\otimes \nabla_{j}\sigma^{{k}}dx^{{j}}$ et
   $\nabla' s'=s'_{l}\otimes \nabla_{j}\sigma'^{{l}}dx^{{j}}.$
 On a donc en utilisant l'argument de la démonstration de la proposition \ref{totau}, les égalités $\nabla s\otimes s'= s_{k}\otimes s'_{l} \sigma'^{l}\otimes\nabla_{j}\sigma^{k} dx^{{j}} $
 et $ s\otimes \nabla's'= s_{k}\otimes s'_{l} \sigma^{k}\otimes\nabla'_{j}\sigma'^{l} dx^{{j}} .$ On en déduit 
 $ (\nabla\otimes \nabla')_{j}\lambda^{kl}=\sigma'^{l}\nabla_{j}\sigma^{k} 
+\sigma^{k}\nabla'_{j}\sigma'^{l}=
\sigma'^{l}\big(\mfrac{\partial \sigma^{k}}{\partial x^{j}}+\omega^{{k}}_{js}\sigma^{s}\big)+\sigma^{k}\big(\mfrac{\partial \sigma'^{l}}{\partial x^{j}}+\omega'^{{l}}_{js}\sigma'^{s}\big)=\mfrac{\partial\lambda^{{kl}}}{\partial x^{{j}}}+\omega^{k}_{js}\lambda^{sl}+\omega'^{l}_{js}\lambda^{ks}$                             
   \end{preuve}
   Cette proposition nous permettra d'écrire sous sa forme la plus générale l'équation de Schrödinger en présence d'un champ électromagnétique (voir sous-section \ref{eqschrochma}).
\begin{remark}\label{decor}

\noindent Si $\tau$ est un fibré vectoriel sur la variété différentiable $M$, muni d'une connexion $\nabla$ et $\{s_{i}\}_{i}$ une base locale de sections de $\tau$, toute section $s$ s'écrit localement $s=\lambda^{i}s_{i}$. Une base locale de $\tau\otimes \tau^*(M)$ s'écrit $\{s_{i}\otimes dx^j\}$. D'après les équations (\ref{der}) et (\ref{dersec}), on a 
$\nabla s=\lambda^{i}_{/j}s_{i}\otimes dx^j= (\nabla s)^{i}_{j }s_{i}\otimes dx^j$ : les scalaires $\lambda^{i}_{/j}$ apparaissent comme les composantes de $\nabla s$ dans la base locale 
 $\{s_{i}\otimes dx^j\}$ et $\nabla$ définit un opérateur $\lambda^{i}\mapsto \lambda^{i}_{/j}$ sur les composantes $\lambda^{i}$ de $s$. \emph{La proposition  \ref{calculasse} montre en fait que cet opérateur est une dérivation sur les fonctions.}
 
 \noindent Pour justifier ceci considérons deux fibrés vectoriels sur $M$, $\tau_{1}$ et $\tau_{2}$ muni respectivement des connexion $\nabla_{1},\nabla_{2}$. Notons encore $\{s^1_{i}\}_{i}$ et 
 $\{s^2_{j}\}_{j}$ deux bases locales de sections respectivement de $\tau_{1}$ et $\tau_{2}$. Si $s_{1}$ est une section de $\tau_{1}$, $s_{2}$ est une section de $\tau_{2}$, on peut écrire $s_{1}=\lambda^{i} s^1_{i}$ et $s_{2}=\mu^j s^2_{j}$. L'équation (\ref{contessa}) s'écrit directement sous la forme :
 \begin{equation}\label {derche}
 (\lambda^k\mu^l)_{/j}=\lambda^k_{/j}\mu^l+\lambda^k\mu^l_{/j}
\end{equation}
 où $\lambda^k_{/j}$ correspond à l'opérateur induit par $\nabla_{1}$ et $\mu^l_{/j}$ à l'opérateur induit par $\nabla_{2}$.
 En prenant $\tau_{1}=\tau_{2}=\tau$ et $\nabla_{1}=\nabla_{2}=\nabla$, on voit que $\nabla$ induit une dérivation sur les composantes des sections de $\tau$.
\end{remark}
\subsubsection{Changement de cartes locales}
Les composantes d'un tenseur dans une carte locale changent selon des formules bien connues lors d'un changement de carte. Inversement une famille de nombre lié à un système de coordonnées locales qui se transforment selon ces mêmes formules sont les composantes d'un tenseur. Ces formules constituent donc un \emph{critère de tensorialité} (voir voir par exemple [\ref{lichné}] ou [\ref{rinrin}]). Nous allons, de la même fa\c con déterminer un ''critère de  connexivité'' en établissant les formules de changement des matrices de connexion. 

\noindent Considérons sur $\tau$ deux cartes locales $(U,\Phi_{U})$,
$(V,\Phi_{V})$ de sorte que $U\cap V\not=\emptyset$. 
Notons $\{s_{1,U},s_{2,U},\cdots,s_{n,U}\}$ une base de sections sur $U$ et $\{s_{1,V},s_{2,V},\cdots,s_{n,V}\}$ une base de sections sur $V$. Avec la notation matri\-cielle
$e_{U}=(s_{1,U},s_{2,U},\cdots,s_{n,U})$ et $e_{V}=(s_{1,V},s_{2,V},\cdots,s_{n,V})$, l'égalité (\ref{loli}) nous assure que $\nabla e_{U}=e_{U}\otimes \omega_{U}$ et 
$\nabla e_{V}=e_{V}\otimes \omega_{V}$. 

\noindent Soit $s$ une section de $\tau$.
Pour tout   $x\in U\cap V, s(x)\in p^{-1}(\{x\})$; on pose :  $\Phi_{U}(s(x))=\big(x,(\lambda^{1}_{U},\cdots,\lambda^{n}_{U})\big)=(x,^{t}\Lambda_{U})$ où $\Lambda_{U}=^{t}(\lambda^{1}_{U},\cdots,\lambda^{n}_{U}).$  Avec les mêmes notations, on a : $\Phi_{V}(s(x))=(x,^{t}\Lambda_{V})$. Le changement de coordonnées s'écrit d'après  l'équation (\ref{chacord}) : $\Lambda_{U}=C_{UV}\Lambda_{V}$; ce qui nous donne également la relation entre les deux bases : $e_{V}=e_{U}C_{UV}$ (équation (\ref{superchapo})). On peut en utilisant la propriété de Leibnitz et l'égalité (\ref{loli}), écrire :
 
 \noindent$\nabla (e_{V})=\nabla (e_{U}c_{UV})=\nabla e_{U}C_{UV}+e_{U}\otimes dc_{UV}=e_{U}\otimes (\omega_{U}C_{UV}+dc_{UV})=e_{V}\otimes \omega_{V}=e_{U}\otimes C_{UV}\;\omega_{V}$, d'où la formule de changement de cartes :
 \begin{equation}\label{chacal}
 \omega_{V}=C_{VU}\omega_{U}C_{UV}+C_{VU}dC_{UV}
 \end{equation}\index{formule de changement de cartes pour une matrice de connexion}
 
\noindent Inversement, supposons l'égalité (\ref{chacal})  vérifiée et considérons dans un domaine de carte $U$ la section  $s=e_{U}\Lambda_{U}$. Définissons sur $U$ l'opérateur $\nabla^{{U}}$ par 
 
 \begin{equation}\label{loca}\nabla^{{U}} s=e_{U}\otimes (d\Lambda_{U}+\omega_{U}\Lambda_{U}).
  \end{equation} 
 Alors si dans un domaine de carte $V$ non disjoint de $U$, on a $s=e_{V}\Lambda_{V}$, nous avons les égalités
 \begin{align}
 \nabla^{{U}}(fs)&=f \nabla^{{U}}s+s\otimes df\label{sucre}\\
 e_{U}\otimes (d\Lambda_{U}+\omega_{U}\Lambda_{U})&=e_{V}\otimes (d\Lambda_{V}+\omega_{V}\Lambda_{V})\label{sel}
 \end{align}
 \begin{exo}\label{susucre}
 Démontrer les égalités (\ref{sucre}) et (\ref{sel}).
 \end{exo}
  L'égalité (\ref{loca}) définit donc de manière intrinsèque une con\-nexion.
  Une autre conséquence intéressante concernant l'expression  locale (\ref{loca}) est la suivante : les matrices colonnes $\nabla \Lambda_{U}=d\Lambda_{U}+\omega_{U}\Lambda_{U}$ se transforment comme les composantes d'un vecteur :
 \begin{equation}\label{chacaux}
 \nabla \Lambda_{U}=C_{UV} \nabla \Lambda_{V}
 \end{equation} 
 En effet, d'après l'équation (\ref{dedecor}), on a :
 
  \noindent $ \nabla \Lambda_{U}=\nabla (C_{UV}\Lambda_{V})=d(C_{UV}\Lambda_{V})+\overbrace{(C_{UV}\omega_{V}C_{VU}+c_{UV}dC_{VU})}^{\omega_{U}}C_{UV}\Lambda_{V}=
 C_{UV}(d\Lambda_{V}+\omega_{V}\Lambda_{V})+(dC_{UV}+C_{UV}dC_{VU}C_{UV})\Lambda_{V}= C_{UV}(d\Lambda_{V}+\omega_{V}\Lambda_{V})=C_{UV}\nabla \Lambda_{V},$
 
 \noindent car $dC_{UV}+C_{UV}dC_{VU}C_{UV}=0$ du fait que $d(C_{UV}C_{VU})=d\I=0$.
 
 \begin{proposition}\label{jojo}
Les équations (\ref{chacal}) et (\ref{chacaux}) sont équivalentes.
\end{proposition}
Nous avons déjà vu que (\ref{chacal}) entraine (\ref{chacaux}).
\begin{exo}\label{pimpim}
Démontrer que l'équation (\ref{chacaux}) entraine l'équation (\ref{chacal})
\end{exo}

\begin{remark}\label{pressi}

\noindent Pour être plus précis dans l'équation (\ref{chacaux}), le terme de gauche  désigne un opérateur $\nabla$ qui agit sur les coordonnées de $s(x)$ dans la carte $(U,\Phi_{U})$. Il faudrait le distinguer du $\nabla$ qui appara\^{i}t  dans le terme de droite, et écrire l'équation (\ref{chacaux}) sous la forme $\nabla^{{U}}\Lambda_{U}=C_{UV} \nabla^{{V}} \Lambda_{V}.$
 Mais alors \begin{gather}\nabla^{{U}}C_{UV} \Lambda_{V}=C_{UV} \nabla^{{V}} \Lambda_{V}\label{chiote}\\
 \intertext{et l'équation (\ref{chacaux}) est équivalente à l'équation }
 \nabla^{{U}}C_{UV}=C_{UV} \nabla^{{V}}.\label{chachacha}
\end{gather} 
\emph{L'équation (\ref{chachacha}) équivaut donc au caractère intrinsèque de la dérivation covariance donnée par l'égalité (\ref{loca})}.
\end{remark}
L'équation (\ref{chacal}) a une conséquence remarquable décrite dans la proposition suivante.
\begin{proposition}\label{antisim}
On considère un fibré $\tau$ défini par des cocycles à valeurs dans $O(n)$ et $\nabla$ une connexion sur $\tau$. Si la matrice de connexion de $\nabla$ dans une carte locale est antisymétrique, alors elle est antisymétrique dans toute carte du fibré.
\end{proposition}
\begin{preuve}{}
On considère deux cartes locales $(U,\Phi_{U})$ et $(V,\Phi_{V})$ d'intersection non vide, et on note $\omega_{U},\omega_{V}$ les matrices de connexion dans ces cartes.
On suppose que $\omega_{U}$ est antisymétrique. On a $ \omega_{V}=C_{VU}\omega_{U}C_{UV}+C_{VU}dC_{UV}$. 

D'où :
$^{{t}} \omega_{V}=C_{VU}\;\;^{{t}}\omega_{U}\;\;C_{UV}-C_{VU}dC_{UV}$, puisque $^{t}C_{UV}=C_{VU}$ et $dC_{VU}\;C_{UV}=-C_{VU}dC_{UV}$. Ainsi, 
$^{{t}} \omega_{V}=-C_{VU}\omega_{U}C_{UV}-C_{VU}dC_{UV}=-\omega_{V}.$
\end{preuve}
Une autre conséquence surprenante de la relation (\ref{chacal}) est énoncée dans la proposition suivante.
\begin{proposition}\label{globalfocou}
Considérons le fibré tangent $\tau(M)$ d'une variété différentiable $M$, équipé d'une connexion. Sur toute carte locale $(U,\Phi)$, on peut définir une matrice de connexion $\omega_U$ et donc une matrice de $2$-formes \; $\theta_U=d\omega_U+\omega_U\wedge\omega_U$. Il existe sur $\tau(M)$ une forme (matricielle) $\theta$ sur $M$ dont la restriction à $U$ est $\theta_U$. 
\end{proposition}
\begin{preuve}{}
Soit $(V,\Phi_V)$ une carte locale telle que $U\cap V\not=\emptyset$. On a : 

\noindent$\theta_V=d\omega_V+\omega_V\wedge\omega_V=d(C_{VU}\omega_{U}C_{UV}+C_{VU}dC_{UV})+(C_{VU}\omega_{U}C_{UV}+C_{VU}dC_{UV})\wedge(C_{VU}\omega_{U}C_{UV}+C_{VU}dC_{UV}).$ Mais d'une part:

\noindent$d(C_{VU}\omega_{U}C_{UV}+C_{VU}dC_{UV})=\left(dC_{VU}\wedge\omega_UC_{UV}+C_{VU}d\omega_UC_{UV}+C_{VU}\omega_U\wedge dC_{UV}\right.$
\noindent $\left.+dC_{VU}\wedge dC_{UV}\right)=-C_{VU}dC_{UV}C_{VU}\wedge\omega_UC_{UV}+C_{VU}d\omega_UC_{UV}+C_{VU}\omega_U\wedge dC_{UV}-C_{VU}dC_{UV}C_{VU}\wedge dC_{UV}$

\noindent et d'autre part : $(C_{VU}\omega_{U}C_{UV}+C_{VU}dC_{UV})\wedge(C_{VU}\omega_{U}C_{UV}+C_{VU}dC_{UV})=..$

\noindent $
C_{VU}\omega_UC_{UV}\wedge C_{VU}\omega_UC_{UV}+C_{VU}\omega_UC_{UV}\wedge C_{VU}\omega_UdC_{UV}+C_{VU}(\omega_U\wedge\omega_U )C_{UV}+C_{VU}dC_{UV}\wedge C_{VU}dC_{UV}=
C_{VU}(\omega_U\wedge \omega_U) C_{UV}+..$

\noindent$..+C_{VU}\omega_U\wedge C_{UV}+C_{VU}dC_{UV}\wedge C_{VU}\omega_UC_{UV}+C_{VU}dC_{UV}\wedge C_{VU}dC_{UV}$.
En additionnant les deux termes on obtient:

 \noindent $\theta_V=C_{VU}\left( d\omega_U+\omega_U\wedge\omega_U\right)C_{UV}=C_{VU}\theta_UC_{UV}$, ce qui montre l'existence d'une forme globale sur $M$.
\end{preuve}
Pour des raisons qui apparaîtront dans la section \ref{mollusk}, $\theta$ est appelée \emph{matrice des formes de de courbure}.\index{forme de courbure}
 \subsection{Exemples de connexions}
 \subsubsection{La connexion préhistorique}\label{prehis}
 Il s'agit de l'opérateur de dérivation des fonctions différentiables sur une variété. Il a été détaillé comme motivation de la définition \ref{connex} p.\pageref{connex}.
 Pour cette connexion l'équation locale (\ref{dersec}) s'écrit
  \begin{equation}\label{croma}
  df=f_{/i}dx^{{i}} \;\;\mathrm{où}\;\; f_{/i}=\mfrac{\partial f}{\partial x^{{i}}}.
  \end{equation}
 La connexion préhistorique est donc la connexion sur le fibré trivial $(M\times \rr,p_{1},M)$ pour laquelle les formes de connexions dans tout système de coordonnées  est nulle.
 \begin{exo}\label{suichac} (cohérence avec l'équation (\ref{chacal}))
 
 Vérifier que si les formes de connexion sont nulles dans un système de coordonnées pour la connexion préhistorique, elles le sont dans toutes, en se référant à l'équation (\ref{chacal}).

 \end{exo}
\subsubsection{La connexion de Levi-Civita sur $\tau(M)$}\index{connexion de Levi-Civita sur le fibré tangent}
On suppose que $M$ est de dimension $n$ et
soit $\{e_{1},e_{2},\cdots,e_{n}\}$ une base de champs de vecteurs sur un ouvert de $M$. D'après l'équation (\ref{caroule}), pour tout champ de vecteurs sur $M$, on a localement une écriture de la forme
$\nabla X=e_{i}\otimes \nabla X^{i}$ si $X=X^{{i}}e_{i}$, avec $\nabla X^{{i}}=dX^{{i}}+\omega^{{i}}_{j}X^{{j}}$. Ainsi pour tout  vecteur $Y$ en un point $b$ de cet ouvert , $\nabla X(Y)=\nabla X^{{i}}(Y) e_{i}=(L_{Y}X^{i}+X^{{j}}\omega^{{i}}_{j}(Y))e_{i}$. On note habituellement en géométrie $\nabla_{Y}X=\nabla X (Y)$. Le vecteur $\nabla_{Y}X$ est appelé \emph{ la dérivée covariance de $X$ dans la direction $Y$ au point $b$}.
 \begin{definition}\label{sysy}\index{dérivée covariante d'un champ de vecteurs dans la direction d'un vecteur}
 Une connexion $\nabla$ sur $\tau(M)$ est symétrique si $\nabla_{X}Y-\nabla_{Y}X=[X,Y].$
\end{definition}\index{connexion symétrique sur le fibré tangent}
 
On munit maintenant $M$ d'une structure riemannienne $\la\;\ra.$ \emph{Tullio Levi-Civita a montré qu'il existe une unique connexion symétrique sur 
$\tau(M)$ telle que localement :}
\begin{equation}\label{leci}
d\la e_{i},e_{j}\ra(e_{k})=\la \nabla_{e_{k}}e_{i},e_{j}\ra+\la e_{i},\nabla_{e_{k}}e_{j}\ra.
\end{equation}
Cette connexion est \emph{la connexion de Levi-Civita.} Ce résultat de Levi-Civita est la justification des définitions adoptées pour une connexion puisque cet objet existe naturellement dans le cadre des variétés riemanniennes avec la propriété supplémentaire de compatibilité avec la métrique évoquée par l'égalité (\ref{leci}). On trouvera la preuve de l'existence dans la démonstration du théorème \ref{levita} à l'appendice $4$. L'unicité découle de ce qui suit.
Prenons le cas où $U$ est le domaine d'une carte locale de $M$, munie de coordonnées locales $(x_{1},x_{2},\cdots,x_{n})$ et où la base locale est
 $\left\{\frac{\partial}{\partial x^{{i}}}\right\}_{i}$. Les formes de connexion de la connexion de Levi-Civita s'écrivent $\omega^{{i}}_{j}=\Gamma^{i}_{kj} dx^{k}$ où les coefficients $\Gamma^{{i}}_{kj}$ sont les \emph{coefficients de Christoffel}. On note $\Gamma$ la matrice de connexion correspondante. 
 
  \noindent\textbf{\'{E}criture locale}
 
De la définition \ref{sysy} il découle immédiatement que
 la propriété de symétrie de la connexion de Levi-Civita équivaut à $\Gamma^{l}_{ij}=\Gamma^{l}_{ji}$ pour tout $(i,j)$.

\noindent L'équation (\ref{leci}) s'écrit dans des coordonnées locales si on note $g_{ij}=\la \frac{\partial}{\partial x^{{i}}},\frac{\partial}{\partial x^{{j}}}\ra : $ 
\begin{equation}\label{christo}
\frac{\partial g_{ij}}{\partial x^{k}}=\Gamma^{s}_{ki} g_{sj}+\Gamma^{s}_{kj} g_{is}.
\end{equation}
En résolvant ce système, on obtient  une expression des coefficients de Christoffel en fonction de la métrique :
\begin{equation}\label{cristo}
\Gamma^{\tau}_{\alpha\beta}=\frac{1}{2}   g^{\tau\mu}\left(
\frac{\partial g_{\alpha\mu}}{\partial x^{\beta}}+\frac{\partial g_{\mu\beta}}{\partial x^{\alpha}}-\frac{\partial g_{\alpha\beta}}{\partial x^{\mu}}
\right).
\end{equation}
L'équation (\ref{cristo}) montre l'unicité de la connexion de Levi-Civita.
 \begin{remark}\label{pseudo}
 L'existence d'une connexion symétrique vérifiant l'équation (\ref{leci}) est également vraie sur une variété pseudo-riemanienne, c'est à dire lorsque la métrique en chaque point de l'espace tangent est une forme bilinéaire symétrique non dégénérée (i.e. $\la X,Y\ra=0$ pour tout $Y$ entraine $X=0$), comme c'est le cas de l'espace de Minkovski en relativité restreinte ou de l'espace temps en relativité générale.
L'égalité (\ref{cristo}) est bien sûr vérifiée dans le cadre d'une variété pseudo-riemannienne.
\end{remark}
Dans une variété riemannienne,
on a une autre remarque  sur la symétrie quand on se place localement dans un repère orthonormé $\mathcal{R}=\{e_{i}\}$.
On note $\{\sigma^{i}\}$ le repère dual de $\mathcal{R}$. Ceci permet d'écrire $\omega^{{k}}_{i}=\Gamma^{{k}}_{li}\sigma^{{l}}$.
 Alors l'équation (\ref{leci}) dit que pour tout $i,j,k$, on la la relation de symétrie
 $\Gamma^{j}_{ki}=-\Gamma^{i}_{kj}$ et donc \[\omega^{j}_{i}=-\omega^{i}_{j},\]
égalité dont la version covariante  est
\[\omega_{ij}=-\omega_{ji}\]
où $\omega_{ij}=g_{ik}\omega^{{k}}_{j}$.
Résumons ceci dans la proposition \ref{kelestbel}.
\begin{proposition}\label{kelestbel}En repère orthonormé la matrice des formes de connexion de Levi-Civita est antisymétrique. Elle est donc antisymétrique dans tout repère si les cocyles sont à valeurs dans $O(n)$ selon la proposition \ref{antisim}.
\end{proposition}
\noindent \textbf{Un cas particulier important :
La connexion de Levi-Civita de $\tau(\rr^n)$.}

\noindent On considère $\rr^n$ muni de sa structure euclidienne standard et son fibré tangent $\tau(\rr^n)$. L'application $L :\mathcal{S}_{\tau(\rr^n)}\mapsto \mathcal{S}_{\tau(\rr^n)\otimes \tau^*(\rr^n)}$ définie par 
\[\forall Y\in \mathcal{S}_{\tau(\rr^n)}\; \; LY=\frac{\partial Y^{i}}{\partial x^{j}}\frac{\partial}{\partial x^{i}}\otimes dx^j\]
si $Y=Y^j\frac{\partial}{\partial x^j}$ vérifie pour tout 
 $X=X^j\frac{\partial}{\partial x^j}$ : \[LY(X)=\frac{\partial Y^{i}}{\partial x^{j}}X^j\frac{\partial}{\partial x^{i}}=L_XY.\]

Les deux premiers items de l'exercice \ref{kon} montrent que $L$ est une connexion sur $\tau(\rr^n)$, les deux derniers items de ce même exercice montrent qu'il s'agit de la connexion de Levi-Civita. Les coefficients de Christoffel associés sont nuls.
 \begin{exo}\label{conlevcicris}
 \begin{enumerate}
 \item Soit $A$ une matrice carrée réelle \;$\rr$-diagonalisable et inversible dont les termes sont des fonctions dérivables d'une variable $x$. La matrice $A'(x)$ est la matrice dont les termes sont les dérivées des termes de $A$. Montrer que $\tr(A'A^{-1})=(\ln\abs{\det A})'$.
 
\item \textbf{Application} : Si $G$ est la matrice qui représente le tenseur métrique d'une variété pseudo-riemannienne dans une carte locale munie de coordonnées $x^1,\ldots,x^n$, montrer que :
 $$\frac{\partial}{\partial x^{i}}(\ln\abs{\det G})=\tr\left(\frac{\partial G}{\partial x^j}G^{-1}\right).$$

\item Déduire  de l'égalité (\ref{cristo}) que dans une variété pseudo-riemanienne les coefficients de Christoffel vérifient : \begin{equation}\label{goodform}\Gamma^{i}_{ij}=\frac{\partial\ln \sqrt{\abs{\det G}}}{\partial x^j}.\end{equation}
\end{enumerate}
 \end{exo}
 \subsubsection{Une application à la géométrie}
  Soit $S$ une sous-variété d'une variété $M$.
 On considère $X$ un champ de vecteurs tangent à $S$ et $Y$ un champ de vecteurs le long de $S$, non nécessairement tangent à $S$ comme $X$. On suppose que $\tau(M)$ est muni d'une connexion $\nabla$. On peut prolonger le champ$Y$ en un champ défini sur un  ouvert de $M$, constituant un voisinage de $S$. Ce prolongement n'est a priori pas unique. Cependant, ce prolongement permet de définir $\nabla_XY$, au moins localement sur $S$. C'est l'objet du lemme suivant.
\begin{lemma}\label{taco}
Soient $Y_1$ et $Y_2$ deux champs de vecteurs de $\tau(M)$ définis sur un voisinage  ouvert $U$ de $S$ qui prolongent $Y$ sur ce voisinage. Alors $\nabla_XY_1=\nabla_XY_2 $ sur $U$.  Cette égalité définit le champ
$\nabla_XY\in \tau(M)$ sur l'ouvert $U$.
\end{lemma}
\begin{preuve}{}
Soit $x\in S$. Plaçons-nous sur un voisinage $V$ ouvert de $x$ dans $S$, inclus dans  le domaine de carte $U$ muni de coordonnées locales 
$(x^1,\ldots,x^{n-1},x^n)$ sur lequel existe un chemin différentiable $\gamma$ de coordonnées $(\gamma^1,\ldots,\gamma^n)$ sur $U$,  tel que les égalités  :  $X=\frac{d\gamma}{dt}$ et $x=\gamma(t)$ aient lieu sur $V$.

Ainsi on peut écrire :

\noindent $\nabla_XY_1(x)=X^\al(x)\nabla_\al(Y_1^{i}\frac{\partial}{\partial x^{i}}) =X^\al(x)
\left(\frac{\partial Y_1^{i}}{\partial x^\al}\frac{\partial}{\partial x^{i}}
+\omega^{i}_{\al k}Y_1^{k}\frac{\partial}{\partial x^{i}}\right)=$

$=\left(\frac{d(Y_1^{i}\circ\gamma)}{dt}+\frac{d\gamma^\al}{dt}Y_1^{i}(x)\omega^{i}_{\al k}\right)\frac{\partial}{\partial x^{i}}=
\left(\frac{d(Y_2^{i}\circ\gamma)}{dt}+\frac{d\gamma^\al}{dt}Y_2^{i}(x)\omega^{i}_{\al k}\right)\frac{\partial}{\partial x^{i}}
=\nabla_XY_2.$ 
 \end{preuve}
  Nous allons utiliser ce lemme lorsque $S$ est une courbe paramétrée de $M$.
  
 \noindent Supposons que $X$ soit un champ de vecteurs le long d'un chemin différentiable $t\to \gamma(t)=(x^1(t),\cdots,x^{n}(t))$ en coordonnées locales. Les composantes de $X$ sont alors des fonctions de $t$ et on définit à partir de l'équation (\ref{meca}) la dérivée covariante de $X$ par rapport à $t$ : on a un champ de vecteurs naturel le long de ce chemin, le vecteur vitesse $v(t)=\mfrac{d\gamma}{dt}$.
  \begin{equation}\label{decoco}\mfrac{\nabla X}{dt}=:\nabla X(v)=X^{i}_{/k}\mfrac{dx^{k}}{dt}\mfrac{\partial}{\partial x^{{i}}}.\end{equation} En développant le terme de droite on a aussitôt :
  \begin{equation}\label{devcov}
  \mfrac{\nabla X}{dt}=\mfrac{\nabla X^{i}}{dt}\mfrac{\partial}{\partial x^{i}} \;\mathrm{où}\;\mfrac{\nabla X^{i}}{dt}=\mfrac{d X^{i}}{dt}+\Gamma^{i}_{kj}X^{j}\mfrac{dx^{k}}{dt}.
  \end{equation}
  On vérifie qu'on a une relation de Leibnitz sur les champs de vecteurs le long d'une courbe. Si $f$ est une fonction différentiable sur $M$ on a immédiatement 
  \[\mfrac{\nabla fX}{dt}=\mfrac{df}{dt}X+f\mfrac{\nabla X}{dt}.\]
  Soit $\gamma :[0,1]\mapsto M$ un chemin différentiable sur $M$ et $X$ un champ de vecteurs sur 
la variété riemannienne $(M,\la\;,\;\ra)$ de dimension $n$.
On dit que le champ de vecteurs $X$ est \emph{(auto-)parallèle }sur $\gamma$ si $\frac{\nabla X}{dt}=0$, où $\nabla$ est la connexion de Levi-Civita sur $\tau(M)$.\index{champ de vecteurs autoparallèle}

\noindent Traduction en coordonnées locales : Soit $(x^1,\ldots,x^n)$ un système de coordonnées locales sur un voisinage de $\gamma(0)$. 

\noindent \'{E}crivons $X=X^{i}\frac{\partial}{\partial x^{i}}, \gamma(t)=(\gamma^1(t),\ldots,\gamma^n(t)),
v^k=\frac{d\gamma^k}{dt},\left(X(\gamma(t))\right)^j=X^j(t)$. La condition de parallélisme se traduit par les $n$ équations suivantes :
\begin{equation}\label{paral}
\frac{dX^{i}}{dt}+\Gamma^{i}_{kj}v^kX^j=0, \;i=1,2,\ldots,n.
\end{equation}

\noindent Le système linéaire (\ref{paral}) admet une unique solution \ $t\mapsto X(t)$ pour une condition initiale $X(0)$ donnée définie pour tout $t\in[0,1]$. On peut en effet recouvrir $\img\gamma$ par un nombre fini de domaines de cartes locales.
\begin{definition}\label{accel}\index{géodésique}
Soit $\gamma$ une courbe différentiable de champ des vitesses $v=\frac{d\gamma}{dt}$ et $X$ un champ de vecteurs défini sur un voisinage ouvert de l'image de $ \gamma$ tel que $X\circ \gamma =v$.
   Si $X$ est  parallèle sur $\gamma$, on dit $\gamma$ est une courbe est une géodésique.  \end{definition}
  Examinons une conséquence \og naturelle\fg\; de la définition.  Supposons que $\gamma$ soit une géodésique.
    Alors,  $$\frac{d\la X,X\ra}{dt}=\frac{\partial \la X,X\ra}{\partial x^{{i}}}\frac{dx^{{i}}}{dt}=d\la X,X\ra (X)=2\la \nabla_{X}X,X\ra=2\la \frac{\nabla X}{dt},X\ra=0.$$ Ceci montre que 
 $ \Vert\frac{d\gamma}{dt}\Vert$ est constant. Si on considère une autre paramétrisation  de la géodésique $\gamma$,  de paramètre $u$,  on a :
 $\la \frac{d\gamma}{dt}, \frac{d\gamma}{dt}\ra=(\frac{du}{dt})^{2}\la \frac{d\gamma}{du}, \frac{d\gamma}{du}\ra.$ D'où $u=\alpha t+\beta$. Le changement de paramétrisations dans une géodésique est une fonction affine, comme on le voit en première année pour le changement de paramètres dans la représentation paramétrique d'une droite de $\rr^{n}$.
 En particulier l'abscisse curviligne $s$ d'une géodésique paramétrée par $t$ nulle pour  $t=0$, est égale à la paramétrisation  $s=t\Vert \frac{d\gamma}{dt}\Vert$. C'est donc le paramètre admissible pour lequel $\Vert \frac{d\gamma}{ds}\Vert=1$.
 \begin{remark}\label{derien}
 Cette remarque prolonge l'application géométrique précédente et donne une version de l'équation (\ref{derche}) pour des sections le long d'un chemin. On considère une section $\sigma$ d'un fibré vectoriel $\tau$ sur la variété différentiable $\tau$, muni d'une connexion $\nabla$, un chemin différentiable exprimé en coordonnées locales par $t\to \gamma(t)=(x^1(t),\cdots,x^{n}(t))$  et  de vecteur vitesse $v=\frac{d\gamma}{dt}$.
 Par définition $\frac{\nabla \sigma}{dt}=\nabla \sigma(v)$. Ainsi en supposant que $\{s_{i}\}$ soit une base locale de sections de $\tau$ et que\; $\sigma=\lambda^{i}s_{i} $, on a 
$\frac{\nabla \sigma}{dt}=\lambda^{i}_{/j}v^js_{i}= \frac{\nabla \lambda^{i}}{dt}s_{i}$ si on pose $ \frac{\nabla\lambda^{i}}{dt}=\lambda^{i}_{/j}v^j$. On en déduit alors comme dans la remarque \ref{decor} que si $\mu=\mu^js_{j}$ est une autre section de $\tau$, alors 
\begin{equation}\label{derichon}
\frac{\nabla(\lambda^{i}\mu^k)}{dt}=\lambda^{i}\frac{\nabla\mu^k}{dt}+\frac{\nabla\lambda^{i}}{dt}\mu^k.
\end{equation}
 \end{remark}
 Définissons maintenant ce qu'est le \emph{transport parallèle} ou {déplacement parallèle} d'un vecteur le long de la courbe $\gamma$.
\begin{definition}\label{trapar}\index{transport parallèle}\index{déplacement parallèle}
Soit $X_0$ un vecteur de $T_{\gamma(0)}M$ et $X$ l'unique solution de $\frac{\nabla X}{dt}=0$  
telle que $X(0)=X_0.$ Le vecteur $X(t)$ est le vecteur résultant du transport parallèle de $X_0$ le long de $\gamma $ du point $\gamma(0)$ au point $\gamma (t)$. On écrira : $X(t)=\tau_{ot}(X_0)$.
\end{definition}
Ce transport parallèle a des propriétés immédiates qui résultent de l'unicité des solutions de l'équation 
différentielle linéaire $\frac{\nabla X}{dt}=0$ . En particulier 
\begin{align*}
 &i)\forall X_0,Y_0\in T_{\gamma(0)}M, \forall \lambda\in \rr,\;\tau_{ot}(X_0+\lambda Y_0)=\tau_{ot}(X_0)+
\lambda\tau_{ot}(Y_0),\\
 &ii)\tau_{ot}\circ\tau_{t0}=I_{ T_{\gamma(0)}M,}.
\end{align*}

\noindent Si on considère alors une base orthogonale $(E_1,\ldots,E_n)$ de  $T_{\gamma(0)}M$, on a pour tout 
$(\lambda^1,\ldots,\lambda^n)\in \rr^n$ :
\[\tau_{0t}( \lambda^{i}E_i)=\lambda^{i}E_i(t)\]
où $E_i(t) =\tau_{0t}(E_i).$

\noindent Une autre conséquence de la linéarité du transport parallèle est que la différentiation covariante $\frac{\nabla}{dt}$ est comme la dérivée habituelle obtenue comme limite d'un taux d'accroissement. Plus précisément, soit $X(t)$ un champ de vecteurs le long de $\gamma$, \emph{qui n'est pas 
nécessairement parallèle}. Pour tout $t\in[0,1],\;\tau_{t0}(X(t))\in T_{\gamma(0)}M$ et de même $\dfrac{ \tau_{t0}(X(t))-X(0)}{t}\in T_{\gamma(0)}M$. Alors 
\begin{proposition}
\begin{equation}\label{sedov}\frac{\nabla X}{dt}(0)=\lim_{t\to 0}\dfrac{ \tau_{t0}(X(t))-X(0)}{t}.\end{equation}
\end{proposition}
\begin{preuve}{}
On a avec les notations qui précèdent $X(t)=\alpha^{i}(t)E_i(t)$. La linéarité du transport parallèle permet 
d'écrire $\dfrac{ \tau_{t0}(X(t))-X(0)}{t}=\dfrac{\alpha^{i}(t)-\alpha^{i}(0)}{t}E^{i}(0)$.
Ainsi \[\lim_{t\to 0}\dfrac{ \tau_{t0}(X(t))-X(0)}{t}={\alpha^{i}}'(0)E_{i}(0)=\dfrac{\nabla (\alpha^{i}E_i)}{dt}(0)=\frac{\nabla X}{dt}(0).\]
\end{preuve}
\begin{exo}\label{trescool}
Soit $t\in [0,1]\mapsto \gamma(t)$ une courbe différentiable, $X_0$ un vecteur de $T_{\gamma(0)}M$
et $X(t)\in T_{\gamma(t)}M$ obtenu par transport parallèle le long de $\gamma$.
On pose $\gamma^{-1}(t)=\gamma(1-t)$ appelé chemin inverse de $\gamma$. Montrer que le transport parallèle de $X(t)$ le long de $\gamma^{-1}$ en $\gamma(0)$ est le vecteur $X_0$.
\end{exo}
\begin{exemple}
Considérons le cas particulier $M=\rr^n$. On considère sur $\rr^n$ le champ de vecteurs différentiable $X$. Soit $a\in \rr^n$ et $v\in T_a\rr^n=\{a\}\times \rr^n$ et $\gamma(t)=tv+a$ la géodésique issue de $a$ avec la vitesse $v$. En considérant le transport parallèle habituel sur $\rr^n$ (c'est à dire en fait défini par la connexion de Levi-Civita de $\tau(\rr^n)$ pour laquelle les coefficients de Christoffel sont nuls), 
l'équation (\ref{sedov}) s'écrit 
\[\frac{\nabla X}{dt}(0)=\lim_{t\to 0}\frac{X(t)-X(0)}{t}=L_vX(a)=\nabla_vX(a).\] 
Le programme annoncé dans l'introduction du chapitre \ref{cacudeux} est réalisé.
\end{exemple}
  \subsubsection{La connexion de Levi-Civita sur $\tau^{*}(M)$}\index{connexion de Levi-Civita sur le fibré cotangent}
  \begin{proposition}
 $M$ est toujours une variété différentiable. Si $\nabla$ est une connexion sur $\tau(M)$, associée dans un domaine de carte locale à une matrice de connexion $\omega$, il existe sur 
 $\tau^{*}(M)$ une connexion associée dans cette carte à la matrice $-\omega$. Elle définit la connexion de Levi-Civita sur $\tau^{{*}}(M)$ si $\nabla$ définit la connexion de Levi-Civita sur $\tau(M)$.
  \end{proposition}
 \begin{preuve}{}
 Notons $\{x^{1},
x^{2},\cdots,x^{{n}}\}$ les coordonnées locales associées à  une carte locale $(U,\varphi)$ de $M$ dans laquelle la connexion $\nabla$ est exprimée par la matrice de connexion $\omega$.
 Si $\alpha$ est une $1$-forme sur $U$ et $X$ un champ de vecteurs sur $U$, on peut écrire localement $\alpha=\alpha_{i}dx^{i}$ et $X=X^{i}\frac{\partial}{\partial x^{i}}$. Mais alors 
 $\alpha(X)=\alpha_{i}X^{i}$ qui est un scalaire dont la dérivée partielle par rapport à $x^{j}$ s'écrit : 
  \[\frac{\partial\alpha(X)}{\partial x^{j}}=X^{{i}}\left(\frac{\partial\alpha_{i}}{\partial x^{j}}-\Gamma^{{s}}_{ji}\alpha_{s}\right)+\alpha_{i}\left(\frac{\partial X^{{i}}}{\partial x^{j}}+\Gamma^{{i}}_{jk}X^{k}.\right)\]
 Or $X^{{i}}_{/j}=\frac{\partial X^{{i}}}{\partial x^{j}}+\Gamma^{{i}}_{jk}X^{k}$ sont les composantes d'un tenseur $1$ fois contravariant, donc $\alpha_{i}X^{{i}}_{/j}$ est un scalaire. Ainsi
 $X^{{i}}\left(\frac{\partial\alpha_{i}}{\partial x^{j}}-\Gamma^{{s}}_{ji}\alpha_{s}\right)$ est un scalaire. Le critère de tensorialité nous permet de conclure que \begin{equation}\label{civicot}{\alpha_{i}}_{/j}=\frac{\partial\alpha_{i}}{\partial x^{j}}-\Gamma^{{s}}_{ji}\alpha_{s}\end{equation} sont les composantes d'une $1$-forme. Posons $\nabla{\alpha}=dx^{i}\otimes {\alpha_{i}}_{/j}dx^{{j}}$ et $\tau^{*}(M)=(T^{*}(M),p,M)$. On définit ainsi une 
 $1$-forme à valeurs dans $T^{*}(M)$ c'est à dire une section du fibré vectoriel $\tau^{*}(M) \otimes \tau^{*}(M)$ et l'opérateur $\nabla$ vérifie la relation de Leibnitz : la vérification est directe. Ce qui achève la démonstration.
  \end{preuve}
\begin{remark}\label{rabio}
Pour une forme différentielle $\alpha$ le long d'une courbe différentiable $t\to (x^{{1}}(t),\cdots,x^{n}(t))$ on a comme dans le cas d'un champ de vecteurs, la notion de dérivée covariance par rapport à $t$ :
\begin{equation}\label{rose}
\mfrac{\nabla \alpha}{dt}=:\nabla \alpha(v)=\mfrac{\nabla \alpha_{i}}{dt}dx^{{i}} \;\mathrm{où}\;\mfrac{\nabla \alpha_{i}}{dt}=\mfrac{d \alpha_{i}}{dt}-\Gamma^{k}_{ji}\alpha_{k}\mfrac{dx^{j}}{dt}.
\end{equation}
La dérivée  exprimée par l'équation (\ref{rose}) servira à écrire une relation fondamentale covariante de la dynamique newtonienne dans la section \ref{colagene}.
\end{remark}
\subsubsection{Connexion de Levi-Civita sur $\tau^{r}_{s}(M)$}\label{tiptop}
Par définition $\tau^{r}_{s}(M)=\overbrace {\tau(M)\otimes\cdots\otimes\tau(M)}^{r fois}\otimes \overbrace {\tau(M^{*})\otimes\cdots\otimes\tau^{*}(M)}^{sfois}.$
Si $T$ est une section de $\tau^{r}_{s}(M)$ (on dit que $T$ est un \emph{ champ de tenseurs $r$ fois contrariant et $s$ fois covariant}), on a en coordonnées locales $$T=T^{i_{1}\cdots i_{r}}_{j_{1}\cdots j_{s}}\mfrac{\partial}{\partial  x^{i_{1}}}\otimes\cdots\otimes \mfrac{\partial}{\partial  x^{i_{r}}}\otimes dx^{j_{1}}\otimes\cdots\otimes dx^{j_{s}}.$$ On suppose $\tau(M)$ muni d'une connexion représentée localement par la matrice $\omega$. En généralisant  le procédé de l'exemple précédent, on obtient que
\begin{equation}\label{pim}
{T^{i_{1}\cdots i_{r}}_{j_{1}\cdots j_{s}}}_{/j}
=\mfrac{\partial T^{i_{1}\cdots i_{r}}_{j_{1}\cdots j_{s}}}{\partial x^{j}}+\Gamma^{i_{1}}_{j\alpha}T^{\alpha i_{2}\cdots i_{r}}_{j_{1}\cdots j_{s}}
+\cdots+\Gamma^{i_{r}}_{j\alpha}T^{i_{1}\cdots i_{r-1}\alpha}_{j_{1}\cdots j_{s}}
-\Gamma^{\alpha}_{jj_{1}}T^{ i_{1}\cdots i_{r}}_{\alpha j_{2}\cdots j_{s}}-\cdots\Gamma^{\alpha}_{jj_{s}}T^{i_{1}\cdots i_{r}}_{ j_{1}\cdots j_{s-1}\alpha}
\end{equation}
sont les composantes d'un tenseur de $\tau^{r}_{s}(M)$ et qu'il existe sur $\tau^{r}_{s}(M)$ une con\-nexion $\nabla$ telle que $(\nabla_{j}T)^{i_{1}\cdots i_{r}}_{j_{1}\cdots j_{s}}={T^{i_{1}\cdots i_{r}}_{j_{1}\cdots j_{s}}}_{/j}.$ La connexion de Levi-Civita  $\nabla$ sur $\tau^{r}_{s}(M)$ est celle qu'on obtient par ce procédé à partir de la connexion de Levi-Civita sur $\tau(M)$. Ainsi en utilisant les égalités (\ref{der}) et (\ref{dersec}), on peut écrire :
\[\nabla T={T^{i_{1}\cdots i_{r}}_{j_{1}\cdots j_{s}}}_{/j}\;\frac{\partial}{\partial x^{i_{1}}}\otimes\ldots\otimes \frac{\partial}{\partial x^{i_{r}}}\otimes dx^{j_{1}}\otimes dx^{j_{s}}\otimes dx^j.\]

En particulier si $\nabla$ est la connexion de Levi-Civita sur $\tau(M)$ et $g$ le tenseur métrique alors on a le théorème de Levi-Civita : \begin{equation}\label{pam}
\nabla g=0
\end{equation} 
\begin{exo}\label{poum}
Démontrer l'égalité (\ref{pam}).
\end{exo}
\begin{proposition}\label{leileilei}
Si $T$ est un tenseur $p$ fois covariant sur $M$ munie de sa connexion de Levi-Civita, et si $X, X_1, \ldots,X_p$ sont des champs de vecteurs sur $M$, alors $$X.T(X_1,\ldots,X_p)=\nabla_XT+\sum_{s=1}^pT(X_1,\ldots,\nabla_XX_s,\ldots,X_p).$$
\end{proposition}
\begin{preuve}{}
En généralisant l'équation(\ref{derche}), on peut écrire :

\noindent$\frac{\partial}{\partial x^k}T(X_1,\ldots,X_p)=T(X_1,\ldots,X_p)_{/k}=(T_{i_1,\ldots,i_p}X^{i_1}_1\ldots,X^{i_p})_{/k}={T_{i_1,\ldots,i_p}}_{/k}X_1^{i_1}\ldots X_p^{i_p}+..$

\noindent$+T_{i_1,\ldots,i_p}\sum_{s=1}^pX_1^{i_1}\ldots\ldots {X_{s}^{i_s}}_{/k} \ldots X_p^{i_p}=$

\noindent $=(\nabla_{k}T)(X_1,\dots,X_p)+\sum_sT(X_1,\ldots,\nabla_{k}X_s,\ldots,X_p)$. L'égalité est vérifiée pour $X=\frac{\partial}{\partial x^k}$ et par suite pour tout champ $X=X^k\frac{\partial}{\partial x^k}$.
\end{preuve}
\begin{remark}\label{contderi}
Remarquons une propriété utile : si on note $c$ la contraction de deux indices et $D_{j}$ la dérivation ${T^{i_{1}\cdots i_{r}}_{j_{1}\cdots j_{s}}}\mapsto {T^{i_{1}\cdots i_{r}}_{j_{1}\cdots j_{s}}}_{/j}$, alors $c\circ D_{j}=D_{j}\circ c$. On le vérifiera à titre d'exercice avec la contraction des indices $i_{1}$ et $j_{1}$ dans l'égalité (\ref{pim}).
\end{remark}

\begin{exo}[Divergence d'un tenseur]\label{divdivergence}\hspace{10.5cm}\index{divergence d'un tenseur}
\begin{enumerate}
 \item Soit $T$ un tenseur défini dans une carte locale d'une variété par ses coordonnées $T^{i_1
i_2\ldots i_n}_{j_1\ldots j_p}$ symétrique dans ses indices contravariants. Montrer que les scalaires 
$(\Div T)^{i_2\ldots i_n}_{j_1\ldots j_p}=:T^{i
i_2\ldots i_n}_{{j_1\ldots j_p}_{/i}}$  définit par ses composantes un tenseur également symétrique en ses indices contravariants. Ce tenseur est la \emph{divergence du tenseur $T$}.
\item On suppose que le tenseur $T$ est un champ de vecteurs sur  $\rr^4$ dont on note  $(x^0,x^1,x^2,x^3)$ les coordonnées canoniques. Soit 
$ \omega= dx^0\wedge dx^1\wedge dx^2\wedge dx^3$ la forme volume canonique de $\rr^4$.
 Alors $d(i_T\omega)=(\Div T)\omega$.
\end{enumerate}
\end{exo}\index{divergence d'un tenseur}
\subsubsection{Connexion de Levi-Civita induite sur le fibré tangent d'une hypersurface d'une variété pseudo-riemannienne.}\label{conlehyp}
Soit $(M,\la\;,\;\ra)$ une variété pseudo-riemannienne de dimension $n$ et $S$ une sous-variété de $M$ de dimension $n-1$, c'est à dire une hypersurface de $M$. Elle hérite de $M$ par la métrique induite d'une structure de variété pseudo-riemannienne. On considère $X$ un champ de vecteurs tangent à $S$ et $Y$ un champ de vecteurs le long de $S$, non nécessairement tangent à $S$ comme $X$. On suppose que $\tau(M)$ est muni d'une connexion $\nabla$.
Le lemme \ref{taco} donne un sens au champ de vecteur sur $M$, $\nabla_XY$.
Soit $N$ un champ de vecteurs unitaires le long de $S$, orthogonal à $S$. On a donc $\la N,N\ra\in\{1,-1\}$. Notons 
$\epsilon(N)=\la N,N\ra$. Alors $\nabla_X Y -\epsilon(N)\la \nabla_XY,N\ra N$ est un champ de vecteurs tangent à $S$ : en effet, on vérifie immédiatement qu'il est orthogonal à $N$. Ce champ est la projection orthogonale du champ le long de $S$ 
$\nabla_XY$. On le note $\nabla^{S}_XY$.
\begin{proposition}\label{conind}
L'opérateur $\nabla^S=\nabla-\epsilon(N)\la \nabla,N\ra N$ qui à tout champ de vecteurs $Y\in\tau(S)$ associe la forme $\nabla^SY$ à valeurs dans $\tau(S)$, définie par  $$\nabla^SY=\nabla Y-\epsilon(N)\la \nabla Y,N\ra N$$  telle que pour tout champ $X\in\tau(S)$ par  $(\nabla^SY)(X)=\nabla_X Y-\epsilon(N)\la \nabla_X Y,N\ra N$ est une connexion sur $\tau(S)$ appelée la connexion induite par $\nabla$ sur $S$. Si $\nabla$ est la connexion de Levi-Civita de $M$, $\nabla^S$ est la connexion de Levi-Civita de $S$.
\end{proposition}
Bien entendu  $(\nabla^SY)(X)$ se notera essentiellement $\nabla^S_XY$.
\begin{preuve}{}
Le fait que $\nabla^S$ soit une connexion provient d'une part de la linéarité pour la structure de module sur les fonctions différentiables pour l'argument $X$, d'autre part de l'additivité en l'argument $Y$ et de la vérification  de la propriété de Leibnitz (\ref{leibi}) pour ce même argument $Y$. Toutes ses propriétés sont directement vérifiables.

Supposons maintenant que $\nabla$ soit la connexion de Levi-Civita.  on a  pour tous champs de vecteurs $X,X,Z $ de $\tau(S)$ : $Z.\la X,Y\ra= \la \nabla_ZX,Y\ra+ \la X,\nabla_ZY\ra= $
 $ =  \la \nabla^S_ZX+\epsilon(N)\la \nabla_ZX,N\ra N,Y\ra +
\la  X,\nabla^S_ZY+\epsilon(N)\la \nabla_ZY,N\ra N\ra=..$

\hspace{3.5cm}$..=\la \nabla^S_ZX,Y\ra+ \la X,\nabla^S_ZY\ra$. 
\end{preuve}
On peut exprimer la connexion induite à l'aide de la fonction de Weingarten que l'on trouve dans l'étude des surfaces de $\rr^n$ (voir par exemple [\ref{rinrin}] ou [\ref{franki}]).
\begin{proposition}\label{weini}
On suppose $\tau(M)$ muni de la connexion de Levi-Civita, notée $\nabla$.
Avec les notations précédentes concernant $S$, l'application  $\cal{W}$ qui à tout $X\in \tau(S)$ associe $\mathcal{W}(X)=-\nabla_XN$ définit un morphisme du fibré $\tau(S)$ vers lui-même au dessus de l'identité. 
La forme bilinéaire associée définie sur $\tau(S)$ par $w(X,Y)=\la X,\mathcal{W}(Y)\ra$ est symétrique et on a :
\begin{equation}\label{conwein}
\nabla_XY=\nabla^S_XY+\epsilon(N)w(X,Y)N.
\end{equation}
\end{proposition}
\begin{definition}\label{weinensie?}\index{application de Weingarten}\index{courbure de Gauss}
$\mathcal{W}$ est appelée l'application de Weingarten de la sous-variété $S$. Les valeurs propres (réelles) de $\mathcal{W}(x) : T_xS\longrightarrow T_xS$ sont les courbures principales de $S$ au point $x$. La courbure de Gauss de $S$ est le produit des courbures principales.
\end{definition}
\begin{preuve}{de la proposition \ref{weini}}
Remarquons tout d'abord $\nabla_XN$ est un champ de vecteurs tangent à $S$.  En effet $\la \nabla_XN,N\ra =\frac{1}{2}X.\la N,N \ra=0$. Soit $x\in S$.
On sait que $(\nabla_{X(x)}N)(x)\in T_xS$ et donc $X(x)\mapsto \nabla_{X(x)}N(x)$ est une application de $T_xS$ dans lui-même qui est clairement $\rr$-linéaire. 

\noindent Compte tenu de de la définition d'une connexion,  $w$ est bilinéaire au sens que pour toute fonction différentiable $f$, on a :

\noindent $\la fX,\mathcal{W}(Y)\ra=f \la X,\mathcal{W}(Y)\ra,
\la X,\mathcal{W}(fY)\ra =f \la X,\mathcal{W}(Y)\ra$, $w$ étant de plus additive par rapport à ses deux arguments. De plus :

$w(X,Y)=\la X,-\nabla_YN \ra=-\la \nabla_YX ,N\ra$ (Car $\nabla $ est la connexion de Levi-civita)

\noindent $=-\la \nabla_XY+[X,Y],N\ra$ (car la connexion de Levi-Civita est symétrique)

\noindent $=-\la \nabla_XY,N\ra=w(Y,X)$ (car $[X,Y]$ est tangent à $S$ comme $X$ et $Y$).

Cela signifie que la fonction de Weingarten de la fibre $T_xS$ dans lui-même est symétrique et donc admet des valeurs propres réelles associées à une base de vecteurs propres qui constitue une base $w$-orthogonale.

\noindent L'équation (\ref{conwein}) résulte directement de la définition de $\nabla^S_XY$ qui apparaît dans la proposition \ref{conind}.

\end{preuve}
\begin{definition}\label{souvargeo}\index{sous-variété géodésique}
Une sous-variété $S$ d'une variété pseudo-riemanienne $M$ est géodésique en un point si toute géodésique de $S$ passant par ce point est également une géodésique de $M$. Si $S$ est géodésique en tout point de $S$ on dit que $S$ est une sous-variété (totalement) géodésique de $M$.
\end{definition}
Par exemple une  courbe fermée de $S^2$ est géodésique si et seulement s'il s'agit d'un grand cercle donc d'une géodésique de $S^2$.
On peut facilement caractériser la propriété de sous-variété géodésique lorsqu'on a affaire à une hypersurface. 
\begin{proposition}\label{hypgeo}
On se donne une hypersurface $S$ de $M$. C'est une sous-variété géodésique en un point si et seulement si la fonction de Weingarten associée à $S$ en ce point est nulle. 
\end{proposition}
\begin{preuve}{}
Soit $X$ un vecteur tangent à $S$ en $x$, $S$ étant géodésique en $x$. On sait qu'il existe sur un voisinage de $x$ dans $S$ une unique géodésique issue de $x$ à l'instant $0$ et tangente en $x$ à $X(x)$. Cette géodésique est habituellement notée $t\to\exp_xtX(x)$ (voir [\ref{mil}]). Elle est définie sur un certain intervalle $]-\epsilon,\epsilon[$. Par hypothèse cette géodésique est également une géodésique de $M$. On a donc en particulier 
$\nabla_{X(x)}X(x)=\nabla^S_{X(x)}X(x)=0$ donc $w(X,X)(x)=0$ (et ceci en fait pour tout $x$ de $S$ si $S$ est géodésique).
Mais la propriété de symétrie de $w$ montre que si $X,Y$sont deux champs de vecteurs tangents à $S$, alors :

\noindent$w(X,Y)=\frac{1}{2}\left(W(X+Y,X+Y)-w(X,X)-w(Y,Y)\right)$ et ainsi $w(X,Y)(x)=0.$
Et on conclut.
\end{preuve}
\subsection{Connexion de Simon sur un fibré vectoriel en droites.} \label{ciment}
\index{connexion de Simon}
Dans le cas du fibré de Hopf la base $S^{2}$ est plongée dans $\rr^{4}$ : par abus de langage on dit que $S^{2}$ est plongée dans $
\cc^{2}$ en pensant que $\cc^{2}=\varphi^{{-1}}(\rr^{4})$ où $\varphi$ est la bijection $\rr$-linéaire de $\cc^{2} $ sur $\rr^{4}$ de la définition \ref{aze}.
Supposons plus généralement une variété réelle $M$  de dimension $p$ plongée dans $\cc^{n}$ de sorte que l'on ait un fibré en droites complexes $\tau=(\cc^{n},\pi,M)$. 

Pla\c cons-nous sur un domaine de carte $U$ muni de paramètres $\alpha=(\alpha^{1},\cdots,\alpha^{p})$.
En tout point de $U$ considérons un vecteur unitaire $e^{U}(\alpha)$ qui définit une section de $\tau$ sur $U$. Alors $\mfrac{\partial e^{U}(\alpha)}{\partial \alpha^{k}}d\alpha^{k}$ est une forme sur $U$ à composantes dans $\cc^{n}$ et $\omega_{1}^{{U}}=\la e^{U}(\alpha), \mfrac{\partial e^{U}(\alpha)}{\partial \alpha^{k}}\ra d\alpha^{k}$ est une $1$-forme sur $U$. Montrons qu'elle définit  une $1$- forme de connexion sur $\tau$. 
On considère une autre  carte  $V$ munie de paramètres $\beta=(\beta^{1},\cdots,\beta^{p})$ de sorte que $\omega_{1}^{V}(\beta)=\la e^{V}(\beta),\mfrac{\partial e^{V}(\beta)}{\partial \beta^{l}}\ra d\beta^{l}$ sur $V$. Soit $c_{UV}\in U(1)$ tel que $e^{U}(\alpha)=e^{V}(\beta)c_{VU}$. On a :

$\omega_{1}^{V}(\beta)=\la e^{U}(\alpha) c_{UV}, \mfrac{\partial e^{U}(\alpha)}{\partial \beta^{l}} c_{UV}+ e^{U}(\alpha)\mfrac{\partial c_{UV}}{\partial \beta^{l}}\ra d\beta^{l}=$

$\la e^{U}(\alpha) , \mfrac{\partial e^{U}(\alpha)}{\partial \beta^{l}}\ra d\beta^{l} +\la e^{U}(\alpha) c_{UV}, e^{U}(\alpha) \mfrac{\partial c_{UV}}{\partial \beta^{l}}\ra d\beta^{l}=$

$ \la e^{U}(\alpha) , \mfrac{\partial e^{U}(\alpha)}{\partial \alpha^{l}}\ra d\alpha^{l} +c_{UV}dc_{UV}=c_{VU}\;\omega_{1}^{U}(\alpha)c_{UV}+c_{UV}dc_{UV}.$
Il existe donc (souvenons-nous de l'équation (\ref{loli})) une connexion sur $\tau$ définie sur $U$ par 
\begin{equation}\label{cosim}
\nabla^{U}e^{{U}}(\alpha)=e^{{U}}\otimes \la e^{U}(\alpha),\mfrac{\partial e^{U}(\alpha)}{\partial \alpha^{k}}\ra d\alpha^{k}
\end{equation} ou encore symboliquement 
\begin{equation}\label{cosimsy}
\nabla e=e\otimes \la e,de\ra 
\end{equation}


\subsection{Complexifiée de la connexion de Levi-Civita : le cas général pour une variété de dimension $2n$}\label{cru}

Soit $\tau$ le fibré tangent à une variété différentiable de dimension $2n$. On suppose que $\nabla$ est la connexion de Levi-civita sur $\tau$. Considérons deux cartes locales 
$(U,\Phi_{U})$ et $(V,\Phi_{V})$ telles que :
\begin{enumerate}
\item $U\cap V\not=\emptyset,$
\item les changements de cartes sont dans $S0(2n),$
\end{enumerate}
Soit $b\in U\cap V$. Si $C_{UV}(b)$ désigne la matrice associée au cocycle $c_{UV}(b)$ ( dans la base canonique de $\rr^{2n}$), la matrice  $\tilde C_{UV}(b)$
 est associée (dans la base canonique de $\cc^{n}$) au cocycle $\tilde c_{UV}$ correspondant sur le fibré $\tau^{c}$ (voir notation suivant la définition \ref{fico} p.\pageref{fico}).
On considère les matrices de connexion $\omega_{U}$ et $\omega_{V}$ associées à $\nabla$ dans ces cartes. Ce sont des matrices carrées d'ordre $2n$. Alors 
$ \omega_{V}=C_{VU}\omega_{U}C_{UV}+C_{VU}dC_{UV}$

 L'application $ \varphi(z_{1},\cdots,z_{n})=(x_{1},x_{2},\cdots,x_{2n-1},x_{2n})$ nous permet d'écrire 
 
\[\omega_{V}=(\varphi\tilde C_{VU}\varphi^{{-1}})\omega_{U}(\varphi \tilde C_{UV}\varphi^{{-1}})+\varphi\tilde C_{VU}\varphi^{{-1}}d\left(\varphi\tilde{C}_{UV}\varphi^{{-1}}\right).\]
La linéarité de $\varphi$ donne l'égalité $d\left(\varphi\tilde{C}_{UV}\varphi^{{-1}}\right)=\varphi d\tilde{C}_{UV}\varphi^{{-1}}$.
D'où 
\[\omega_{V}=(\varphi\tilde C_{VU})(\varphi^{{-1}}\omega_{U}\varphi )(\tilde C_{UV}\varphi^{{-1}})+\varphi\tilde C_{VU}d\left(\tilde C_{UV}\right)\varphi^{{-1}}\]
où $\varphi^{{-1}}\omega_{U}\varphi $ est une matrice carrée d'ordre $n$  dont les éléments sont des $1$-formes sur $M$ à coefficients complexes.
On peut donc écrire

\[ \varphi^{-1}\omega_{V}\varphi=\tilde{C}_{VU} \left(\varphi^{-1}\omega_{U}\varphi\right)\tilde{C}_{UV}+\tilde{C}_{VU}d\tilde{C}_{UV}.\]
La famille $\{\tilde\omega_{V}= \varphi^{-1}\omega_{V}\varphi\}$ constitue une famille de matrices de $1$-formes sur $M$ à coefficients complexes. Elle définit une connexion $\nabla^{c}$ sur $\tau^{c}$ appelée la\emph{ connexion complexifiée de $\nabla$}.\index{complexifiée d'une connexion}
\begin{exemple}
Soit $\tau(M)$ le fibré tangent à une variété riemannienne $M$ de dimension $2$: $\tau=\tau(M)$ et on demande que les cocycles soient à valeur dans $SO(2)$. On sait alors (voir proposition \ref{kelestbel}) que la matrice de connexion est de la forme $\omega=\left(\begin{matrix}0&-\varpi\\\varpi&0\end{matrix}\right)$, où $\varpi$ est une $1$-forme sur $M$. 
Posons  $z=x+iy$, on a $\omega\varphi(z)=\left(\begin{matrix}0&-\varpi\\\varpi&0\end{matrix}\right)\left(\begin{matrix}x\\y\end{matrix}\right)=
 \left(\begin{matrix}-y\varpi\\x\varpi\end{matrix}\right)$. D'où $\varphi^{{-1}}\omega\varphi (z)=i\varpi z$.
 
\end{exemple}
 
\subsection{Dérivations covariantes d'ordre 2 et opérateur laplacien}\label{lapilapin}
On introduit la dérivation covariante d'ordre $2$ et à sa suite l'opérateur laplacien, avec deux cas particuliers remarquables, en cinq points :

\begin{enumerate}
\item On considère une variété $M$ pseudo-riemannienne et on note $g_{ij}$ le tenseur métrique dans les cartes locales que l'on considèrera. Soit un fibré vectoriel $\tau$  sur $M$, muni d'une connexion $\nabla$ et le fibré $\tau^{*}(M)$, muni de sa connexion de Levi-Civita. Si $s$ est une section sur $\tau$
qui s'écrit localement dans une base $s=\lambda^{i}s_{i}$, $\nabla s$  s'écrit localement, d'après l'équation (\ref{dersec}),
$\nabla s=\lambda^{i}_{/j}s_{i}\otimes dx^{j}$ et c'est une section de $\tau\otimes \tau^*(M)$.

\item De fa\c con générale si $\sigma$ est une section sur $\tau\otimes \tau^{*}(M)$, avec les notations du point précédent, $\sigma$ s'écrit sous la forme 
$\sigma=\zeta^{i}_{j}s_{i}\otimes dx^{{j}}$. En considérant $\tau^{*}(M)$ muni de la connexion de Levi-Civita $\nabla_{\ma{LC}}$, nous avons de façon analogue au résultat de la proposition \ref{calculasse} une connexion $\nabla''=\nabla\otimes \nabla_{\ma{LC}}$ sur 
$\tau\otimes \tau^{*}(M)$  qui vérifie : 
\begin{equation}\label{intermede}
\nabla''_{k}(\zeta^{i}_{j})=\mfrac{\partial \zeta^{i}_{j}}{\partial x^{k}}+\omega^{i}_{ks}\zeta^{s}_{j}-\Gamma^{s}_{kj}\zeta^{i}_{s}
\end{equation}
\item  Dans le point précédent, considérons la section de $\tau\otimes \tau^*(M)$, $\sigma=\nabla s$, où $s=\lambda^{i}s_{i}$. Alors $(\nabla\otimes \nabla_{\ma{LC}})(\nabla s)$ est une section de $\tau\otimes \tau^*(M)\otimes \tau^*(M)$. Notons 
$$\nabla^2=\left(\nabla\otimes \nabla_{\ma{LC}}\right)\circ\nabla .$$ 
Attention $\nabla^2\not=\nabla\circ\nabla$ qui n'a aucun sens! L'opérateur $\nabla^2$ (\emph{opérateur dérivée seconde  covariante})\index{opérateur  de dérivée seconde covariante} associe à toute section $s$ de $\tau$ une section de
$\tau\otimes \tau^*(M)\otimes \tau^*(M)$ l'on peut noter $\lambda^{k}_{/ij}s_{k}\otimes d x^{i}\otimes dx^j$.
 L'égalité (\ref{intermede}) et l'équation (\ref{contessa}) de la proposition \ref{calculasse} nous donnent l'expression de $\lambda^{k}_{/ij} $
\begin{equation}\label{fucklap}
\lambda^{k}_{/ij}:=\nabla''_{i}(\lambda^{k}_{/j})=\mfrac{\partial\lambda^{k}_{/j}}{\partial x^{{i}}}+\omega^{k}_{is}\lambda^{s}_{/j}-\Gamma_{ij}^{s}\lambda^{k}_{/s}
\end{equation}
où
\[\lambda^{k}_{/j}=\mfrac{\partial\lambda^{k}}{\partial x^{{j}}}+\omega_{js}^{k}\lambda^{s}.\]

\item {E}xaminons l'égalité (\ref{fucklap}). Si on y développe les dérivations covariantes d'ordre $1$, on obtient :

\begin{equation}\label{lapilapi}
\lambda^{k}_{/ij}=\big(\mfrac{\partial^{2}\lambda^{k}}{\partial x^{i}\partial x^{j}}-\Gamma^{s}_{ij}\mfrac{\partial \lambda^{k}}{\partial x^{s}}\big)+A^{k}_{ijs}\lambda^{s}\end{equation}
où \begin{center}$A^{k}_{ijs}=\mfrac{\partial \omega^{k}_{js}}{\partial x^{i}}+\omega^{k}_{is}\mfrac{\partial }{\partial x^{j}}+\omega^{k}_{js}\mfrac{\partial }{\partial x^{i}}+\omega^{k}_{it}\omega^{t}_{js}-\omega^{k}_{us}\Gamma^{u}_{ij}.$\end{center}

\noindent La définition habituelle du laplacien sur une variété riemannienne $M$ est $$\Delta_{M} f=\mathrm{div}(\grad f).$$\index{laplacien d'une fonction}
Le champ de vecteurs $\grad f$ est défini par l'égalité $\la \grad f,\;\ra=df$.\index{gradient d'un champ de vecteurs} Il s'écrit donc en coordonnées locales
 \begin{equation}\label{gradus}
\grad f=g^{ij}\mfrac{\partial f}{\partial x^{j}} \mfrac{\partial}{\partial x^{i}}.
\end{equation}
On a donc, en utilisant les équations  (\ref{derche}),  (\ref{croma}),(\ref{civicot}) et (\ref{pam}) :

\noindent $\mathrm{div}(\grad f)=\big(g^{{ij}}\mfrac{\partial f}{\partial x^{j}}\big)_{/i}=
g^{{ij}}\big(\mfrac{\partial f}{\partial x^{j}}\big)_{/i}=g^{{ij}}\big(\mfrac{\partial^{2} f}{\partial x^{i}\partial x^{j}}-\Gamma_{ij}^{k}\mfrac{\partial f}{\partial x^{k}}\big)= 
g^{ij}{(f_{/j})}_{/i}=g^{ij}{(f_{/i})}_{/j}$.
On écrira ceci sous la forme :
\begin{equation}\label{lapone}
\Delta_M f=g^{{ij}} f_{/ij}
\end{equation}   pour toute fonction $f$  suffisamment différentiable sur $M$. 

On définit le laplacien de toute section $s=\lambda^{i}s_{i}$ de $\tau$ (voir premier point) par :\index{laplacien d'une section}
\begin{equation}\label{lapine}
\Delta_{\tau}(s)=g^{{ij}}\lambda^{k}_{/ij}s_{k}
\end{equation}
où $\lambda^{k}_{/ij}$ est donné par l'égalité (\ref{lapilapi}).
 L'équation (\ref{lapone}) montre que $\Delta_{\tau}(s)=(\Delta_{M}\lambda^{k})s_{k}.$ 

Par exemple si on prend $\tau=(\rr^{n}\times \rr,p_{1},\rr^{n})$, et $f$ une fonction différentiable sur $\rr^{n}$ à valeur dans $\rr$, $\Delta_{\tau}f=\sum_{i=1}^{n}\mfrac{\partial^{2}f}{{\partial x^{i}}^{2}}$.
\item \textbf{Deux cas particuliers simples mais non triviaux}\index{connexion sur un fibré en droite}
\begin{enumerate}
\item
\emph{On suppose que $\tau$ est un fibré en droites (complexes éventuellement) sur $\rr^{3}$ euclidien. }

Alors $s$ est localement représenté par une fonction (éventuellement complexe) $\psi$. Selon l'équation (\ref{loloo}), la matrice de la connexion $\nabla$ est une forme $\omega=\omega_{i}dx^{{i}}$ où les $\omega_{i}$ sont éventuellement des complexes. Dans l'équation (\ref{fucklap}) on a si on s'en réfère à l'équation (\ref{coefcon}) : $\omega^{k}_{ij}=\omega_{i}$ pour tout $(j,k)$. 
On peut donc écrire à partir de l'égalité (\ref{lapilapi}) :

$\Delta_{\tau}s=g^{{ii}}\psi_{/ii}=\sum_{i=1}^{3}\left(\mfrac{\partial^{2}\psi}{{\partial x^{i}}^{2}}\right)+\mfrac{\partial \omega_{i}}{\partial x^{{i}}}\psi+2\omega_{i}
\mfrac{\partial\psi}{\partial x^{i}}+\omega_{i}^{2}\psi,$ ou encore
\begin{equation}\label{schon}
\Delta_{\tau}s=\sum_{i=1}^{3}\left(\mfrac{\partial}{\partial x^{{i}}}+\omega_{i}\right)^{2}\psi.
\end{equation}
\item \emph{On suppose que $\tau$ est un fibré en droites (complexes)  sur une variété pseudo-riemannienne $M$.}

Les hypothèses et notations sur $\omega$ sont les mêmes que dans le cas précédent.
L'équation (\ref{fucklap}), donne de fa\c con identique au cas (a) :
\begin{equation}\label{schonschon}
\Delta_{\tau}s=g^{{ij}}\left( \big(\mfrac{\partial}{\partial x^{i}}+\omega_{i}\big) \big(\mfrac{\partial}{\partial x^{j}}+\omega_{j}\big) -\Gamma^{s}_{ij}\big(\mfrac{\partial}{\partial x^{s}}+\omega_{s}\big)\right)\psi
\end{equation}
\end{enumerate}
\end{enumerate}
\begin{remark}\label{sioux}

\noindent $1$) Pour être précis il faudrait noter $\nabla^{U}$ la connexion $\nabla$ du fibré $\tau$, exprimée dans les coordonnées du domaine de carte $U$ et $\Delta^{U}_\tau$ le laplacien qui intervient dans l'équation (\ref{schon}).

\noindent $2$) La remarque \ref{decor} nous dit que la connexion $\nabla$ sur $\tau$ induit une dérivation sur les composantes des sections de ce fibré, dérivation dont on a précisé les différentes notations dans les égalités (\ref{nonotes}). Lorsque $\tau$ est un fibré en droites complexes sur $\rr^3$
euclidien alors l'égalité (\ref{lapilapi}) montre après un calcul direct que 
\begin{equation}\label{der2}
\psi_{/ij}=\left(\frac{\nabla}{\partial x^{i}}\circ\frac{\nabla}{\partial x^{j}}\right)(\psi).
\end{equation}
Le laplacien donné par l'équation (\ref{schon}) s'écrit dans ce cas 
\begin{equation}\label{chapichapi}
\Delta_\tau=\sum_{i=1}^3\left(\frac{\nabla}{\partial x^{i}}\right)^2.
\end{equation}
Par contre si $\tau$ est un fibré en droites complexes sur une variété riemannienne $M$ dont la connexion de Levi-Civita n'est pas nulle, alors l'équation (\ref{der2}) n'est plus vérifiée car la dérivée
première ne fait pas intervenir la connexion de Levi-Civita.
\end{remark}
\section{Les connexions en mécaniques classique et quantique}\label{conclaqua}
\subsection{Connexions de Levi-Civita et équations de Lagrange}\label{colagene}\index{Lagrange (équations de)}
On obtient classiquement les équations de Lagrange en mécanique classique à partir d'un principe variationnel, le principe des travaux virtuels de Hamilton. Décrivons ce principe. On considère une particule de masse $m$ qui se déplace avec un certains nombre de contraintes, soumise à des forces dérivant d'un potentiel $\ma V$.
Ces contraintes traduisent le déplacement de la particule  sur une variété $M$ de dimension $n$. On peut trouver le détail de cette idée dans [\ref{poi}]. 
Soit une fonction $L$   définie sur l'espace $TM$ du fibré tangent à M ($M$ \emph{l'espace de configuration} de la particule).\index{espace de configuration} Cette fonction définit l'intégrale  $S=\int_{\gamma_{\alpha}}L(q,\dot{q},t)dt$ où 
 $\gamma_{\alpha} (t)$ est une variation à un paramètre $\alpha$  de la trajectoire $\gamma_{0}(t)$ paramétrée par le temps de la particule telle qu'elle est déduite des équations de Newton,   entre deux instants
  $t_{0}$ et $t_{1}$ de sorte que 
\begin{enumerate}
\item Pour tout $\alpha$, $\gamma(\alpha,t_{0})=a$ et $\gamma(\alpha,t_{1})=b$
\item  pour tout $\alpha$, on a $\mfrac{\partial\gamma}{\partial\alpha}(t_{0})=\mfrac{\partial\gamma}{\partial\alpha}(t_{1})=0$
\end{enumerate}
Le principe des travaux virtuels dit qu'il existe une fonction $L$ sur $TM$ telle que le fait que $S(\gamma_{\alpha})$ admette $\gamma_{0}$ comme point critique lorsqu'on considère les variations de $\gamma_{0}$ dans la famille décrite ci-dessus équivaut à la relation fondamentale de la dynamique. La fonction $L$ a été trouvée par Lagrange et se nomme \emph{le lagrangien}. Plus précisément $L=E_{c}-\ma V$ où $E_{c}$ est l'énergie cinétique de la masse $m$. L'écriture de $\mfrac{dS}{d\alpha}(0)$ aboutit aux équations de Lagrange données par l'égalité (\ref{lag}) ci-dessous.
 Pour plus de détails, consulter [\ref{poi}] ou [\ref{franki}]. 
 
 \noindent Nous allons montrer que l'écriture covariante des équations de Newton  aboutit directement aux équations de Lagrange, sans principe variationnel.
 Notons que l'énergie cinétique munit  la variété $M$   d'une structure riemannienne :  exprimée en coordonnées locales, c'est une forme quadratique définissant une métrique $\la,\ra$ de sorte que si l'espace de configuration  est muni des coordonnées locales $(q^{1},\cdots q^{n},\dot{q}^{1},\cdots,\dot{q}^{n})$,  l'énergie cinétique s'écrit dans ce système de coordonnées : $$E_{c}=\mfrac{1}{2}
g_{ij}\dot{q}^{{i}} \dot{q}^{j}.$$
le tenseur métrique défini par $E_c$ est le tenseur deux fois covariant $g=(g_{ij})$.

\noindent Le fibré tangent est alors naturellement muni de la connexion de Levi-Civita $\nabla$ associée à $g$.
Sup\-po\-sons que la particule décrive une courbe 
paramétrée par le temps $$t\to(q^{1}(t),\cdots q^{n}(t),\dot{q}^{1}(t),\cdots,\dot{q}^{n}(t)).$$
 Sa vitesse au point de coordonnées $q(t)=(q^{1},\cdots,q^{n})$ est $v=\frac{d{q}^{i}}{dt}\frac{\partial}{\partial {q}^{{i}}}=:\dot{q}_{i}(v)\frac{\partial}{\partial {q}^{{i}}}$ et son accélération est d'après la définition \ref{accel} : $\frac{\nabla v}{dt}=:\nabla v (v).$ L'équation de Newton s'écrit 
\begin{equation}\label{nounou}
m\frac{\nabla v}{dt}=-\grad \vv.
\end{equation}Compte tenu de l'équation (\ref{devcov}) les équations du mouvement s'écrivent en coordonnées locales 

\begin{equation}\label{mouv}
\frac{\nabla \dot{q}^{i}}{dt}=\frac{d\dot{q}^{i}}{dt}+\Gamma^{i}_{jk}\dot{q}^{j}\dot{q}^{k}=-g^{ik}\frac{\partial \vv}{\partial q^{k}},
\end{equation}
où pour tout indice $i$ on a $\dot{q}^{i}=\mfrac{d q^{i}}{dt}.$
Les équations (\ref{mouv}) pour $i\in\{1,\ldots,n\}$ sont les \emph{équations de Newton covariantes}. \index{Newton (équations de) }En les transformant nous allons en donner la forme classique des \emph{équations de Lagrange} que l'on obtient directement à partir
du principe variationnel rappelé ci-dessus.

Remarquons que dans l'équation (\ref{mouv}), la masse $m$ est contenue dans le terme $g^{ik}$.
Le lagrangien de la particule dont il a été question au début de la section est la fonction $L=E_{c}-\vv$. En coordonnées locales : $L=\frac{1}{2}g_{ij}\dot{q}^{{i}}\dot{q}^{j} -\vv$.
On définit à partir du lagrangien   la $1$-forme \emph{moment} $p=\frac{\partial L}{\partial \dot{q}^{i}}dq^{i}=g_{ij}\dot{q}^{j}dq^{i}=:p_{i}dq^{i}.$ En utilisant l'équation de Newton ainsi que les égalités (\ref{derichon}) et (\ref{poum}),
on a :
\[\frac{\nabla p_{i}}{dt}=g_{ij}\frac{\nabla\dot{q}^{j}}{dt}=-g_{ij}\left(g^{{jk}}\frac{\partial \vv}{\partial q^{k}}\right)=-\frac{\partial \vv}{\partial q^{i}}.\]
Par ailleurs, d'après l'équation (\ref{rose}) on a : 
\begin{equation}\label{nunu}
\frac{\nabla p_{i}}{dt}=\frac{d p_{i}}{dt}-\Gamma^{k}_{ij}p_{k}\dot{q}^{j}.
\end{equation} 

\noindent On a donc $\frac{d p_{i}}{dt}=\Gamma^{k}_{ij}p_{k}\dot{q}^{j}-\frac{\partial \vv}{\partial q^{i}}$.
En considérant l'égalité (\ref{cristo}), il vient :
\begin{gather*}
\frac{d p_{i}}{dt}=\mfrac{1}{2} g^{k\alpha}\left(
\frac{\partial g_{\alpha j }}{\partial q^{i}}+\frac{\partial g_{i \alpha}}{\partial q^{j}}-\frac{\partial g_{ij}}{\partial q^{k}}\right)g_{ks}\dot{q}^{s}\dot{q}^{j}-\frac{\partial \vv}{\partial q^{i}}=\\
=\mfrac{1}{2}\left(\frac{\partial g_{s j }}{\partial q^{i}}+\frac{\partial g_{si }}{\partial q^{j}}-\frac{\partial g_{i j }}{\partial q^{s}}\right)\dot{q}^{s}\dot{q}^{j}-\frac{\partial \vv}{\partial q^{{i}}}
=\mfrac{1}{2}\frac{\partial g_{s j }}{\partial q^{i}}\dot{q}^{s}\dot{q}^{j}-\frac{\partial \vv}{\partial q^{i}}=
\frac{\partial L}{\partial q^{{i}}}\\
\mathrm{car}\;\left(\frac{\partial g_{si }}{\partial q^{j}}-\frac{\partial g_{i j }}{\partial q^{s}}\right)\dot{q}^{s}\dot{q}^{j}=0
\end{gather*}


\noindent On obtient ainsi les équations de Lagrange classiques :
 \begin{equation}\label{lag}
\frac{d}{dt}\left(\frac{\partial L}{\partial \dot{q}^{{i}}}\right)-\frac{\partial L}{\partial q^{{i}}}=0
\end{equation}
On mesure à partir de cet exposé le progrès que constituent les équations de Lagrange. Lorsque l'espace de configuration est une variété, la mise en oeuvre des équations de Newton 
nécessite le calcul des coefficients de Christoffel, là où les équations de Lagrange ne nécessite que l'expression du tenseur métrique dans les coordonnées locales choisies. Illustrons ceci par un exemple.
\begin{exemple}

$1.$ Supposons que deux particules de masse $1$ circulent chacune dans le plan $\rr^2$ soumises à un potentiel $\ma{V}$. On peut considérer que l'espace des configurations pour le système constitué des deux particules est la variété de dimension $4$, $\rr^2\times \rr^2$, munies des coordonnées $(x^1,x^2,x^3,x^4)$. La particule $1$ a pour coordonnées $(x^1,x^2,0,0)$ et elle est soumise au potentiel $\ma{V}(x^1,x^2)$. La particule $2$ a pour coordonnées $(0,0,x^3,x^4)$ et elle est soumise au potentiel $\ma{V}(x^3,x^4)$.
Le lagrangien associé au système est 
$L=\frac{1}{2}\big((\dot{x}^1)^2+(\dot{x}^2)^2+(\dot{x}^3)^2+(\dot{x}^4)^2\big)-\ma{V}(x^1,x^2)-\ma{V}(x^3,x^4).$
Les équations de Newton s'écrivent $\frac{d^2 x^{i}}{dt^2}=-\frac{\partial \ma{V}}{\partial x^{i}}$ pour $i\in\{1,2,3,4\}$ et elles sont identiques (sans calcul supplémentaire) aux $4$ équations de Lagrange $\frac{d}{dt}\left(\frac{\partial L}{\partial \dot{x}^{i}}\right)=\frac{\partial L}{\partial x^{i}}.$

$2.$ Nous allons rajouter une contrainte au système précédent, en supposant les deux particules liées entre elles par une barre rigide sans masse, de longueur $1$. Cette contrainte relie les coordonnées des deux particules et plus précisément on a : \[(x^1-x^3)^2+(x^2-x^4)^2-1=0.\]
Posons $f(x^1,x^2,x^3,x^4)=(x^1-x^3)^2+(x^2-x^4)^2-1$. Cette fonction est de rang $1$ sur l'ouvert de $\rr^4$, $\ma{U}=\{(x^1,x^2,x^3,x^4)\in\rr^4\;;\;x^1\not=x^3\;\mathrm{et}\;x^2\not=x^4\} $. Par conséquent l'équation $f=0$ représente une hypersurface $M$  de $\rr^4$. Si on considère $M$ comme espace de configuration du système constitué par les deux particules, nous pouvons équiper cette variété de la métrique induite par la métrique standard de $\rr^4$ et évaluer dans une carte locale son tenseur métrique.
Pla\c cons-nous maintenant dans l'ouvert $\tilde{\ma{U}}$ de $M$ défini par  $\tilde{\ma{U}}=M\cap \{(x^1,x^2,x^3,x^4)\;;\; x^4>x^2\}$. Sur cet ouvert $M$ est le graphe de la fonction 
$S(x^1,x^2,x^3)=x^2+\sqrt{1-(x^1-x^3)^2}$. Ainsi l'ouvert $\tilde{\ma{U}}$ est le domaine d'une carte locale équipée des coordonnées\\ $(x^1,x^2,x^3)$.

\noindent Notons tout d'abord que le vecteur de $\rr^4$, $N(x^1,x^2,x^3)=\left(\frac{\partial S}{\partial x^1},\frac{\partial S}{\partial x^2},\frac{\partial S}{\partial x^3},-1\right)$ est normal à $M$ au
point de $\tilde{\ma{U}}$ de coordonnées $(x^1,x^2,x^3)$ et que par conséquent la famille de trois vecteurs $\left\{T_{1}=(1,0,0,\frac{\partial S}{\partial x^1}), T_{2}=(0,1,0,\frac{\partial S}{\partial x^2}), T_{3}=(0,0,1,\frac{\partial S}{\partial x^3})\right\}$ est une base de l'espace tangent à $M$ en ce point.
Nous allons exprimer le tenseur métrique sur $\tilde{\ma{U}}$ dans les coordonnées $(x^1,x^2,x^3)$.
Pour cela remarquons que sur l'espace tangent à $M$ en un point de coordonnées $(x^1,x^2,x^3)$ de $\tilde{\ma{U}}$,  on a : $$T_{1}=\frac{\partial}{\partial x^1},T_{2}=\frac{\partial}{\partial x^2},T_{3}=\frac{\partial}{\partial x^3}.$$

\noindent En effet si $h$ est une fonction sur $\rr^4$, sa restriction $h_{\vert M}$ à $M$ est sur l'ouvert $\tilde{\ma{U}}$ la fonction $(x^1,x^2,x^3)\mapsto 
h(x^1,x^2,x^3,S(x^1,x^2,x^3)$. 
De plus $\frac{\partial h_{\vert M}}{\partial x^1}=\frac{\partial h}{\partial x^1}+\frac{\partial h}{\partial x^4}\frac{\partial S}{\partial x^1}=T_{1}.h$, la dérivée de $h$ dans la direction $T_{1}$. Idem pour les deux autres égalités. Notons $g_{ij}=\la\frac{\partial}{\partial x^{i}},\frac{\partial}{\partial x^{j}}\ra_{M}$ les composantes du tenseur métrique, le crochet désignant la métrique induite sur $M$ par $\la.,.\ra$, le crochet de $\rr^4$. Alors $g_{ij}=\la T_{i},T_{j}\ra$. 

\noindent Le calcul donne immédiatement 

\noindent $g_{11}=1+\left(\frac{\partial S}{\partial x^1}\right)^2=\frac{1}{1-(x^1-x^3)^2},\;g_{22}=2,\;g_{33}=1+\left(\frac{\partial S}{\partial x^3}\right)^2=g_{11},$

\noindent $g_{12}=g_{21}=\frac{\partial S}{\partial x^1}\frac{\partial S}{\partial x^2}=-\frac{x^1-x^3}{\sqrt{1-(x^1-x^3)^2}}=-g_{23}=-g_{32}$

\noindent $g_{13}=g_{31}=\frac{\partial S}{\partial x^1}\frac{\partial S}{\partial x^3}=-\frac{(x^1-x^3)^2}{\sqrt{1-(x^1-x^3)^2}}.$

Si on note $g$ le tenseur métrique, on vérifie bien que $\det g=2\frac{1-(x^1-x^3)^2}{1+(x^1-x^3)^2}>0$ puisque sur $\tilde{\ma{U}}$, on a $(x^1-x^3)^2<1$.

Ceci nous donne le lagrangien associé au système des deux particules liées : $L=\frac{1}{2}g_{ij}\dot{x}^{i}\dot{x}^j-\ma{V}$. La résolution des équations de Lagrange donne accès à l'orbite $t\mapsto (x^1(t),x^2(t),x^3(t))$, qui est une courbe paramétrée sur $M$ par le temps et qui indique le mouvement des deux particules liées dans le plan $\rr^2$.
\end{exemple}
\subsection{Premier groupe des équations de Maxwell et connexion électromagnétique.}\label{erer}\index{Maxwell (premier groupe d'équations)}
\noindent 

\subsubsection{Rapide rappel des principes fondamentaux de la relativité restreinte}

La mécanique de Newton privilégie les référentiels galiléens, référentiels en translation rectiligne uniforme par rapport à un 
un référentiel absolu dont on pose l'existence par principe. Les lois de la mécanique ne peuvent mettre en évidence un référentiel galiléen par rapport à un autre. \index{référentiel galiléen}

Le premier principe de la relativité restreinte affirme que \emph{la vitesse de la lumière par rapport à tout référentiel galiléen est une constante $c$.}

De ce fait on peut définir dans chaque référentiel galiléen un temps $t$ : un rayon lumineux issu de l'origine d'un référentiel galiléen de coordonnées $(x,y,z)$ met un temps $t=\frac{\sqrt{x^2+y^2+z^2}}{c}$ pour atteindre ce point.
Les coordonnées $(ct,x,y,z)$ sont appelées \emph{coordonnées galiléennes }ou \emph{inertielles} de l'espace $\rr^4$. Ces quatre coordonnéees ont la dimension d'une longueur.

Le principe de la mécanique de Newton s'étend maintenant relativité restreinte :

\noindent\textbf{principe de relativité d'Einstein}\\
  \emph{Les lois de la mécanique et de l'électromagnétisme  ne peuvent mettre en évidence un référentiel galiléen par rapport à un autre. }

Autrement dit, les transformations qui échangent deux systèmes de coordonnées inertielles laissent les lois de la mécanique et de l'électromagnétisme invariantes.
Supposons une ligne d'univers d'un photon (son orbite dans $\rr^4$). Considérons pour la décrire deux systèmes de coordonnées
inertielles $(ct,x,y,z)$ et $(ct',x',y',z')$. 
Alors $$\left(\frac{dx}{dt} \right)^2+\left(\frac{dy}{dt} \right)^2+\left(\frac{dz}{dt} \right)^2=\left(\frac{dx'}{dt'} \right)^2+\left(\frac{dy'}{dt'} \right)^2+\left(\frac{dz'}{dt'} \right)^2=c^2.$$
Ainsi les formes quadratiques $Q=-c^2dt^2+dx^2+dy^2+dz^2$ et $Q'=-c^2dt'^2+dx'^2+dy'^2+dz'^2$ s'anullent en même temps.

Nous admettons en relativité restreinte \emph{le principe d'inertie de Galilée} : 

\noindent\emph{un point matériel libre de toute action extérieures  admet dans des coordonnées galiléennes comme trajectoire une  ligne droite. }

\noindent Ce principe implique que les transformations qui échangent les coordonnées inertielles sont linéaires. 
En conséquence , un calcul direct montre que les formes $Q$ et $Q'$ doivent être proportionnelles. La coïncidence des référentiels inertiels à un instant donné fait de cette proportionnalité une égalité :  \begin{equation}\label{lolo}
-c^2dt^2+dx^2+dy^2+dz^2=-c^2dt'^2+dx'^2+dy'^2+dz'^2.
\end{equation}
Les transformations qui échangent les coordonnées inertielles sont donc celles qui laissent invariantes la forme quadratique
$ds^2=-c^2dt^2+dx^2+dy^2+dz^2$. Elles forment un groupe  de $\mathcal{G}l(\rr^4)$: le \emph{groupe de Lorentz}.
La métrique de $\rr^4$ associée à cette forme sera notée $\la,\ra_L$. L'exercice \ref{clebard}  donne un exemple de transformation de Lorentz. 
\begin{exo}\label{clebard}
\begin{enumerate}
\item On pose dans cet item $c=1$. On considère le plan $\rr^2$ muni de la forme quadratique $q(t,x)=-c^2t^2+x^2=-t^2+x^2$.
Montrer que le groupe orthogonal de $(\rr^2,q)$ est constitué de deux types de transformations représentées dans la base canonique de $\rr^2$ par des matrices de déterminant $1$ de la forme
$\left(\begin{matrix}
\cosh\theta&\sinh\theta\\
\sinh\theta&\cosh\theta
\end{matrix}\right)$  et des matrices de déterminant $-1$ de la forme $\left(\begin{matrix}
\cosh\theta&\sinh\theta\\
-\sinh\theta&-\cosh\theta
\end{matrix}\right)$ .
On parlera pour le premier type de rotations hyperboliques et pour le second type de symétries hyperboliques.

\item On considère dans $\rr^2$ la droite $D=\vect\{(0,1)\}$ ainsi que deux points $O$ et $O'$ de $D$, puis deux  référentiels galiléens $\mathcal{R}$ d'origine $O'$ et de coordonnées $(ct',x')$ et $\mathcal{R}_0$ d'origine $O$ et de coordonnées $(ct,x)$ de telle manière qu'un observateur lié au repère affine $(O,x)$ voit s'éloigner l'origine $O'$ du repère affine
$(O',x')$  à la vitesse constante $v$ portée par $D$. On suppose que $\mathcal{R}$ et $\mathcal{R}_0$ coïncident 
à un moment donné.
\begin{enumerate}
\item Montrer que la transformée de Lorentz qui fait passer des coordonnées $(ct',x')$ aux coordonnées $(ct,x)$ est une rotation hyperbolique.
\item \'{E}tablir les relations
$\left\{
\begin{matrix}
t'&=&\gamma(t-\frac{v}{c^2}x)\\
x'&=&\gamma(-vt+x)
\end{matrix}\right.
$

\noindent où \;$\gamma =\frac{1}{\sqrt{1-\frac{v^2}{c^2}}}$ ( $\gamma $ est appelé le \emph{facteur de Lorentz})
\item Extension de la question précédente : on considère $\rr^4$, la droite $D=\vect\{(0,1,0,0)\}$, $\mathcal{R}=(O',t',x',y',z')$ et $\mathcal{R}_0=(O,t,x,y,z)$,  de telle manière qu'un observateur lié au repère affine $(O,xy,z)$ voit s'éloigner l'origine $O'$ du repère affine
$(O',x'y',z')$  à la vitesse constante $v$ portée par $D$. Alors
\[\left\{
\begin{array}{ll}
t'&=\gamma(t-\frac{v}{c^2}x)\\
x'&=\gamma(-vt+x)\\
y'&=y\\z'&=z
\end{array}\right.
\]
\end{enumerate}
\item Montrer que la forme volume $cdt\wedge dx$ est invariante par la transformée de Lorentz.
\end{enumerate}
\end{exo}

\noindent Les lois de la mécanique ou de l'électromagnétisme s'exprimeront donc sous la forme d'égalité de (champs de) vecteurs invariants par transformées de Lorentz ( appelés \emph{quadrivecteurs}) ou  sous forme covariante, par l'égalité de formes différentielles invariantes par transformées de Lorentz
\subsubsection{Dynamique du point en relativité restreinte}\label{diporr}

Nous voulons, préalablement à l'exposé des lois de l'électromagnétisme, étendre à l'espace de Minkowski la dynamique newtonienne en une dynamique invariante par le groupe de Lorentz. Cela permettra 
\begin{enumerate}
\item d'introduire le quadrivecteur force de Lorentz qui est la base de la théorie de l'électromagnétisme,
\item d'avoir un guide pour établir les lois du mouvement en mécanique relativiste restreinte à partir des lois classiques exposées dans la section \ref{dynemicon} de l'appendice $6$.
\end{enumerate}
Introduisons les notations de Minkowski.
 L'espace de Minkowski est la variété $\rr^4$ équipée de la forme quadratique de Lorentz. Il est recouvert par une carte locale dont les coordonnées sont $\{x^{i}\}_{i\in\{0,1,2,3,4\}}$ avec $x^0=ct, x^1=x,x^2=y,x^3=z$,  appelées classiquement \emph{coordonnées réduites}. Dans ces coordonnées la forme de Lorentz s'écrit $ds^2=-(dx^0)^2+\sum_{i=1}^3(dx^{i})^2$. La forme bilinéaire associée à cette forme quadratique se notera  $\la,.\ra_L$.
 
 \noindent Un repère affine est déterminé par une origine et un système de coordonnées dans un domaine de carte locale. Un tel repère est appelé rappelons-le un référentiel en physique que l'on peut noté sous la forme 
 $\mathcal{R}=\{O, (x^0,x^1,x^2,x^3)\}$. Un point $E$ de l'espace de Minkowski est aussi appelé \emph{événement} et si on l'écrit par ses coordonnées dans $\mathcal{R}$ on parlera  d'un événement observé par un observateur lié à $\mathcal{R}$. La $i$-ième coordonnée de $E$ dans $\mathcal{R}$ se note alors $x^{i}(E)$. Par exemple $x^{i}(O)=0$ pour tout $i\in\{0,1,2,3\}$.
 Deux événements $E_1$ et $E_2$  sont \emph{simultanés dans $\mathcal{R}$ (ou pour un observateur lié à $\mathcal{R}$) }
  si $x^0(E_1)=x^0(E_2)$.
 \begin{exo}\label{brut} \hspace{1cm}\\
\noindent Considérons l'espace de Minkowski muni des coordonnées réduites. On considère dans $\rr^4$ la droite $D=\vect\{(0,1,0,0)\}$, deux points $O$ et $O'$ de $D$, puis deux  référentiels galiléens $\mathcal{R}$ d'origine $O'$ et muni  des coordonnées $((x^0)',(x^1)',(x^2)',(x^3)')$ et $\mathcal{R}_0$ d'origine $O$ muni des coordonnées $((x^0),(x^1),(x^2), (x^3))$de telle manière qu'un observateur lié au repère affine $\mathcal{R}_0$ voit s'éloigner l'origine $O'$ du repère affine
$\mathcal{R}$  à la vitesse constante $v$ portée par $D$. On suppose que $\mathcal{R}$ et $\mathcal{R}_0$ coïncident 
à un moment donné. 
\begin{enumerate}
\item Montrer que la transformée de Lorentz s'écrit en coordonnées réduites :
\begin{equation}\label{rafu}
\left\{
\begin{array}{rcl}
(x^{0})'&=&\gamma(x^0-\frac{v}{c}x^1)\\
(x^{1})'&=&\gamma(-\frac{v}{c}x^0+x^1)\\
(x^2)'&=&x^2\\
(x^3)'&=&x^3
\end{array}\right.
\end{equation}
 Comme les deux dernières coordonnées dans $\mathcal{R}_0$ et $\mathcal{R}$ sont égales deux à deux on se restreint dans la suite de l'exercice  aux référentiels $\{O, (x^0,x^1)\}$ et  $\{O, ((x^0)',(x^1)')\} $ encore nommés $\mathcal{R}_0$ et $\mathcal{R}$.
 \item Montrer que dans le référentiel $\mathcal{R}_0$ les axes $(x^0)'$ et $(x^1)'$ du référentiel $\mathcal{R}$ ont pour équations respectivement $x^0=\frac{c}{v}x^1$ et $x^0=\frac{v}{c}x^1$.
On notera que l'axe temporel $(x^0)'$ a une pente $>1$ et l'axe spatial $(x^1)'$ une pente $<1$. 
\item Considérons un événement $E$ de coordonnées $(a,b)$ dans $\mathcal{R}$ et $(a',b')$ dans $\mathcal{R}_0$. Les événements simultannés à $E$ pour un observateur de $\mathcal{R}_0$ a pour équation dans les coordonnées de $\mathcal{R}_0$ : \begin{equation}\label{symu1}x^0=a.\end{equation} Montrer que dans ces mêmes coordonnées les événements simultanés à $E$ pour un observateur lié à $\mathcal{R}$ a pour équation dans les coordonnées de $\mathcal{R}_0$ :
\begin{equation}\label{symu2}x^0=a+\frac{v}{c}(x^1-b)\end{equation}
Les ensembles déterminés par les équations (\ref{symu1}) et (\ref{symu2}) sont distincts.
\noindent Ceci précise le fait que la notion de simultanéité est lié à l'observateur et n'est plus un absolu comme dans la mécanique newtonnienne.
\item On considère une horloge liée à $\mathcal{R}_0$. Cette horloge a, à deux instants $t^1$ et $t^2$ les coordonnées réduites $( \tau^1, a)$ et $( \tau^2, a)$ dans le référentiel $\mathcal{R}_0$ (où $\tau^{i}=ct^{i}$) et repésentent deux événements $E_1$ et $E_2$ observés par un observateur lié à $\mathcal{R}_0$. La durée séparant ces deux événements $E_1$ et $E_2$ dans $\mathcal{R}_0$ est égale à $(t^2)-(t^1)=\frac{1}{c}((\tau^1)-(\tau^2))=T_0$. Montrer qu'un observateur lié à $\mathcal{R}$ évalue cette durée à $T$ de sorte que $T=\gamma T_0$. On a donc $T> T_0$
\item On considère deux événements simultanés $E_1=(a,b)$ et $E_2=(a,\be)$ pour un observateur de $\mathcal{R}_0$  Le segment de droite $[E_1,E_2]$ apparaît comme une règle rigide pour un observateurs de $\mathcal{R}$, de longueur $l=\abs{b'-\be'}$. Montrer que la longueur $l$ évaluée par un observateur lié à $\mathcal{R}$ à un instant donné $t'$ est égale à $l=\frac{ l_0}{\gamma}$. On a donc $l< l_0$
\item On considère un parallélépipède rectangle dont un des cotés est porté par l'axe des $x^1$. Soit $V$, $V_0$ son volume estimé respectivement par un observateur de $\mathcal{R}$ et de $\mathcal{R}_0$. Monter que $V_0=\gamma V$. En déduire que la relation entre les mesures de Lebesgue de $\rr^3$, $dV$ et $dV_0$, évaluées dans $\mathcal{R}$ et $\mathcal{R}_0$ est
 \begin{equation}\label{begue}
dV_0=\gamma dV
\end{equation}
\end{enumerate}
\end{exo}

 \noindent Sur un ouvert de $\rr^4$ où cette métrique est définie négative, notons $:d\sigma=\sqrt{-ds^2}$ (qui a la dimension d'une longueur). Aussi définissons-nous sur cet ouvert le temps propre par 
\begin{equation}\label{temtrepro}
d\tau=\frac{1}{c}d\sigma
\end{equation}

\noindent Le quadrivecteur vitesse\index{quadrivecteur vitesse} d'une particule  se déplaçant à la vitesse $v=\sum_{i=1}^3v^{i}\frac{\partial}{\partial x^{i}}$ dans un référentiel galiléen ( de la mécanique newtonienne) de coordonnées $(x,y,z)$ se définit  par 
\begin{equation}\label{qvi}u=\left(\frac{dx^0}{d\sigma}, \frac{dx^1}{d\sigma},\frac{dx^2}{d\sigma},\frac{dx^3}{d\sigma}\right)=\sum_{i=0}^3 u^{i}\frac{\partial}{\partial x^{i}}.\end{equation}
Notant que $d\sigma=\frac{c}{\gamma} dt$ où $\gamma^{-1}=\sqrt{1-\frac{\norm{v}^2}{c^2}}$ et $\Vert.\Vert$ désigne la norme euclidienne standard de $\rr^3$, on a donc immédiatement les égalités :
\begin{equation}\label{quavi}
u^0=\gamma, u^{i}=\frac{\gamma}{c} v^{i}\;\mathrm{où}\; i\in\{1,2,3\} \;\mathrm{et}\;d\tau=\frac{dt}{\gamma}.
\end{equation}

 \noindent Remarquons que les composantes de $u$ sont sans dimension. Le \emph{quadrivecteur impulsion}\index{quadrivecteur impulsion} qui devrait intervenir dans dans la loi fondamentale doit avoir la dimension d'une masse par une vitesse. On le définit alors  par 
\[P=m_0cu,\] 
que l'on peut écrire encore :
 \begin{equation}\label{qas}P=m_0 c\gamma\frac{\partial}{\partial x^0}+\sum_{i=1}^3m_0 v^{i}\gamma\frac{\partial}{\partial x^{i}}=m_0 c\gamma\frac{\partial}{\partial x^0}+m_0v\gamma.\end{equation}
 Notons que $m_0v\gamma$ est précisément la correction de l'impulsion classique dans laquelle $\gamma=1$. Le quadrivecteur impulsion conduit au quadrivecteur $\frac{dP}{d\tau}$ appelé \emph{quadrivecteur force de Minkowski}.\index{quadrivecteur force de Minkowski}
 La loi fondamentale de la dynamique du point sera par analogie avec la loi de Newton une relation du type 
\begin{equation}\label{ainsifond}
\frac{dP}{d\tau}=F
\end{equation} 
où $F$ est un quadrivecteur qui doit contenir dans ses composantes la correction relativiste de la force classique de la mécanique de Newton. 
\begin{exo}\label{secret}
Montrer que $u$ est unitaire avec $\la u,u\ra_L=-1$. En déduire que $\la u,F\ra_L=0$ où $F$ est défini par l'équation (\ref{ainsifond}).
\end{exo}
\noindent Examinons l'équation (\ref{ainsifond}). Cette relation s'écrit encore compte tenu de l'équation (\ref{temtrepro}) : $m_0c^2\frac{du}{d\sigma}=F$. 
On note $F=\sum_{i=0}^3F^{i}\frac{\partial}{\partial x^{i}}$.  Cette loi fondamentale, grâce aux équations (\ref{quavi}), implique
en particulier les trois relations 
\begin{equation}\label{promesse}
\frac{d}{dt}(m_0\gamma v^{i})=\frac{F^{i}}{\gamma}=f^{i}, i=1,2,3.
\end{equation}
Or le membre de gauche de cette égalité est la correction relativiste de la dérivée temporelle de la quantité de mouvement en mécanique newtonienne, dans laquelle $\gamma =1$. Il faut donc, si on accepte la relation fondamentale (\ref{ainsifond}), interpréter $f^{i}$ dans l'égalité (\ref{promesse})   pour $i\in\{1,2,3\}$ comme les composantes de la force extérieure agissant sur la particule.
L'orthogonalité de $u$ et $\frac{dP}{d\tau}$ (cf. l'exercice \ref{secret}) permet d'interpréter la composante $F^0$. Commençons par écrire la composante $0$ de l'équation vectorielle (\ref{ainsifond}). Elle s'écrit directement :
\begin{equation}\label{techos}
m_0 \frac{d}{dt}(c\gamma)=\frac{F^0}{\gamma}.
\end{equation}
De l'équation $\la F,u\ra_{L}=0$ on a: $F^0\gamma=\sum_{i=1}^3F^{i}\frac{\gamma}{c}v^{i}=\frac{\gamma^2}{c}\sum_{i=1}^3f^{i}v^{i}.$ D'où :
\begin{equation}\label{unipro}
F^0=\frac{\gamma}{c} \sum_{i=1}^3 f^{i}v^{i}=(\frac{\gamma}{c} )f.v 
\end{equation}
Cette relation et l'égalité (\ref{techos}) permettent d'écrire 
\begin{equation}\label{enerj}
\frac{d}{dt}(m_0 c^2\gamma)=f.v
\end{equation} 
En mécanique classique, $f.v$ est la dérivée de l'énergie cinétique, ce qui incite à interpréter $E=m_0c^2\gamma$ comme l'énergie de la particule. \`{A} la différence de l'énergie cinétique, $E$ n'est pas nulle quand $v=0$ : elle est alors égale à $E_0=m_0 c^2$.
Si l'on examine $T=E-E_0$, on observe que $T=\frac{1}{2}m_0 \norm{v}^2 + \eta$ où $\eta$ est d'autant plus négligeable devant
$\frac{1}{2}m_0 \norm{v}^2 $ que le quotient $\frac{\norm{v}}{c}$ est petit. $T$ est donc un correctif relativiste de l'énergie cinétique classique. On l'appellera également \emph{énergie cinétique}.
\`{A} la lumière de ceci nous pouvons réécrire la composante $P^0$ de l'impulsion : \begin{equation}\label{wert}P^0=\frac{E_0}{c} \gamma.\end{equation}
\noindent Nous venons d'établir qu'une particule de masse au repos $m_0$ a une énergie au repos $E_0=m_0c^2$.
Réciproquement, on peut énoncer le théorème suivant.
\begin{theorem}
Supposons qu'une particule au repos dans un référentiel galiléen $\ma{R}_0$ ait une énergie $E_0$. Alors dans tout référentiel galiléen $\ma{R}$, cette particule a l'impulsion d'une particule de masse au repos $m_0=\frac{E_0}{c^2}$.
\end{theorem}
\begin{preuve}{}
On notera $(x^0,x^1,x^2,x^3)$ les coordonnées dans $\mathcal{R}_0$ et $((x^0)',(x^1)',(x^2)',(x^3)')$) les coordonnées dans $\mathcal{R}$.
 On supposera  que l'axe des $x$ des deux référentiel sont confondus et que l'origine de $\mathcal{R}_0$ se déplace sur l'axe des $x$ avec une vitesse $v$ par rapport au référentiel $\mathcal{R}$. La transformée de Lorentz passant des coordonnées de $\mathcal{R}_0$ aux coordonnées de $\mathcal{R}$ a pour matrice, d'après l'exercice \ref{brut} :
\[\left(
\begin{matrix}
\gamma&\gamma v/c&0&0\\
\gamma v/c&\gamma&0&0\\
0&0&1&0\\
0&0&0&1
\end{matrix}
\right)
\]
Avec les notations de l'énoncé du théorème, le quadrivecteur impulsion de la particule dans le référentiel $\mathcal{R}_0$ est 
$P_0=\frac{E_0}{c}\frac{\partial}{\partial x^0}$.
Le quadivecteur impulsion s'écrit donc dans $\mathcal{R}$ : $P=\left(\frac{E_0}{c^2}\right)c \gamma\frac{\partial}{\partial x^0}+
\left(\frac{E_0}{c^2}\right)v\gamma\frac{\partial}{\partial x^{1}}$. L'équation (\ref{qas}) montre que la masse au repos de cette particule est $\frac{E_0}{c^2}$.
\end{preuve}
Ce théorème traduit l'équivalence entre la masse et l'énergie. On comprend le sens de cette équivalence dans l'expérience suivante.
Supposons qu'une particule au repos dans un référentiel galiléen de masse au repos $m_0$, perd tout en restant au repos avec une énergie $\mathcal{E}_0$. Cette énergie équivaut à une masse au repos $\frac{\mathcal{E}_0}{c^2}$ et sa nouvelle masse au repos est $m_0- \frac{\mathcal{E}_0}{c^2}$.
\begin{remark}
Nous avons parlé de masse au repos, qu'il faut distinguer de la masse relativiste que l'on peut voir apparaître  dans la composante spatiale (c'est à dire le vecteur déterminé par les $3$ dernières composantes) de $P$, dans l'équation (\ref{qas}). Cette composante spatiale peut s'écrire $mv$ si l'on pose 
$m=m_0\gamma$ qu'on appelle aussi la \emph{masse relativiste } de la particule.\index{masse relativiste}
\end{remark}
\begin{remark}\label{inertierr}
Un observateur qui veut vérifier expérimentalement la loi de (Newton)-Minkowski (\ref{ainsifond}), se placerait dans un référentiel pour faire des mesures, puisqu'alors il disposerait de coordonnées. Supposons qu'il soit lié à un référentiel galiléen et que le quadrivecteur force agissant sur une particule test soit nul. Alors la loi de Minkowski montre que le le quadrivecteur vitesse $u$ de la particule vérifie l'équation $\frac{du}{ds}=0$. Ce qui montre que la ligne d'univers de la particule est une droite, donc indépendamment des coordonnées une géodésique de l'espace de Minkowski. Et on retrouve dans le cadre de la relativité restreinte le principe d'inertie de Galilée dont l'énoncé précis est le suivant.
\begin{principe}[d'inertie de Galilée en relativité restreinte]\label{pringarr}
\hspace{1cm}\\
Un point matériel isolé admet comme ligne d'univers dans l'espace de Minkowski une géodésique. Dans un système de coordonnées galiléenne, la métrique $ds^2$ est strictement négative le long de cette ligne pour une particule de masse non nulle (voir exercice \ref{secret}). La métrique de Lorentz est nulle le long  de la ligne d'univers d'un photon. 
\end{principe}
\end{remark}
\subsubsection{Première équation tensorielle de Maxwell}\index{première équation de Maxwell}

En théorie classique de l'électromagnétisme dans le vide  on dispose d'un champ électrique $E$ et d'un champ magnétique $B$ définis sur l'espace euclidien $(\rr^{3},\la\;\ra)$. Ces champs sont mis en évidence expérimentalement à partir  de la force de Lorentz $f$\index{force de Lorentz} qu'ils créent sur une charge $e$ en mouvement avec une vitesse $v$ 
(vitesse relative au référentiel $\rr^{3}$) . Cette force s'écrit :
 \[f_\ma{L}
=e(E+\frac{v}{c}\times B).\] On associe au champ électrique $E$ la $1$-forme 
\begin{equation} \label{elco}
\mathcal{E}^{1}=\la  E,\;\ra=E_{i}dx^{i}=E_1 dx+E_2 dy+E_3 dz
\end{equation}
et au champ magnétique $B$ la $2$-forme
 \begin{equation}\label{themag}\mathcal{B}^{2}=i(B)(dx\wedge dy\wedge dz)=B^{1}dy\wedge dz+B^{2 }dz\wedge dx+B^{3}dx\wedge dy.
 \end{equation}

\noindent Par symétrie on peut également également définir une $2$-forme champ électrique $\mathcal{E}^2$ et une $1$-forme champ magnétique $\mathcal{B}^1$, en posant
\begin{align}
\mathcal{E}^2&=i(E)dx\wedge dy\wedge dz=E^{1}dy\wedge dz+E^{2 }dz\wedge dx+E^{3}dx\wedge dy,\label{lasyme}\\
\mathcal{B}^1&=\la B,\;\ra\label{lasymb}
\end{align}
La théorie relativiste de l'électromagnétisme peut se fonder sur le principe physique suivant. 
\begin{principe}\label{lolorenz}
 Le vecteur force de Lorentz $f$ se prolonge dans l'espace de Minkowski en un quadrivecteur $\ma{F}_\ma{ML}$ vérifiant l'équation fondamentale (\ref{ainsifond}). 
\end{principe}

\`{A} partir de ce principe le prolongement est unique. En effet si  $\ma{F}_\ma{ML}=F^0\frac{\partial}{\partial x^0} +F^{i}\frac{\partial}{\partial x^{i}}=F_0\frac{\partial}{\partial x^0}+F$, on a d'après les équations (\ref{promesse}) et (\ref{unipro}) $F=\gamma f_\ma{L}$ et $F^0=\frac{\gamma e}{c}\la E,v\ra$.
\begin{definition}\index{quadrivecteur de Minkowski-Lorentz}
 Le quadrivecteur $F_\ma{ML}=\frac{\gamma e}{c}\la E,v\ra\frac{\partial}{\partial x^0}+\gamma e(E+\frac{v}{c}\times B)$ est le quadrivecteur de Minkowski-Lorentz. On a donc :
 \begin{equation}\label{emel}
 F_\ma{ML}=\gamma e\left(\frac{1}{c^2}\la E,v\ra\frac{\partial}{\partial t}+ E+\frac{v}{c}\times B\right).
 \end{equation}
 \end{definition}
 
\noindent On associe au quadrivecteur force de Minkowski-Lorentz  une $1$-forme $f^{1}$ par l'intermédiaire de la métrique de Lorentz : $$f^{1}=\la F_{\ma{ML}},.\ra_{\mathrm{L}}.$$ 
Il convient ici de noter, sous forme d'une remarque, un aspect de la démarche mathématique de l'étude des champs de la physique. 
\begin{remark}\label{vecoufor?}
De façon générale si $X$ est un champ de vecteurs sur une variété pseudo-riemmanienne $(M,g)$, on peut lui associer canoniquement la $1$-forme $\al=g(X,.)$ Soit $h$ une isométrie  de $M$ sur lui-même. Dire que $X$ est invariant par $h$ signifie que $dh(X)=X\circ h$ ou encore : pour tout $x\in M$, $dh(x)(X(x))=X(h(x))$.
que $\al$ est invariante par $h$ signifie que $h^*\al=\al$.
Alors il y a équivalence entre les deux items suivants :
\begin{enumerate}
\item $X$ est invariant par $h$ .
\item $\al$ est invariante  par $h$.
\end{enumerate}
La machinerie du calcul différentiel sur les formes est plus aisée que celle sur les champs de vecteurs, ce qui explique pourquoi on travaille avec les formes covariantes des champs dans les structures pseudo-riemaniennes.
\end{remark}
\begin{exo}\label{falco}
Démontrer l'équivalence des deux items de la remarque \ref{vecoufor?}.
\end{exo}
\noindent Le principe \ref{lolorenz} énoncé ci-dessus  va permettre d'établir la transformation des composantes les champs électriques et magnétiques par changement de coordonnées galiléennes. Conformément à la remarque \ref{vecoufor?} nous traiterons cette question à partir de la forme covariante $f^1$ du quadrivecteur de Minkowski-Lorentz (voir exercice \ref{gloups})

\noindent En s'inspirant de la théorie de l'élasticité (voir la section \ref{dynemicon} de l'appendice $6$), on définit le  \emph{tenseur (des contraintes) électromagnétique} par \index{tenseur (des contraintes) électromagnétique}
\begin{equation}\label{tenconmag}
 \mathcal{F}^{2}=c\mathcal{E}^{1}\wedge dt+ \mathcal{B}^{2}=\ma{E}^1\wedge dx^0+\ma{B}^2
 \end{equation}
  qui permet de calculer en chaque point du milieux la force de Lorentz. En effet, comme le montre l'exercice \ref{gral}, l'équivalent de la relation $f_{}=e(E+\frac{v}{c}\times B)$ en relativité restreinte s'écrit :
\begin{equation}\label{lor} f^{1}=-ei(u)\mathcal{F}^{2}
\end{equation}
Et maintenant une série de quatre exercices pour ne pas s'endormir.
 \begin{exo}\label{gral}
  Démontrer l'égalité (\ref{lor}).
  \end{exo}
 \begin{exo}\label{quadteconmag}
 \hspace{1cm}\\
  Montrer qu'il y a équivalence entre les deux items :
 \begin{enumerate}
  \item $F_\ma{LM}$ est un quadrivecteur.
 \item $\ma{F}^2$ est invariant par transformation de Lorentz.
 \end{enumerate}
 \end{exo}
 \begin{exo}\label{ortint}
 \hspace{1cm}\\Orthogonalité et produit intérieur.
 \begin{enumerate}
 \item Soit $X$ et $Y$ deux champs de vecteurs de l'espace de Minkowski. Soit $\al$ la $1$-forme définie par  $\al=\la X,.\ra_L$. Montrer que $X$ et $Y$ sont orthogonaux dans l'espace de Minkowski si et seulement si $i(Y)\al=0$.
 \item Retrouver à partir de l'égalité (\ref{lor}) que le quadrivecteur force de Minkowski-Lorentz est orthogonal au quadrivecteur vitesse en chaque point de la trajectoire de la particule chargée.
 \end{enumerate}
 \end{exo}
 \begin{exo}\label{gloups}
 En utilisant l'invariance de $\ma{F}^2$ par toute transformation de Lorentz, déterminer les relations entre les composantes des champs électrique et magnétique de  deux référentiels galiléens.
 Pour préciser les hypothèses,  on considère $\rr^4$, la droite $D=\vect\{(0,1,0,0)\}$, $\mathcal{R}_0=(O',ct',x',y',z')$ et $\mathcal{R}=(O,ct,x,y,z)$,  de telle manière qu'un observateur lié au repère affine $(O,xy,z)$ voit s'éloigner l'origine $O'$ du repère affine
$(O',x'y',z')$  à la vitesse de norme  constante $v$ portée par $D$. On note $E'^{i}$ et $B'^{i}$ les composantes des champs électriques et magnétiques dans $(O',x',y',z')$ et $E^{i}$ et $B^{i}$ les composantes de ces mêmes champs dans $(O,x,y,z)$. \'{E}tablir  les relations :
\[\left\{
\begin{array}{rcl}
E'^1=E^1,&
E'^2=\gamma(E^2-\frac{v}{c}B^3),&
E'^3=\gamma(E^3+\frac{v}{c}B^2),\\
B'^1=B^1,&
B'^2=\gamma(\frac{v}{c}E^3+B^2),&
B'^3=\gamma(-\frac{v}{c}E^2+B^3).
\end{array}
\right.
\]
 \end{exo}

  \noindent Le tenseur électromagnétique permet, en dehors de son interprétation mécanique exprimée par l'égalité (\ref{lor}) d'exprimer synthétiquement le premier groupe des équations de Maxwell.

\noindent \textbf{Notation}

\noindent$\mathbf{d}$ désigne la différentiation extérieure sur $\rr^{3}$ et $d$ la différentiation exté\-rieure sur $\rr^{4}$
\begin{exo}\label{rougi}
On se place sur $\rr^4$ munie des coordonnées $(t,x^0,x^1,x^2)$.
Soit $\alpha=\sum_{i<j}\alpha_{ij}dx^{i}\wedge dx^j+\alpha_{0i} dt\wedge dx^{i}$ une $2$-forme sur $\rr^4$.

\noindent On pose $\frac{\partial \alpha}{\partial t}=\frac{\partial \alpha_{ij}}{\partial t}dx^{i}\wedge dx^j +\frac{\partial \alpha_{0i}}{\partial t}dt\wedge dx^{i} $. Vérifier que $d\alpha=\mathbf{d}\alpha+dt \wedge \frac{\partial \alpha}{\partial t} .$
\end{exo}
\begin{principe}[Premier groupe des équations de Maxwell]  \hspace{10cm}
 
\noindent Le tenseur des con\-traintes électromagnétique vérifie l'égalité :
 \begin{equation}\label{maxwi}
d\mathcal{F}^{2}=0 
\end{equation}
 Ce qui équivaut à 
 \begin{equation}\label{max}
 \begin{dcases}
 \mbf{d}\mathcal{E}^{1}=-\frac{1}{c}\frac{\partial \mathcal{B}^{2}}{\partial t} \\
\mbf{d}\mathcal{B}^{2}=0
\end{dcases}
\end{equation}
où on a posé suite à l'égalité (\ref{themag}) et à l'exercice \ref{rougi} :\[\mfrac{\partial \mathcal{B}^{{2}}}{\partial t}=\mfrac{\partial B^{{1}}}{\partial t} dy\wedge dz+
\mfrac{\partial B^{{2}}}{\partial t} dz\wedge dx+\mfrac{\partial B^{{3}}}{\partial t} dx\wedge dy\]
Les équations (\ref{max}) sont les équations de Maxwell covariantes du premier groupe des équations de Maxwell.
\end{principe}
\noindent L'équation (\ref{maxwi}) équivaut à  (\ref{max}) puisque l'on a, d'après l'exercice \ref{rouge} : \[d(\mathcal{E}^{1}\wedge dt)=\mbf{d}\mathcal{E}^{1}\wedge dt, \; \mathrm{et}\;d\mathcal{B}^{2}=\mfrac{\partial \mathcal{B}^{2}}{\partial t}\wedge dt+\mbf{d}\mathcal{B}^{2}.\]
Notons que les équations de Maxwell obtenues sont bien invariantes par toute transformée de Lorentz dès que l'on admet le principe \ref{lolorenz}.
\begin{corollary}\label{lepote}
Sur toute partie simplement connexe de $\rr^{4}$ où $\mathcal{F}^{2}$ est différentiable, il exis\-te une $1$-forme \emph{potentiel} $\ma A=\Phi dt +A_{\alpha}(t,x^{1},x^{2},x^{3})dx^{{\alpha}}$ telle que $\mathcal{F}^{2}=d \ma A$. On note $\bf{ A}$ la $1$-forme de $\rr^{3}$ définie par $\mbf{A}=A_{\alpha}(t,x^{1},x^{2},x^{3})dx^{{\alpha}}.$ 
On a donc :
\[\mathcal{E}^{1}=\frac{1}{c}\left(\mbf{d}\Phi-\frac{\partial \mbf{A}}{\partial t}\right)\; \;\mathrm{et}\;\; \mathcal{B}^{2}=\mbf{d}\mbf{A}.\]
 \end{corollary}
 La fonction $\Phi$ est appelée \emph{potentiel scalaire}\index{potentiel scalaire} et le vecteur $(A_{1},A_{2},A_{3})$ est appelé \emph{potentiel vecteur}.\index{potentiel vecteur}
 \begin{exo}\label{klaciko}
 Soient $X$ et $Y$ deux champs de vecteurs différentiables sur $\rr^3$ muni du produit scalaire euclidien standard $\la,\ra$, $\omega=dx\wedge dy\wedge dz$ la forme volume canonique de $\rr^3$. On considère la $1$-forme $\zeta= \la X,.\ra$ et la $2$-forme $\xi=i(X)\omega$.
 
\begin{enumerate}
 \item Montrer qu'il existe un unique champ de vecteurs différentiable $Y$ sur $\rr^3$ tel que $d\zeta=i(Y)\omega$. 
 Par définition $Y=\rot (X)$ : $Y$ est le rotationnel de $X$.
 \item Monter qu'il existe une unique fonction $\lambda$ sur $\rr^3$ telle que $d\xi=\lambda\omega$.
 Par définition $\lambda=\Div X$ : $\lambda$ est la divergence $X$.
 \item Montrer que le premier groupe des équations de Maxwell s'écrit (classiquement) dans $\rr^3$
 : \[\rot E=-\frac{1}{c}\frac{\partial B}{\partial t},\;\;\;\;\Div B=0.\]
 \end{enumerate}
 \end{exo}
 \subsubsection{Connexion électromagnétique et principe d'invariance de jauge.}
 \noindent\emph{Dans le reste de la section \ref{conclaqua}, l'exposé des principes d'invariance de jauge \ref{conek} et \ref{coconek} se fera en posant $c=1$}
 
 \noindent Le tenseur des contraintes électromagnétique est invariant par transformée de Lorentz. Si $\mathcal{L}$ est une transformée de Lorentz, elle fait passer d'une carte $U$ de $\rr^{4}$ à une carte $V$. En terme de physique elle transforme un référentiel galiléen en un référentiel galiléen. L'invariance de $\mathcal{F}^{2}$ par $\mathcal{L}$  s'écrit $\mathcal{L}^{*}\mathcal{F}^{2}=\mathcal{F}^{2}$.
 On a donc à partir du corollaire \ref{lepote}  l'égalité $d(\mathcal{L}^{*}\ma A-\ma A)=0$, ce qu'on peut écrire $\ma A_{U}-\ma A_{V}={d}f_{UV}$, si $\ma A_{U}$ représente $\ma A$ dans les coordonnées du référentiel $U$, et $\ma A_{V}$ représente $\ma A$ dans les coordonnées du référentiel $V$.
 Si l'on note $M^{4}$ l'espace de Minkowski, on  considère cet espace les deux fibrés triviaux : $M^{4}\times \rr^{2}$ et son complexifié $M^{4}\times \cc^{}$. Considérons la $1$-forme 
 \begin{equation}\label{sliman}
\omega=-\frac{ie}{\hbar}\ma A.
 \end{equation}
  La normalisation $-\frac{ie}{\hbar}$ sera justifiée par la physique dans la suite du texte. Elle est représentée dans les cartes $U$ et $V$ par les formes $\omega_{U}$ et $\omega_{V}$ de sorte que $\omega_{U}-\omega_{V}=-\frac{ie}{\hbar} df_{UV}$. Si on considère pour toutes les cartes $U$ et $V$ de $M^{4}$, d'intersection non vide, la famille de complexes \begin{equation}\label{cuve}C_{UV}=\exp(-\frac{ie}{\hbar}f_{UV}),\end{equation} on a immédiatement \begin{equation}\label{chacra}\omega_{U}-\omega_{V}=C_{UV}^{{-1}}dC_{UV}\end{equation}
  où l'on reconnait l'égalité (\ref{chacal}).
 On en déduit que \textit{ si les $c_{UV}$ constituent une famille de cocycles}, alors les formes $\omega$ définissent sur $M^{4}\times \cc$ une con\-nexion. Cette connexion s'appelle  naturellement la \textbf{con\-nexion électromagnétique}.
 \index{connexion électromagnétique}La condition est donc que l'on ait chaque fois que cela a un sens $C_{UV}C_{VW}=C_{UW}$. \begin{principe}\label{conek}(Principe d'invariance de jauge. \'{E}noncé géométrique)
 
\noindent Les fonctions $f_{UV}$ sont telles que  $C_{UV}$ forment une famille de cocycles pour tous les ouverts de cartes $U$, $V$, d'intersection non vide.
\end{principe}
Cette condition entraine  un autre  principe de l'électrodynamique quantique, \textit{ le principe d'invariance de jauge}. Nous allons en effet montrer que la connexion électromagnétique est  la justification géométrique  de l'équation de Schrödinger. Le principe d'invariance de jauge décrit comment change la fonction d'onde solution de l'équation de Schrödinger lorsqu'on perturbe le potentiel par une forme exacte. 

 \subsection{Equations de Lagrange, équations d'Hamilton et équation de Schrödinger en coordonnées cartésiennes de $\rr^{3}$}
 \subsubsection{Du lagrangien au hamiltonien}\index{lagrangien}\index{hamiltonien}
 
 \noindent Rappelons succintement comment passer du lagrangien au hamiltonien, des équations de Lagrange aux équations d'Hamilton.
 Le lagrangien est défini sur l'espace $TM$ du fibré tangent à l'espace \textit{de configuration} représenté par la variété différentiable $M$. Le hamiltonien est défini sur l'espace $T^{*}M$
 du fibré cotangent sur $M$ à partir de la donnée du lagrangien $L$, et ceci de la fa\c con suivante. On se donne des coordonnées locales $(q^1,\cdots,q^{n})$ sur $M$. 
 Sur ces coordonnées se définissent des coordonnées locales $(q^1,\cdots,q^{n},\dot{q}^{1},\cdots,\dot{q}^{n})$ 
 (symboliquement notées $ (q,\dot{q})$) sur $TM$ et $(q^1,\cdots,q^{n},p_{1},\cdots,p_{n})$ sur $T^{*}M$
  (symboliquement notées $ (q,p)$). On considère l'application $\mathfrak{L}$ de $TM$ vers $T^{*}M$ définie 
  en coordonnées locales par $(q,\dot{q})\to (q,p=\frac{\partial L}{\partial \dot{q}})$.  L'application $\mathfrak{L}$ est appelée \textit{transformée de Legendre}.\index{transformée de Legendre} Lorsque $\det(\frac{\partial^{2}L}{\partial \dot{q}^{i} \dot{q}^{j}})\not=0$ ( $\mathfrak{L}$ est alors dite \textit{régulière})
la transformée de Legendre  est un difféomorphisme local et on peut exprimer $\dot{q}$ en fonction
   de $(q,p)$. On pose dans ce cas $H(q,p)=p\dot{q}-L(q,\dot{q})$ 
   (symboliquement $p\dot{q}=p_{i}\dot{q}^{{i}}$). Alors \begin{equation}\label{enfin}\frac{\partial H}{\partial q}=-\frac{\partial L}{\partial q}=-\frac{dp}{dt}\end{equation} d'après les équations (\ref{lag}) et on a directement
   $\frac{\partial H}{\partial p}=\frac{dq}{dt}$
   \begin{definition}\label{ham}		
 Les équations 
\begin{equation}\label{hamec}
\begin{dcases}
\mfrac{\partial H}{\partial q}=-\mfrac{dp}{dt}\\
\mfrac{\partial H}{\partial p}=\mfrac{dq}{dt}
\end{dcases}
\end{equation}
 sont appelées les équations d'Hamilton.
   \end{definition}
  Il découle immédiatement des équations d'Hamilton  que $\frac{dH}{dt}=0$, ce qui signifie que $H$ est constante sur l'orbite de la particule, paramétrée par le temps. $H$ représente l'énergie de la particule. 
  Par exemple si $M=\rr^{3}$ et si $L=\frac{1}{2} m (\dot{x}^{2}+\dot{y}^{2}+\dot{z}^{2})-\vv(x,y,z,t)=E_{c}-\vv$, on est dans le cadre d'une application de Legendre régulière et un calcul direct donne $H=E_{c}+\vv$.
  \begin{remark}\label{renewton}
  Cette remarque complète la réflexion de la section \ref{colagene} sur les équations de Newton.
  Nous y avons montré l'équivalence des équations de Newton intrinsèques exprimées par l'équation (\ref{nounou}) (ou encore l'équation (\ref{mouv}) qui traduit l'égalité (\ref{nounou}) en cordonnées locales) et des équations de Lagrange. Dans le cas où la transformée de Legendre est régulière, les équations de Lagrange sont elles même équivalentes aux équations de Hamilton. La structure riemannienne sur l'espace de configuration permet de donner une version duale des équations de Newton. Dans cette version duale de l'équation (\ref{nounou}) la connexion de Levi-civita sur le fibré tangent devient la connexion de Levi-Civita sur le fibré cotangent, le champ des vitesses $v$ devient 
  la $1$-forme $v^{*}=\la v,.\ra$. Notons-la en coordonnées locale $v^{*}=v^{*}_idq^{i}$. Alors c'est un petit exercice de montrer que $v^{*}_i=g_{ij}\dot{q}^j=p_i$ avec les mêmes notations que dans la section \ref{colagene}. Par ailleurs la forme duale du champ $\grad \mathcal{V}$ est la $1$-forme 
  $d\mathcal{V}$. Les équations de Newton duales sont alors exactement les équations (\ref{nunu}). 
  Alors en suivant la démonstration de la section \ref{colagene}, on aboutit à partir de (\ref{nunu}) à l'égalité (\ref{enfin}) et donc aux équations de Hamilton.
 En conclusion, les équations de Hamilton, dans le cas régulier sont équivantes aux équations de Newton sur le fibré cotangent muni de la connexion de Levi-Civita.
  \end{remark}
 
 \subsubsection{Du hamiltonien  à l'équation de Schrödinger. Cas d'une particule soumise à des forces dérivant d'un potentiel (il n'y a pas de champ électromagnétique).} 
 
 
 \noindent On sait qu'en mécanique quantique l'impulsion, le potentiel sont modélisés par des opérateurs qui opèrent sur la fonction d'onde associée à \og la particule\fg\- pour peu que ce mot désigne une réalité représentable à notre esprit \footnote{Comme par exemple un minuscule chat à la fois mort et vivant}.
  Ces opérateurs ont été inspirés par les grandeurs classiques à partir de recettes de quantification. Décrivons ce processus qui aboutira à un opérateur hamiltonien.  
  Reprenons les notations de la section précédente. L'énergie cinétique sur $M$ muni d'une métrique $\{g_{ij}\}$ s'écrit $E_{c}=\frac{1}{2}mg_{ij}\dot{q}^{{i}}\dot{q}^{{j}}$
   si $\dot{q}^{{i}}=\frac{dq^{i}}{dt}$. D'où \begin{equation}\label{duc}
   p_{i}=\frac{\partial L}{\partial \dot{q}^{{i}}}=mg_{ij}\dot{q}^{{j}}
   \end{equation}

     L' égalité entraine (\ref{duc}) permet d'écrire l'énergie cinétique en fonctions des moments (c'est à dire des $p_{i}$) :
 $E_{c}=\frac{1}{2m}g^{{ji}}p_{i}p_{j}$. On en déduit l'expression explicite de $H$ en fonction des moments :
    \begin{equation}\label{hamham}
   H=\frac{1}{2m}g^{{ji}}p_{i}p_{j}+\vv(q^{{i}},t).
   \end{equation}
   Considérons maintenant $\rr^{3}$ muni de sa structure euclidienne standard. L'équa\-tion (\ref{hamham}) s'écrit 
   \begin{equation}\label{lavraie}
   H=\frac{1}{2m}(p_{1}^{2}+p_{2}^{2}+p_{3}^{2})+\vv 
   \end{equation}

   La recette de quantification proposée par Schrödinger est de remplacer $p_{j}$ par l'opérateur $-i\hbar\frac{\partial}{\partial x^{j}}$ et $\vv$ par l'opérateur $\psi\to \vv. \psi$ où $\psi$ est la fonction d'onde. 
   Le facteur $i$ de l'opérateur impulsion, vient de la nécessité d'avoir un opérateur auto-adjoint, dont les valeurs propres, alors réelles, sont des grandeurs mesurables expérimentalement. Ceci est donc la première justification liée à la quantification du facteur $\frac{ie}{\hbar}$ (voir par exemple [\ref{riri}]). La géométrie va être en cohérence avec cette nécessité
   physique!

   Avec cette recette et dans le cas où la particule n'est soumise qu'à un potentiel $\ma{V}$, $H$ devient \emph{l'opérateur hamiltonien} \index{opérateur hamiltonien}\[\mathcal{H} : \psi\to-\frac{\hbar^{2}}{2m}\sum_{i=1}^{{3}}\frac{\partial^{2}\psi}{{\partial x^{i}}^{2}}+\vv\psi.\]
  Schrödinger proposa, sous ces hypothèses  le principe dynamique suivant.
   \begin{principe}\label{prish}(\'{E}quation de Schrödinger)\index{Schrödinger (équation, sans champ électromagnétique)}
   
   \noindent La fonction d'onde $\psi$ évolue au cours du temps en vérifiant l'équation 
   \begin{equation}
   i\hbar \frac{\partial \psi}{\partial t}=\mathcal{H}(\psi)
   \end{equation}
   \end{principe}
   Dans $\rr^{3}$, l'équation de Schrödinger s'écrit donc :
   \begin{equation}\label{eqschepla}
   i\hbar\frac{\partial \psi}{\partial t}=-\frac{\hbar^{2}}{2m}\sum_{i=1}^{{3}}\frac{\partial^{2}\psi}{{\partial x^{i}}^{2}}+\vv\psi.
      \end{equation}
      
    \subsubsection{Hamiltonien, lagrangien, hamiltonien en présence d'un champ électromagnétique.}  
    
 Dans le cas d'une loi décrite par des opérateurs différentiels, ceux-ci ne doivent pas être liés à un système de coordonnées mais avoir un sens intrinsèque. Mais dans ce cas ils sont formés à partir de dérivations covariantes et on doit pouvoir mettre en évidence une connexion qui sert à les écrire (voir par exemple la loi de Newton (équation (\ref{nounou})). Pour ce qui concerne l'équation de Schrödinger, c'est précisément la connexion électromagnétique qui intervient. Ceci est un argument supplémentaire en faveur du principe  \ref{conek}.
   Rappelons que si $\nabla$ est la connexion électromagnétique sur le fibré $M^{4}\times \cc$, on a d'après les équations (\ref{der}) et (\ref{dersec}) :
   
    $$\nabla \psi=d\psi-\frac{ie}{\hbar}\psi \ma A=\sum_{i=0}^{4}\Big(\frac{\partial \psi}{\partial x^{i}}-\frac{ie}{\hbar}A_{i}\psi\Big)dx^{i}=\sum_{i=0}^{4}\frac{\nabla \psi}{\partial x^{{i}}}dx^{{i}}$$ où
   
  \begin{equation}\label{rapcon}
\mfrac{\nabla \psi}{\partial x^{i}}=\mfrac{\partial \psi}{\partial x^{i}}-\mfrac{ie}{\hbar}A_{i}\psi .
 \end{equation} 
 Les coefficients $A_{i}$ sont définis dans le corollaire \ref{lepote} pour $i\in\{1,2,3\}$,  $A_{0}=\Phi$.
   On suppose maintenant qu'une particule de charge $e$ est soumise à des forces dérivant d'un potentiel. On lui associe dans $\rr^{3}$ un hamiltonien $H(q_{\alpha},p_{\alpha},t)$. Puis on rajoute un champ électromagnétique lié à un potentiel de forme covariante la $1$-forme du corollaire \ref{lepote}. Ceci conduit à introduire la perturbation du hamiltonien $H(q,p,t)$, définie par la fonction
   \[ H^{*}(q,p^{*},t)=H(q_{\alpha},p^{*}_{\alpha}-eA_{\alpha},t)-e\Phi.\]\index{opérateur hamiltonien avec champ électromagnétique}
   \begin{proposition}\label{nouham} La fonction $H^*$ est le \emph{hamiltonien associé à la particule chargée en présence d'un champ électromagnétique}. Elle vérifie les équations :
    \[
 \left\lbrace
 \begin{array}{l}
 
 \frac{\partial H^{{*}}}{\partial q}=-\frac{dp^{{*}}}{dt}\\
 \\
  \frac{\partial H^{{*}}}{\partial p^{{*}}}=\frac{dq}{dt}
 \end{array}
 \right.
\]

   \end{proposition}
  La démonstration de ce résultat peut être lue dans [\ref{franki}] au chapitre $XVI$. Je la retranscris dans l'appendice $1$ pour l'autonomie de ce document.
   On en déduit par le principe \ref{prish} l'équation de Schrödinger que vérifie la fonction d'onde attachée à la \og particule\;\fg :
   \begin{equation}\label{macbouc}
   i\hbar \frac{\partial \psi}{\partial t}=\mathcal{H^{{*}}}(\psi)
   \end{equation}
   où, conformément à l'équation (\ref{lavraie})
    \begin{equation}\label{nouvham}
    \mathcal{H^{{*}}}=\mfrac{1}{2m}\sum_{i=1}^{3}(p^{*}_{i}-eA_{i})^{{2}}+\vv-e\Phi 
   \end{equation}
   On a donc à partir de (\ref{nouvham}) :
   \begin{equation}\label{newhamlook}\mathcal{H^{{*}}}(\psi)=\mfrac{1}{2m}\sum_{i=1}^{3}(-i\hbar \frac{\partial}{\partial x^{i}}-eA_{i})^{{2}}(\psi)+\vv\psi-e\Phi \psi=-\frac{\hbar^{2}}{2m}\Delta_\tau\psi+\vv\psi-e\Phi \psi,\end{equation}  
   où $\Delta_\tau$ désigne le laplacien sur le fibré  trivial en droites complexes  sur l'espace de Minkowski, équipé de la connexion électromagnétique (voir équation (\ref{schon})).
   D'où par la même recette qui a conduit au principe \ref{prish},  l'équation de Schrödinger en présence du champ électromagnétique va s'écrire :
   
\noindent $ i\hbar \mfrac{\partial \psi}{\partial t}=-\frac{\hbar^{2}}{2m}\Delta_\tau\psi+\vv\psi-e\Phi\psi$
ou $i\hbar \left(\mfrac{\partial \psi}{\partial t}- \frac{ie}{\hbar}\Phi\right)\psi=-\frac{\hbar^{2}}{2m}\Delta_\tau\psi+\vv\psi$. Ainsi,
  \begin{equation}\label{schretwo}
   i\hbar \mfrac{\nabla \psi}{\partial t}=-\frac{\hbar^{2}}{2m}\Delta_\tau\psi+\vv\psi.
  \end{equation}
  Cette équation permet de donner un sens physique au principe d'invariance de jauge géométrique \ref{conek}
  
   Nous avons vu que si l'on adopte le principe \ref{conek}, alors passer du potentiel $A$ au potentiel $A+df$, revient à un changement de cartes : on  passe d'un domaine de carte $U$ à un domaine de carte V de telle manière à ce que
   \begin{enumerate}
   \item les dérivées covariantes $\nabla^{U}$ et $\nabla^{V}$ exprimées dans chacune des cartes vérifient la relation (\ref{chachacha}), dans laquelle $c_{UV}=\exp(-\frac{ie}{\hbar}f)$ (équation (\ref{cuve})),
    \item la fonction d'onde $\psi_{U}$ exprimée dans la carte $U$ s'écrit dans la carte $V$ : $\psi_{V}=C_{VU}\psi_{U}$ conformément à l'équation (\ref{chacord}).
    \end{enumerate}
   L'équation de Schrödinger (\ref{schretwo}) correspondant au potentiel $\ma A$ peut s'écrire dans la carte $U$ (voir remarques \ref{sioux}) :
    \[ i\hbar\mfrac{\nabla^{U}\psi_{U}}{\partial t}
    =-\frac{\hbar^{2}}{2m}\sum_{\alpha=1}^{3}\big(\frac{\nabla^{U}}{\partial x^{\alpha}}\big)\big(\frac{\nabla^{U}}{\partial x^{\alpha}}\big)(\psi_{U})
    +\vv\psi_{U}.\]
    On a donc, en utilisant successivement les relations (\ref{chiote}) et (\ref{chachacha}) qui sont les conséquences du principe \ref{conek} :
   \begin{align*}
   i\hbar \mfrac{\nabla^{V}\psi_{V}}{\partial t}&= i\hbar C_{VU}\nabla_{0}^{U}\psi_{U}\\
    &=-\frac{\hbar^{2}}{2m}C_{VU}\left(\sum_{\alpha=1}^{3}\big(\frac{\nabla^{U}}{\partial x^{\alpha}}\big)\big(\frac{\nabla^{U}}{\partial x^{\alpha}}\big)(\psi_{U}) +\vv\psi_{U}\right)\\
    &=-\frac{\hbar^{2}}{2m}C_{VU}
  \left(  \sum_{\alpha=1}^{3}\big(C_{UV}\frac{\nabla^{V}}{\partial x^{\alpha}}C_{VU}\big)\big(C_{UV}\frac{\nabla^{V}}{\partial x^{\alpha}}C_{VU}\big)(\psi_{U})+\vv\psi_{U} \right) \\
  &=-\frac{\hbar^{2}}{2m} \sum_{\alpha=1}^{3}\biggl(\frac{\nabla^{V}}{\partial x^{\alpha}}\Bigr)\Bigl(\frac{\nabla^{V}}{\partial x^{\alpha}}\Bigr)(\psi_{V})
    +\vv\psi_{V}.
    \end{align*}
   Le principe d'invariance de jauge s'énonce donc ainsi, si l'on n'évoque pas la connexion électromagnétique sous-jacente :
   \begin{principe}(Principe d'invariance de jauge. \'{E}noncé physique)\label{coconek}\index{invariance de jauge (principe d')}
   
   \noindent Si $\psi$ vérifie l'équation de Schrödinger  (\ref{macbouc}) sous un potentiel $A$, c'est à dire avec un opérateur hamiltonien donné par l'égalité (\ref{nouvham}), alors $\exp(-\frac{ie}{\hbar}f)\psi$ vérifie l'équation de Schrödinger sous le potentiel $\mathcal{A}+df$, c'est à dire avec un opérateur hamiltonien donné par (\ref{nouvham}), en y remplaçant les termes $A_{i}$ par 
   $A_{i}+\mfrac{\partial f}{\partial x^{{i}}}$ pour $i\in\{1,2,3\}$ et $A_0=\Phi$ par\; $\Phi+\frac{\partial f}{\partial t}$.
   \end{principe}
   \subsubsection{\'{E}quation de Schrödinger dans un champ électromagnétique en coordonnées locales d'une variété pseudo-rieman\-nienne }\label{eqschrochma}
Dans le cas où n'existe aucun champ électromagnétique sur $\rr^{3}$, l'équation de Schrödinger (\ref{eqschepla})) s'écrit :
\[  \frac{\partial \psi}{\partial t}=\frac{i\hbar^{}}{2m}\Delta \psi+\vv\psi\]
où $\Delta$ est le laplacien ordinaire sur $\rr^{3}$ et  $\mfrac{\nabla}{\partial t}=\mfrac{\partial}{\partial t}$.
Lorsqu'il existe un champ électromagnétique, l'équation de Schrödinger (\ref{schretwo}))
  s'écrit 
   \begin{equation}\label{schregéné}
 \mfrac{\nabla\psi}{\partial t }=\mfrac{i\hbar}{2m}\Delta_{\tau}\psi+\vv\psi
   \end{equation}
   où $\Delta_{\tau}$ est le laplacien sur le fibré en droites complexes $M\times \cc$ muni de la connexion électromagnétique. Ce laplacien est celui défini dans l'équation (\ref{schon}) avec
$\omega_{i}=\mfrac{-ie}{\hbar}A_{i}$.

Dans le cas général envisagé dans cette section, l'équation (\ref{schregéné}) doit encore être l'équation de Schrödinger qui décrit l'évolution de la fonction d'onde si  $\Delta_{\tau}$ est le laplacien donné par l'équation (\ref{schonschon}) dans laquelle on a toujours $\omega_{i}=\mfrac{-ie}{\hbar}A_{i}$ et les $\Gamma^{{s}}_{ij}$ les coefficients de Chritoffel de la connexion de Levi-Civita de la variété $M$, qui traduit l'existence physique d'un champ de gravitation.

\section {Connexion et courbures d'une variété pseudo-rie\-man\-nienne. }\label{mollusk}

\noindent L'objet de cette section est de définir le tenseur qui est à la base des équations du champ de gravitation en relativité générale, à savoir le tenseur de Cartan-Einstein-Hilbert. Pour cette raison nous ne développons ici la notion de courbure dans un fibré vectoriel quelconque. On se bornera au fibré tangent d'une variété différentiable que l'on supposera équipé d'une connexion. Nous distinguerons le cas d'une connexion quelconque, situation où on n'invoque   pas de métrique sur la variété et le cas plus spécifique de la connexion de Levi-Civita sur une variété pseudo-riemmannienne, liée à sa métrique et déterminée par elle.
\subsection {Courbures et torsion.}
\subsubsection{Le tenseur de courbure.}
Soit $M$ une variété différentiable, $\nabla$ une connexion sur $\tau(M)$. $\mathcal{F}$ désigne l'anneau des fonctions réelles différentiables sur $M$. Considérons l'application 
\begin{align}
\tau(M)\times\tau(M)\times&\tau(M)\overset{\mathcal{R}}{\longrightarrow}\tau(M) \nonumber\\
&(X,Y,Z)\mapsto R(X,Y)Z=\nabla_X\nabla_YZ-\nabla_Y\nabla_XZ-\nabla_{[X,Y]}Z\label{courbue}
\end{align}
\begin{proposition}\label{tc}
L'application $\mathcal{R}$ est $\mathcal{F}$-trilinéaire. Pour tout $x\in M, \mathcal{R}(X,Y,Z)(x)$ ne dépend que $X(x),Y(x),Z(x)$.
\end{proposition}
\begin{preuve}{}
L'additivité par rapport à chaque argument est évidente. Du fait de l'égalité $R(X,Y)Z=-R(Y,X)Z$, il reste encore à démontrer les deux items suivants pour tous champs $X,Y,Z$ et toute fonctions différentiable $f$ :
\begin{enumerate}
\item $R(fX,Y)Z=fR(X,Y)Z$ 
\item $R(X,Y)(fZ)=fR(X,Y)Z$.
\end{enumerate}
Pour le premier point, on a :

\noindent$R(fX,Y)Z=\nabla_{fX}\nabla_YZ-\nabla_Y\nabla_{fX}Z-\nabla_{[fX,Y]}Z=$

\noindent$=\left(f\nabla_X\nabla_YZ\right)-\left((Y.f)\nabla_XZ+f\nabla_Y\nabla_XZ\right)-\left(-(Y.f)\nabla_XZ+f\nabla_{[X,Y]}Z\right)=f\nabla_X\nabla_YZ$,
en utilisant au passage l'exercice \ref{nok} p.\pageref{nok} pour évaluer le terme $[fX,Y]$.
\end{preuve}

\noindent Pour le deuxième point, 

\noindent $R(X,Y)(fZ)=\nabla_X\nabla_YfZ-\nabla_Y\nabla_XfZ-\nabla_{[X,Y]}(fZ)=$

\noindent$=\left(X.(Y.f)Z+(Y.f)\nabla_XZ+(X.f)\nabla_YZ+f\nabla_X\nabla_YZ\right)-\cdots$

\noindent$\cdots-\left(Y.(X.f)Z+(X.f)\nabla_YZ+(Y.f)\nabla_XZ+f\nabla_Y\nabla_XZ\right)-\left(([X,Y].f)Z+f\nabla_{[X,Y]}Z\right)=$

\noindent$=f\nabla_X\nabla_YZ.$

\begin{definition}
\hspace{1cm}\\
Le tenseur $\mathcal{R}$ est le \emph{tenseur de courbure de $M$ pour la connexion $\nabla$}. \index{tenseur de courbure}
\end{definition}
\noindent\textbf{Expression  du tenseur de courbure dans une carte locale.}

\noindent Sur un ouvert $U$ d'une  carte locale munie de coordonnées$(x^1,\ldots,x^n)$. Posons  $X=X^{i}\frac{\partial}{\partial x^{i}},Y=Y^j\frac{\partial}{\partial x^{j}},Z=Z^k\frac{\partial}{\partial x^{k}}$ et  $\omega=(\omega^{i}_j)$  la matrice locale de la connexion $\nabla$ . Alors
$\omega^{i}_j=\omega^{i}_{sj} dx^s$.
On a : $R(X,Y)Z=R^l_{kij}X^{i}Y^jZ^k \frac{\partial}{\partial x^{l}}$ avec 
\begin{equation}\label{courloca}
R_{kij}^l=\frac{\partial\omega^l_{jk}}{\partial x^{i}}-\frac{\partial\omega^l_{ik}}{\partial x^{j}} +\omega^r_{jk}\omega^l_{ir}
-\omega^r_{ik}\omega^l_{jr}.
\end{equation}
En effet,

\noindent $\nabla_i(\nabla_j)\frac{\partial}{\partial x^k}=\nabla_i(\omega^s_{jk}\frac{\partial}{\partial x^s})=
\left(\frac{\partial}{\partial x^s}\otimes d\omega^s_{jk}+\omega^s_{jk}\nabla\frac{\partial}{\partial x^s}\right)(\frac{\partial}{\partial x^{i}})=\left(\frac{\partial \omega^t_{jk}}{\partial x^i}+\omega^{s}_{jk}\omega^t_{is}\right)\frac{\partial}{\partial x^t}$.

\noindent De même $\nabla_j(\nabla_i)\frac{\partial}{\partial x^k}=\left(\frac{\partial \omega^t_{ik}}{\partial x^j}+\omega^{s}_{ik}\omega^t_{js}\right)\frac{\partial}{\partial x^t}$. Et on conclut en se souvenant que 
$[\frac{\partial }{\partial x^i},\frac{\partial }{\partial x^j}]=0.$

\subsubsection{Le tenseur de Riemann-Christoffel d'une variété pseudo-riemannienne.}

On suppose que $M$ est une variété pseudo-riemannienne munie de sa connexion de Levi-Civita $\nabla$.
On note $g$ le tenseur métrique.

\begin{definition}\label{tenricri}\index{tenseur de Riemann-Christoffel}
Avec les notations de la section précédente, le tenseur de Riemann-Christoffel est défini pour tous champs de vecteurs 
$X,Y,Z,T$ par l'égalité :
\[R(X,Y,Z,T)=\la R(X,Y)Z,T\ra.\]
\end{definition}
\noindent Ce tenseur est intérressant pour ses nombreuses symétries. Deux d'entre elles sont  exposées dans la proposition  \ref{simet}. Deux autres sont exposées comme corollaire de la première identité de Bianchi (équation (\ref{bian1})).

\noindent{\textbf{\'{E}criture dans des coordonnées locales}

\noindent Sur un ouvert muni des coordonnées $(x^1,\ldots,x^n)$, on a :

\noindent $$R_{lkij}=:R\left(\frac{\partial}{\partial x^{i}},\frac{\partial}{\partial x^{j}},\frac{\partial}{\partial x^{k}},\frac{\partial}{\partial x^{l}}\right)=\left\la R^s_{kij}\frac{\partial}{\partial x^{s}},\frac{\partial}{\partial x^{l}}\right\ra=g_{sl}R^s_{kij}.$$
On a de façon équivalente :$$R^{a}_{kij}=g^{as}R_{skij}.$$

L'égalité exprime que  $R_{lkij}$ est obtenu à partir de $R^s_{kij}$ par \emph{ abaissement de l'indice contravariant}.
\begin{proposition}\label{simet}
On suppose ici la variété $M$ munie d'une structure pseudo-riemannienne et de sa connexion de Levi-Civita.	 
On a pour tous champs $X,Y,Z,T$ :

\begin{align}
 R(X,Y,Z ,T)&=- R(Y X,Z,T)\label{mis1}\\
R(X,Y,Z,T)&=-R(X,Y,T,Z)\label{mis2}
\end{align}
ce qui se traduit sur les composantes par 
\begin{enumerate}
\item $ R_{lkij}=-R_{lkji}$,
 \item $R_{lkij}=-R_{klij}$
\end{enumerate} 
 \end{proposition}
\begin{preuve}{}
L'égalité (\ref{mis1}) est immédiate. Montrons l'égalité (\ref{mis2}).

\noindent Comme $\nabla$ est la connexion de Levi-Civita, elle vérifie pour tous champs $X,Y,Z$
$X.\la Y,Z\ra=\la \nabla_XY,Z\ra+\la X,\nabla_XZ\ra$. On vérifie avec cette propriété directement l'identité:
\[\la \nabla_X\nabla_YZ,T\ra=X.\la\nabla_YZ,T\ra-Y.\la Z,\nabla_XT\ra+\la Z,\nabla_Y\nabla_XT\ra.\]
En permutant $X$ et $Y$ on a: 
\[\la \nabla_Y\nabla_XZ,T\ra=Y.\la\nabla_XZ,T\ra-X.\la Z,\nabla_YT\ra+\la Z,\nabla_X\nabla_YT\ra.\]
De plus :
\[\la \nabla_{[X,Y]}Z,T\ra=[X,Y].\la Z,T\ra-\la Z,\nabla_{[X,Y]}T\ra.\]
Ainsi

$R(X,Y,Z,T)=\la \nabla_X\nabla_YZ,T\ra-\la \nabla_Y\nabla_XZ,T\ra-\la \nabla_{[X,Y]}Z,T\ra=$

$=-\la Z,\nabla_X\nabla_YT\ra+\la Z,\nabla_Y\nabla_XT\ra+\la Z,\nabla_{[X,Y]}T\ra$
car \[X.\la\nabla_YZ,T\ra-Y.\la Z,\nabla_XT\ra-\left(Y.\la\nabla_XZ,T\ra-X.\la Z,\nabla_YT\ra\right)-[X,Y].\la Z,T\ra=0.\]
On en déduit que $R(X,Y,Z,T)=-R(X,Y,T,Z).$
\end{preuve}

\subsubsection{Définition de la torsion.}
On considère l'application 
\begin{align}
\tau(M)\times\tau(M)&\overset{\mathcal{T}}{\longrightarrow}\tau(M) \nonumber\\
(X,Y)&\mapsto \nabla_XY-\nabla_YX-[X,Y]\label{tordue}
\end{align}
\begin{proposition} \label{teto}
L'application $\mathcal{T}$ est $\mathcal{F}$-bilinéaire.
\end{proposition}
\begin{preuve}{}
L'additivité par rapport à chaque argument est évidente. En remarquant que $\tau(X,Y)=-\tau(Y,X)$, il reste à montrer que pour toute fonction différentiable $f$, on a : $\tau(fX,Y)=f\tau(X,Y)$. Or,
 $\tau(fX,Y)=\nabla_{fX}Y-\nabla_YfX-[fX,Y]=$

\noindent$=f\nabla_XY-(f\nabla_YX+(Y.f)X)-(-(Y.f)X+f[X,Y])$ ( d'après l'exercice \ref{nok}) 

\noindent$=f(\nabla_XY-\nabla_YX-[X,Y])=f\tau(X,Y)$.
\end{preuve}
\begin{definition}\index{torsion}
Le tenseur $\mathcal{T}$ est le tenseur de torsion de $M$ pour la connexion $\nabla$.
\end{definition}
\noindent\textbf{Expression  de la torsion dans une base locale.}

\noindent Soit sur un ouvert $U$ une base locale $(e_1,\ldots,e_n)$. On pose $X=X^{i}e_i, Y=Y^je_j$.
Alors  $\tau(X,Y)=T_{ij}X^{i}Y^j$ avec :
\begin{equation}\label{torloca}
T_{ij}=(\omega^k_{ij}-\omega^k_{ji})e_k-[e_i,e_j].
\end{equation}
La preuve en est immédiate. On note bien sûr l'antisymétrie $T_{ij}=-T_{ji}$.

L'équation (\ref{torloca}) montre (en considérant une base locale liée à des coordonnées locales) que si $\nabla$ est la connexion de Levi-Civita alors le tenseur de torsion est nul. 
\subsubsection{Formalisme d'\'{E}lie Cartan.}
On a vu en électromagnétisme tout l'intérêt qu'il y a à manipuler des formes différentielles (des tenseurs covariants)
plutôt que des tenseurs contravariants comme les champs de vecteurs : la machinerie du calcul différentiel est plus aisée avec les formes. De plus dans le cas d'une structure pseudo-riemannienne le passage aux formes est facilité par le fait qu'on passe canoniquement de l'aspect contravariant à l'aspect covariant par l'intermédiaire du tenseur métrique.
Il y a une idée analogue dans le cas du calcul différentiel sur une variété munie d'une connexion où la torsion et la courbure interviennent respectivement dans le calcul de la dérivée extérieure et d'une dérivée covariante seconde des champs de vecteurs, comme nous allons le voir.
\'{E}lie Cartan a introduit des $2$-formes de torsion et de courbure qui sont des outils extrêmement efficaces pour ce  ce calcul. Nous développons ci-dessous les formules de base du calcul différentiel de Cartan (voir [\ref{franki}] ou [\ref{eca}] ).

\noindent Commençons par réinterpréter le tenseur de torsion. Avec les notations de la sections précédente, plaçons-nous sur un ouvert $U$ muni d'une base locale $e=(e_1,\ldots,e_n)$ dont la base duale est $\varepsilon=(\varepsilon^1,\ldots,\varepsilon^n)$.  Si l'on pose $T_{ij}=T^k_{ij}e_k$ et qu'on tient compte de l'antisymétrie $T^k_{ij}=-T^k_{ji}$, on peut écrire :
$\mathcal{T}(X,Y)=T^{i}_{jk}X^jY^ke_i=\tau^{i}(X,Y)e_i$ où \begin{equation}\label{focuavaleur}\tau^{i}=T^{i}_{jk}\epsilon^{j}\wedge\epsilon^k. \end{equation} 
D'où l'égalité 
\begin{equation}\label{focuaval}
\mathcal{T}=e_i\otimes \tau^{i}.
\end{equation}
Dans cette égalité, $\mathcal{T}$ est vu comme une $2$-forme à valeurs dans $TM$.
\begin{definition}\label{tordons}\index{formes de torsion}\index{matrice des formes de torsion}
Les $2$-formes $\tau^{i}$ sont les formes de torsion.  La matrice colonne $\tau=^t(\tau^1,\ldots, \tau^n)$ est la matrice des formes de torsion. 
\end{definition}
\noindent Pour toute $1$-forme $\al$ et tous champs de vecteurs $X,Y$, on a sur $U$ les écritures locales 
$X=X^ke_k, Y=Y^le_l, \al=\lambda^{i}\epsilon_i$ et par suite on peut  écrire : $d\al (X,Y)=(d\lambda_i\wedge \epsilon^{i})(X,Y)+\lambda_i X^kY^l d\epsilon^{i}(e_k,e_l) $. La proposition suivante donne ainsi un formulaire  de calcul de $d\al$ pour toute $1$-forme $\al$. Soit $\omega$ la matrice de connexion $\nabla$ sur $U$ liée à la base $e$ : on sait que $\nabla e= e\otimes \omega$. De plus,
\begin{proposition}\label{strucar}
Pour tout $\in\{1;\ldots,n\}, d\varepsilon^{i}=-\omega^{i}_k\wedge \varepsilon ^k+\tau^{i}$ ce qui équivaut à l'écriture  matricielle
\begin{equation}\label{eqstr1}
d\varepsilon=-\omega\wedge\varepsilon+\tau
\end{equation}
\end{proposition}
\begin{preuve}{}
 On a :
 
 \noindent$d\epsilon^{i}(e_j,e_k)=e_j.\epsilon^{i}(e_k)-e_k.\epsilon^{i}(e_j)-\epsilon^{i}([e_j,e_k])=-\epsilon^{i}([e_j,e_k])$=
 
 \noindent$=-\epsilon^{i}\left(\big(T^s_{jk}-(\omega^s_{jk}-\omega^s_{kj})\big)e_s\right)=T^{i}_{jk}-(\omega^{i}_{jk}-\omega^{i}_{kj}).$ Par ailleurs,
 
  \noindent$\left(-\omega^{i}_s\wedge \varepsilon ^s+\tau^{i}\right)(e_j,e_k)=\left(-\omega^{i}_{jk}+T^{i}_{jk}\right)\epsilon^j\wedge \varepsilon ^k(e_j,e_k)=\omega^{i}_{kj}-\omega^{i}_{jk}+T^{i}_{jk}$ et on conclut.
\end{preuve}
\noindent La base du formalisme d'Elie Cartan pour le calcul de la courbure d'une variété pseudo-riemannienne  $M$ consiste à prolonger l'opérateur de connexion $\nabla$ qui opère sur les champs de vecteurs en un opérateur sur les formes différentielles à valeurs dans  le fibré tangent $\tau(M)$. Exposons cela en détails.
\begin{definition}\label{pfo}\index{formes à valeurs dans le fibré tangent}
Soit $M$ une variété différentiable et $p$ un entier naturel.
Une $p$-forme à valeurs dans $TM$ est une section du fibré $F^p(M)=\tau(M)\otimes \Lambda^p(M)$.
\end{definition}
\noindent Remarquons que $F^0(M)$ est isomorphe à $\tau(M)$.
Si $U$ est un ouvert sur lequel existe une base de champs de vecteurs $e=(e_1,\ldots,e_n)$, dont la base duale est $\epsilon=(\epsilon^1,\ldots,\epsilon^n)$, toute $p$-forme $\al$ à valeurs dans $TM$ peut s'écrire comme somme de termes sous la forme :$$ \al= e_i\otimes \al^{i}.$$En effet $\al$ est une somme de termes de la forme $(\lambda^{i}e_i)\otimes (\mu_j\epsilon^j)=e_i\otimes \al^{i}$ en posant $\al^{i}=\lambda^{i}\mu_j\epsilon^j$. Dans le cas où $p=0$, une base des sections de $F^0(M)$ s'écrit $(e_i=e\otimes \bf{1}_i)$ avec ${\bf{1}_i}=^t(0\ldots,0,1,0\ldots,0)$ le $1$ étant en i-ème position.

\noindent  On note $\mathcal{S}(F^p(M))$ l'espace vectoriel des sections de $F^p(M)$ et  $\mathcal{S}\left(F(M)\right)=\bigoplus_{p\in \nn}\mathcal{S}\left(E^p(M)\right)$.

On munit $\mathcal{S}\left(F(M)\right)$ d'une structure d'algèbre (graduée) en le munissant de la multiplication interne suivante :
\[
\begin{array}{ccc}
\mathcal{S}\left(F^p(M)\right)\times \mathcal{S}\left(F^q(M)\right)&\longrightarrow&\mathcal{S}\left(F^{p+q}(M)\right)\\
(e_i\otimes \al^{i},e_i\otimes \be^{i})&\mapsto&e_i\otimes (\al^{i}\wedge\be^{i})=:(e_i\otimes \al^{i})\wedge(e_i\otimes \be^{i})
\end{array}
\]
l'addition et la multiplication externe naturellement définies.
On suppose $\tau(M)$ muni d'une connexion $\nabla$ de matrice $\omega$ sur l'ouvert $U$ précédent.
Considérons sur l'ouvert $U$  l'opérateur $D_U$ défini pour tout $p\in\nn$ sur $\mathcal{S}\left(F^p(U)\right)$ par \begin{equation}\label{dercolo}
D_U\al=D_U(e_i\otimes \al^{i})=e_i\otimes (d\al^{i}+\omega^{i}_k\wedge\al^k)\in \mathcal{S}\left(F^{p+1}(U)\right)
\end{equation}
\begin{proposition}\label{dercoglo}
\hspace {1cm}\\
L'égalité (\ref{dercolo}) définit un opérateur global $D$ sur l'algèbre $\mathcal{S}\left(F(M)\right)$.
Cet opérateur prolonge $\nabla$ à $\mathcal{S}\left(F(M)\right)$.
\end{proposition}
\begin{preuve}{}
\noindent \emph{Premier point}

\noindent On considère deux ouverts $U$ et $V$ d'intersection non vide, sur lesquels existent respectivement les bases de champs de vecteurs $e_U$ et $e_V$ et on pose $C_{UV}$ la matrice de changement de bases telle que $e_V=e_UC_{UV}$ (voir égalité (\ref{superchapo}) p.\pageref{superchapo}). La $p$-forme à valeurs dans $TM$ s'écrit 
$e_U\otimes \al^{U}=e_V\otimes \al^V$. On a donc :

\noindent$D_U(e_U\otimes \al^{U})=e_VC_{VU}\otimes \left(d(C_{UV}\al^V)+(C_{UV}\omega_VC_{VU}+C_{UV}dC_{VU})\wedge C_{UV}\al^V\right)=$

\noindent$e_V\otimes \left(C_{VU}dC_{UV}+dC_{VU}C_{UV})\wedge \al^V+d\al_V+\omega_V\wedge\al^V\right)=e_V\otimes (d\al^V+\omega_V\wedge \al^V)=D_V(e_V\otimes \al^V)$ car 
$C_{VU}dC_{UV}+dC_{VU}C_{UV}=0$. Et on conclut quant à la globalité de $D$.
 
 \noindent\emph{Deuxième point}
 
\noindent Plaçons-nous sur l'ouvert $U$. On a vu (égalité (\ref{loloo}) p. \pageref{loloo}) que $\nabla e_i=e_j\otimes \omega^{i}_j$ et par ailleurs
 $D(e\otimes {\bf{1}_i})=:e\otimes (\omega\wedge{\bf{1}_i})=e_j\otimes \omega^j_i=\nabla e_i$. Or les opérateurs $D$ et $\nabla$ sont déterminés par leurs valeurs sur une base de \emph{$\tau(U)$ identifié à $\tau(U)\otimes \Lambda^0(M)$.} Ceci prouve que 
 $D$ et $\nabla$ coïncident sur $\mathcal{S}(F^0(M))$ et démontre le deuxième point.
\end{preuve}
\noindent Notons l'écriture matricielle suivante de l'équation (\ref{dercolo}).
Posons $\al=e_i\otimes \al^{i}$, $e=(e_1,\ldots,e_n)$, $d\al+\omega\wedge \al=^t(d\al^1+\omega^1_k\wedge \al^k,\ldots,d\al^n+\omega^n_k\wedge \al^k)$. Alors :
\begin{equation}\label{dercolomat}
D(\al)=e\otimes (d\al+\omega\wedge \al)
\end{equation}
\noindent L'opérateur $D$ évoque immanquablement la dérivée extérieure sur $\bigoplus _{p\in \nn}\Lambda^p(M)$ dont le carré nul permet d'obtenir des invariants topologiques. Quelle information le carré de $D$ fournit-il?
\begin{proposition}\label{matfocu}
On se place sur un ouvert $U$ d'une variété équipée d'une connexion $\nabla$, ouvert sur lequel existe une base de champs de vecteurs  $e=(e_1,\ldots,e_n)$. On note $\omega$ la matrice de la connexion $\nabla$ associée à la base $e$.
Alors :
 $$D\circ D(e)=e\otimes \theta$$
où\; $\theta=d\omega+\omega\wedge\omega$.
\end{proposition}
\begin{definition}\label{courgette}\index{formes de courbures}
la matrice $\theta$ de la proposition \ref{matfocu} est la matrice des formes de courbure sur l'ouvert $U$ associée à la connexion $\nabla$ et à la base $e$.
\end{definition}
\begin{preuve}{}
On a $De=\nabla e=e\otimes \omega$ d'après la proposition \ref{dercoglo} et l'égalité (\ref{loloo}).
D'où $D\circ D(e)=e\otimes (d\omega+\omega\wedge\omega)$ à partir de la définition (\ref{dercolo}).
\end{preuve}
\noindent\textbf{Cas particulier d'un ouvert muni de coordonnées locales $(x^1,\ldots,x^n)$.}

\noindent On a : 

\noindent $\theta^t_k=d\omega^t_k+\omega^t_{s}\wedge\omega^s_{k}=\left(\frac{\partial \omega^t_{jk}}{\partial x^{i}}+\omega^t_{is}\omega^s_{jk}\right)dx^{i}\wedge dx^j=\left(\frac{\partial \omega^t_{ik}}{\partial x^{j}}+\omega^t_{js}\omega^s_{ik}\right)dx^{j}\wedge dx^{i}$

\noindent$=\frac{1}{2}\left(\frac{\partial \omega^t_{jk}}{\partial x^{i}}-\frac{\partial \omega^t_{ik}}{\partial x^{j}}+\omega^t_{is}\omega^s_{jk}-\omega^t_{js}\omega^s_{ik}\right)dx^{i}\wedge dx^j.$
D'où, d'après l'égalité (\ref{courloca}) la relation :
\begin{equation}\label{focutencu}
\theta^{i}_j=\frac{1}{2}R^{i}_{j\al\be}dx^\al\wedge dx^\be.
\end{equation}
Ce qui justifie la dénomination de \og forme de courbure\fg \;pour $\theta^k_l$.

\subsubsection{Identités de Bianchi}\index{Bianchi (identités)}
\begin{theorem}[Identités de Bianchi]\label{ijn}
\hspace{1cm}\\
On se donne sur un ouvert $U$ d'une variété différentiable,  une base locale de champs de vecteurs $e=(e_i)_{i}$ et $\epsilon$ sa base duale. On suppose $\tau(M)$ muni d'une connexion $\nabla$. Soit sur $U$, $\omega$ la matrice des formes de connexion, $\theta$ la matrice des \;$2$-formes de courbure  et $\tau$ la matrice des $2$-formes de torsion. Alors 
\begin{align}
 D\tau&=\theta\wedge \epsilon\label{bian1}\\
 d\theta&=\theta\wedge\omega-\omega\wedge\theta\label{bian2}
\end{align}
\end{theorem}
\begin{preuve}{}
Montrons l'égalité (\ref{bian1}).
On part de l'équation (\ref{eqstr1}): $d\epsilon=-\omega\wedge\epsilon+\tau$. En appliquant la dérivée extérieure, on obtient $0=-d\omega \wedge\epsilon+\omega\wedge d\epsilon+d\tau=-d\omega \wedge\epsilon+
\omega\wedge(-\omega\wedge\epsilon+\tau)+d\tau=-\theta\wedge\epsilon+d\tau+\omega\wedge \tau$
que les équations (\ref{focuaval}) et  (\ref{dercolomat}) permettent d'écrire sous la forme $D\tau=\theta\wedge\epsilon$.

\noindent Montrons l'égalité (\ref{bian2}). On se rappelle que $\theta=d\omega+\omega\wedge\omega$. En appliquant la dérivée extérieure, on obtient $d\theta=d\omega\wedge\omega-\omega\wedge d\omega=
(\theta-\omega\wedge\omega)\wedge\omega-\omega\wedge(\theta-\omega\wedge\omega)
=\theta\wedge\omega-\omega\wedge\theta$.
\end{preuve}
Des identités de Bianchi, nous allons déduire les cinq corollaires suivants.
\begin{corollary}\label{sisim1}

Dans les hypothèses du théorème \ref{ijn}, on suppose : 

\noindent$1)$ que la connexion $\nabla$ est symétrique, c.à d. que la torsion est nulle (voir défi\-nition \ref{sysy} p. \pageref{sysy}).

\noindent$2)$ qu'il existe des coordonnées locales $(x^1,\ldots,x^n)$ telles 
que $e_i=\frac{\partial}{\partial x^{i}}$.

\noindent Alors les coordonnées du tenseur de courbure vérifie la symétrie:
\begin{equation}\label{trisikl}
R^{i}_{jkr}+R^{i}_{rjk}+R^{i}_{krj}=0.
\end{equation}
Ceci traduit l'égalité intrinsèque suivante. Pour tous champs de vecteurs $X,Y,Z$, on a lorsque la connexion est symétrique :
\begin{equation}\label{sitencou}
R(X,Y)Z+R(Z,X)Y+R(Y,Z)X=0
\end{equation}
\end{corollary}
\begin{preuve}{}
L'équation (\ref{bian1}) s'écrit,sous les hypothèses du corollaire :
 $$R^{i}_{jkr}dx^k\wedge dx^r\wedge dx^j=0.$$
l'équation (\ref{trisikl}) se déduit directement de l'égalité  $$dx^k\wedge dx^r\wedge dx^j=dx^j\wedge dx^k\wedge dx^r=dx^r\wedge dx^j\wedge dx^k$$
\end{preuve}
\begin{corollary}\label{simtenricristo}
\hspace{1cm}\\
Lorsque la connexion est symétrique, le tenseur de Riemann-Christoffel vérifie les deux symétries suivantes :
\begin{align}
&R(X,Y,Z,T)+R(Z,X,Y,T)+R(Y,Z,X,T)=0\label{sim3}\\
&R(X,Y,Z,T)=R(Z,T,X,Y)\label{sim4}
\end{align}
ce qui se traduit sur les composantes par 
\begin{enumerate}
\item $R_{lkij}+R_{ljki}+R_{lijk}=0$
\item $R_{lkij}= R_{ijlk}$
\end{enumerate}
\end{corollary}
\begin{preuve}{}
L'équation (\ref{sim3}) résulte immédiatement de l'équation (\ref{sitencou}). Démontrons l'égalité (\ref{sim4}). C'est plus sportif. En fait nous allons montrer que l'équation (\ref{sim4}) est une conséquence des trois équations (\ref{mis1}),(\ref{mis2}) et (\ref{sim3}).
On commence par démontrer quatre identités qui résultent toutes des équations (\ref{mis1}) et (\ref{mis2}) et (\ref{sim3}).
\'{E}nonçons ces quatre égalités :
\begin{enumerate}
\item[i)] $R(X,Y,Z,T)+R(Z,X,Y,T)+R(Y,Z,X,T)=0$
\item[ii)] $R(X,Y,Z,T)+R(Y,T,Z,X)+R(X,T,Y,Z)=0$
\item[iii)] $R(Y,T,Z,X)+R(Y,Z,X,T)+R(Z,T,X,Y)=0$
\item[iv)] $R(X,T,Y,Z)+R(Z,X,Y,T)+R(Z,TX,Y)=0$
\end{enumerate}
l'égalité $i)$ est directement l'égalité (\ref{sim3}). En utilisant les équations (\ref{mis1}) et (\ref{mis2}), l'équation  $(ii)$ s'écrit :

\noindent$-R(X,Y,T,Z)-R(Y,T,X,Z)-R(T,X,Y,Z)=0$ d'après $(\ref{sim3})$. Elle est donc vérifiée.

\noindent Les identités $iii)$ et $iv)$ se vérifient de la même manière. En faisant $i)$+$ii)$-$iii)$-$iv)$, on obtient l'équation (\ref{sim4}).

\noindent On en déduit que

\noindent$R_{lkij}=R(\frac{\partial}{\partial x^{i}},\frac{\partial}{\partial x^{j}},\frac{\partial}{\partial x^{k}},\frac{\partial}{\partial x^{l}})=R(\frac{\partial}{\partial x^{k}},\frac{\partial}{\partial x^{l}},\frac{\partial}{\partial x^{i}},\frac{\partial}{\partial x^{j}})
=R(\frac{\partial}{\partial x^{l}},\frac{\partial}{\partial x^{k}},\frac{\partial}{\partial x^{j}},\frac{\partial}{\partial x^{i}})=
R_{ij,l,k}$.

\end{preuve}
\begin{exo}\label{contrac}
Supposons que $T^{i}_{jkl}$ soient les composantes en coordonnées locales d'un tenseur $1$ fois contravariant et $3$ fois covariant. 
\begin{enumerate}
\item Montrer que $T_{jl}=T^{i}_{jil}$ sont les composantes d'un tenseur $2$ fois covariant. \emph{C'est le tenseur obtenu par contraction des indices $i$ et $k$}.
\item  Montrer que la contraction commute avec la dérivée covariante : si $c$ désigne la contraction des indices $i$ et $k$ alors $\left(c(T^{i}_{jkl}\right)_{/s}=c\left({T^{i}_{jkl}}_{/s}\right).$
\end{enumerate}
\end{exo}
\subsubsection{Tenseur de Ricci, courbure scalaire.}
\begin{definition}\label{ricci}\index{tenseur de Ricci}
Soit $M$ une variété différentiable dont le fibré tangent est muni d'une connexion $\nabla$.
Soit $\mathcal{R}$ le tenseur de courbure de composantes $R^l_{kij}.$
Il y a  équivalence entre les deux définitions suivantes.
\begin{enumerate}
\item Le tenseur  de Ricci est le tenseur deux fois covariant  de composantes locales  $R_{kj}$ obtenues par contraction des indices $l$ et $i$ dans $R^l_{kij}$. On a donc $$R_{kj}=:R^s_{jsk}\;\mathrm{et}\;\;
\mathrm{Ricc}(X,X)=R_{ij}X^{i}X^j.$$
\item Le tenseur de Ricci est la forme quadratique sur les champs de vecteurs notée $\mathrm{Ricc}$ telle que pour tout champ de vecteurs $X$ : 
\[\mathrm{Ricc}(X,X)=\tr \;r_X\]
où $r_X$ est l'endomorphisme défini par
\[\forall Y\in \mathcal{S}(\tau(M)),\;\;r_X(Y)=R(Y,X)X.\]
\item Si $M$ est une variété pseudo-riemannienne munie de sa connexion de Levi-Civita, cette définition équivaut encore à la suivante.

 Le tenseur de Ricci est le tenseur $\mathrm{Ricc}$ deux fois covariant vérifiant sur un ouvert muni d'une base $(e_1,\ldots,e_n)$ de champs de vecteurs 
\[\forall X,Y\in \mathcal{S}(\tau(M)),\;\;\mathrm{Ricc}(X,Y)=g^{jl}R(X,e_j,e_l,Y).\]
En particulier on peut écrire 
\begin{equation}\label{tetec}\mathrm{Ricc}(e_i,e_j)=R_{ij}=g^{kl}R(e_i,e_k,e_l,e_j)=g^{kl}R_{jlik}.
\end{equation}
\end{enumerate}
On appelle également $\mathrm{Ricc}$ le \emph{tenseur de courbure de Ricci}.
\end{definition}
\begin{preuve}{des équivalences}
Examinons la deuxième définition. Soit $(e_1,\ldots,e_n)$ une base locale de champs de vecteurs. On a :

\noindent $r_X(e_i)=R(e_i,X)X=R^{s}_{lik}X^kX^l e_s$ D'où $\tr (r_X)=R^{i}_{lik}X^kX^l=R_{lk}X^kX^l$ le terme $R_{lk}$ étant obtenu par contraction comme dans la première définition. Ainsi $\tr r_X= \mathrm{Ricc}(X,X)$ au sens de la première définition.

\noindent Examinons la troisième définition.

\noindent On a :
$g^{jl}\la R(X,e_j)e_l,Y\ra=g^{jl}X^{i}Y^k\la R(e_i,e_j)e_l,e_k\ra=g^{jl}X^{i}Y^k\la R(e_j,e_i)e_k,e_l\ra$

(d'après les symétries de la proposition \ref{simet})

$=X^{i}Y^{k} g^{jl} g_{sl}R^{s}_{kji}=R^{s}_{ksi}X^{i}Y^k=R_{ki}X^{i}Y^k$. L'équation (\ref{tetec}) en découle.
En particulier on a : $$\mathrm{Ricc}(X,X)=R_{ki}X^{i}X^k$$ qui correspond à la deuxième définition.
\end{preuve}
\begin{corollary}\label{sisim2}
Si $\nabla$ est la connexion  de Levi-Civita, le tenseur de Ricci est symétrique.
\end{corollary}
\begin{preuve}{}
En considérant la définition du troisième item et ses notations, on a : 
$$\mathrm{Ricc}(X,Y)=g^{jl}R(X,e_j,e_l,Y)=g^{jl}R(e_l,Y,X,e_j)=
g^{lj}R(Y,e_l,e_j,X)=\mathrm{Ricc}(Y,X).$$
\end{preuve}
\begin{corollary}\label{sisim3}
On se place sur un ouvert d'une carte locale, munie des coordonnées $(x^1,\ldots,x^n)$. On suppose la connexion $\nabla$ symétrique. Alors pour tout $\{i,j,t,l,s\}\subset\{1,\ldots, n\}$ on a l'identité :

\[{R^{i}_{jtl}}_{/s}+{R^{i}_{jst}}_{/l}+{R^{i}_{jls}}_{/t}=0.\]
\end{corollary}
\begin{preuve}{}
La deuxième identité de Bianchi s'écrit $d\theta^{i}_j-\theta^{i}_t\wedge\omega^t_j+\omega^{i}_t\wedge\theta^t_j=0$.
En utilisant l'égalité (\ref{focutencu}), cette relation s'écrit directement 
\[\left(\frac{\partial R^{i}_{jtl}}{\partial x^s}-\omega ^\al_{sj}R^{i}_{\al tl}+\omega^{i}_{s\al}R^\al_{jtl}\right)dx^s\wedge dx^t\wedge dx^l=0.\]
Cette équation peut se récrire à partir de la dérivation covariante des composantes du tenseur de courbure, en utilisant l'égalité (\ref{pim}) p. \pageref{pim}. Cela aboutit à l'écriture suivante :
\[ \left({R^{i}_{jtl}}_{/s}+\omega^\al_{st}R^{i}_{j\al l}+\omega^{\al}_{sl}R^{i}_{jt\al}\right)dx^s\wedge dx^t\wedge dx^l=0.\]
De l'égalité $dx^s\wedge dx^t\wedge dx^l=dx^l\wedge dx^s\wedge dx^t=dx^t\wedge dx^l\wedge dx^s,$ on déduit 
\begin{align*}
&{R^{i}_{jtl}}_{/s}+\omega^\al_{st}R^{i}_{j\al l}+\omega^{\al}_{sl}R^{i}_{jt\al}+\\
&{R^{i}_{jst}}_{/l}+\omega^\al_{ls}R^{i}_{j\al t}+\omega^{\al}_{lt}R^{i}_{js\al}+\\
&{R^{i}_{jls}}_{/t}+\omega^\al_{tl}R^{i}_{j\al s}+\omega^{\al}_{ts}R^{i}_{jl\al}=0
\end{align*}
La symétrie de la connexion fait que la somme des termes contenant les coefficients de connexion est nulle
(par exemple $\omega^\al_{st}R^{i}_{j\al l}=-\omega^{\al}_{ts}R^{i}_{jl\al}$).
Ce qui achève la démonstration.
\end{preuve}

\begin{definition}\label{couscal}\index{courbure scalaire}
Si $R_{ij}$ désigne en coordonnées locales les composantes du tenseur de Ricci d'une variété pseudo-riemannienne, la courbure scalaire $R$
 se définit par  $R=g^{ij}R_{ij}$.
\end{definition}
\begin{exo}\label{cuscacon}
Sur une variété pseudo-riemannienne dont le tenseur de Ricci a pour composantes locales $R_{ij}$, on considère la  version mixte du tenseur de Ricci de composantes locales $R^{i}_j=g^{jk}R_{ik}$. Alors $R=R^{i}_i$.
\end{exo}
\begin{corollary}\label{sisim4}
On se place sur un ouvert $U$ d'une carte locale d'une variété pseudo-riemannienne munie de sa connexion de Levi-Civita. Soient $(x^1,\ldots,x^n)$ les coordonnées sur $U$. Alors la courbure scalaire $R$ vérifie 
\begin{equation}
\frac{\partial R}{\partial x^s}=2 {R^{i}_s}_{/i}
\end{equation}
\end{corollary}
\begin{preuve}{}
Partons du corollaire \ref{sisim3}. on a : ${R^{i}_{jtl}}_{/s}+{R^{i}_{jst}}_{/l}+{R^{i}_{jls}}_{/t}=0$.
Contractons les indices $i$ et $t$. En utilisant la commutation de la dérivation covariante et de contraction (voir exercice \ref{contrac}), on obtient : ${R^{i}_{jil}}_{/s}+{R^{i}_{jsi}}_{/l}+{R^{i}_{jls}}_{/i}=0$, c'est à dire 
${R_{jl}}_{/s}-{R_{js}}_{/l}+{R^{i}_{jls}}_{/i}=0$.  Ainsi 

\noindent $g^{\al j}\left({R_{jl}}_{/s}-{R_{js}}_{/l}+{R^{i}_{jls}}_{/i}\right)=
{(g^{\al j}{R_{jl}}})_{/s}-{(g^{\al j}{R_{js}}})_{/l}+{(g^{\al j}R^{i}_{jls}})_{/i}={R^\al_l}_{/s}-{R^\al_s}_{/l}
+{(g^{\al j}R^{i}_{jls}})_{/i}=0$. On contracte $\al$ et $s$ dans la dernière égalité. On obtient :
${R^\al_l}_{/\al}-{R^\al_\al}_{/l}+(g^{\al j} R^{i}_{jl\al})_{/l}$ . Or $g^{\al j} R^{i}_{jl\al}=g^{\al j}g^{\be i}R_{\be jl\al}=
g^{\be i}R_{\be l}$ ( d'après l'équation (\ref{tetec})

\noindent $=R^{i}_l$. On a donc ${R^\al_l}_{/\al}-{R^\al_\al}_{/l}+{R^{i}_l}_{/i}=0.$ On conclut sachant que $R=R^\al_\al$.
\end{preuve}
\subsubsection{Courbure sectionnelle}\index{courbure sectionnelle}
\begin{proposition}\label{cousec}
On se donne une variété pseudo-riemannienne munie de sa connexion de Levi-Civita et sur un ouvert $U$ deux champs de vecteurs $X,Y$ linéairement indépendants en tout point de $U$. Soit $\mathcal{P}$ la famille de plans
au dessus de $U$ engendrée par $X$ et $Y$ : pour tout $x\in U,\; \mathcal{P}(x)=\left\{aX(x)+bY(x)\;a,b\in\rr\right\}$.
Alors \[K(X,Y)=\frac{R(X,Y,Y,X)}{\la X,X\ra\la Y,Y\ra-\la X,Y\ra^2}\]
ne dépend que de $\mathcal{P}$.
\end{proposition}
\begin{preuve}{}
Un calcul élémentaire montre que si $X'=aX+bY, Y=cX+dY, ad-bc\not=0$, alors 
\begin{enumerate}
\item $R(X',Y',Y',X')=(ad-bc)^2R(X,Y,Y,X),$
\item $\la X',X'\ra \la Y',Y'\ra -{\la X',Y'\ra}^2=(ad-bc)^2(\la X,X\ra\la Y,Y\ra-\la X,Y\ra^2)$.
\end{enumerate}
ce qui montre que $K(X',Y')=K(X,Y)$.
\end{preuve}
\noindent Le tenseur $K$ est symétrique : $K(X,Y)=K(Y,X)$. Si $X$et $Y$ sont deux vecteurs orthogonaux et unitaires de $\mathcal{P}$, alors
$K(X,Y)=R(X,Y,Y,X)$.

\noindent Choisissons une base orthonormée de $\mathcal{P}$,  $\{e_1,\ldots,e_n\}$. Alors $\la e_i,e_i\ra\in\{-1,1\}$ et $\la e_i,e_j\ra=0$ pour $i\not= j$. On pose $\epsilon_i=\la e_i,e_i\ra$.
\begin{corollary}\label{seccourisca}
En considérant la base orthonormée précédente, on a :
\begin{enumerate}
\item $\mathrm{Ricc}(e_i,e_i)=\epsilon_i\displaystyle\sum_{\{j;j\not=i\}}K(e_i,e_j)$
\item Si $R$ désigne la courbure scalaire, $R=\displaystyle\sum_i\displaystyle\sum_{\{j;j\not=i\}}K(e_i,e_j)$
\end{enumerate}
\end{corollary}
\begin{preuve}{}
Pour le premier item, on a $\mathrm{Ricc}(X,X)=\sum_{j}\epsilon_j\la r_X(e_j),e_j\ra=\sum_{j}\epsilon_j\la R(e_j,X)X,e_j\ra$.
D'où $\mathrm{Ricc}(e_i,e_i)=\displaystyle\sum_{j,j\not=i}\epsilon_j\la R(e_j,e_i)e_i,e_j\ra$ car $\la R(e_i,e_i)e_i,e_i\ra=0$.

\noindent Et comme $K(e_i,e_j)=\epsilon_i\epsilon_j\la R((e_i,e_j)e_i,e_i)\ra$, on conclut.

\noindent Pour le deuxième item, d'après la définition \ref{couscal} et l'hypothèse d'orthonormalité, on a : $R=\displaystyle\sum_j \epsilon_j\mathrm{Ricc}(e_j,e_j)$ et on conclut avec le premier item.
\end{preuve}

\subsubsection{Courbure induite sur une hypersurface d'une variété pseudo-rieman\-nien\-ne.}
\noindent Les notations sont celles du paragraphe \ref{conlehyp} : $S$ est une hypersurface de la variété pseudo-riemannienne $M$, munie de sa connexion de Levi-Civita $\nabla$  et $\nabla^S$ est la connexion induite sur $S
$. On sait que $\nabla^S$ est la connexion de Levi-Civita de $S$.
On définit naturellement le tenseur de courbure induit sur $S$ par 
\begin{align}\index{courbure induite sur une hypersurface}
\tau(M)\times\tau(M)\times&\tau(M)\overset{\mathcal{R}^S}{\longrightarrow}\tau(M) \nonumber\\
&(X,Y,Z)\mapsto R^S(X,Y)Z=\nabla^S_X\nabla^S_YZ-\nabla^S_Y\nabla^S_XZ-\nabla^S_{[X,Y]}Z\label{courbue}
\end{align}\index{courbure induite}
Quel est le lien entre $\mathcal{R}$ et $\mathcal{R}^S$? Pour le savoir on se servira du lien entre $\nabla$ et $\nabla^S$ donné par l'égalité (\ref{conwein}) p. \pageref{conwein}, en exprimant $\nabla$ en fonction de $\nabla^S$ dans l'expression de 
$R(X,Y)Z$. L'exposé du calcul est facilité par le lemme technique suivant.
\begin{lemma}\label{theq}
On se place sur un ouvert  d'une carte locale de $S$ munie des coordonnées $(x^1,\ldots,x^n)$.
On considère la fonction de Weingarten $\mathcal{W}$ de $S$ et le tenseur symétrique deux fois covariant $w$ associé (voir définition \ref{weinensie?} p. \pageref{weinensie?}). On a l'égalité :

\noindent$\frac{\partial}{\partial x^\al}w\left(\frac{\partial}{\partial x^\be},\frac{\partial}{\partial x^\gamma}\right)$
$-w\left(\frac{\partial}{\partial x^\be},\nabla^S_\al\frac{\partial}{\partial x^\gamma}\right)-\frac{\partial}{\partial x^\be}w\left(\frac{\partial}{\partial x^\al},\frac{\partial}{\partial x^\gamma}\right)$
$+w\left(\frac{\partial}{\partial x^\al},\nabla^S_\be\frac{\partial}{\partial x^\gamma}\right)=$

\noindent$\left(\nabla^S_{\al}w\right)\left(\frac{\partial}{\partial x^\be,}\frac{\partial}{\partial x^\gamma}\right)-$
$\left(\nabla^S_{\be}w\right)\left(\frac{\partial}{\partial x^\al},\frac{\partial}{\partial x^\gamma}\right)$
\end{lemma}
\begin{preuve}{}
En utilisant la proposition \ref{leileilei} p. \pageref{leileilei}, on écrit les deux égalité suivantes:
\begin{align}
\frac{\partial}{\partial x^\al}w\left(\frac{\partial}{\partial x^\be},\frac{\partial}{\partial x^\gamma}\right)&=
\left(\nabla^S_{\al}w\right)\left(\frac{\partial}{\partial x^\be,}\frac{\partial}{\partial x^\gamma}\right)+w\left({\nabla^S_\al} \frac{\partial}{\partial x^\be},\frac{\partial}{\partial x^\gamma}\right)+w\left(\frac{\partial}{\partial x^\be},\nabla^S_\al \frac{\partial}{\partial x^\gamma}\right)\label{ed}\\
&\empty\nonumber\\
\frac{\partial}{\partial x^\be}w\left(\frac{\partial}{\partial x^\al},\frac{\partial}{\partial x^\gamma}\right)&=
\left(\nabla^S_{\be}w\right)\left(\frac{\partial}{\partial x^\al,}\frac{\partial}{\partial x^\gamma}\right)+w\left({\nabla^S_\be} \frac{\partial}{\partial x^\al},\frac{\partial}{\partial x^\gamma}\right)+w\left(\frac{\partial}{\partial x^\al},\nabla^S_\be \frac{\partial}{\partial x^\gamma}\right)\label{eded}
\end{align}
En remarquant que ${\nabla^S_\be} \frac{\partial}{\partial x^\al}={\nabla^S_\al} \frac{\partial}{\partial x^\be}$ puisque la connexion est symétrique, on obtient le résultat annoncé en faisant (\ref{ed})-\ref{eded}).
\end{preuve}
Ce lemme a pour conséquence la proposition \ref{courbin}
\begin{proposition}\label{courbin}
Avec les notations qui précèdent on a pour tous champs $X,Y,Z$ tangents à l'hypersurface $S$ admettant un champ de vecteurs normal $N$, on a :
\begin{align}
R(X,Y)Z=R^S(X,Y)Z+\epsilon_N\big(w(X,Z)\mathcal{W}(Y)&-w(Y,Z))\mathcal{W}(X)\big)+\nonumber\\
&+\epsilon_N\left((\nabla^S_X w)(Y,Z)-(\nabla^S_Yw)(X,Z)\right)N.\label{coubin}
\end{align}
\end{proposition}
\begin{preuve}{}
\noindent On notera dans cette démonstration $\partial_\al$ le champ de vecteurs $\frac{\partial}{\partial x^\al}$.
Plaçons-nous sur la carte locale du lemme \ref{theq}. 

\noindent Puisque $R(X,Y)Z=X^\al,Y^\be Z^\ga R(\partial_\al,\partial_\be)\partial_\ga$, il suffit d'établir la proposition pour $X=\partial_\al, Y=\partial_\be, Z=\partial_\ga$.
\`{A} partir de la relation (\ref{conwein}), on peut écrire :

\noindent$\nabla_\al(\nabla_\be\partial_\ga)=\nabla^S_\al(\nabla_\be\partial_\ga)+\epsilon_N w(\partial_\al,\nabla_\be\partial_\ga)N = \nabla^S_\al(\nabla_\be\partial_\ga)+\epsilon_N w(\nabla_\be\partial_\ga,\partial_\al)N
=..$

$..=\nabla^S_\al(\nabla_\be\partial_\ga)-\epsilon_N\la \nabla_\be\partial_\ga,\nabla_\al N\ra N$.
\noindent Or,

\noindent $\nabla^S_\al(\nabla_\be\partial_\ga)=\nabla^S_\al\left(\nabla^S_\be\partial_\ga+\epsilon_N w(\partial_\be,\partial_\ga)N\right)$. En écrivant 

\noindent $w(\partial_\be,\partial_\ga)=w(\partial_\ga,\partial_\be)=\la \partial_\ga,-\nabla_\be N\ra=\la \nabla_\be\partial_\ga,N\ra$ (puisque $\partial_\ga$ est tangent à $S$), on a :

\noindent$\nabla^S_\al(\nabla_\be\partial_\ga)=\nabla^S_\al\nabla^S_\be\partial_\ga+\epsilon_N\left(\dfrac{\partial}{\partial x^\al}\la\nabla_\be\partial_\ga,N\ra N+\la\nabla_\be\partial_\ga,N\ra\nabla^S_\al N\right)$ et par conséquent

\noindent$\nabla_\al(\nabla_\be\partial_\ga)=\nabla^S_\al\nabla^S_\be\partial_\ga+\epsilon_N\left(\dfrac{\partial}{\partial x^\al}\la\nabla_\be\partial_\ga,N\ra N+\la\nabla_\be\partial_\ga,N\ra\nabla^S_\al N-\la \nabla_\be\partial_\ga,\nabla_\al N\ra N\right)$.

\noindent Ainsi, en évaluant de la même manière $\nabla_\be(\nabla_\al\partial_\ga)$ et en faisant la différence avec $\nabla^S_\al(\nabla_\be\partial_\ga)$, on obtient :

\noindent$R(\partial_\al,\partial_\be)\partial_\ga=..$

\noindent$R^S(\partial_\al,\partial_\be)\partial_\ga
+\epsilon_N \left(\dfrac{\partial}{\partial x^\al}\la\nabla_\be\partial_\ga,N\ra +\la \nabla_\be\partial_\ga,\nabla_\al N\ra -\dfrac{\partial}{\partial x^\be}\la\nabla_\al\partial_\ga,N\ra -\la \nabla_\al\partial_\ga,\nabla_\be N\ra \right)N$

\noindent$+\epsilon_N\left(\la\nabla_\be\partial_\ga,N\ra\nabla^S_\al N-\la\nabla_\al\partial_\ga,N\ra\nabla^S_\be N\right)$. En utilisant le lemme \ref{theq} on peut écrire :

\noindent$R(\partial_\al,\partial_\be)\partial_\ga=R^S(\partial_\al,\partial_\be)\partial_\ga+\epsilon_N\left((\nabla^S_\al w)(\partial_\be,\partial_\ga)-(\nabla^S_\be w)(\partial_\al,\partial_\ga)\right)N+..$

\hspace{5cm} $..+\epsilon_N\left(w(\partial_\be,\partial_\ga)\mathcal{W}(\partial_\al)-w(\partial_\al,\partial_\ga)\mathcal{W}(\partial_\be)\right)$.
\end{preuve}
\begin{corollary}\label{cousecind}
Notons pour deux champs de vecteurs unitaires et orthogonaux $X,Y$ tangents à l'hypersurface $S$ de la variété pseudo-riemannienne $M$,
$K^S(X,Y)$ la courbure sectionnelle correspondante. Alors 
\begin{equation}\label{cousecindeq}
K(X,Y)=K^S(X,Y)+\epsilon_N\epsilon_X\epsilon_Y\left(w(X,Y)^2-w(X,X)w(Y,Y)\right)
\end{equation}
où $\epsilon_X=\la X,X\ra\in\{-1,1\}$  et $\epsilon_Y=\la Y,Y\ra\in\{-1,1\}$
\end{corollary}
\noindent La preuve est immédiate à partir de la proposition \ref{courbin} et de la définition de la courbure sectionnelle.
\begin{corollary}[théorème egregium de Gauss]\label{thega}
Si $S$ est une surface différentiable de $\rr^3$, alors en tout point de $S$ la courbure sectionnelle, déterminée par la métrique induite de la structure euclidienne de $\rr^3$ est égale à la courbure de Gauss de $S$.
\end{corollary}
\begin{preuve}{}
Sous les hypothèses du corollaire, le terme $K(X,Y)$ de l'équation (\ref{cousecindeq}) est nul. On a donc 
$K^S(X,Y)=w(X,X)w(Y,Y)-w(X,Y)^2=\det\mathcal{W}$ qui est la courbure de Gauss de $S$.
\end{preuve}
L'intérêt de ce théorème est le suivant : il montre que la courbure de Gauss est invariante par toute isométrie de $\rr^3$ car  c'est le cas pour la courbure sectionnelle (voir [\ref{rinrin}] p. 105). En effet $\mathcal{W}$ dépend a priori de $N$, champ de vecteurs  unitaire normal  à la surface et l'invariance par isométrie de $\det\mathcal{W}$ n'est pas immédiate.
\subsection{Le tenseur d'Hilbert-Einstein-Cartan}
\begin{definition}\label{hec}\index{tenseur d'Hilbert-Einstein-Cartan}
On se donne une variété pseudo-riemannienne $M$ munie de sa connexion de Levi-Civita.  Comme ci-dessus la courbure scalaire de $M$
 est désignée par $R$. Le tenseur de Hilbert-Einstein-Cartan (tenseur $\mathcal{HEC}$) est le tenseur $G$ deux fois covariant sur $M$ défini par
\begin{equation}
\forall X\in\mathcal{S}(\tau(M))\;\;G(X,X)=\mathrm{Ricc}(X,X)-\frac{1}{2}\la X,X\ra R
\end{equation}
\end{definition}
En coordonnées locales, si on pose $G(X,X)=G_{ij}X^{i}X^j$, on a la relation
\[G_{ij}=R_{ij}-\frac{1}{2}g_{ij}R\]
et sa version mixte :
\[G^{i}_j=R^{i}_j-\frac{1}{2} g^{i}_j R.\]
\begin{proposition}\label{divtehec}
La divergence du tenseur $\mathcal{HEC}$ est nulle.
\end{proposition}
\begin{preuve}{}
En se référant à l'exercice \ref{divdivergence}, il s'agit de montrer que ${G^{i}_j}_{/i}=0$. Ceci est une conséquence 
des identités de Bianchi par l'intermédiaire du corollaire \ref{sisim4}. En effet,
${G^{i}_{j}}_{/i}={R^{i}_j}_{/i}-\frac{1}{2}{g^{i}_j}_{/i}R-\frac{1}{2}g^{i}_jR_{/i}=\frac{1}{2}\frac{\partial R}{\partial x^j}-\frac{1}{2}g^{i}_j\frac{\partial R}{\partial x^{i}}=\frac{1}{2}\frac{\partial R}{\partial x^j}-\frac{1}{2}\frac{\partial R}{\partial x^j}=0$.
\end{preuve}
Si l'on doit attribuer en partie  à Hilbert ce tenseur, c'est parce qu'il en obtint le premier l'expression exacte et 
donna ainsi les bonnes équations du champ de gravitation de la relativité générale. Si on doit attribuer en partie ce tenseur à Einstein, c'est parce qu'il en avait compris le rôle dans les équations du champ de gravitation, jouant ainsi un rôle moteur dans l'élaboration de la relativité générale. Si on doit
l'attribuer en partie à \'{E}lie Cartan, c'est pour le théorème que l'on va énoncer ci-après et qui démontre que les choix d'Einstein et Hilbert étaient à des constantes multiplicatives et additives près le seul choix possible ce qui 
donne une cohérences aux équations du champ établies par Hilbert-Einstein (en dehors de leur adéquation avec les observations, bien sûr). Les références concernant cette justification sont dans l'ordre historique  [\ref{Ein}], 
[\ref{Hil}], [\ref{Car}].
Voici, en substance le théorème de Cartan.
\begin{theorem}\label{tdca}\index{Cartan (théorème de)}
On se place sur une variété pseudo-riemannienne munie d'un tenseur métrique de signature $(3,1)$ (comme la métrique de Lorentz).
On considère un tenseur deux fois covariant $T$, de composantes $T_{ij} $ dans des coordonnées locales et on suppose que ces composantes $T_{ij}$ sont des fonctions continues $x\mapsto T_{ij}(g_{kl}(x),\frac{\partial g_{kl}(x)}{\partial x^s},\frac{\partial^2g_{kl}(x)}{\partial x^s\partial x^t})$, linéaires par rapport au troisième argument. Alors il existe trois constantes $\lambda,\mu,\nu$
telles que 
\begin{equation}\label{K}
T_{ij}=\lambda R_{ij}+\mu g_{ij}R+\nu g_{ij}.
\end{equation}
\end{theorem}
\begin{corollary}\label{fou}
\hspace{1cm}\\
Si on suppose en outre que la divergence du tenseur  $T$ du théorème \ref{tdca} est nulle alors $T=\lambda G+\nu g$, ce qui s'écrit en coordonnées pour la tversion mixte du tenseur $T$ :
\begin{equation}\label{foudre}
T^{i}_j=\lambda \left(R^{i}_j-\frac{1}{2}g^{i}_j(R+\nu)\right)
\end{equation}
\end{corollary}
Nous ne donnerons pas ici la démonstration du théorème de Cartan. Sa lecture pour un lecteur d'aujourd'hui est difficile (et longue). Il existe sur internet un document de Joël Merker [\emph{Sur les équations de la gravitation d'Einstein, d'après \'{E}lie Cartan}, 2010] qui reconstitue le raisonnement de Cartan sous une forme accessible avec le formalisme actuel. 
Montrons par contre le corollaire.
\begin{preuve}{du corollaire \ref{fou}}
En partant de l'équation (\ref{K}), on a : $T^{i}_j=\lambda R^{i}_j+\mu g^{i}_j R+\nu g^{i}_j$. 
Ainsi,

\noindent${T^{i}_j}_{/i}=\lambda {R^{i}_j}_{/i}+\mu g^{i}_j R_{/i}=\lambda\frac{1}{2}\frac{\partial R}{\partial x^j}+\mu \frac{\partial R}{\partial x^j}=0$. D'où $\mu=-\frac{\lambda}{2}$ et on conclut.

\end{preuve}
Dans le corollaire \ref{seccourisca}, nous avons vu que pour une base locale orthonormée $\{e_i\}$ du fibré tangent à une variété pseudo-riemannienne l'on peut exprimer  les valeurs $\mathrm{Ricc}(e_i,e_i)$ ainsi que la courbure scalaire $R$ en fonction des courbures sectionnelles $K(e_i,e_j)$. Il en est donc de même pour $G(e_i,e_i)$. Le résultat est l'objet de la proposition suivante.
\begin{proposition}\label{versrelat}
Avec les notations du corollaire \ref{seccourisca} on a :
\begin{equation}\label{eincoursec}
G(e_i,e_i)=-\epsilon_i\sum_{\{(k,l);k<l,k\not= i,l\not= i\}}K(e_k,e_l).
\end{equation}
\end{proposition}
\begin{exo}\label{versrelati}
Démontrer la proposition \ref{versrelat}.
\end{exo}
Remarquons que la somme du membre de droite de l'équation (\ref{eincoursec}) est la somme des courbures sectionnelles  de tous les plans orthogonaux à $e_i$. En dimension $n$,  elle contient donc $\frac{(n-1)(n-2)}{2}$ termes. \emph{Par la suite on notera cette somme $\mathcal{K}(e_i^\perp)$.}

\section{Les équations du champ de gravitation}\label{equachation}
\subsection{Un théorème de Levi-Civita}\index{Levi-Civita (un théorème de)}
\begin{theorem}[Levi-Civita]\label{li}
\hspace{1cm}\\
Soit $M$ une variété pseudo-riemannienne de dimension $4$ munie sur une carte locale des coordonnées $(x^0,x^1,x^2,x^3)=(t,r,\theta,\varphi)$, de sorte que sur cette carte la métrique soit de signature $(3,1)$ exprimée par  $ds^2=g_{00}(r)dt^2+dr^2+r^2d\theta^2+r^2\sin^2\theta d\varphi^2$, avec $g_{00}<0$. Dans ces coordonnées, on note $R^{i}_j$ les composantes du tenseur mixte de Ricci.
Alors \begin{equation}\label{lilimarlene}
\Delta(\sqrt{-g_{00}})=-R^0_0\sqrt{-g_{00}}.
\end{equation}
\end{theorem}
\begin{preuve}{}
On se souvient ( voir équation (\ref{lapone})) que pour une fonction $f$ deux fois différentiable on a :
\[\Delta f=g^{ij}\left(\frac{\partial^2 f}{\partial x^{i}\partial x^j}-\Gamma^k_{ij}\frac{\partial f}{\partial x^k}\right),\]
 où $g_{11}=1,g_{22}=r^2$ et $g_{33}=r^2\sin^2\theta$ et en utilisant l'égalité (\ref{cristo}) :
 $$\Gamma^1_{00}=-\frac{1}{2}\frac{\partial g_{00}}{\partial r},\Gamma^1_{11}=0,\Gamma^1_{22}=-r,\Gamma^1_{33}=-r\sin^2\theta.$$ 

Ce qui donne ici :
\begin{align*}
\Delta (\sqrt{-g_{00}})=&g^{ii}\left(\frac{\partial^2\sqrt{-g_{00}}}{{\partial x^{i}}^2}-\Gamma^1_{ii}\frac{\sqrt{-g_{00}}}{\partial r}\right)=\frac{1}{g_{11}}\frac{\partial^2\sqrt{-g_{00}}}{\partial r^2}-\frac{\Gamma^1_{ii}}{g_{ii}}\frac{\partial\sqrt{-g_{00}}}{\partial r}\\
&=\frac{1}{\sqrt{-g_{00}}}\left( \frac{1}{4g_{00}}\big(\frac{\partial g_{00}}{\partial r}\big)^2-\frac{1}{2}\frac{\partial^2g_{00}}{\partial r^2}\right)-\frac{1}{r\sqrt{-g_{00}}}\frac{\partial g_{00}}{\partial r}.
\end{align*}
\noindent Par ailleurs $R_{ij}=R^l_{ilj}$ et $R^{i}_j=g^{is}R_{sj}=g^{is}R^l_{slj}.$ D'où $R^0_0=g^{0s}R^l_{sl0}=g^{00}R^l_{0l0}.$
Les termes $R^l_{0l0}$ s'évaluent à partir de l'équation (\ref{courloca}) p. \pageref{courloca}.  On a  de façon précise :

\noindent$R^0_{000}=\Gamma^s_{00}\Gamma^0_{0s}-\Gamma^s_{00}\Gamma^0_{0s}=0, R^1_{010}=
\frac{\partial\Gamma^1_{00}}{\partial r}+\Gamma^s_{00}\Gamma^1_{1s}-\Gamma^s_{10}\Gamma^1_{0s}
=-\frac{1}{2}\frac{\partial^2 g_{00}}{\partial r^2}+\frac{1}{4}\left(\frac{\partial g_{00}}{\partial r}\right)^2$ après avoir établi que $\Gamma^s_{00}\Gamma^1_{1s}-\Gamma^s_{10}\Gamma^1_{0s}=-\Gamma^0_{10}\Gamma^1_{00}=
\left(\frac{1}{2g_{00}}\frac{\partial g_{00}}{\partial r}\right)\left(\frac{-1}{2}\frac{\partial g_{00}}{\partial r}\right)$.

\noindent De la même façon, on établit que $R^2_{020}=R^3_{030}=-\frac{1}{2r}\frac{\partial g_{00}}{\partial r}$.
On en déduit que $-\sqrt{-g_{00}}R^0_0=\frac{1}{\sqrt{-g_{00}}}\left( \frac{1}{4g_{00}}(\frac{\partial g_{00}}{\partial r})^2-\frac{1}{2}\frac{\partial^2g_{00}}{\partial r^2}-\frac{1}{r}\frac{\partial g_{00}}{\partial r}\right).$
\end{preuve}
\begin{corollary}\label{li1}
\hspace{1cm}\\
Si on pose $U=1-\sqrt{-g_{00}}$, on a :  $\Delta U=R^0_0 \sqrt{-g_{00}}$.
\end{corollary}
\begin{corollary}\label{li2}
\hspace{1cm}\\
én métrique $ds^2$ du théorème \ref{li} sous la forme $g_{00}(r)c^2dt^2+dr^2+r^2d\theta^2+r^2\sin^2\theta d\varphi^2$, en renormalisant $g_{00}(r)$, avec $g_{00}$ négatif et voisin de $-1$ et  $\lim_{r\to\infty}g_{00}(r)=-1$.
 ce qui la fait apparaître  comme une perturbation de la métrique de Lorentz. 
 
\noindent Avec cette normalisation de $g_{00}$ l'équation (\ref{lilimarlene}) est encore vérifiée.
\end{corollary}
\subsection{Le principe d'équivalence d'Einstein et principes fondamentaux de la relativité générale}\label{princece}
\subsubsection{Les principes d'équivalence et d'inertie}\label{peipei}
La loi fondamentale de la dynamique de Newton indique que la résultante $F$ des forces extérieures qui s'appliquent sur une particule est proportionnelle à son accélération $\gamma$ calculée dans un référentiel galiléen : $F=m_i\gamma.$ Si la seule force qui s'exerce sur cette particule est son poids, il résulte de la loi de gravitation universelle que celle ci s'écrit $F=m_p g_p$ où $g_p$ est un vecteur dirigé vers le centre de la Terre\footnote{Si le centre de la terre concentre sa masse la loi de gravitation universelle dit que la terre exerce sur la masse $m_p$ une force d'intensité $F=\frac{\kappa m_T}{r^2}m_p$, où $r$ est la distance entre la masse et le centre de la Terre. Si on pose $g=\frac{\kappa m_T}{r^2}$, on voit que $g$ n'est pas constant puisque $r$ diminue dans la chute mais varie si peu dans une expérience de chute libre à partir du sommet de la tour de Pise qu'on la mesure constante}  , qu'on appelle \og accélération de la pesanteur\;\fg. Newton a admis comme principe l'égalité $m_i=m_p$ égalité, ce qui rend sa loi conforme aux observations de Galilée sur la chute libre. Cette égalité a été confirmée expérimentalement fin du 19-ième siècle par Lor\'{a}n Eötvös. On posera $m=m_i=m_p$.

Plaçons nous dans un référentiel qui est une bonne approximation d'un référentiel galiléen, $\mathcal{R}$ d'origine $O$ situé au pôle nord muni de coordonnées $x,y,z$ associées à trois axes orthogonaux, l'axe des $z$ étant vertical parallèle au vecteur $g_p$. Comme d'habitude on note $\mathcal{R}=(O,x,y,z)$. Imaginons l'expérience suivante : un ascenseur dont le cable est parallèle à l'axe des $z$ se retrouve suite à un accident en chute libre le long de cet axe (le cable a laché). Dans cet ascenseur se trouve par hasard un physicien qui profite de l'aubaine pour étudier cette chute. Le référentiel de mesure du physicien est un référentiel $\mathcal{R}'=(O', x,y,z)$. Le point $O'$ situé sur le plancher de l'ascenseur subit par rapport au référentiel $\mathcal{R}$ une accélération $g$. Le physicien lance verticalement dans l'ascenseur une petite balle de masse $m$ avec une vitesse initiale $v_0$. 
 Soit $\gamma$ l'accélération de la balle dans le référentiel $\mathcal{R}$ et $\gamma'$ dans le référentiel $\mathcal{R}'$. On a la relation $\gamma=\gamma'+g_p$ (l'accélération de Coriolis est nulle). L'écriture de la loi fondamentale de Newton dans le référentiel galiléen $\mathcal{R}$ aboutit à l'égalité $\gamma'=0$ ce qui signifie que la balle a comme trajectoire dans le référentiel lié à l'ascenseur une droite parcourue à vitesse $v_0$.
 Le physicien applique le principe d'inertie de Galilée et conclut qu'il est dans un référentiel galiléen dans lequel la balle n'est soumise à aucune force extérieur et décrit en coordonnées galiléennes une droite.
 
 \noindent Généralisons le scénario précédent. Supposons que $\mathcal{R}$ soit un référentiel galiléen dans lequel l'accélération de la pesanteur est $g_p$ et considérons un référentiel $\mathcal{R}'$ déduit de $\mathcal{R}$ par une translation accélérée d'accélération le vecteur $g_p$ : si $O$ et $O'$ sont les origines respectivement de $\mathcal{R}$ et $\mathcal{R}'$ alors $\frac{d^2 \overrightarrow{OO'}}{dt^2}=g_p.$ Si $M$ est un point matériel lancé dans avec une vitesse initiale par rapport à $\mathcal{R}'$ égale à $v_0$ et  soumis uniquement à son poids, on obtient par la loi fondamentale que l'accélération de $M$ par rapport à $\mathcal{R}'$ est nulle est que $M$ décrit pour un observateur lié à $\mathcal{R}'$ une droite avec la vitesse constante $v_0$. Inversement, si $\mathcal{R}$ est un référentiel galiléen, si $\mathcal{R}'$ est déduit de $\mathcal{R}$ par une translation accélérée d'accélération $g_p$, si $M$ a pour un observateur de $\mathcal{R}'$ une orbite à vecteur vitesse constante $v_0$, un observateur lié à $\mathcal{R}$ en déduira par la loi fondamentale de la dynamique newtonienne qu'il est soumis à la force extérieur 
 $F=mg_p$ qu'il pourra librement interpréter comme le poids de $M$ dans le champ de gravité d'accélération $g_p$.
 D'où le principe d'équivalence énoncé par Einstein :
 \begin{principe}[d'équivalence]\index{principe d'équivalence}
 \hspace{1cm}\\
Les forces de gravitation sont fictives.
 On peut les éliminer localement en se mettant dans un référentiel en translation accélérée par rapport à un référentiel galiléen et inversement traiter un tel référentiel  comme doté d'un champ de gravitation.
\end{principe}
Nous devons conserver un principe d'inertie à la Galilée et étendre le principe  \ref{pringarr} p.\pageref{pringarr} de la relativité restreinte de manière en tenant  compte du principe d'équivalence.
 Tout référentiel en translation rectiligne accélérée par rapport à un référentiel galiléen de la relativité restreinte sera appelé référentiel galiléen ou référentiel d'inertie. Voici son énoncé.

\begin{principe}[d'inertie de la relativité générale]
\hspace{1cm}\\
Si une particule test, avec ou sans masse n'est soumise qu'à l'attraction d'une distribution de masses, elle décrit pour un observateur liée à un référentiel d'inertie une géodésique de la métrique attachée à ce référentiel.
Si $(x^0,x^1,x^2,x^3)$ sont les coordonnées locales du référentiel d'inertie, celui-ci est équipée d'une métrique 
$ds^2=g_{ij}dx^{i}\otimes dx^j$ de signature $(-,+,+,+)$ de sorte que le long de la géodésique on ait $ds^2\le 0$,
avec la condition $ds^2=0$ si la géodésique correspond à la ligne d'univers de propagation de la lumière.
\end{principe}
Quitte à faire un changement de coordonnées $ds^2$ peut s'exprimer sous la forme diagonalisée 
$ds^2=g_{ii}{dx^{i}}^2$. L'expérience montre que les coefficients $g_{ii}$ ne sont pas constants car cela impliquerait une trajectoire rectiligne (voir équation (\ref{paral} p.\pageref{paral})) ce qui n'est pas conforme aux observations de déviation de la lumière en présence d'une planète. Ceci confirme que la relativité restreinte ne peut traiter le problème de la gravitation, ce que l'on a déjà mis en évidence dans la section \ref{dimicorr} dans les hypothèses du théorème \ref{eqmvmcrr}.

On peut considérer par exemple une métrique comme celle proposée dans les hypothèses du théorème \ref{li} dans la version du corollaire \ref{li2}  appelée \emph{métrique statique} pour modéliser l'existence d'un champ faible (cette métrique est proche de la métrique de Lorentz au sens où $g_{00}(r)$ est proche de $-1$.

\noindent Les principes précédents mettent en évidence 
\begin{enumerate}
\item que l'espace temps de la relativité générale doit être une variété pseudo-rie\-ma\-nienne,
\item que la trajectoire d'une particule test est liée à la métrique de cette variété.
\end{enumerate}
En mécanique newtonienne, l'orbite d'une particule pesante dans un champ gravitationnel est déterminée par le potentiel de gravitation tel qu'il apparaît dans la loi de Poisson. Aussi une théorie relativiste  de la gravitation doit-elle nécessairement réaliser trois choses.
\begin{enumerate}
\item \'{E}tablir un lien entre les coefficients de la métrique galiléenne et le potentiel de gravitation newtonien.

\noindent Nous donnerons effectivement une évaluation de ce lien  en section \ref{heuri}
\item \'{E}tablir une loi du mouvement qui étend, en présence de gravité, celle de la mécanique des milieux continus exposée dans le corollaire \ref{eqmeumeu}.

\noindent Ces lois appelées seront exposées en section \ref{eh} sous le nom d'\og équations du champ de gravitation\;\fg. 
\item Retrouver à partir de cette loi une correction relativiste de la loi de Poisson rappelée par l'équation (\ref{poiclas}) de la section suivante. 

\noindent Cette correction est exposée dans le cadre particulier d'un champ de gravitation faible dans la section \ref{poigra}.
\end{enumerate}
\subsubsection{Potentiels de gravitation et coefficients de la métrique galiléenne. Une heuristique.}\label{heuri}
Exposons ici un calcul approché du potentiel par les coefficients de la métrique emprunté à [\ref{franki}].

\noindent On se place dans le cadre de la métrique statique $ds^2=g_{00}(r)c^2dt^2+dr^2+r^2d\theta^2+r^2\sin^2\theta d\varphi^2$ c'est à dire dans le cadre d'un champ faible pour lequel on supposera que $g_{00}$ est proche de $-1$
 et on s'intéresse à une particule test dont la vitesse $v$ est négligeable devant la vitesse de la lumière au point que l'on puisse négliger $\frac{v}{c}$ devant $1$ et $\grad \frac{g_{00}}{2}$.
 On se rappelle que si $\tau$ est le temps propre on a $d\tau^2=-ds^2$. Les coordonnées locales sont ici $(x^0,x^1,x^2,x^3)=(ct,r,\theta,\varphi)$. Un calcul direct montre que 
 
 $$\left(\frac{d\tau}{dx^0}\right)^2=-g_{00}+\underbrace{\left(\frac{dr}{dx^0}\right)^2+r^2\left(\frac{d\theta}{dx^0}\right)^2
 +r^2\sin^2\theta\left(\frac{d\varphi}{dx^0}\right)^2}_{(\frac{v}{c})^2\sim 0 \;\mathrm{d'après \;notre \;hypothèse}}.$$ 
 On a donc $\frac{dx^0}{d\tau}\sim\frac{1}{\sqrt{-g_{00}}}\sim 1$.
Pour $i\in\{1,2,3\}$, on a $\frac{dx^{i}}{d\tau}=\frac{dx^{i}}{dx^0}\frac{dx^0}{d\tau}=\frac{1}{c\sqrt{-g_{00}}}v^{i}\sim\frac{v^{i}}{c}$ où 
$v^{i}=\frac{dx^{i}}{dt}$. Les coordonnées de la particule test vérifient selon le principe d'inertie les égalités 
\[\frac{d^2x^{i}}{d\tau^2}=-\Gamma^{i}_{jk}\frac{dx^j}{d\tau}\frac{dx^k}{d\tau}=-\Gamma^{i}_{00}\left(\frac{dx^0}{d\tau}\right)^2-\sum_{j\ge 1,k\ge 1}\Gamma^{i}_{jk}\left(\frac{v^j}{c}\right)\left(\frac{v^k}{c}\right).\]
Or d'après l'égalité ( \ref{cristo}) p.\pageref{cristo}, on a : $\Gamma^{i}_{00}=-\frac{1}{2}g^{is}\frac{\partial g_{00}}{\partial x^s}=-\left(\grad\big(\frac{g_{00}}{2}\big)\right)^{i}$. Donc d'après nos hypothèses sur les ordres de grandeurs, on peut écrire : 
\[\frac{d^2x^{i}}{d\tau^2}=\left(\grad\big(\frac{g_{00}}{2}\big)\right)^{i}=(\grad U)^{i}\]
si $U$ désigne le potentiel de gravitation de la loi de Poisson. Sachant qu'à l'infini $U$ s'annule quand $g_{00}=-1$, on obtient l'estimation 
\begin{equation}\label{estime}
g_{00}\sim(2U-1).
\end{equation}

\subsubsection{Loi d'Einstein-Hilbert du mouvement d'un fluide relativiste}\label{eh}
Les principes d'équivalence et d'inertie nous montrent que la gravité est liée à la géométrie de l'espace-temps : 
 l'équation (\ref{estime}) qui en découle montre que l'énergie (ici dûe à la gravitation) et la géométrie de l'espace-temps sont liées. Ainsi la relativité générale fait apparaître une nouvelle synthèse non traitée par la relativité restreinte qui elle même avait réalisé une synthèse entre  la masse et  l'énergie étrangère à la physique newtonienne.
 Il s'agit maintenant de donner la loi du mouvement d'un fluide soumis à toute sorte d'énergie dont celle de la gravitation. 
 
 \noindent Pour établir les lois du mouvement, il faut s'appuyer sur le principe fondamental suivant :
  les lois de cette physique relativiste doivent avant toute chose \emph{être en conformité avec la relativité restreinte}, au sens où les solutions des équations du mouvement doivent être celles de la relativité restreinte lorsqu'il n'y a pas de gravitation. Nous partirons pour décrire les équations du mouvement de la théorie des milieux continus de la relativité restreinte. 
  Aussi tout comme en relativité restreinte :
  \begin{enumerate}
  \item l'espace temps a une structure de variété pseudo-riemmanienne $M$ de signature $(-,+,+,+)$ (voir section \ref{peipei}),
  \item l'énergie dûe aux pressions et aux charges électriques est décrite par un tenseur \emph{impulsion-énergie}
  $\mathcal{E}=\mathcal{P}+\mathcal{M}$ vérifiant sur $M$ l'égalité de conservation $\Div\;\mathcal{E}=0$.
  Par exemple dans le cadre d'un fluide parfait le tenseur  $\mathcal{P}$ est donné par l'équation (\ref{impenparf}) p. \pageref{impenparf}.
  Le tenseur $\mathcal{M}$ conserve l'expression donnée par l'équation (\ref{toneuneu}) p. \pageref{toneuneu}.
  \item La synthèse masse-énergie mise en évidence par la relativité restreinte doit demeurer, ce qui impose que la totalité de l'énergie est liée à la géométrie de l'espace-temps.
 \end{enumerate}
 Nous avons vu par le corollaire \ref{fou} du théorème de Cartan, qu'il existe, moyennant une hypothèse peu restrictive une unique famille de tenseurs à deux paramètres $G(\lambda,\nu)$ liée à la géométrie de $M$ qui soit de divergence nulle. Le tenseur d'Hilbert-Einstein-Cartan est par définition le tenseur $G(1,0)=:G$.
 La synthèse énergie-géométrie nous impose nécessairement l'égalité entre un tenseur $G(\lambda,\nu)$ et le produit  $\mathrm{ C}.\mathcal{E}$ où $\mathrm{C}$ est une constante et $\mathcal{E}=\mathcal{P}+\mathcal{M}$.
 Les réflexions initiées par Einstein, ont conduit à poser comme équation du mouvement en présence d'un champs de gravitation (et d'un champ électromagnétique) une égalité de la forme :
 \[G=8\pi\kappa\mathcal{E}\]
 où $\kappa$ est la constante de Newton.\footnote{$\kappa \sim 6,67407. 10^{-11}  m^3/kgs^2$}
 Le choix $\mathrm{C}=8\pi\kappa$ permet de retrouver dans le cadre newtonien la loi de Poisson  (voir section \ref{poigra}).
 D'où l'équation tensorielle, appelée \emph{équation du champ de gravitation}\index{équation du champ de gravitation}
 \begin{equation}\label{eqch}
 \mathrm{Ricc}-\frac{1}{2}gR=8\pi\kappa \mathcal{E},  \end{equation}
 où $g$ désigne le tenseur métrique,  $R$ la courbure scalaire. 
 \begin{remark}
 Le fait que $\Div\;\mathcal{E}=0$, montre que les solutions issues de cette équation sont un cas particulier de celles de l'équation (\ref{eqch}), ce qui est bien conforme à ce qu'on attendait.
 \end{remark}
 \subsection{\'{E}quation du champ de gravitation et loi de Poisson}\label{poigra}

On considère une région de l'espace dans laquelle la masse est répartie avec une densité $\rho$. Cette répartition crée en chaque point un potentiel newtonien $U$. La loi de Poisson de la mécanique newtonienne donne le lien entre ce potentiel et la densité de masse. Elle se traduit par l'équation :
\begin{equation}\label{poiclas}
\Delta U=-4\pi \kappa \rho
\end{equation}
  
 Dans la remarque \ref{impclarela}, nous avons vu que pour passer du tenseur impulsion classique au tenseur énergie-impulsion relativiste il fallait remplacer la densité de masse par la densité de masse-énergie, plus précisément on remplace la densité de masse $\rho$ par la densité $\rho_0+\frac{p}{c^2}$.
 Ainsi la correction relativiste de l'équation (\ref{poiclas}), dans le cadre de la métrique du théorème \ref{li1} qui doit tenir compte de la densité de masse et de la densité d'énergie dûe aux forces de pression est de la forme 
\begin{equation}\label{poirela}
\Delta U=-4\pi \kappa F(\rho_0,\frac{p}{c^2})\sqrt{-g_{00}}
\end{equation}
où $F$ est une fonction réelle continue.
Le principe d'équivalence et l'heuristique qui en a suivi dans la section \ref{princece} montre qu'il y a une relation entre le coefficient $g_{00}$ de la métrique du corollaire \ref{li2} et le potentiel de gravitation $U$. Une approximation ponctuelle de $U$ est donnée par $\frac{1+g_{00}}{2}$ (sans rien préciser sur le sens mathématique de cette approximation).
Notons maintenant deux résultats rigoureux :
\begin{enumerate}
\item Avec les hypothèses du corollaire \ref{li2}, on a : $1-\sqrt{g_{00}}=\frac{1}{2}+\frac{1}{2}g_{00}+o(1+g_{00})$
(au sens où $\lim_{r\to\infty}\frac{o(1+g_{00})}{1+g_{00}}=0$)
\item Le théorème \ref{li} de Levi-Civita.
\end{enumerate}
Ces deux faits, associés à l'estimation (\ref{estime}) incitent à penser que 
\begin{enumerate}
\item $U=1-\sqrt{-g_{00}}$ est le potentiel de gravitation
\item Le théorème de Levi-Civita est la correction relativiste de la loi de Poisson \end{enumerate}
Alors, compte tenu des équations (\ref{lilimarlene}) et (\ref{poirela}), l'équation de Poisson relativiste implique l'égalité 
\begin{equation}\label{poirelageo}
R^0_0=-4\pi\kappa \;F(\rho_0,\frac{p}{c^2})
\end{equation}
Considérons maintenant le théorème suivant.
\begin{theorem} \label{poietein}
\hspace{1cm}\\
Supposons que nous soyons en présence d'un fluide parfait immobile, sans interaction électromagnétique.
On note toujours $\mathcal{P}$ le tenseur impulsion-énergie.
Les équations du champ impliquent l'équation
\begin{equation}\label{vraiepoire}\index{loi de Poisson relativiste}
R^0_0=-4\pi\kappa\left(\rho_0+\frac{3p}{c^2}\right)
\end{equation}
\end{theorem}
\noindent L'équation (\ref{poirela}) est  bien alors une correction de la loi de Poisson (\ref{poiclas}). On peut la nommer \og loi de Poisson relativiste\fg. Elle s'écrit sous forme :
$$\Delta U=-4\pi\kappa\left(\rho_0+\frac{3p}{c^2}\right)\sqrt{-g_{00}}.$$ 
On retrouve la loi de Poisson classique en négligeant le terme $\frac{3p}{c^2}$ et en considérant que $g_{00}=-1$.

\noindent La démonstration de la relation (\ref{vraiepoire}) n'utilise pas la totalité des $10$ équations scalaires contenues dans les équations du champ de gravitation mais seulement $4$ d'entre elles comme on va le voir. Les informations des équations du champ sont donc plus riches que l'équation de Poisson. On développe des aspects géométriques nouveaux de ces équations dans la section suivante.
\begin{preuve}{du théorème \ref{poietein}}
On commence par énoncer un lemme concernant le tenseur impulsion-énergie.
\begin{lemma}Sous les hypothèses du corollaire \ref{li2} p. \pageref{li2}, on a  les égalités
\begin{align}
\mathcal{P}^0_0&=-\rho_0,\\
\tr\;\mathcal{P}&=-\rho_0+\frac{3p}{c^2}.
\end{align}
\end{lemma}
\begin{preuve}{du lemme}
Le quadrivecteur vitesse $u$ qui entre dans la définition du tenseur impulsion-énergie a pour norme $-1$ relativement à la métrique du théorème \ref{li}.  Or dans les hypothèses d'un fluide parfait immobile on a :$u=u^0\frac{\partial}{\partial u^0}$ car pour $i\in\{1,2,3\}$ on a  $u^{i}=\frac{dx^{i}}{d\sigma}=0$ où $d\sigma =\sqrt{-ds^2}$. Ce qui implique $u^0=\frac{1}{\sqrt{-g_{00}}}$. On a donc $\mathcal{P}^0_0=g_{0i}\mathcal{P}^{0i}=g_{00}\mathcal{P}^{00}=g_{00}\left((\rho_0+\frac{p}{c^2})(u^0)^2+\frac{p}{c^2}g^{00}\right)=-\rho_0.$

\noindent Par ailleurs $\tr \mathcal{P}=\mathcal{P}^{i}_i=(\rho_0+\frac{p}{c^2})\langle u,u\rangle+4\frac{p}{c^2}=
-\rho_0+\frac{3p}{c^2}.$
\end{preuve}

\noindent Des équations du champ $R^{i}_j-\frac{1}{2}\delta^{i}_jR=8\pi\kappa \mathcal{P}^{i}_j$, on déduit d'une part que  $R^{i}_i -2R=8\pi\kappa \tr\mathcal{P}$ ou encore $R=-8\pi\kappa \tr\mathcal{P}$ et d'autre part que 
$R^0_0=\frac{1}{2}R+8\pi\kappa\mathcal{P}^0_0$. D'où en utilisant le lemme, on obtient : 
$$R^0_0=-4\pi\kappa(-\rho_0+\frac{3p}{c^2})-8\pi\kappa\rho_0=-4\pi\kappa(\rho_0+\frac{3p}{c^2}).$$
\end{preuve}
\section{Géométrie des équations du champ.}\label{geoequachation}
Les relations (\ref{eincoursec}) p. \pageref{eincoursec} permettent de donner une écriture équivalente aux équations du champ de gravitation (\ref{eqch}) faisant intervenir les courbures sectionnelles.
\begin{proposition}\label{geau}
En tout point de l'espace temps, notons $(e_0,e_1,e_2,e_3)$ une base orthonormée du fibré tangent où $e_0$ est du type temps et $e_1,e_2,e_3$ sont du type espace ( i.e. $\la e_0,e_0\ra=-1, \la e_i,e_i\ra=1$ pour $i\in\{1,2,3\}$).
Rappelons que $\mathcal{E}$ désigne le tenseur impulsion-énergie. Alors les équations (\ref{eqch}) sont équivalentes à
\begin{equation}\label{pifo}
\forall i\in\{0,1,2,3\}\;\;\mathcal{K}(e_i^\perp)=-8\pi\kappa \epsilon_i\mathcal{E}(e_i,e_i).
\end{equation}
\end{proposition}
\begin{preuve}{}
Il est clair que l'équation (\ref{eqch}) implique l'équation (\ref{pifo}) à partir de (\ref{eincoursec}).
Inversement si l'équation(\ref{pifo}) est vérifiée, alors  $G(e_i,e_i)=8\pi\kappa\mathcal{E}(e_i,e_i)$ pour tout $i\in\{0,1,2,3\}$ grâce à (\ref{eincoursec}). Soit $x$ une point de l'espace-temps $M$ et $X$ un vecteur non nul de 
$T_xM$. On peut compléter $X/\norm{X}$ en une base de $T_xM$. Mais alors, par hypothèse on a :
$G\left(\frac{X}{\norm{X}},\frac{X}{\norm{X}}\right)=8\pi\kappa\mathcal{E}\left(\frac{X}{\norm{X}},\frac{X}{\norm{X}}\right)$, donc $G(X,X)=8\pi\kappa\mathcal{E}(X,X)$. Ce qui montre que $G$ est entièrement défini sur la diagonale
$\Delta=\{(v,v);v\in T_xM\}$ de $T_xM$. $G$ étant symétrique, il est défini partout et on a :
\begin{align*}
G(e_i,e_j)&=\frac{1}{2}\left(G(e_i+e_j,e_i+e_j)-G(e_i,e_i)-G(e_j,e_j)\right)\\
&=\frac{1}{2}8\pi\kappa\left(\mathcal{E}(e_i+e_j,e_i+e_j)-\mathcal{E}(e_i,e_i)-\mathcal{E}(e_j,e_j)\right)\\
&=8\pi\kappa \;\mathcal{E}(e_i,e_j),
\end{align*}
ce qui implique l'équation (\ref{eqch}).
\end{preuve}
On considère dans l'espace temps une hypersurface $S$, un point $x$ de $S$. La fonction de Weingarten $\mathcal{W}$ associée à $S$ en $x$ est symétrique donc diagonalisable. Notons $(e_j,e_k,e_l)$ une base orthonormée de vecteurs propres 
associés respectivement aux valeurs propres $\lambda_j,\lambda_k,\lambda_l$. Dans la théorie classique des surfaces ces valeurs propres sont appelées les \emph{courbures principales }en $x$, le déterminant de $\mathcal{W}$ est la \emph{courbure de Gauss} que nous avons déjà vue dans le théorème egregium (corollaire \ref{thega})\index{courbure de Gauss}\index{courbure moyenne}\index{courbure de Wheeler}
la trace de $\mathcal{W}$ est la \emph{courbure moyenne} de $S$ en $x$. Le polynôme caractéristique de 
$\mathcal{W}$ s'écrit $X^3-\tr(\mathcal{W})X^2+W(S)X-\det(\mathcal{W})$. Il met en évidence trois invariants : la courbure de Gauss, la courbure moyenne et $W(S)=\lambda_j\lambda_k+\lambda_j\lambda_l+\lambda_k\lambda_l$ que nous appellerons la \emph{courbure de Wheeler} de $S$ en $x$ dans cet ouvrage.
\begin{corollary}\label{geaugeau}
\hspace{1cm}\\
On considère une hypersurface $S$ et $x\in S$. On note $e_i$ un vecteur unitaire de $T_xM$ normal en $x$ à $S$.
On désigne par $R^S$ la courbure scalaire de $S$ en $x$. Alors \begin{equation}\label{deouf}
G(e_i,e_i)=-\frac{\epsilon_i}{2}R^S+W(S).
\end{equation}
\end{corollary}

\begin{preuve}{}
On considère  la base $(e_0,e_1,e_2,e_3)$ de la proposition \ref{geau}. On suppose que 
\begin{enumerate}
\item $\{e_i,e_j,e_k,e_l\}=\{e_0,e_1,e_2,e_3\}$,   
\item $S$ admet $e_i$ comme vecteur normal en $x$,
\item $(e_j,e_k,e_l)$ est une base orthonormée de vecteurs propres de $\mathcal{W}$ en $x$.
\end{enumerate}
Alors le corollaire \ref{cousecind} permet d'écrire que 

\noindent$K(e_j,e_k)=K^S(e_j,e_k)+\epsilon_i\epsilon_j\epsilon_k\left(\la \mathcal{W}(e_j),e_k\ra^2-\la\mathcal{W}(e_j),e_j\ra\la\mathcal{W}(e_k),e_k\ra\right)$ ou encore $K(e_j,e_k)=K^S(e_j,e_k)-\epsilon_i\lambda_j\lambda_k.$

\noindent Ainsi $\mathcal{K}(e_i^\perp)=K(e_j,e_k)+K(e_j,e_l)+K(e_k,e_l)=\frac{1}{2}R^S-\epsilon_i W(S)$ en appliquant le deuxième item du corollaire \ref{seccourisca} p.\pageref{seccourisca}. On a donc 

\noindent $\mathcal{K}(e_i^\perp)=-8\pi\kappa\epsilon_i\mathcal{E}(e_i,e_i)=-\epsilon_i G(e_i,e_i)$
\end{preuve}
Il est clair que le vecteur normal à $S$ en un point ne détermine pas la courbure scalaire d'une hypersurface $S$ en ce point ni la courbure de Wheeler en ce point. Cependant dans l'espace temps de la relativité générale 
$-\frac{\epsilon_i}{2}R^S+W(S)$ est déterminé par vecteur unitaire normal au point considéré par l'équation (\ref{deouf}).
\begin{corollary}
Avec les mêmes notations, dans l'espace de Minkowski, on a :$$W(S)=\frac{\epsilon_i}{2} R^S.$$
\end{corollary}
\begin{corollary}\label{relhypgeo}
Si $S$ est une hypersurface totalement géodésique de l'espace-temps dont on note $N$ un champ de vecteurs normal unitaire alors le tenseur impulsion-énergie détermine sa courbure scalaire de sorte que :
\[R^S=-16\epsilon_N\pi\kappa\mathcal{E}(N,N).\]
\end{corollary}
\begin{preuve}{du corollaire \ref{relhypgeo}}
D'après le corollaire \ref{geaugeau}, on a : $G(N,N)=-\frac{\epsilon_N}{2}R^S$. En effet, $W(S)=0$ puisque $S$ est géodésique (voir corollaire \ref{hypgeo} p. \pageref{hypgeo}). Ainsi $8\pi\kappa\mathcal{E}(N,N)=-\frac{\epsilon_N}{2}R^S$. D'où la relation annoncée.
\end{preuve}
\begin{definition}\label{etreoupas}
Si $S$ est une hypersurface de type espace (c'est à dire que $N$ dans le corollaire précédent est du type temps), $\mathcal{E}(N,N)$ est appelée densité d'énergie de courbure.
\end{definition}
La définition \ref{etreoupas} est justifiée par le corollaire suivant.
\begin{corollary}\label{relhypgeoesp}
Si $S$ est une hypersurface géodésique de type espace 
de l'espace temps dont le tenseur impulsion-énergie est produit par un fluide parfait, alors la courbure scalaire de $S$, $R^S$, en tout point où la normale unitaire est $N$  vérifie :
\[R^S=16\pi\kappa \mathcal{E}(N,N).\]
\end{corollary}

 \section{Appendice 1 : démonstration du théorème \ref{cossi}}\label{cosssi}

Tout d'abord quelques rappels généraux.
\subsubsection{Somme topologique et topologie quotient}
 Soit $\{E_i\}_{i\in I}$ une famille d'espaces topologiques, $E$ un ensemble  et $\{f_i\}$ une famille d'applications $f_i : E_i\longrightarrow E$. L'ensemble des parties $A\subset E$ telles que pour tout 
 $i\in I, \;f_i^{-1}(A)$ est un ouvert de $E_i$, forment une topologie sur $E$.\index{topologie finale}
 
 Cette topologie de $E$ est appelée la \emph{topologie finale} sur $E$ définie par la famille $\{(E_i,f_i)\}$.
 \begin{exo}\label{topfi}
 Démontrer cette assertion.
 \end{exo}
 Pour tout $i\in$ on munit $\{i\}\times E_i$ de la topologie naturelle suivante :
 $\{i\}\times A_i$  est un ouvert de $\{i\}\times E_i$  si $A_i$ est un ouvert de $E_i$.
On définit alors pour tout $j\in I$ l'injection canonique $i_j$ de $E_j$ dans $\bigcup_{i\in I}\{i\}\times E_i$ : pour tout $x\in E_j$, $i_j(x)=(j,x)$.
\begin{definition}\label{sotop}
L'ensemble $\Sigma=\bigcup_{\alpha\in I}\{\alpha\}\times E_\alpha$ muni de la topologie finale définie par la famille 
$( E_\alpha,i_\alpha)\}$ est appelé la somme topologique des espaces $E_\alpha$ ou encore la réunion disjointe des $E_\alpha$ et parfois notée $\coprod_{\alpha\in I} E_\alpha$.\index{somme topologique}\index{réunion disjointe}
\end{definition}
\begin{remark}
Il faut distinguer réunion et réunion disjointe : $E\coprod E$ est la réunion de deux ensembles distincts alors que $E\cup E=E$. Il est commode lorsqu'on veut distinguer $E$ de lui-même de l'écrire sous deux formes : par exemple $\{1\}\times  E$ et $\{2\}\times E$.
\end{remark}
Avec la notation de la définition \ref{sotop}, une partie $A$ de $\Sigma$ est de la forme 
\[\bigcup_{\alpha\in I}\{\alpha\}\times A_\alpha,\;\mathrm{avec}\;A_\alpha\subset E_\alpha\]
et on a $i_\alpha^{-1}(A)=A_\alpha.$
Ainsi, $A$ est un ouvert si pour tout $i\in I$, $A_i$ est un ouvert de $ E_i$. Autrement dit les ouverts de 
$\coprod E_i$ sont de la forme $\coprod O_i$ où $O_i$ est un ouvert de $E_i$.
\subsubsection{Topologie quotient}\index{topologie quotient}
Soit $E$ un ensemble et $\sim$ une relation d'équivalence sur $E$. On note $E/{\sim}$ l'ensemble quotient et $q:E\longrightarrow E/{\sim}$ la surjection canonique qui à tout élément de $E$ associe sa classe d'équivalence.
Si $E$ est un espace topologique, l'ensemble des parties $[A]$ de $E/{\sim}$ telles que $q^{-1}([A])$ est un ouvert de $E$ forment une topologie de $E/{\sim}$ appelée \emph{topologie quotient}.
\begin{exo}\label{toqo}
Démontrer l'assertion précédente.
\end{exo}
Si $A$ est une partie de $E$, la partie $q^{-1}\big(q(A)\big)$ contient $A$ ainsi que tous les éléments de $E$ qui sont équivalents à un élément de $A$.  On l'appelle de ce fait la \emph{partie saturée de $A$.} On la notera parfois $\sat(A)$. On dit que la \emph{relation d'équivalence est ouverte} si l'image par $q$ de tout ouvert de $E$ est un ouvert de $E/{\sim}$. \index{saturé d'un ensemble}

\noindent Alors $\sim$ est ouverte si et seulement le saturé de tout ouvert de $E$ est un ouvert.
\begin{exo}\label{sature}
Démontrer l'assertion précédente.
\end{exo}
\subsubsection{Démonstration de la proposition \ref{cossi}}
\noindent En deux points. 

\noindent \emph{Premier point : construction d'un fibré vectoriel $(E,p,M)$ de fibre $F$ (sur laquelle on a une topologie définie par une norme), de groupe $G$ ayant pour changements de cartes locales un cocycle donné $\{c_{ij}\}$ sur $M$.}

\noindent On suppose donc un recouvrement ouvert $\{U_i\}$ de $M$ et on précise que pour tout $x\in U_i\cap U_j, c_{ij}(x)$ est un automorphisme de $F$ appartenant à $G$.

\noindent On définit sur la somme topologique $\Sigma=\coprod_i U_i\times F$ la relation d'équivalence : 

\noindent $\forall  (x,e)\in U_i\times F,\forall (x',e')\in U_j\times F,\; (i,x,e)\sim (j,x',e') \;\mathrm{si}\; x=x'\in U_i\cap U_j\;\mathrm{ et}\; e'=c_{ij}(x) e$.
On vérifie directement qu'il s'agit bien d'une relation d'équivalence. Soit $E$ l'ensemble quotient
et $q$ la surjection canonique de $\Sigma$ sur $E$. 

\noindent On remarque deux propriétés de cette relation : elle est ouverte et elle est compatible avec l'application continue
$\Phi :\Sigma\longrightarrow M$ définie  pout tout $\alpha\in I$ et tout $(x,e)\in U_\alpha\times F$ par $\Phi \big((\alpha,x,e)\big)=x.$
Justifions ces deux remarques.

\noindent Pour montrer que $\sim$ est ouverte, on considère un ouvert $\{\alpha\}\times \mathcal{O}_\alpha\times F_\alpha $ de $\Sigma$ où $ \mathcal{O}_\alpha$ est un ouvert de $ U_\alpha$ et $F_\alpha $ est un ouvert de $F$. Soit 
$I_\alpha =\{\beta\in I\;;\;\mathcal{O}_\alpha\cap U_\beta\not=\emptyset\}$. Alors
 \[\sat\Big(\{\alpha\}\times \mathcal{O}_\alpha\times F_\alpha \Big)=
 \bigcup_{\beta\in I_\alpha}\{\beta\}\times (\mathcal{O}_\alpha\cap U_\beta)\times \big(\bigcup
 _{x\in U_\beta\cap \mathcal{O}_\alpha}c_{\beta\alpha}(x)F_\alpha\big).\]
Le saturé de $\{\alpha\}\times \mathcal{O}_\alpha\times F_\alpha $  est  une réunion d'ouverts (en effet les $c_{\beta\alpha}(x)$ sont aussi des homéomorphismes de $F$ dans $F$) donc un ouvert. Comme tout ouvert de $\Sigma$ est une réunion de tels ouverts et que le saturé d'une réunion est la réunion des saturés (petit exercice), on peut conclure que $\sim$ est une relation ouverte.
L'application $\Phi$ est continue car si $\mathcal{O}$ est un ouvert de $M$, alors 
$\Phi^{-1}(\mathcal{O} )= \coprod_{i\in I}(U_i\cap \mathcal{O} )\times F$ qui est ouvert dans $\Sigma$ et clairement $(x,e)\sim (x',e')$ entraine $\Phi((x,e))=\Phi((x',e'))=x$.

Une conséquence de ces deux remarques est l'existence d'une application $p$ continue de $E$ vers $M$ telle que $\Phi=p\circ q$ : $\forall (\alpha,x,e)\in \Sigma, \;p(q(\alpha,x,e))=\Phi ((\alpha,x,e))$. Cette application $p$ est continue puisque pour tout ouvert $A$ de $M$, on a  $p^{-1}(A)=q(\phi^{-1}(A))$ qui est ouvert.

Notons $\Psi_\alpha$ la restriction de $q$ à $\{\alpha\}\times U_\alpha\times  F$ noté désormais $U_\alpha\times  F$  . Les couples $(U_\alpha,\Phi_\alpha)$
possèdent les propriétés suivantes :
\begin{enumerate}
\item $\Psi_\alpha $ est un homéomorphisme de $U_\alpha\times F$ vers $p^{-1}(U_\alpha)$.

On pose $\Phi_\alpha=\Psi_\alpha^{-1}$.
\item si $U_\alpha\cap U_\beta\not=\emptyset$, alors 
\[\forall x\in U_\alpha\cap U_\beta, \forall e\in F,\;\Phi_\beta^{}\Phi_\alpha^{-1}(x,e)=\big(x,c_{\beta\alpha}(x)e\big).\]
\end{enumerate}
L'établissement de ces deux propriétés confèrera à $(E,p,M)$ une structure de fibré vectoriel de fibre $F$ de groupe $G$ admettant $\{c_{ij}\}$ comme changement de  cartes locales.
Pour le premier item, si $\Psi_\alpha(x,e)=\Psi_\alpha(x',e')$ pour deux éléments $(x,e)$ et $( x',e')$
de $U_\alpha\times F$, alors $\Phi (x,e)=\Phi(x',e')$ et donc $x=x'$ et $e'=c_{\alpha\alpha}e=e.$
Ainsi $\Psi_\alpha$ est une injection, ouverte (puisque $q$ est ouverte) et continue. C'est donc un homéomorphisme de $U_\alpha$ sur son image qui est clairement $p^{-1}(U_\alpha)$.

\noindent Pour le deuxième item, il résulte directement de l'égalité 
\[ \forall x\in U_\alpha\cap U_\beta\forall e\in F\;q(x,e)=q(x,c_{\beta\alpha}(x)e).\]

\noindent \emph{Deuxième point : unicité du fibré déterminé par un cocycle.}

\noindent On se donne un deuxième fibré vectoriel $(E',p',M)$ de fibre $F$ muni d'un altlas $\{U_\alpha,\Phi'_\alpha\}$ tel que
$\forall x\in U_\alpha\cap U_\beta, \forall e\in F,\;{\Phi'}_\beta^{}{\Phi'}_\alpha^{-1}(x,e)=\big(x,c_{\beta\alpha}(x)e\big).$ On considère l'application $\tilde{H}$ de $\Sigma$ vers $E'$
qui est égale à ${\Phi'}_\alpha^{-1}$ sur $U_\alpha\times F$. La clef de la démonstration est que $\tilde{H}$ est compatible avec la relation $\sim$ définie sur $\Sigma$. 
En effet, si $(x,e)\in U_\alpha\times F$, $(x',e')\in U_\beta\times F$ et si $(x,e)\sim ((x',e')$ alors $x=x'$ et
$\Phi_\beta\Phi_\alpha^{-1}(x,e)=(x, c_{\beta\alpha}(x)e)=(x,e')={\Phi'}_\beta{\Phi'}_\alpha^{-1}(x,e)$, c'est à dire que ${\Phi'}_\alpha^{-1}(x,e)={\Phi'}_\beta^{-1}(x',e')$. 

Ainsi $\tilde{H}$ définit une application $H$ de $E=\Sigma/{\sim}$ vers $E'$ vérifiant $p'\circ H=p$. Dans les cartes 
$(U_\alpha,\Phi_\alpha)$ (sur $E$) et $(U_\beta, {\Phi'}_\beta)$ ( sur $E'$), l'application $H$ est s'exprime conformément à l'équation (\ref{morfer}) par l'application $h_{\beta\alpha}=c_{\beta\alpha}$ et donc $H$ est selon l'exercice \ref{job} un isomorphisme de fibré au-dessus de l'identité. Ceci achève la démonstration de l'unicité à un isomorphisme près.
\section{Appendice 2 : fibré localement trivial}\label{ququ}
\subsection{Groupe opérant sur un espace topologique}\index{groupe topologique}
Un groupe topologique est un groupe muni d'une topologie pour laquelle la loi de groupe est continue ainsi que le passage à l'inverse. Par exemple $U(1)=\left\{e^{i\theta}\;;\;\theta\in[0,2\pi[\right\}$ muni de la multiplication habituelle est un groupe topologique. En effet si on pose $d(e^{i\theta},e^{i\alpha})=\abs{\theta-\alpha}$ on définit une distance sur $U(1)$ pour laquelle la multiplication est continue ainsi que l'application $e^{i\theta}\mapsto e^{-i\theta}$.
\begin{definition}\label{optop}\index{opération de groupe sur un espace topologique}
Soit $G$ un groupe topologique, d'élément neutre $e$, soit $E$  un espace topologique $E$. 
On dit que $G$ opère sur $E$ s'il existe une application continue
\begin{eqnarray*}
G\times E&\longrightarrow&E\\
(g,x)&\mapsto g.x
\end{eqnarray*}
telle que 
\begin{enumerate}
\item
$\forall x\in E,\;e.x=x$
\item $ \forall x\in E, \forall g_1,g_2\in G,\; g_1.(g_2.x)=(g_1g_2).x$.
\end{enumerate}
\end{definition}
Si $G$ opère sur $E$ et pour tout $x\in E$, l'ensemble $G_x=\{g\in G\;;\;g.x=x\} $ est un sous-groupe de $G$ : en effet, il n'est pas vide car $e\in G_x$, il est stable par la loi de groupe et par l'inversion, à partir de la définition \ref{optop}.
\begin{definition}\label{effir}\index{opération de groupe effective}
On dit que l'opération de $G$ sur $E$ est effective si $$\bigcap_{x\in E}G_x=\{e\}$$
\end{definition}
\noindent L'intérêt de cette définition est donné par la proposition \ref{efiso}. Notons tout d'abord que si $G$ opère sur $E$, alors pour tout $g\in G$, $x\mapsto g.x$ est un homéomorphisme de $E$ sur lui-même. Notons $\mathcal{H}(E)$ le groupe des homéomorphismes de $E$ sur lui-même. 
\begin{proposition}\label{efiso}
Si l'opération de $G$  sur $E$ est effective, alors $G$ s'identifie à un sous-groupe de $\mathcal{H}(E)$ 
\end{proposition}
\begin{exo}\label{efisoso}
Démontrer la proposition \ref{efiso}.
\end{exo}

\subsection{Fibré localement trivial}
Nous sommes en mesure de généraliser la définition \ref{fifibre} d'un fibré vectoriel ainsi que les définitions qui en découlent.
\begin{definition}\label{fififibre}
On considère un espace topologique $E$, une variété $M$ (qui sera ici différentiable), $p$ une surjection continue de $E$ vers $M$, $F$ un espace topologique   et enfin $G$ un groupe topologique  qui opère de façon effective sur $F$. Le triplé $\tau=(E,p,M)$  est
\emph{un fibré de fibre} $F$ et de \emph{groupe} $G$  si on dispose d'un recouvrement ouvert de $M$, $\{U_{i} \}_{i}$ de sorte que :
\begin{enumerate}
\item Il existe un homéomorphisme $\Phi_i$ de $p^{{-1}}(U_{i})$ sur $U_{i}\times F$, tel que 
si  $p_{1}$ désigne la projection de $ U_{i}\times F$ sur $ U_{i}$ alors 
\begin{equation}\label{restoto}p_1\circ\Phi_i=p.\end{equation}
Le couple $(U_{i},\Phi_{i})$ est appelé \emph{une carte locale} du fibré ou aussi une trivialisation de $\tau$. L'ensemble des cartes $\{(U_{i},\Phi_{i})\}_{i}$ est appelé \emph{un atlas} du fibré.
\item Si $U_{i}$ et $U_{j}$ sont des domaines de cartes non disjoints les changements de cartes $\Phi_{i}\Phi_{j}^{{-1}}$ sont de la forme :
\begin{equation}\label{changchang}
 \begin{array}{cccc}
\Phi_{i}\Phi_{j}^{{-1}} :& U_{j}\cap U_{i}\times F&\longrightarrow&U_{i}\cap U_{j}\times F\\
&(x,f)&\longmapsto&\bigl(x, c_{ij}(x)f\bigr)
\end{array}
\end{equation}
\noindent où $c_{ij}$ est une application continue de $U_i\cap U_j$ vers $G$.
\end{enumerate}
 La variété $M$ est \emph{la base du fibré} $\tau$. Si seul le premier item est réalisé, on dit que $\tau$ est un fibré localement trivial (sans référence au groupe $G$).
 \end{definition}
 \noindent Comme dans le cas des fibrés vectoriels, la famille  $\{c_{ij}\}$ est un cocycle sur $M$.
 Notons que
la proposition \ref{efiso} montre qu'on peut remplacer dans la définition \ref{fififibre} $G$ par un sous-groupe du groupe des homéomorphismes de $F$.
Voici comment se modifie la définition de morphisme d'in fibré vectoriel. 
\begin{definition}\label{morficus}
Soient $\tau=(E,p,M)$ et $\tau'=(E',p',M')$ deux fibrés localement triviaux de  fibres respectivement   $F$ et $F'$
 et d'atlas respectifs $\{(U_{i},\Phi_{i})\}_{i}$ et  $\{(U'_{j},\Phi'_{j})\}_{j}.$  Soit $h$ une application différentiable de $M$ vers $M'$. Un morphisme $H$ de $\tau$ vers $\tau'$ au-dessus de $h$ est une application continue de $E$ vers $E'$ telle que :
\begin{enumerate}
\item si $e\in p^{-1}(b)$, alors $H(e)\in p'^{-1}\big(h(b)\big)$,
\item si $e$ s'écrit dans la carte locale $(U_{i},\Phi_{i})$ : $e=\Phi_{i}^{-1}(b,f)$, alors $H(e)$ s'écrit dans une carte locale  $(U'_{j},\Phi'_{j})$ telle que $h(b)\in U'_{j}$ :
$H(e)=\Phi'^{-1}_{j}\big(h(b),h_{ji}(b)f\big)$
où $h_{ji}$ est une application continue de $U_{i}\cap h^{{-1}}(U'_{j})$ dans $\mathcal{C}_0(F,F')$, ensemble des applications continues de $F$ vers $F'$.
\end{enumerate}
\end{definition}
Les applications $\{h_{ij}\}$ vérifient les relations (\ref{concon}) et la réciproque énoncée dans la proposition \ref{crac} est encore vraie :
\begin{proposition}\label{drac}
si $h$ est une application de $M$ vers $M'$ et si une famille d'applications continues $\{h_{ij}\}$ vérifie les équations (\ref{concon}), où $h_{ij}$  est une application continue de  $h^{-1}(U'_{i})\cap U_{j}$ vers $\mathcal{C}_0(F,F')$, alors elles définissent par l'équation (\ref{morfer}) un morphisme de fibré $\tau$ vers le fibré $\tau'$.
\end{proposition}
\begin{preuve}{}
Soit $e\in p^{-1}(b)$. Soient $(U_i,\Phi_i)$ et $(U_k\Phi_k)$ deux cartes locales en $b$ et $(U'_j,\Phi'_j)$,$(U'_l\Phi'_l)$ deux cartes locales en $h(b)$. On a $e=\Phi_i^{-1}(b,f)=\Phi_k^{-1}(b,s)$.
Montrons que si $H\Phi_i^{-1}(b,f)={\Phi'_j}^{-1}\left(h(b),h_{ji}(b)f)\in {p'}^{-1}(h(b)\right))$ alors on a également
$H{\Phi_k}^{-1}(b,s)={\Phi'_l}^{-1}\left(h(b),h_{lk}(b)s\right)$, ce qui montrera que l'écriture de $H(e)$ dans les cartes locales défini un opérateur intrinsèque qui applique chaque fibre $p^{-1}(b)$ sur la fibre ${p'}^{-1}(h(b))$.

Supposons donc $H\Phi_i^{-1}(b,f)={\Phi'_j}^{-1}\left(h(b),h_{ji}(b)f\right)$. On a : 

$H\Phi_i^{-1}(b,f)=H\Phi_k^{-1}\big(\Phi_k\Phi_i^{-1}(b,f)\big)=H\Phi_k^{-1}(b,c_{ki}(b)f)=H\Phi_k^{-1}(b,s)=$

$={\Phi'_l}^{-1}\left(\Phi'_l{\Phi'_j}^{-1}\right)\big(h(b),h_{ji}(b)f\big)={\Phi'_l}^{-1}\left(h(b),c'_{lj}(h(b))h_{ji}(b)f\right)=$

$={\Phi'_l}^{-1}\big(h(b),h_{li}(b)f\big)={\Phi'_l}^{-1}\big(h(b),h_{li}(b)c_{ik}(b)(s)\big)={\Phi'_l}^{-1}\big(h(b),h_{lk}(s)\big)$.
\end{preuve}
Si $H$ est un homéomorphisme alors on dit que c'est un\emph{ isomorphisme} de fibrés au-dessus de $h$.
\begin{exo}\label{gibus}
Dans  la définition \ref{morficus}, on suppose les deux items suivants : \begin{enumerate}\item Les fibres $F$ et $F'$ sont homéomorphes  et
les applications $h_{ij}$   sont à valeurs dans le groupe des homéomorphismes de $F$ vers $F'$,\item $h$ est un homéomorphisme.\end{enumerate}Montrer que $H$ est un isomorphisme de fibré.
\end{exo}

Reste le problème de la caractérisation \emph{à une équivalence près} des fibrés par ses cocycles. Avant cela nous allons restreindre la classe des isomorphismes entre deux fibrés localement triviaux pour définir précisément ce qu'on entend par \emph{fibrés localements triviaux équivalents}.
\begin{definition}\label{eqfilotri}
On considère les deux fibrés localement triviaux $\tau=(E,p,M)$ et $\tau'=(E',p',M)$ de même base $M$, de même groupe \;$G$, de fibres repectives $F$ et $F'$. On suppose que $G$ opère de façon effective sur $F$ et sur $F'$. Une équivalence  de $\tau$ et $\tau'$ est un morphisme $H$ de $\tau$ vers $\tau'$ \emph{au-dessus de l'identité de $M$} tel que les applications $h_{ij}$ de la définition \ref{fififibre} soient à valeurs dans $G$.

\noindent Dans ce cas on dit que $\tau$ et $\tau'$ sont équivalents.\index{fibrés localement triviaux équivalents}
\end{definition}
On peut énoncer maintenant la caractérisation annoncée, d'un fibré localement trivial par ses cocycles.
\begin{theorem}\label{supercoco}
Un cocycle sur un espace topologique $M$ à valeurs dans un groupe $G$ qui opère de façon effective sur un espace topologique $F$ définit un fibré $(E,p,M)$ de groupe $G$ et de fibre $F$.
Ce fibré est unique à une équivalence près.
\end{theorem}
La preuve de ce théorème se calque sur la preuve  du théorème \ref{cossi}, donnée en détail dans l'appendice $1$. On la trouvera toutefois dans [\ref{steen}].
 \subsection{Un exemple}\label{ququ}
 On considère sur la sphère $S^{2}$ la structure de variété différentiable définie par l'atlas à deux cartes décrit ci-après.
 
 \noindent \`{A}  tout complexe $z$ on associe le point de $\rr^{3}$ de coordonnées $\dfrac{(1-\lvert z_{}\rvert^{2} ,2\bar{z_{}})} {1+\lvert z_{}\rvert^{2}}$ où $\dfrac{1-\lvert z_{}\rvert^{2}}{1+\lvert z_{}\rvert^{2}}$ est la troisième coordonnée  dans $\rr^{3}$ et  $\dfrac{2\bar{z_{}}} {1+\lvert z_{}\rvert^{2}}$ est le complexe représentant naturellement les deux premières coordonnées dans le plan $0xy$ de $\rr^3$. On vérifie directement que ce point appartient à $S^{2}$ et que l'ensemble 
$ V=\left\{\dfrac{(1-\lvert w_{}\rvert^{2} ,2\bar{w_{}})} {1+\lvert w_{}\rvert^{2}}\;;\;w\in\cc\right\}$ est égal à l'ouvert $S^{2}\setminus{(0,0,-1)}$ de $S^{2}$ (pour la topologie induite) . Posons $\varphi_{V}^{{-1}}(w)=\dfrac{(1-\lvert w_{}\rvert^{2} ,2\bar{w_{}})} {1+\lvert w_{}\rvert^{2}}$. On vérifie que $\varphi_{V}^{{-1}}$ définit une bijection de $\cc$ sur $V$, dont on note bien sûr $\varphi_{V}$ la réciproque.
\noindent De même posons $U=\left\{\dfrac{(\lvert z_{}\rvert^{2}-1 ,2{z_{}})} {1+\lvert z_{}\rvert^{2}}\;;\;z\in\cc\right\}$. On vérifie que $U=S^{2}\setminus\{(0,0,1)\}$ et notant $\varphi_{U}^{{-1}}(z)=\dfrac{(\lvert z_{}\rvert^{2}-1 ,2{z_{}})} {1+\lvert z_{}\rvert^{2}}$, que $\varphi_{U}^{{-1}}$ réalise une bijection de $\cc$ sur $U$, de réciproque $\varphi_{U}$.

 \noindent On obtient ainsi un atlas $\{(U,\varphi_{U}), (V,\varphi_{V})\}$ sur $S^2$.

\noindent \textbf{Petite remarque} :

\noindent on peut se demander pourquoi avoir choisi $ V=\left\{\dfrac{(1-\lvert w_{}\rvert^{2} ,2\bar{w_{}})} {1+\lvert w_{}\rvert^{2}}\;;\;w\in\cc\right\}$ et non par exemple
$ V=\left\{\dfrac{(1-\lvert w_{}\rvert^{2} ,2{w_{}})} {1+\lvert w_{}\rvert^{2}}\;;\;w\in\cc\right\}$ et idem pour $U$. La réponse est donnée par l'exercice suivant.
\begin{exo}\label{hopfi}
Montrer que $\varphi_{U}\circ\varphi_{V}^{-1}$ est l'inversion $w\mapsto 1/w$ sur  $\cc\setminus{\{0\}}$.
\end{exo}
 Avec les notations ainsi introduites, énon\c cons un théorème inspiré des travaux de Hopf.
 \begin{theorem} (Hopf)\label{hop}
 
\noindent  Pour tout entier relatif  non nul $n$, il existe un unique fibré (à une équivalence près) $\ma{H}_n=\left(\cc^{2}\setminus \{(0,0)\},p,S^{2}\right) $ de fibre $\cc^*$ déterminé  par le cocycle $c_{VU}$ à valeurs dans $U(1)$  défini par l'égalité 
\begin{equation}\label{hoho}\forall \varphi_U^{-1}(z)\in U\cap V,\;c_{VU}\left(\varphi_U^{-1}(z)\right)= e^{in\theta}\: \mathrm{si} \;z=\abs{z}e^{i\theta}.\end{equation}
Autrement dit, $c_{VU}\left(\varphi_U^{-1}(z)\right)(\lambda) =e^{in\theta}\lm$ pour tout $\lm\in\cc^*$.
 \end{theorem}
 \begin{preuve}{}
 D'après la proposition \ref{cossi}, il existe une unique à un isomorphisme près structure de fibré sur $S^{2}$, de fibre $\cc$, de groupe $U(1)$ définie par le cocycle $c_{UV}$  de l'énoncé. 
 \end{preuve}
 \noindent\textbf{Construction d'un exemplaire du fibré de Hopf $\ma{H}_n$ où $n\in\zz ^*$}

 \noindent On a sur $\cc^{2}\setminus \{(0,0)\}$ la relation d'équivalence $(z,z')\sim(w,w')$ s'il existe $\lambda\in\cc$ tel que $(z,z')=\lambda(w,w')$. L'ensemble quotient est l'espace projectif $\cc P^{1}$. C'est l'ensemble des droites complexes de $\cc^{2}$ passant par l'origine, privées de cette origine. Notons $[(z,z')]$ la classe de $(z,z')$ et $\pi'$ la surjection canonique de $\cc^{2}\setminus \{(0,0)\}$ sur $\cc P^{1}$. On considère l'application 
\[
 \begin{array}{ccc}
 \pi:\cc P^{1}&\longrightarrow &S^{2}\\
 \lbrack(w_{0},w_{1})\rbrack&\mapsto&\dfrac{(\lvert w_{1}\rvert^{2}-\lvert w_{0}\rvert^{2} ,2\bar{w_{0}}w_{1})} {\lvert w_{0}\rvert^{2}+\lvert w_{1}\rvert^{2}}
 \end{array}
 \]
 On vérifiera que $\pi$ est bien définie et qu'elle est surjective et continue.
 On pose $\ma{V}=\pi^{{-1}}(V)$. Alors $$\ma{V}=\left\{\lbrack(w,1)\rbrack\;;\;w\in\cc\right\}.$$ Supposons en effet que $\lbrack(w_{0},w_{1})\rbrack\in \ma{V}$, alors 
 $\dfrac{(\lvert w_{1}\rvert^{2}-\lvert w_{0}\rvert^{2} ,2\bar{w_{0}}w_{1})} {\lvert w_{0}\rvert^{2}+\lvert w_{1}\rvert^{2}}\not=(0,0,-1)$ et donc $w_{1}\not=0$. D'où $\lbrack(w_{0},w_{1})\rbrack=\lbrack(w,1)\rbrack$ avec $w=w_{0}/w_{1}$. On a vérifié l'inclusion $\ma{V}\subset
\left\{\lbrack(w,1)\rbrack\;;\;w\in\cc\right\}$. L'inclusion réciproque est claire.

\noindent De même on pose $\ma{U}=\pi^{{-1}}(U)$ et comme ci-dessus on a l'égalité $$\ma{U}=\{\lbrack(1,z)\rbrack\;;z\in\cc\}.$$
Notons $p$ la surjection $p=\pi\circ \pi'$. On a alors \[p^{-1}(V)=\{(w_{0},w_{1})\in \cc^{2}\setminus\{(0,0)\}\;;\;w_{1}\not=0\}. \]
Notons que  $p^{-1}(V)=\left\{\frac{\lm}{\sqrt{1+\abs{w}^2}}(w,1),\; \lm\in\cc^{*},\;w\in\cc\right\}$.
De même on a: 
$p^{-1}(U)=\left\{\frac{\lm}{\sqrt{1+\abs{z}^2}}(1,z),\; \lm\in\cc^{*},\;z\in\cc\right\}$.
Observons  que $p^{-1}(V)$ et $p^{-1}(U)$ sont deux ouverts de $\cc^2\setminus\{(0,0)\}$.
Soit $m$ un point appartenant à l'intersection $U\cap V.$  Il existe deux complexes $W$ et $Z$ tels que  $m=\f_V^{-1}(W)=\f_U^{-1}(Z)$ et $WZ=1$.

Posons enfin $e_V(m)=\frac{(W^n,1)}{\sqrt{1+\abs{W^n}^2}}\in p^{-1}(V)$ et $e_U(m)=\frac{(1,Z^n)}{\sqrt{1+\abs{Z^n}^2}}\in p^{-1}(U)$. 
 On définit les applications $\Phi_V : p^{-1}(V)\longrightarrow  V\times \cc^*$ et $\Phi_U : p^{-1}(U)\longrightarrow  U\times \cc^*$  par : 
 
 $\Phi_V\big(\lm e_V(m)\big)=\left(\frac{(1-\abs{W}^2,2\bar{W})}{1+\abs{W}^2},\lm\right)=(m,\lambda)$ et 
  $\Phi_U\big(\lm e_U(m)\big)=\left(\frac{(\abs{Z}^2-1,2{Z})}{1+\abs{Z}^2},\lm\right)=(m,\lambda)$.
  Ces deux homéomorphismes définissent deux cartes locales  par lesquelles $(\cc^2\setminus\{(0,0)\},p,S^2)$ est un fibré localement trivial de fibre $\cc^*$. Déterminons le cocycle de ce fibré.
  
  On a : 
  
  \noindent$\Phi_V{\Phi_U}{^{-1}}(m,\lm)=\Phi_V(\lm e_U(m))=\Phi_V\left(\lm \frac{Z^n\bigl(\tfrac{1}{Z^n},1\bigr)}{ \abs{Z^n}\sqrt{ 1+\tfrac{1}{\abs{Z^n}^{2}} } }\right)=\Phi_V\left(\lm \frac{Z^n}{\abs{Z^n}}e_V(m)\right)=..$
  
\noindent$  =(m,\lm \frac{Z^n}{\abs{Z^n}})=(m,C_{VU}(m)\lm)$. Ainsi pour $m=\f_U^{-1}(Z)$ et $Z=\abs{Z}e^{i\theta}$ , on a : $$c_{VU}(m).\lm=e^{in\theta}.\lm.$$
  Et on conclut que $(\cc^{2}\setminus\{(0,0)\},p,S^{2})=\ma{H}_{n}$.
\begin{corollary}(Hopf)
\hspace{1cm}\\
Pour tout entier relatif non nul $n$, il existe un unique fibré (à une équivalence près) 
$\tilde{\ma{H}}_n=\left(S^3,\tilde{p},S^2\right)$ de fibre $S^1$ déterminé par le cocycle $c_{VU}$ à valeurs dans $U(1)$ défini par l'équation (\ref{hoho}).
\end{corollary}
\begin{preuve}{}
Dans la construction du fibré de Hopf $(\cc^2\setminus\{(0,0)\},p S^2)$, on considère la restriction de $p$ à $S^3=\{(z_1,z_2)\in \cc^2\setminus\{(0,0)\}\;;\;\abs{z_1}^2+\abs{z_2}^2=1\}$. On la note $\tilde{p}$. Alors en reprenant les notations de la démonstration du théorème \ref{hop}, on a :
\[\tilde{p}^{_1}(V)= \left\{\frac{\lm}{\sqrt{1+\abs{w}^2}}(w,1),\; \lm\in S^1,\;w\in\cc\right\}\]
et 
\[\tilde{p}^{_1}(U)= \left\{\frac{\lm}{\sqrt{1+\abs{z}^2}}(1,z),\; \lm\in S^1,\;z\in\cc\right\}\]
Les cartes $\Phi_U$ et $\Phi_V$ sont définies de façon identique mais sont à valeurs dans respectivement $U\times S^1$ et $V\times S^1$.
\end{preuve}
L'application $\tilde{p} : S^3\longrightarrow S^2$ définie par $\tilde{p}(z_1,z_2)=\left(\abs{z_1}^2-\abs{z_0}^2,2\bar{z_0}z_1\right)$ est appelée \emph{la fibration de Hopf}.\index{fibration de Hopf}
\section{Appendice 3 : démonstration de la proposition \ref{nouham}}\label{museio}
On aura besoin de quelques égalités conséquences du corollaire \ref{lepote}. Donnons-les dans le lemme suivant.
\begin{lemma}\label{ugh}
On est dans $\rr^3$ mais pour respecter les notations de la proposition \ref{nouham}, on notera 
$x=q^1,y=q^2,z=q^3$ les coordonnées canoniques.
Soit une particule de vitesse $v=\frac{dq^{i}}{dt} \frac{\partial}{\partial q^{i}}$. 
Soit $\ma{B}$ la $2$-forme champ magnétique, et $\mathbf{A}=A_\alpha dx^{\alpha}$ la $1-$forme potentiel vecteur. Alors 
\begin{enumerate}
\item $B^{i}=B_i=\frac{\partial A_{i+2}}{\partial q^{i+1}}-\frac{\partial A_{i+1}}{\partial q^{i+2}}$ où l'addition des indices est l'addition dans $\{1,2,3\}$ modulo $3$. (exemple $2+2=1$)
\item Si $\times $ est le produit vectoriel dans $\rr^3$ euclidien, alors pour tout $i,j\in \{1,2,3\}$, on a :
$(v\times B)_j=\frac{dq^{i}}{dt}\left(\frac{\partial A_i}{\partial x^j}-\frac{\partial A_j}{\partial x^{i}}\right).$
\end{enumerate}
\end{lemma}
\begin{preuve}{du lemme.}
Le premier item résulte directement de l'égalité $\ma{B}^2=\mathbf{d}\mathbf{A}$.
Pour le deuxième item, vérifions l'égalité pour $j=1$.
On a : 

\noindent$(v\times B)_1=\frac{dq^2}{dt}B_3-\frac{dq^3}{dt}B_2=\frac{dq^1}{dt}\left(\frac{\partial A_1}{\partial q^1}-
\frac{\partial A_1}{\partial q^1}\right)+\frac{dq^2}{dt}\left(\frac{\partial A_2}{\partial q^1}-
\frac{\partial A_1}{\partial q^2}\right)+\frac{dq^3}{dt}\left(\frac{\partial A_3}{\partial q^1}-
\frac{\partial A_1}{\partial q^3}\right)=\frac{dq^{i}}{dt}\left(\frac{\partial A_i}{\partial q^1}-
\frac{\partial A_1}{\partial q^{i}}\right)$. Idem pour $j=2$ et $j=3$.
\begin{preuve}{de la proposition \ref{nouham}}

\end{preuve}
On pose $p_i^*=p_i+eA_i$ pour tout $i\in \{1,2,3\}$ (on écrira symboliquement en omettant les indices $p^*=p+eA$) et on définit $H^*(q,p^*,t)$ par :
\[H^*(q,p^*,t)=H(q,p,t)-e\Phi=H(q,p^*-eA,t)-e\Phi.\]
Alors on a immédiatement $\frac{\partial H^*}{\partial p^*}=\frac{dq}{dt}$.
Pour démontrer la première égalité de la proposition \ref{nouham}, on imagine la particule tout d'abord soumise à une force $f$ qui dérive d'un potentiel, de sorte que $\frac{\partial H}{\partial q}=-\frac{dp}{dt}
=-f$ (selon Newton). Puis on installe en plus un champ électromagnétique qui fait apparaitre un potentiel scalaire $\Phi$ et un potentiel vecteur $A$. On évalue alors la quantité $Q_i=\frac{dp^*_i}{dt}+\frac{\partial H^*}{\partial q^{i}}$ pour $i=1,2,3$. On a :

$Q_i=(\frac{dp_i}{dt}+e\frac{dA_i}{dt})+\frac{\partial H}{\partial q^{i}}-e\frac{\partial H}{\partial p_j}\frac{\partial A_j}{\partial q^{i}}-e\frac{\partial \Phi}{\partial q^{i}}=(\frac{dp_i}{dt}+e\frac{dA_i}{dt})
-f_i-e(\frac{dq^j}{dt}\frac{\partial A_j}{\partial q^{i}})-e\frac{\partial\Phi}{\partial q^{i}}
=(\frac{dp_i}{dt}+e\frac{dA_i}{dt})
-f_i-e\frac{dq^j}{dt}(\frac{\partial A_j}{\partial q^{i}}-\frac{\partial A_i}{\partial q^{j}})
-e\frac{dq^{j}}{dt}\frac{\partial A_i}{\partial q^{j}}-e\frac{\partial\Phi}{\partial q^{i}}
=(\frac{dp_i}{dt}+e\frac{dA_i}{dt})
-f_i-e(v\times B)_i-e(\frac{d A_i}{dt}-\frac{\partial A_i}{\partial t})-e\frac{\partial\Phi}{\partial q^{i}}
=\frac{dp_i}{dt}-f_i-e\left((v\times B)_i+\frac{\partial\Phi}{\partial q^{i}}-\frac{\partial A_i}{\partial t}\right)
=\frac{dp_i}{dt}-f_i-e\left((v\times B)+E\right)_i=0$, d'après Newton et le corollaire \ref{lepote}. Ce qui achève la démonstration.
\end{preuve}
\section{Appendice 4 :
orientation d'un fibré vectoriel et con\-nexion de Levi-Civita}\label{appquatro}
\subsection{Connexion de Levi-Civita sur $\tau(M)$}\label{demlevci}

\begin{theorem}\label{levita}\index{connexion de Levi-Civita (construction)}
Soit $M$ une variété différentiable munie d'une structure riemannienne $\la\;,\;\ra$.
Il existe sur $\tau(M)$ une unique connexion symétrique $\nabla$ telle que pour tous champs de vecteurs $X,Y,Z$ sur $M$ on a :
\[L_X\la Y,Z\ra=\la\nabla_XY,Z\ra+\la Y,\nabla_XZ\ra.\]
\end{theorem}
\begin{preuve}{}
L'unicité a déjà été montrée avec l'égalité (\ref{cristo}).
Pour l'existence on évalue la fonction $A(X,Y,Z)$ définie par :
\[A(X,Y,Z)=L_X\la Y,Z\ra+L_Y\la Z,X\ra-L_Z\la X,Y\ra.\]
Compte tenu de la symétrie de $\nabla$ ( voir définition \ref{sysy}), on a directement :
\[A(X,Y,Z)=\la [X,Z],Y\ra+\la [Y,Z],X\ra+\la [Y,X],Z\ra+2\la \nabla_XY,Z\ra.\]
Ainsi
 \[2\la \nabla_XY,Z\ra=A(X,Y,Z)-(\la [X,Z],Y\ra+\la [Y,Z],X\ra+\la [Y,X],Z\ra)\]
 Cette égalité détermine complètement le champ de vecteurs $\nabla_XY$. Reste à 
 montrer deux points pour conclure :
 
 $i) \mathrm{Pour\; toute\; fonction \;différentiable} \;f , \nabla_{fX}Y=f\nabla_XY$
 
 $ii)\mathrm{Pour \;toute\; fonction\; différentiable} \;f , \nabla_X{fY}=(X.f)Y+f\nabla_XY.$
 
 \noindent Ces deux points ont une vérification directe à partir de l'exercice \ref{nok}.
 
 \end{preuve}

\subsection{Orientation d'un fibré vectoriel, orientation d'une variété différentiable}
\subsubsection{L'intuition que donne le cas d'un espace vectoriel}
Considérons le cas simple d'une droite (vectorielle). Une droite peut être parcourue dans un sens ou dans l'autre. Chacun de ces choix correspond à l'idée d'une orientation de la droite. En élaborant un peu cette idée, une orientation de la droite vectorielle est le choix d'une base $\{e\}$. Une autre base $\{e'\}$ correspond à la même orientation si $e'=a e$ avec $a>0$, sinon à l'orientation opposée.
Pour un espace vectoriel $E$ de dimension $n$, on peut reprendre l'intuition de la dimension $1$, de la façon suivante : soit $\ma{B}=(e_1,\ldots,e_n)$ une base de $E$, l'espace des $n$-formes sur $E$ est de dimension $1$, une base étant $\omega=\varepsilon^1\wedge\ldots\wedge\varepsilon^n$ avec $(\varepsilon^1,\ldots,\varepsilon^n)$ la base duale de $(e_1,\ldots,e_n)$. Si $\ma{B'}=(e'_1,\ldots,e'_n)$ est une autre base de base duale $(\varepsilon'^1,\ldots,\varepsilon'^n)$, posons 
$\omega'=\varepsilon'^1\wedge\ldots\wedge\varepsilon'^n$. 
Alors, de façon analogue à la droite,  le choix de $\ma{B}$  correspond à une orientation de $E$ et un autre choix $\ma{B}'$ correspond à la même orientation si \begin{equation}\label{cho}
\omega'=\lambda\omega,\;\;
\mathrm{avec}\;\lambda>0.
\end{equation}
On a évoqué ici les $n$- formes pour se ramener à la dimension $1$. 

\noindent Cependant on voit que de façon équivalente on peut écrire que $\ma{B}$ et $\ma{B}'$ correspondent à la même orientation si en notant 
$f$ l'isomorphisme défini par $f(e_i)=e'_i$, ces deux bases correspondent à la même orientation de $E$ si et seulement si $\det f>0$. En effet le coefficient $\lambda$ de l'équation (\ref{cho}) est égal 
à $\det f.$

\noindent Considérons maintenant un fibré vectoriel (au sens de la définition \ref{fifibre}) $(E,p,M)$ où $M$ est une variété différentiable. Pour tout $x\in M$ la fibre  $p^{-1}(x)$ est un espace vectoriel (voir proposition \ref{ev}). On peut l'orienter par le choix d'une base. Une façon naturelle de comparer les orientations de fibres voisines est donnée par la notion d'\emph{orientation locale. }
\begin{definition}\label{orloc}\index{orientation locale d'un fibré vectoriel}
Soit $\tau=(E,p,M)$ un fibré vectoriel de fibre $F$ espace vectoriel orienté de dimension $n$.
Une orientation locale au voisinage d'un point $x\in M$ est la donnée 
\begin{enumerate}
\item[i)] d'un voisinage $V$ de $x$, 
\item[ii)] d'un $n$-uplet de sections $(s_1,\ldots,s_n)$ définies sur $V$,
\end{enumerate}
de sorte que 
pour tout $y\in V$, pour toute carte locale $(U,\Phi)$ de $\tau$ telle que $y\in U$, $(\Phi\circ s_1(y),\ldots,\Phi\circ s_n(y))$ soit une base $\mathcal{B}_y$ définissant l'orientation de $F$.
\end{definition}
\begin{definition}\label{replo}\index{référentiel local d'un fibré vectoriel}
Une base de sections d'un fibré vectoriel $\tau$ définie sur un ouvert est un \emph{référentiel local} de ce fibré sur cet ouvert.
\end{definition}
\begin{exemple}\label{reflocvar}
Soit  $\tau=\tau(M)$, de fibre $\rr^n$  munie de son orientation canonique.  La donnée d'un référentiel local $(X_1,\dots,X_n)$ défini sur un ouvert $V$ inclus dans le domaine d'une carte locale munie de coordonnées $(x_1,\ldots,x_n)$ de sorte que l'orientation de la base $\left(\frac{\partial}{\partial x^1},\ldots, \frac{\partial}{\partial x^n} \right)$ soit l'orientation canonique de $\rr^n$   et tel que si $X_j=X_j^{i}\frac{\partial}{\partial x^{i}}$, alors $\det \left(X_j^{i}\right)>0$ sur $V$ définit une orientation locale sur $\tau(M)$ (on parlera d'orientation locale sur $M$).
\end{exemple}
Une idée intuitive d'orientabilité peut être l'existence d'une orientation locale en chaque point de $M$.

L'exemple de l'espace vectoriel suggère une autre idée. Un espace vectoriel $F$ équipé d'une base fournit des coordonnées. On peut le voir comme une carte locale $(F,\Phi)$ où pour tout vecteur $u$,
$\Phi(u)$ est le $n$-uplet des composantes de $u$ dans la base choisie. Le choix d'une autre base définit alors la carte $(F,\Psi)$ et le fait que les deux bases aient même orientation signifie que $\det (\Psi\Phi^{-1})>0$. Donc une idée également naturelle de l'orientabilité d'un fibré vectoriel est l'existence  d'un atlas $\{(U_i,\Phi_i\}_i\in I$ tel que pour tout couple $(i,j)$ tel que $U_i\cap U_j\not=\empty$, on ait $\det(\Phi_{i}\Phi_{j}^{-1})>0$.
Reste enfin l'intuition de la dimension $1$. Si $F=\rr^n$ et si $\{c_{ij}\}$ est un cocycle de $(E,p,M)$
alors $\{(c_{ij}^*)^{-1}\}$  où pour tout $x\in M$  $(c_{ij}^*)^{-1}(x)$ détermine le fibré des formes volume sur $M$ dont le fibre est de dimension $1$ comme ceci est décrit dans l'exemple \ref{true} p.\pageref{true}.
L'exemple des espaces vectoriels suggère qu'on puisse décrire l'orientabilité par  l'existence d'une section du fibré des formes volume qui ne s'annule en aucun point de $M$ (de sorte que sur chaque fibre on ait une forme volume qui donne la même orientation sur les fibres voisines

La section suivante reprend rigoureusement  ces notions  et montre leur équivalence.

\subsubsection{Orientation d'un fibré vectoriel}\label{cricri}
\begin{theorem}\label{orifi}
Soit $\tau=(E,p,M)$ un fibré vectoriel de base $M$ une variété différentiable et de fibre un espace vectoriel $F$. 
Il y a équivalence entre les trois items suivants.
\begin{enumerate}
\item Le fibré $\tau$ possède un atlas $\{U_i,\Phi_i)\}_{i\in I}$ dont les changements de cartes sont définis, conformément à l'équation (\ref{chang}), par le cocycle $\{c_{ij}\}$
qui vérifie la propriété de compatiblité suivante:
\[\forall (i,j)\in I^2 \; \;U_i\cap U_j\not=\emptyset \Rightarrow\forall x\in U_i\cap U_j\;\det \left(c_{ij}(x)\right)>0.\]
\item Au voisinage de tout point de $M$ existe une orientation locale de $\tau$.
\item Il existe sur $\tau$ une forme volume qui ne s'annule en aucun point de $M$.
\end{enumerate}
\end{theorem}
\begin{definition}\label{orib}\index{fibré vectoriel orienté}
Si un des  trois items du théorème \ref{orifi} est vérifié, on dit que le fibré $\tau$ est orientable. 
 Dans le premier item, le choix d'un atlas $\{U_i,\Phi_i)\}_{i\in I}$ tel que $\det c_{ij}$ ait un signe donné est \emph{une orientation du fibré}. Il correspond également au choix d'une forme volume ou au choix d'une orientation locale en chaque point de $M$.
 \end{definition}
 \begin{definition}\label{orivardi}\index{variété orientable}
 Si le fibré $\tau$ de la définition\ref{orib} est le fibré tangent à $M$, on dit que $M$ est orientable, une orientation de $M$ correspondant à une orientation de son fibré tangent. Une orientation de $M$ est définie par le choix d'une forme volume sans zéro sur $M$.
 \end{definition}
 \begin{preuve}{du théorème \ref{orifi}}
$1)$ \emph{Montrons l'équivalence des deux premiers items.}
 
 Supposons l'item $1.$ vérifié. Soit $x\in M$,  $(U_i,\Phi_i)$ une carte locale contenant $x$ et $(e_1,\ldots,e_n)$ une base positive de $F$. On pose pour tout $y\in U_i\cap U_j, \; s_k(y)=
 \Phi_i^{-1}(y,e_k)=\Phi_j^{-1}\left(y,c_{ji}(y)e_k\right)$.
 On définit ainsi sur $U_i$ un référentiel local $\mathcal{R}=(s_1,\ldots,s_n)$ tel que pour toute carte locale $(U_j,\Phi_j)$ et tout $y\in U_i\cap U_j$, la base $\mathcal{B}_y$ de $F$ définie par $\mathcal{B}_y=(\Phi_j\circ s_1(y),\ldots,\Phi_j\circ s_n(y))$ définit la même orientation que $(e_1,\ldots,e_n)$ puisque $\det c_{ji}>0$.
 
 Inversement supposons qu'on ait une orientation locale sur un voisinage $V$ de tout point $x$ de $M$.  Sur $V$ existe un référentiel local $(s_1,\ldots,s_n)$ tel que si $(U_i,\Phi_i)$ et $(U_j,\Phi_j)$
 sont deux cartes locales telles que $y\in V\cap U_i\cap U_j$, alors $\left(e_{1i},\ldots,e_{ni}\right)=\left(
 \Phi_i(s_1(y)),\ldots, \Phi_i(s_n(y)\right)$ et  $\left(e_{1j},\ldots,e_{nj}\right)=\left(
 \Phi_j(s_1(y)),\ldots, \Phi_j(s_n(y)\right)$ définissent la même orientation qui est celle de $F$.
 Or $\Phi_j\Phi_i^{-1}(y,e_{ki})=(y,e_{kj})=(y,c_{ji}(y)e_{ki})$. On en déduit que $\det c_{ji}>0$ ce qui démontre le premier item.
 
 \noindent $2)$\emph{ Montrons l'équivalence des items $1$ et $3$.}
 
 Supposons donné un atlas $\{U_i,\Phi_i\}_{i\in I}$ vérifiant les hypothèses de l'item $1$. 
 On se donne sur $F$ une forme volume $v$, et une partition de l'unité subordonnée au recouvrement
 $\{U_i\}_{i\in I}$ de $M$, c'est à dire une famille de fonction $\{\varphi_i\}_{i\in I}$ différentiables à valeurs dans $[0,1]$ telles que $\su{\varphi_i}\subset U_i$ et $\sum_{i\in I} \varphi_i=1$. Une telle famille existe sur toute variété différentiable car on peut tout choisir le recouvrement $\{U_i\}_{i\in I}$ localement fini (tout point a un voisinage qui ne rencontre qu'un nombre fini de $U_i$), auquel cas la somme $\sum \varphi_i(x)$ est finie pour tout $x$. On peut trouver une démonstration de l'existence d'une partition de l'unité sur une variété dans [\ref{BG}].
 
 On considère la section $\sigma_i$ définie sur $M$ par $\sigma_i(x)=\varphi_i \Phi_i^*(x,v)$ pour tout $x\in M$. C'est une section à support dans $U_i$. Alors $s=\sum_{i\in I}\sigma_i$ est une section sur $M$ qui ne s'anulle en aucun point de $M$. En effet pour tout $i\in I $ et tout $x\in M$, on a :
 ${\Phi_i^{-1}}^*(s(x))=\sum_{j\in I}\varphi_j(x)(\Phi_j\Phi_i^{-1})^*(x,v)=\Big(x,\underbrace{\big(\sum_{j\in I}\varphi_j (x)\det{c_{ji}}(x)\big)}_{\not=0}(v)\Big).$
 
 \noindent Et on conclut avec l'item $3.$
 
 Supposons l'item $3.$ : il existe sur $M$ une forme volume $s$ qui ne s'annule en aucun point.
 On se donne un atlas $\{U_i,\Phi_i\}_{i\in I}$ de $\tau$, pour lequel on peut toujours supposer les ouverts $U_i$ connexes car $M$ est localement connexe. On se donne sur la fibre $F$ une forme volume $v$ et une structure euclidienne. Toutes les données précédentes ne restreignent pas la généralité de la démonstration. Pour tout point $x\in U_i$ la forme $\Phi_i^*(x,v)$ est une forme volume sur la fibre $p^{-1}(x)$ de $\tau$. Par conséquent il existe une fonction réelle continue non nulle $\lambda_i$ définie sur $U_i$ telle que $\Phi^*(x,v)=\lambda_i s_i(x)$ où $s_i$ est la restriction de $s$ à $U_i$. Du fait de la connexité de $U_i$, la fonction $\lambda_i$ a un signe fixe sur $U_i$ . Quitte  à remplacer $\Phi_i$ par $s\circ\Phi_i$, où $s$ est une isométrie négative de $F$, on peut supposer  $\lambda_i>0$ sur $U_i$. On définit une telle fonction $\lambda_i$ sur tous les domaines de cartes de $\tau$.
 Considérons une carte locale $(U_j,\Phi_j)$ telle que $U_i\cap U_j\not=\emptyset$.
 On a : 
 
 \noindent$\left(\Phi_j\Phi_i^{-1}\right)^*(x,v)=(x,\det c_{ji}(x) v)$, d'où
 $\Phi_j^*(x,v)=\Phi_i^*(x,\det c_{ji}(x) v)$ c'est à dire $ \lambda_j(x)s_j(x)=\lambda_i(x)\det c_{ji}(x)s_i(x).$ Mais $s_i(x)=s_j(x)=s(x)\not=0$ d'où $\det c_{ji}(x)=\frac{\lambda_j(x)}{\lambda_i(x)}>0$.
 D'où l'item $1.$
  \end {preuve}
  \subsubsection{Orientation canonique d'une variété pseudo-riemannienne orientable.}
  Soit $(M,g)$ une variété pseudo-riemannienne connexe orientable. Il existe, d'après  le premier item du théorème \ref{orifi}, un atlas $\mathcal{A}=\{(U_i,\varphi_i)\}$ sur $M$ vérifiant :
  \[\forall i,j,\forall x\in U_i\cap U_j\; \det D(\varphi_i\varphi_j^{-1})(\varphi_j(x))>0.\]
  On suppose $M$ munie de cet atlas $\mathcal{A}$.
 
  \noindent Dans un domaine ouvert $U$ de carte locale munie des coordonnées $(x^1,\ldots,x^n)$, le tenseur métrique $g$ est représenté par une matrice $G_x$ telle que 
  $$(G_x)_{ij}=g(\frac{\partial}{\partial x^{i}},\frac{\partial}{\partial x^{j}}).$$ La matrice $G_x$ représente une forme quadratique non dégénérée et son déterminant n'est pas nul. La variété étant connexe, le signe de ce déterminant est constant. Dans le cas d'une variété riemannienne $G_x$ est définie positive (donc  de déterminant positif). Notons $(U,x)$ la carte locale précédente que nous prenons dans l'atlas $\mathcal{A}$. On considère sur cette carte la forme volume 
  \begin{equation}\label{fovori}
  \sigma_U=\sqrt{\abs{\det G_x}} \;d x^1\wedge\ldots\wedge dx^n.
  \end{equation}
  Avec des notations évidentes, considérons une carte locale $(V,y)\in \mathcal{A}$ telle que $U\cap V\not=\emptyset.$ 
  Le changement de carte $\varphi : U\cap V\longrightarrow U\cap V$, 
  \[(x^1,\ldots,x^n)=\varphi\left((y^1,\ldots,y^n)\right),\]
  a un jacobien de déterminant positif.
  On a bien sûr \begin{equation}\label{qbni}G_y=^tJ(\varphi)G_xJ(\varphi)\end{equation} qui exprime le lien entre deux matrices de formes quadratiques définies dans deux bases ce qui implique la relation
  \[\sqrt{\abs{\det G_y}}=\det\left(J(\varphi)\right)\sqrt{\abs{\det G_x}}.\]
Ainsi, l'égalité (\ref{fovori}) détermine une forme volume globale sur $M$.
En effet,

\noindent $\varphi^*(\sigma_U)=\sqrt{\abs{\det G_x}}\varphi^*(dx^1\wedge\ldots\wedge dx^n)
  =\frac{\sqrt{\abs{\det G_y}}}{\det\left(J(\varphi)\right)} \det\left(J(\varphi)\right)dy^1\wedge\ldots\wedge dy^n=\sigma_V$.
  Cette forme globale est déterminée \emph{au signe près}.  
   \begin{definition}\label{canfori}\index{forme volume riemannienne}
  La \emph{forme volume riemannienne (canonique) }de la variété pseudo-riemannienne orientable  $M$ se définit par son expression dans la carte locale locale
  $(U,x)$ précédente par

 \[ \sigma_U= \sqrt{\abs{\det G_x}} \;d x^1\wedge\ldots\wedge dx^n.\]
  \end{definition}
  \begin{remark}\label{cokette}
Voici une construction alternative de la \emph{forme volume riemannienne}. On garde les notations de la construction précédente.
Il existe sur la carte $(U,x)$une base orthonormée positive $\mathcal{B}=(e_1,\ldots,e_n)$ ( dans le cas d'une structure riemannienne, on peut la construire à partir de la base  $\mathcal{C}=(\frac{\partial}{\partial x^1},\ldots,\frac{\partial}{\partial x^n})$ par le procédé de Gram-Smidth, dans le cas pseudo-riemanienn on peut utiliser le processus de diagonalisation de Gauss). On pose : $\omega^{i}=g(e_i,.)=:\gamma(e_i)$. Ceci définit une $1$-forme sur $U$. Considérons sur $U$ la forme volume 
\[\rho_U=\omega^1\wedge\ldots\wedge\omega^n.\]
Alors  :

$$\rho_U=\sigma_U$$  où $\sigma_U$ est la forme définie par l'égalité (\ref{fovori}).

\noindent En effet, le tenseur métrique $g$ est représenté dans la base $\mathcal{B}$ par une matrice $\tilde{I}$ matrice diagonale contenant sur sa diagonale des $1$ est des $-1$, plus précisément $p$ fois $-1$ si son indice est $p$. Dans le cas riemannien $\tilde {I}$ est la matrice identité.
Si $A$ est la matrice de passage  de la base $\mathcal{C}$ à la base  $\mathcal{B}$, son déterminant est positif. De plus, à partir de la formule de changement de base on a
$\det  G=(-1)^p(\det A^{-1})^2$ et donc $\det A^{-1}=\sqrt{\abs{\det G}}$.

D' après l'exercice \ref{bofbof} p. \pageref{bofbof}, le changement de base se traduit par la relation matricielle $(\omega^1,\ldots,\omega^n)=(dx^1,\ldots,dx^n)^tA^{-1}$. Celle-ci implique l'égalité des $n$- formes
$\omega^1\wedge\ldots\wedge\omega^n=\det A^{-1} \;dx^1\wedge\ldots\wedge dx^n$, d'où l'égalité  
$\rho_U=\sigma_U.$
\end{remark}
\begin{proposition}\label{volerie}
Soit $(M,g)$ une variété  pseudo-riemannienne orientée,
$\epsilon=1$ si $ \det g>0$ et $
-1$\ si $ \det g<0$.
Notons  $\sigma$ la forme volume riemannienne sur $M$. 
Alors $\sigma$ est l'unique forme volume $\omega$ sur $(M,g)$ qui vérifie pour tous champs de vecteurs $X_1,\ldots,X_n,Y_1,\ldots,Y_n$ l'égalité :
\begin{equation}\label{smol}
\omega(X_1,\ldots,X_n)\omega(Y_1,\ldots,Y_n)=\epsilon \det \left(g(X_i,Y_j)\right).
\end{equation}
\end{proposition}
\begin{exo}\label{smutch}
Démontrer la proposition \ref{volerie}
\end{exo}

Nous savons qu'une variété pseudo-riemannienne n'est pas nécessairement orientable, en particulier nous ne savons rien sur l'orientabilité de l'espace-temps de la relativité générale. Si une variété n'est pas orientable, il n'existe pas de forme volume sans zéro sur cette variété. Cependant on peut toujours définir sur une variété riemannienne sur elle une pseudo-forme volume.
\begin{definition}\label{psedofor}
Soit $E$ un espace vectoriel. Une pseudo $p$-forme sur $E$ associe à chaque orientation $o$ de $E
$ une $p$-forme $\al_o$ de sorte que $\al_{-o}=-\al_o$.
Sur une variété, si on choisit une carte locale de domaine $U$ munie de coordonnées $(x^1,\ldots, x^n)$, une pseudo $p$-forme $\al$ associe à chaque $x$ une pseudo-$p$-forme  $\al(x)$ sur $T_xM$ de sorte que si on choisit l'orientation définie par $\mathcal{B}_x=\left(\frac{\partial}{\partial x^1},\ldots,\frac{\partial}{\partial x^n}\right)$, alors $x\mapsto \al(x)$ est différentiable.

Si on choisit  a priori une orientation $o$ de $T_xM$, on posera $o(x)=1$ si $\mathcal{B}_x$ est orientée comme $o$ et $o(x)=-1$ dans le cas contraire.
\end{definition}
\begin{proposition}
Soit $(M,g)$ une variété riemannienne  et $(U,\varphi)$ n'importe quelle carte locale munie de coordonnées locales 
$(x^1,\ldots, x^n)$ d'un atlas $\mathcal{A}$ sur $M$. En posant $\al=o(x)\sqrt{\det G_x}dx^1\wedge\ldots\wedge dx^n$
on définit une pseudo $n$-forme globale sur $M$ appelée pseudo-forme volume riemannienne. Si la variété est orientable orientée par l'alas $\mathcal{A}$ cette pseudo-forme est la forme volume riemannienne définie par l'équation (\ref{fovori}).
\end{proposition}
\begin{preuve}{}
Plaçons-nous sur l'intersection de deux domaines de cartes $U$ et $V$ munis respectivement des cordonnées $(x^1,\ldots, x^n)$ et $(y^1,\ldots, y^n)$. 
On a en reprenant les notations de l'équation (\ref{qbni}) : $\sqrt{\det G_x}=\abs{\det J(\varphi)}\sqrt{\det G_y}$. D'où :

\noindent$o(x)\sqrt{\det G_x}\;dx^1\wedge\ldots\wedge dx^n=o(x)\abs{\det J(\varphi)}\sqrt{\det G_y}\;
dx^1\wedge\ldots\wedge dx^n$

\noindent$=o(y)\det{J(\varphi)}\sqrt{\det G_y}\;
dx^1\wedge\ldots\wedge dx^n=o(y)\sqrt{\det G_y}\;dy^1\wedge\ldots\wedge dy^n$.

\noindent Ce qui permet de  conclure.
\end{preuve}

\subsubsection{Orientation d'une variété et déplacement parallèle}\label{cridep}
\begin{definition}\label{depor}\index{transport d'une orientation}
Soit un chemin $\gamma$  différentiable par morceaux, de la variété différentielle $M$ défini sur $[0,1]$. Soit un référentiel  de $M$, $\mathcal{R}(t)$, le long de $\gamma$, \emph{dépendant continuement de $t$ }. Ce référentiel détermine une orientation $\mathcal{O}(0)$ de $T_{\gamma(0)}M$ et pour tout $t\in [0,1]$ une orientation $\mathcal{O}(t)$ de $T_{\gamma(t)}M$ . On dit que $\mathcal{O}(t)$ est obtenue par transport de $\mathcal{O}(0)$ le long de $\gamma$.
\end{definition}
Dans la définition précédente, on a par exemple $\mathcal{R}(0)=(X_1(0),\ldots,X_n(0))$ où les $X_i(0)$ sont des vecteurs de $T_{\gamma(0)}M$. Si on transporte  ces vecteurs paralèlement le long de $\gamma$, on obtient un transport de l'orientation. Et ce procédé est général pour toute variété différentiable, car il existe toujours sur celle-ci une structure riemannienne à laquelle  se réfèrer naturellement ce transport parallèle.
La définition \ref{depor} est justifiée par le premier item du lemme suivant qui indique que l'orientation transportée à l'instant $t$ ne dépend pas du choix de $\mathcal{R}(0)$ définissant l'orientation $\mathcal{O}(0)$.

\begin{lemma}\label{trumpshit}
On se donne un chemin $\gamma$, différentiable par morceaux sur $M$.
\begin{enumerate}
\item Soit deux bases $\mathcal{R}_0$ et $\mathcal{R}'_0$ de $T_{\gamma(0)}M$ orientées identiquement. Soit $\mathcal{O}(0)$ leur orientation. Le transport de cette orientation en $\gamma(t)$ se fait par l'intermédiaire de deux référentiels $\mathcal{R}(t)$ et $\mathcal{R}'(t)$ tels que $\mathcal{R}(0)=\mathcal{R}_0$ et $\mathcal{R}'(0)=\mathcal{R}'_0$ . L'orientation obtenue à l'instant $t$ ne dépend pas de la base utilisée à $t=0$.
\item Si $\gamma$ et $\mu$ sont deux chemins de mêmes extrémités $a$ et $b$
 homotopes par $F:[0,1]\times [0,1]\longrightarrow M$ telle que  pour tout $t\in [0,1]$ et tout $s\in [0,1]$
 $F(t,0)=\gamma(t), F(t,1)=\mu(t),F(0,s)=a,F(1,s)=b$.
 Soit $\mathcal{O}(0)$ une orientation en $a$. Le transport parallèle de cette orientation en $b$ le long de $\gamma$ ou de $\mu$ donne la même orientation en $b$.
\end{enumerate}
\end{lemma}
\begin{preuve}{}
Montrons le premier item. 

Notons $\mathcal{R}(t)$ et $\mathcal{R}'(t)$ les référentiels dépendant continuement de $t$ issus respectivement de $\mathcal{R}_0$ et de $\mathcal{R}'_0$. Soit $A(t)$ l'automorphisme de 
$T_{\gamma(t)}M$ qui transforme $\mathcal{R}(t)$ en $\mathcal{R}'(t)$. Comme $\det A(t)$ de s'anulle pas et qu'il est une fonction continue de $t$, il garde un signe fixe qui celui de $\det A(0)$ qui est positif.
Et on conclut.

Pour le deuxième item, on considère une base $\mathcal{R}_0$ en $a$ induisant une orientation $\mathcal{O}_0$ sur $T_aM$. Soient $\mathcal{R}(t)$ et $\mathcal{R}'(t)$ deux référentiels continus d'origine $a$ respectivement le long de $\gamma$ et de $\mu$ définissant un déplacement de l'orientation définie par $\mathcal{R}_0$, le long de $\gamma $ et le long de $\mu$. On considère pour tout $t\in [0,1]$, le chemin
$\gamma_t$ défini par $\gamma_t(s)=F(t,s).$ On a $\gamma_t(0)=\gamma(t)$ et $\gamma_t(1)=
\mu(t)$. Notons $\tau_{\gamma(t)\gamma_t(s)}$ le transport parallèle le long de $\gamma_t$ de $\gamma(t)$ à $F(t,s)$(voir définition \ref{trapar} p. \pageref{trapar}).
 En $\mu(t)$, on distingue les deux bases $\tau_{\gamma(t)\mu(t)}\left(\mathcal{R}(t)\right)$ et $\mathcal{R}'(t)$. Soit $A(t)$ l'automorphisme de $T_{\mu(t)}M$ qui échange ces deux bases. Par continuité du transport parallèle, 
 on a pour $t=0, A(0)=Id$ et par suite
 $\det A(t)>0$. En particulier $\det A(1)>0$, ce qui signifie que $\mathcal{R}(1)$ et $\mathcal{R}'(1)$ 
 ont la même orientation.
\end{preuve}
Donnons maintenant un critère d'orientabilité d'une variété différentiable, lié au déplacement parallèle.
\begin{proposition}\label{oripar}
Soit $M$ une variété différentiable connexe. Si le déplacement parallèle le long d'un  chemin fermé ne conserve pas  l'orientation, la variété n'est pas orientable.
\end{proposition}
\begin{preuve}{}
Soit $\gamma $ un chemin fermé défini sur $[0,1]$. On a donc $\gamma(0)=\gamma(1)$. Soit $\mathcal{R}'(t)$ le transport parallèle de l'orientation $\mathcal{R}(0)$. Par hypothèse $\mathcal{R}(1)$ a l'orientation opposée à celle de $\mathcal{R}(0)$.

\noindent Si la variété $M$ était orientable on aurait pour tout $t_0\in ]0,1[$ un $\epsilon >0$ tel que  pour $t\in ]t_0-\epsilon,t_0+\epsilon[$ existerait sur un repère $\mathcal{R}(t)$ restriction d'une orientation locale de $M$ à $\gamma(]t_0-\epsilon,t_0+\epsilon[)$ de sorte que si les orientations de $\mathcal{R}(t_0)$ et de $\mathcal{R}'(t_0)$ coïncident alors $\mathcal{R}(t)$ et de $\mathcal{R}'(t)$ coïncident pour  $t\in ]t_0-\epsilon,t_0+\epsilon[$. Idem pour pour $t=0$ et $t=1$ en remplaçant  l'intervalle $]t_0-\epsilon,t_0+\epsilon[$ par $[0,\epsilon[\cup]1-\epsilon,1]$. Mais alors pour tout $t\in [0,1]$ les orientations de $\mathcal{R}(t)$ et  $\mathcal{R}'(t)$ coïncident. Ceci est en contradiction avec l'hypothèse.
On en déduit que $M$ n'est pas orientable.
\end{preuve}

\subsubsection{Exemples qui illustrent les différents critères d'orientabilité d'une variété  énoncés dans les sections \ref{cricri} et \ref{cridep}}

\noindent\textbf{ i) Orientation des espaces projectifs réels}\index{espaces projectifs réels}

\noindent L'espace projectif réel $\rrp^n$ de dimension $n$ est l'espace quotient obtenu en identifiant  dans $\rr^{n+1}\setminus\{(0,\ldots,0)\}$ (noté encore ${\rr^{n+1}}^*$) tous les vecteurs situés sur une même droite vectorielle. De façon plus précise, on considère la relation binaire sur $\rr^{n+1}\setminus\{(0,\ldots,0)\}$
notée $\sim$ définie par  $(x^1,\ldots,x^{n+1})\sim (y^1,\ldots,y^{n+1})$ s'il existe un réel $\lambda\not=0$ tel que $(x^1,\ldots,x^{n+1})= \lambda (y^1,\ldots,y^{n+1})$. Cette relation est une relation d'équivalence et $\rrp^n=: \rr^{n+1*}/\sim$.

Munissons $\rrp^n$ d'une structure de variété différentiable.
On considère pour tout 
$i\in \{1,2,\ldots,n+1\}$,  l'ouvert de $\rr^{n+1}$ : $\tilde{U_i}=\{(x_1,\ldots,x_{n+1})\in {\rr^{n+1}}^*\;;\;x_i\not=0 \}$ et on pose $\mathcal{U}$ la famille de ces $n+1$ ouverts. Si $q$ est la surjection canonique de $\rr^{n+1}$ sur $\rrp^n$, on a immédiatement
$q^{-1}\left(q(\tilde{U_i)}\right)=\tilde{U_i}$. Ainsi $U_i=q(\tilde{U_i})$ est un ouvert de $\rrp^n$.
L'application continue $\tilde{\varphi_i}$ définie pour tout $(x^1,\ldots,x^{n+1})\in \tilde{U_i}$ par 
$\tilde{\varphi_i}\left((x^1,\ldots,x^{n+1})\right)=\left(\frac{x^1}{x^{i}},\cdots,\frac{x^{i-1}}{x^{i}},\frac{x^{i+1}}{x^{i}},\ldots \frac{x^{n+1}}{x^{i}}\right)\in\rr^n$ est compatible avec la relation $\sim$ : en effet, de l'égalité dans $ \tilde {U_i}$ définie par $(y^1,\ldots,y^{n+1})=\lambda (x^1,\ldots,x^{n+1})$
découle l'égalité $\left(\frac{y^1}{y^{i}},\cdots,\frac{y^{n+1}}{y^{i}}\right)=\left(\frac{x^1}{x^{i}},\cdots,\frac{x^{n+1}}{x^{i}}\right).$ On  pose alors 
\[\forall (x^1,\ldots,x^{n+1})\in \tilde{U_i},\;\; \varphi_i\left(q\left((x^1,\ldots,x^{n+1})\right)\right)=\left(\frac{x^1}{x^{i}},\cdots,\frac{x^{i-1}}{x^{i}},\frac{x^{i+1}}{x^{i}},\cdots\frac{x^{n+1}}{x^{i}}\right).\]
La famille $\{(U_i,\varphi_i)\}_{i\in\{1,\ldots,n+1\}}$ est un atlas conférant à $\rrp^n$ une structure de variété différentiable, conformément à l'exercice suivant.
\begin{exo}\label{proj}

\emph{Adoptons la notation suivante : $(a_1,\ldots ,\widehat{a_i},\ldots, a_{n+1})$ désigne le $n$-uplet 
$(a_1,\ldots ,a_{i-1},a_{i+1},\ldots, a_{n+1})$.}

$i)$ Soit $(\tilde{U_i})$ et $(\tilde{U_j})$ deux ouverts  de $\mathcal{U}$ d'intersection non vide.

\noindent En notant $u=(u^1,\ldots,u^{i},u^{i+1},\ldots,u^{n+1})$  un élément de $\varphi_i(U_i)$ montrer l'égalité :  $$\varphi_j\varphi_i^{-1}(u)=\left(\frac{u^1}{u^j},\cdots,\frac{u^{j-1}}{u^j},\frac{u^{j+1}}{u^j}\ldots,\frac{u^{n+1}}{u^j}\right).$$

$ii)$ Montrer ensuite que $\det D(\varphi_j\varphi_i^{-1})(u)=-\frac{1}{{(u^j)}^{n+1}}$.
\end{exo}
\noindent Donnons une construction équivalente de $\rrp^n$. Tout d'abord, on remarque que l'espace projectif est une variété compacte. En effet si $S^{n+}=\{(x^1,\ldots,n^{n+1})\in S^n\;;\;x^{n+1}\ge 0\}$
alors $q(S^{n+})=\rrp^n$.
 Cette relation inspire la construction suivante. Notons
  \begin{align*}
 D_n&=\{(x^1,\ldots,x^n,0)\in\rr^{n+1}\;;\;(x^1)^2+\ldots +(x^n)^2\le 1\} \\
 \partial D^n&=\{(x^1,\ldots,x^n,0)\;;\;(x^1)^2+\ldots+(x^n)^2=1\}\subset S^{n+}
\end{align*}
La restriction de la relation $\sim$ sur $\rr^{n+1}$ à $S^{n+}$ se définit par :
 pour $x,y\in S^n,\;x\sim y $ si \[\begin{cases} x, y\notin \partial D^n \;\mathrm{et}\; x=y\\
 x,y\in \partial D^n \;\mathrm{et}\; x=-y
  \end{cases}
\]
cette relation d'équivalence définit l'espace topologique quotient $S^{n+}/\sim$. Un atlas est obtenu sur $S^{n+}/\sim$ 
à partir des intersections des $\tilde{U_i}$ avec $S^{n+}$, comme cela a été fait pour $\rr^{n+1}/\sim$
avec les ouverts $\tilde{U_i}$. Notons alors que l'injection canonique $i :S^{n+}\longrightarrow \rr^{n+1}$ induit une bijection
$\bar{i}: S^{n+}/\sim\longrightarrow \rr^{n+1}/\sim$. Cette bijection est un difféomorphisme  entre les structures de variétés différentiables. On peut donc écrire :
\[\rrp^n=S^{n+}/\sim.\]
\begin{proposition}\label{oripro}
L'espace projectif \; $\rrp^n$ est une variété orientable si et seulement si $n$ est impair.
\end{proposition}
\begin{preuve}{}
\emph{Montrons que $\rrp^n$ n'est orientable si $n$ est pair}


On choisit dans $\rr^{2k+1}$ , en $\alpha$ une base $(e_0,\dot{\gamma}(0), e_1,\ldots,e_{2k-1})$ dans laquelle $e_0$ est un vecteur unitaire orthogonal à $S^{2k}$. 
Le transport parallèle de l'orientation le long de cette base le long de $\gamma$ ou de $\delta$ donne la même orientation en $-\alpha$ (voir deuxième item du lemme \ref{trumpshit}). Or à chaque instant le déplacement parallèle dans $\rr^{2k+1}$ de
$(e_0,\dot{\gamma}(0), e_1,\ldots,e_{2k-1})$  est lui-même. Ainsi, en considérant l'identification définie par $\sim$ sur $S^{2k}$, on obtient en $-\alpha$ l'orientation définie par $(\dot{\gamma}(0), -e_1,\ldots,-e_{2k-1})$ que l'on doit comparer à l'orientation de l'espace tangent en ce point à $\rrp^{2k}$ définie par $(\dot{\gamma}(0), e_1,\ldots,e_{2k-1})$. Ces deux orientations sont distinctes. On en déduit, d'après la proposition \ref{oripar} que $\rrp^{2k}$ n'est pas orientable.

\emph{Montrons que $\rrp^n$ est orientable si $n$ est impair}
 
 On a un atlas $\{U_1,U_2,\ldots, U_{n+1}\}$ tel qu'il est décrit ci-dessus. Considérons sur l'ouvert $U_s$, $s\le n-1$, l'application $\psi_s$ définie par :
 \[\psi_s([y^1,\ldots,y^{n+1}])=\left(-\frac{y^1}{y^s},\ldots,-\frac{y^{s-1}}{y^s},-\frac{y^{s+1}}{y^{s}},\frac{y^{s+2}}{y^s},\frac{y^{s+3}}{y^{s}}\ldots,\frac{y^{n+1}}{y^s}\right).\]
 Il suffit de montrer conformément au premier item du théorème \ref{orifi}, qu'il existe pour $n$ pair un atlas de $\rrp^n$ dont les changement de cartes soient tous à déterminant positif.
 Pour $s=1$, un calcul direct montre que 
 
 \[\psi_2\psi_1^{-1}(u^2,u^3,\ldots,u^{n+1})=\left(\frac{1}{u^2},\frac{u^3}{u^2},-\frac{u^4}{u^2},\ldots,-\frac{u^{n+1}}{u^2}\right)\] et que \[\det D(\psi_2\psi_1^{-1})=(-1)^{n-1}\frac{1}{(u^2)^{n+1}}>0.\]
 Pqr un calcul analogue on a : \[\forall  s\in \{1,\ldots n+1\} \;\;\det D(\psi_s\psi_{s-1}^{-1})=(-1)^{n-1}\frac{1}{(u^s)^{n+1}}>0\]
 ce qui permet de déduire que pour tout $(i,j) \det (\psi_i\psi_j^{-1})>0$ et on conclut.
\end{preuve}
\noindent\textbf{ii) La bande de Moebius}\index{bande de Moebius}
Il fallait citer ce grand classique. On définit ici la bande de Moebius comme le quotient 
$\mathcal{M}=:(\rr\times \rr)\big/\sim$ où $(x,y)\sim (x+1,-y)$. 
\begin{proposition}\label{mimo}
La bande de Moebius  est une variété différentiable non orientable. 
\end{proposition}
\begin{preuve}{}
La preuve détaillée nécessite $4$ courtes étapes. Ce résultat est très intuitif et les ouvrages que j'ai pu consulter s'en tirent avec un dessin: si l'on représente $\mathcal{M}$ par un rectangle $[0,1]\times \rr $ où l'on identifie les points $(0,y)$ avec les points $(1,-y)$,
alors le transport parallèle le long  du chemin fermé $\gamma(t)=[(t,0)]\;t\in [0,1]$ de la base en $(0,0)$
$B=\big(e_1=(1,0),e_2=(0,1)\big)$ aboutit en $\gamma(1)=\gamma(0)$ à la base $(e_1,-e_2)$ qui est d'orientation opposée à la base de départ.
Ce qui permet de conclure. Mais nous voulons apporter une justification rigoureuse de ceci, qui utilise tous les concepts fondamentaux introduits dans ce document : les structures de variété, de fibré tangent, de déplacement parallèle lié à la connexion de Levi-Civita sur une variété riemannienne.

\noindent \emph{\'{E}tape $1$ : structure de variété différentiable  de $\mathcal{M}$}

\noindent On note $p$ la surjection canonique de $D=:\rr\times \rr$ sur $\mathcal{M}$.
Pour tout $u= (x,y)\in D$ la projection $p$ réalise un homéomorphisme de $V_{u}=
]x-\frac{1}{2},x+\frac{1}{2}[\times \rr$ sur $U_{u}=p(V_{u})$. On note $\varphi_u$ l'homéomorphisme réciproque. Alors $(U_u,\varphi_u)$ est une carte locale en $u$. 
Si $(U_u,\varphi_u)$ et $(U_{u'},\varphi_{u'})$ sont deux cartes locales de domaines non disjoints
alors la fonction changement de cartes $\varphi_{u'}\circ\varphi_u^{-1}$ est définie par :
\[(x,y)\in \varphi_u(U_u\cap U_{u'})\mapsto (x'=x+1,y'=-y)\in \varphi_{u'}(U_u\cap U_{u'})\] qui est différentiable la différentielle en $(x,y)$ étant représentée par la matrice \[\left(
\begin{matrix}  1&0\\
0&-1
\end{matrix}\right).\]
\noindent\emph{\'{E}tape $2$: description catégorique d'une variété et d'un fibré.}

Dans les définitions \ref{varan} et \ref{vavaran} exposées ici, la structure de variété fournit un ensemble  de cartes locales avec une régularité des fonctions changement de cartes (ou changement de coordonnées). Appelons ces fonctions \emph{fonctions de transition}.
Supposons à l'inverse qu'on ne dispose que d'une famille de cartes locales. Comment peut-on deviner la variété sous-jacente. La réponse est \og naturelle\fg\; : des points on ne perçoit que les coordonnées (dans $\rr^n$ par exemple), qui sont différentes dans chaque carte locale, mais qui se transforment les unes en les autres par les fonctions de transition. Les points seront donc des classes d'équivalence de 
$n$-uplets, deux $n$-uplets $a_i$ et $a_j$ étant équivalents s'il existe une fonction de transition
$\varphi_{ij}$ telle que $a_i=\varphi_{ij}(a_j)$. Nous avons détaillé cette constructiondans [\ref{rinrin}] \footnote{l'adjectif catégorique est lié à cette construction} et montrer l'équivalence des deux constructions. 

Si on reprend le cas du ruban de Moebius $\mathcal{M}$ tel qu'on l'a défini plus haut, la relation d'équivalence sur $\rr\times \rr$
est celle qu'on vient de décrire. Cette définition nous donne les points de $\mathcal{M}$ et nous laisse entrevoir par la définition de la relation $\sim$ les fonctions de transition. Les cartes locales correspondantes existent mais elles ne sont pas explicitement données. Nous les avons trouvées dans la première étape.

Pour les fibrés, il y a la même idée de construction à partir de cartes locales : ici les fonctions de transition sont les cocycles du fibrés dans la définition qu'on en a donné (définitions \ref{fifibre} et \ref{cyclotouriste}). On comprend alors, à la lumière de la définition de $\mathcal{M}$ et de l'étape $1$ qui introduit les cocycle du fibré tangent $\tau(\mathcal{M})$ que l'on peut définir l'espace $T(\mathcal{M})$ du fibré tangent $\tau(\mathcal{M})$ comme 
\begin{equation}\label{ET}
T(\mathcal{M})=\bigcup_{\rr^2\times\rr^2}(x,y,a,b)\big/(x,y,a,b)\sim (x+1,-y,a,-b)
\end{equation}
On pose $\pi : T(\mathcal{M})\longrightarrow\mathcal{M}$ la surjection définie par $\pi([(x,y,a,b)]=[(x,y)]$ où $[(x,y)] =p\big((x,y)\big) $ (voir étape $1$ pour la définition de $p$).

\noindent Notons que sur chaque fibre $\pi^{-1}\big([(x,y)]\big)$ on a bien une structure d'espace vectoriel en posant $[(x,y,a,b)]+[(x,y,\al,\be)]=[(x,y,a+b,\al+\be]$ et $\lm [(x,y,a,b)]=[(x,y,\lm a,\lm b)]$.
On vérifiera que ces deux opérations sont bien définies et qu'elles confèrent à la fibre une structure de plan vectoriel.

\begin{exo}\label{leder}
Montrer que $\big(T(\mathcal{M}),\pi,M\big)$ a une structure de fibré vectoriel de fibre $\rr^2$ au sens de la définition \ref{fifibre},  isomorphe au fibré $\tau(\mathcal{M})$.
\end{exo}

\noindent\emph{\'{E}tape $3$: structure riemannienne sur $\tau(\mathcal{M})$}.

Notons $[(x,y,a,b)]$ la classe de $(x,y,a,b)$ pour la relation d'équivalence décrite dans la définition (\ref{ET}).
On pose lorsque $p(x_1,x_2)=p(y_1,y_2)=x$ : 
\begin{equation}\label{pesc}
\la [x_1,x_2, a_1,a_2],  [y_1,y_2, b_1,b_2]\ra
=a_1b_1+a_2b_2.
\end{equation}
On vérifie que ce produit ne dépend pas du représentant choisi des classes $[x_1,x_2, a_1,a_2]$ et 
$[y_1,y_2, b_1,b_2]$ de $p^{-1}(x)$. Il définit donc une structure riemannienne sur $\tau(\mathcal{M})$.
Les équations (\ref{cristo}) p. \pageref{cristo} et (\ref{paral}) p. \pageref{paral} montrent que si on déplace parallèlement un vecteur tangent le long d'un chemin $\gamma(t)$, ses composantes sont constantes dans une cartes locales (comme pour le déplacement parallèle d'un vecteur de $\rr^2$).

\noindent\emph{\'{E}tape $4$:  $\tau(\mathcal{M})$ n'est pas orientable.}

On considère le chemin de $\mathcal{M}$ définie par $\gamma(t)=[(t,0)]$ pour $t\in [0,1]$. Ce chemin est fermé puisque $\gamma(0)=\gamma(1)$.
On considère les vecteurs $e_1$ et $e_2$ de $p^{-1}([0,0]$ définis par $e_1=[0,0,1,0],e_2=[(0,0 ,0,1)]$. Notons que $e_2=[(1,0,0,-1)]$.
On considère les chemins définis pour $t\in [0,1]$ par $\gamma_1(t)=[(\frac{t}{2},0)]$ et 
$\gamma_2(t)=[(1-\frac{t}{2},0)]$. On a $\gamma_1(0)=\gamma_2(0)$ et $\gamma_1(1)=\gamma_2(1)=[(\frac{1}{2},0)]$. Les deux chemins de mêmes extrémité sont dans le même domaine de carte et ils sont homotopes : $F(t,h)=([\frac{t}{2} h+(1-\frac{t}{2} (1-h),0)]$ décrit une  une homotopie entre $\gamma_1$ et $\gamma_2$.
Supposons qu'au voisinage de $[(0,0)]$ on ait une orientation locale coïncidant avec l'orientation dé finie par $\mathcal{B}=(e_1,e_2)$ sur $p^{-1}([(o,o)])$. D'après l'étape $3$, le transport  de $\mathcal{B}$ en $\gamma_1(1)$ aboutit à $\tilde{\mathcal{B}}=\left(\tilde{e_1},\tilde{e_2}\right)$, avec 
$\tilde{e_1}=[\frac{1}{2},0,1,0)]$ et $\tilde{e_2}=[(\frac{1}{2},0,0,1)]$ et 
le transport  de $\mathcal{B}$ en $\gamma_2(1)$ aboutit à $\check{\mathcal{B}}=\left(\check{e_1},\check{e_2}\right)$, avec 
$\check{e_1}=[\frac{1}{2},0,1,0)]$ et $\check{e_2}=[(\frac{1}{2},0,0,-1)]=-\tilde{e_2}$. Les orientations définies par ces deux bases sont opposées, ce qui contredit le lemme \ref{trumpshit}. Il n'y a donc pas d'orientation locale au voisinage de $[(0,0)]$ donc $\mathcal{M}$ n'est pas orientable d'après le théorème \ref{orifi}.
\end{preuve}

\section{Appendice 5 : opérateur de Hodge et deuxième groupe des équations de Maxwell}\label{maxwiwi}

\subsection{Opérateur de Hodge sur une variété pseudo-riemannienne.}
Soit $(M,g)$ une variété pseudo-riemannienne \emph{orientable} et $\gamma$ le morphisme de fibré
\[ 
\begin{array}{ccc}
\mathcal{S}(\tau(M))&\overset{\gamma}{\longrightarrow}&\Lambda^1(M)\\
X&\mapsto&g(X,.)
\end{array}
\]
Soit $\alpha$ une $p$-forme sur $M$ et $\{X_1,\ldots,X_{n-p}\}$ une famille de $(n-p)$ 
champs de vecteurs indépendants sur $M$. On note $\sigma$ la forme volume riemannienne canonique de $M$.
Les formes volume $\sigma$ et $\alpha \wedge\gamma(X_1)\wedge\ldots\wedge\gamma(X_{n-p})$ sont proportionnelles. On note $*\alpha(X_1,\ldots,X_{n-p})$ le coefficient de proportionnalité :
\begin{equation}\label{hogele}
*\alpha(X_1,\ldots,X_{n-p})\sigma=\alpha \wedge\gamma(X_1)\wedge\ldots\wedge\gamma(X_{n-p}).
\end{equation}
$*\alpha$ est une fonction $(n-p)$ linéaire alternée et définit une $(n-p)$-forme différentielle sur $M$.
\begin{definition}\label{oh!G}\index{opérateur de Hodge}
L'opérateur $*:\Lambda^p(M)\longrightarrow \Lambda^{n-p}(M)$ défini par l'égalité (\ref{hogele}) est l'opérateur de Hodge de la variété 
pseudo-riemannienne $(M,g)$.
\end{definition}
\textbf{Introduisons pour la suite  quelques notations.}

\emph{Plaçons-nous dans une carte locale $(U,x)$ de $(M,g)$ munie des coordonnées $x=(x^1,\ldots,x^n)$.
Soit $I$ un sous-ensemble de $\{1,2,\ldots,n\}$ à $p$ éléments distincts. À partir de $I$ on un construit unique $p$-uplet 
ordonné $ \mathcal{I}=(i_1,\ldots,i_p)$ tel que $I=\{i_1,\ldots,i_p\}$ et $i_1<i_2<\ldots<i_p$. Les majuscules calligraphiques désigneront des arrangements ordonnés dans le sens croissant (on parlera également de $p$-uplet orienté s'il s'agit d'un arrangement à $p$ éléments).\index{p-uplet orienté}
Le sous-ensemble $I$ a un unique complémentaire dans $\{1,\ldots,n\}$, noté $\bar{I}$ et associé à $\bar{I}$ avec la convention précédente, on a l'arrangement orienté $\bar{\mathcal{I}}$.
Notons pour terminer que l'ensemble $I$ admet $p!$ arrangements. On note $\mathcal{A}(I)$ l'ensemble des arrangements de $I$. Par exemple, pour voir si tu suis, $\mathcal{I}\in\mathcal{A}(I)$.
Avec cette notation, on peut écrire une $p$-forme $\alpha$ sous la forme :
\begin{equation}\label{zombi}\alpha=a_\mathcal{I} dx^{\mathcal{I}}\end{equation}
où $\mathcal{I}$ décrit les $p$-uplets ordonnés dans le sens croissant de $\{1,\ldots,n\}$ et où $dx^\mathcal{I} =dx^{i_1}\wedge\ldots \wedge dx^{i_p}$ si  $\mathcal{I}=(i_1,\ldots,i_p)$. }


\noindent Pour calculer $*\alpha$ la définition précédente et quelques conséquences immédiates suffisent.
Parmi ces conséquences notons la proposition suivante.
\begin{proposition}\label{isogi}
Soit $\alpha\in \Lambda^p(M)$.
\begin{enumerate}
\item   $$**\alpha=(-1)^{p(n-p)}\alpha.$$
\item Si $f$ est une fonction différentiable, alors $*(f\al)=f*\al.$
\end{enumerate}
\end{proposition}
\begin{exo}\label{isogigi}
Démontrer la proposition \ref{isogi}
\end{exo}
\begin{corollary}\label{oiso}
Soit $(M,g)$ une variété pseudo-riemannienne orientée, munie de sa forme volume canonique $\sigma$.
Soit $v= f \sigma$ une forme volume sur $M$ où $f$ est une fonction différentiable sur$M$.
On a : $*v=f$ et $*f=v$
\end{corollary}
\begin{preuve}{}
Dans une carte locale munie des coordonnées $(x^1,\ldots,x^n)$, on a :

\noindent $(*f)(\frac{\partial}{\partial x^1},\ldots,\frac{\partial}{\partial x^n})\sigma=f \gamma(\frac{\partial}{\partial x^1})\wedge\ldots\wedge \gamma(\frac{\partial}{\partial x^n})=f g_{1i_1}\ldots g_{ni_n}d x^{i_1}\wedge\ldots\wedge dx^{i_n}=$

\noindent $=f\det g\;dx^1\wedge\ldots\wedge x^n =f\sqrt{\abs{\det g}}\sigma$.

\noindent Par ailleurs il existe une fonction différentiable $\lambda$ telle que $*f=\lambda \sigma$
On a donc $(*f)(\frac{\partial}{\partial x^1},\ldots,\frac{\partial}{\partial x^n})=f\sqrt{\abs{\det g}}=\lambda\sqrt{\abs{\det g}}$ D'où $\lambda =f$. Ainsi $*f=f\sigma.$

D'après la proposition \ref{isogi}, $*(*f)=f$, donc pour toute fonction différentiable $f$, $f=*(f\sigma)=f(*\sigma)$. En prenant $f=1$, on obtient 
$*\sigma =1$ et donc $*v=f$.
\end{preuve}
 \noindent Si maintenant le calcul de $*\alpha$ passe par les coordonnées locales dans une carte, la proposition qui suit donne la recette du calcul des coordonnées de $*\alpha$, en précisant un choix de la forme volume riemannienne (définie a priori au signe près). Dans la proposition suivante, on notera symboliquement $\al=a dx$  une $p$-forme qui s'écrit habituellement dans des coordonnées locales $x=(x^1,\ldots,x^n): \al =a_{i_1,\ldots,i_p}dx^{i_1}\wedge\ldots\wedge dx^{i_p}$. 
 Le cas des fonctions et des formes volume étant réglé, on peut énoncer :
 \begin{proposition}\label{coloduG} 
 Soit $(M,\la\;\ra)$ une variété pseudo-riemannienne de dimension $n$ et
 $\alpha$ une $p$-forme différentielle  donnée par l'égalité (\ref{zombi}). Nous supposons que $p\notin \{0,n\}$. Par ailleurs sur une carte locale $(U, (x^1,\ldots,x^n))$ pour laquelle on note $g_{ij}=\la\frac{\partial}{\partial x^{i}},\frac{\partial}{\partial x^{j}}\ra$, nous considérons la forme volume riemannienne  $\sigma $ restreinte à $U$, $\sigma_U=\sqrt{\abs{\det g}}dx^1\wedge\ldots\wedge dx^n$. On a au signe près :
  \begin{equation}\label{houa}*\alpha=(*a)_\mathcal{K} dx^\mathcal{K} \end{equation} où $\mathcal{K}$ décrit les $(n-p)$-uplets orientés de $\{1,\ldots ,n\}$ avec
  \begin{equation}\label{houaoua}
(*a)_\mathcal{K} =\sqrt{\abs{\det g}}\;\epsilon(\bar{\mathcal{K},}\mathcal{K})a^{\bar{\mathcal{K}}}
\end{equation}
\noindent où \begin{enumerate}
\item
$\bar{\mathcal{K}}$ est le  $p$-uplet orienté complémentaire de $\mathcal{K}$ dans $\{1,\ldots,n\}$  et où $\epsilon(\bar{\mathcal{K},}\mathcal{K})$ désigne 
la signature de la permutation de $\{1,2,\ldots,n\}$ s'écrivant symboliquement $(\bar{\mathcal{K},}\mathcal{K})$.
\item
Si $\bar{\mathcal{K}}=(i_1,\ldots,i_{p})$, alors $a^{\bar{\mathcal{K}}}=g^{i_1,s_1}g^{i_2,s_2}\ldots g^{i_{p},s_{p}}a_{s_1\ldots s_p}$,  qui est une sommation où $(s_1,\ldots,s_p)$ décrit  les  $p$-uplets  de 
$\{1,\ldots,n\}$.
\end{enumerate}
\end{proposition}
\begin{preuve}{}

Quitte à restreindre $U$, nous allons considérer sur $U$ des coordonnées galiléennes $(x^1,\ldots,x^n)$, c'est à dire des coordonnées vérifiant : $g_{ij}= \pm \delta_{ij}$.

\noindent Ainsi $\det g\in \{-1,1\}$. 
 
 \noindent Posons $ \epsilon_i=g_{ii}$. On a l'égalité 
$\gamma(\frac{\partial}{\partial x^{i}})=\epsilon_i dx^{i}$ (sans sommation). Soit la $p$-forme $ \al=a_{i_1,\cdots,i_p}dx^{i_1}\wedge\cdots\wedge dx^{i_p}$ où $i_1<\cdots<i_p$. Notons $\beta=\sum_{\mathcal{K}}(*a)_{\mathcal{K}}dx^{\mathcal{K}}$ où ${\mathcal{K}}$ est un $(n-p)$-uplet orienté. En coordonnées galiléennes, on a :

$(*a)_{\mathcal{K}}=\epsilon(\overline{{\mathcal{K}}},{\mathcal{K}})g^{\overline{{\mathcal{K}}},\overline{{\mathcal{K}}}}a_{\overline{{\mathcal{K}}}}$ (écriture sans sommation, $\overline{{\mathcal{K}}}$ étant le $p$-uplet orienté complémentaire de 
$\mathcal{K}$ dans $\{1,\cdots,n\}$). Ainsi de façon explicite :
\[(*a)_{j_1,\cdots, j_{n-p}}=\epsilon(i_1,\cdots,i_p,j_1,\cdots,j_{n-p})\epsilon_{i_1}\cdots\epsilon_{i_p}a_{i_1,\cdots,i_p},\]
avec 

\noindent $i_1<\cdots<i_p$, $j_1<\cdots <j_{n-p}$, $\{i_1,\cdots,i_p,j_1,\cdots,j_{n-p}\}=\{1,2,\cdots,n\}$ 
 
 et 
 
 \noindent $\epsilon(i_1,\cdots,i_p,j_1,\cdots,j_{n-p})$ désignant la signature de la permutation\\ $(i_1,\cdots,i_p,j_1,\cdots,j_{n-p})$ de $(1,2,\cdots,n)$.
 On peut donc écrire $\beta$ sous la forme 
 \begin{equation}\label{lebeta}
 \beta=\sum_{j_1<\cdots<j_{n-p}}\epsilon(i_1,\cdots,i_p,j_1,\cdots,j_{n-p})\;\epsilon_{i_1}\cdots \epsilon_{i_p}\;a_{i_1,\cdots,i_p}\;dx^{j_1}\wedge\cdots\wedge dx^{j_{n-p}}.
 \end{equation}
 On a donc $\beta(\frac{\partial}{\partial x^{j_1}},\cdots,\frac{\partial}{\partial x^{j_{n-p}}})=\epsilon(i_1,\cdots,i_p,j_1,\cdots,j_{n-p})\epsilon_{i_1}\cdots \epsilon_{i_p}a_{i_1,\cdots,i_p}.$
 Par ailleurs on a directement les égalités :
 
 \noindent$\al\wedge\gamma(\frac{\partial}{\partial x^{j_1}})\wedge\cdots\wedge \gamma(\frac{\partial}{\partial x^{j_{n-p}}})=
 a_{i_1,\cdots,i_p}dx^{i_1}\wedge\cdots\wedge dx^{i_p}\;\epsilon_{j_1}\cdots\epsilon_{j_{n-p}}dx^{j_1}\wedge\cdots\wedge dx^{j_{n-p}}=\epsilon(i_1,\cdots,i_p,j_1,\cdots,j_{n-p})\;\epsilon_{i_1}\cdots \epsilon_{i_p}a_{i_1,\cdots,i_p}\;dx^1\wedge\cdots\wedge d x^n$.
 
 \noindent Si on prend $\sigma=dx^1\wedge\cdots\wedge dx^n$, alors on a montré que 
 \[\beta(\frac{\partial}{\partial x^{j_1}},\cdots,\frac{\partial}{\partial x^{j_{n-p}}})\sigma=\al\wedge\gamma(\frac{\partial}{\partial x^{j_1}})\wedge\cdots\wedge \gamma(\frac{\partial}{\partial x^{j_{n-p}}}).\]
 Ceci démontre la proposition en coordonnées galiléennes. Compte tenu du caractète tensoriel de l'égalité à démontrer, celle-ci est vérifiée dans tout système de coordonnées. 
 De plus, il y a deux choix possibles pour la forme volume riemannienne, ce qui justifie l'égalité au signe près.
\end{preuve}
\begin{remark}
Il est pratique de se souvenir de l'égalité (\ref{lebeta}) pour calculer $*\al$ dans l'espace de Minkovski, ou dans l'espace euclidien $\rr^3$. Par exemple si dans $\rr^3$ on a : $\al =adx+bdy+cdz$, alors par (\ref{lebeta}) on a :
$(*\al)=\epsilon(1,2,3)ady\wedge dz+\epsilon(2,1,3)b dx\wedge dz+\epsilon(3,1,2) dx\wedge dy=ady\wedge dz+bdz\wedge dx+cdx\wedge dy$.
\end{remark}
\begin{proposition}\label{champetre}
On considère l'espace euclidien $\left(\rr^3,\la\;,\;\ra\right)$ standard et l'espace de Minkowski  
$\left(\rr^4,\la\;,\;\ra_{L}\right)$ muni des coordonnées $(x^0=ct ,x,y,z)$. On note $*$ l'opérateur de Hodge de l'espace de Minkowski et $\star$ l'opérateur de Hodge de l'espace euclidien $\rr^3$. 
On reprend les notations des équations (\ref{elco}),(\ref{themag}),(\ref{lasyme}) et 
(\ref{lasymb}) p.\pageref{elco}  de la section \ref{erer}. Alors :
\begin{enumerate}
\item $\star\mathcal{E}^1=\mathcal{E}^2$,
\item $\star \mathcal{B}^2=\mathcal{B}^1$,
\item $*\mathcal{B}^2=-c\mathcal{B}^1\wedge dt,$
\item  $*\left(\mathcal{E}^1\wedge c dt\right)=\mathcal{E}^2$,
\item $*\mathcal{F}^2=\mathcal{E}^2-c\mathcal{B}^1\wedge dt$
\end{enumerate}
\end{proposition}
\begin{exo}\label{maxou}
Démontrer la proposition \ref{champetre}
\end{exo}

\subsection{Deuxième groupe des équations de Maxwell}

On reprend les notations et hypothèses  de l'exercice \ref{brut}, p. \pageref{brut}.
On considère dans $\rr^3$ un fluide chargé en mouvement. Un observateur lié à un référentiel $\mathcal{R}_0$ galiléen qui suit le fluide dans son mouvement observe une densité de charge $\sy_0$ (densité de charge au repos). Un observateur lié au référentiel galiléen $\mathcal{R}$   observe une densité de charge $\sy$. De manière analogue à la masse au repos, $\sy_0$ est une constante qui permet de définir le \emph{quadivecteur  courant électrique} 
$$J=\sy_0c u$$ 
où $u$ est le quadrivecteur vitesse d'une particule chargée du fluide. Le  principe de relativité d'Einstein atteste  que deux observateurs liés à deux référentiels galiléens mesurent la même charge électrique, ce qui se traduit par l'égalité : $\sy_0 dV_0=\sy d V$. On en déduit de la relation (\ref{begue}) p.\pageref{begue} l'égalité $\sy=\gamma\;\sy_0$.
On a donc :
\begin{equation}\label{coucoulele}
J=\sy c\frac{\partial}{\partial x^0}+j\;\mathrm{où}\; j=\sy v
\end{equation}
Le vecteur $j$ est classiquement appelé le \emph{vecteur courant}. Dans des coordonnées galiléennes réduites $(x^0, x^1,x^2, x^3)$ la forme volume riemannienne de l'espace de Minkowski s'écrit ( voir définition \ref{canfori}) $\tau=dx^0\wedge dx^1\wedge dx^2\wedge dx^3$.
On associe au quadrivecteur courant sa version covariante par l'intermédiaire de la forme volume $\tau$ de l'espace de Minkowski : 
\[\mathcal{J}^3=i(J)\tau.\]
Alors \begin{principe}{ Deuxième groupe d'équations de Maxwell }\index{Maxwell (deuxième groupe d'équations)}
\begin{equation}\label{max2}
d*\mathcal{F}^2=\frac{4\pi}{c} \mathcal{J}^3.
\end{equation}
\end{principe}
\begin{exo}\label{max2etc}
\begin{enumerate}
\item Montrer que $\mathcal{J}^3=\sigma ^3-j^2\wedge c dt$ où $\sigma^3=\sy c dx\wedge dy\wedge dz$ et $j^2=i(j)dx\wedge dy\wedge dz$
\item Montrer, en utilisant les exercices \ref{rougi} et la proposition \ref{champetre} que 

$i$) $\bold{d}\star\mathcal{E}^1=\frac{4\pi}{c} \sigma^3$

$ii$) $\bold{d}\star \mathcal{B}^2=\frac{4\pi}{c} j^2+\frac{1}{c}\frac{\partial  \mathcal{E}^2}{\partial t}$.
\item 
En déduire, en utilisant l'exercice \ref{klaciko}, les équations de Maxwell exprimées en termes de  champs de vecteurs:

$i$) $\Div E=4\pi\sy$

$ii$) $\rot B=\frac{4\pi}{c} j+\frac{1}{c}\frac{\partial E}{\partial t}$
\end{enumerate} 
\end{exo}
\subsection{\'{E}criture des équations deMaxwell en coordonnées locales quelconques.
Quadrivecteur densité de force de Lorentz.}
\subsubsection{\'{E}quations de Maxwell}

L'espace de Minkowski $(M,g)$ est un espace pseudo-riemannien muni d'une métrique $g$ de signature $(1,3)$. Dans cet espace les coordonnées galiléennes sont celles qui diagonalisent cette métrique au sens où elle y est représentée par une matrice diagonale 
$\left(\begin{matrix}
-1&0&0&0\\
0&1&0&0\\
0&0&1&0\\
0&0&0&1
\end{matrix}\right)$. Elles sont privilégiées par le principe d'Einstein de la relativité restreinte qui affirme qu'aucune loi de la mécanique et de l'électromagnétisme ne peut mettre en évidence le mouvement d'un référentiel galiléen par rapport à un autre. Et on sait que les changements de cartes locales entre deux cartes munies de coordonnées galiléennes est une transformation de Lorentz. Ces coordonnées sont donc naturellement celles dans lesquelles on écrit les lois de la physique, en particulier les équations de Maxwell. Mais on peut envisager de les écrire dans 
n'importe quelle carte locale de l'espace $(M,g)$. Plaçons nous dans une carte locale  galiléenne munie des coordonnées réduites $(x^0,x^1,x^2,x^3)$ dans laquelle la forme volume riemannienne s'écrit $dx^0\wedge dx^1\wedge dx^2\wedge dx^3$.
Si on pose $\mathcal{F}^2=F_{ij}dx^{i}\wedge dx^j$ où $i<j$, alors avec les notations de l'équation (\ref{houaoua}), on a, d'après la proposition \ref{coloduG} :
$*\mathcal{F}^2=\varepsilon(i,j\alpha,\beta)\mathcal{F}^{ij}dx^{\al}\wedge dx^\be$ où $\alpha<\beta=$ 

 \noindent $*\mathcal{F}^2=\mathcal{F}^{23}dx^0\wedge dx^1-\mathcal{F}^{13}dx^0\wedge dx^2+\mathcal{F}^{12}dx^0\wedge dx^3 
+\mathcal{F}^{03}dx^1\wedge dx^2$

$-\mathcal{F}^{02}dx^1\wedge dx^3+\mathcal{F}^{01}dx^2\wedge dx^3$ que l'on écrit plus naturellement :

\noindent $*\mathcal{F}^2=*F_{01}dx^0\wedge dx^1+*F_{02}dx^0\wedge dx^2+*F_{03}dx^0\wedge dx^3 
+*F_{12}dx^1\wedge dx^2$

$*F_{13}dx^1\wedge dx^3+*F_{23}dx^2\wedge dx^3$

\begin{exo}\label{bebete}
Dans les coordonnées galiléennes $(x^0=ct, x,y,z)$ on a :
\begin{enumerate}
\item ${F}_{01}=-E^1, {F}_{03}=-E^3, {F}_{02}=-E^2,{F}_{12}=B^3,{F}_{13}=-B^2,{F}_{23}=B^1, F_{ij}=-F_{ji}$
\item $*{F}_{01}=B^1, *{F}_{02}=B^2,* {F}_{03}=B^3,*{F}_{12}=-E^3,*{F}_{13}=E^2,*{F}_{23}=-E^1,$

$ *F_{ij}=-*F_{ji}$
\end{enumerate}
\end{exo}
On considère des coordonnées quelconques $(x^0,x^1,x^2,x^3)$ dans lesquelles $\ma{F}^2=F_{ij}dx^{i}\wedge dx^j$ avec $i<j$.
La première équation (tensorielle) de Maxwell $d\mathcal{F}=0$, s'écrit en coordonnées   $\frac{\partial F_{ij}}{\partial x^k}\;dx^k \wedge dx^{i}\wedge dx^j=0$, sommation dans laquelle $i<j$. On retrouve dans cette somme le terme $dx^k\wedge dx^{i}\wedge dx^j$ où
$i\not= j\not=k$ et $i<j$ trois fois à l'ordre près : $dx^k\wedge dx^{i}\wedge dx^j,dx^{i}\wedge dx^{k}\wedge dx^j$,$dx^{i}\wedge dx^{j}\wedge dx^k$. On en déduit que la première équation de Maxwell équivaut en coordonnées locales à :
\begin{equation}\label{primo}
\frac{\partial F_{ij}}{\partial x^k}+\frac{\partial F_{jk}}{\partial x^{i}}+\frac{\partial F_{ki}}{\partial x^j}=0,
\end{equation}
pour toutes les parties à trois éléments distincts deux à deux  $\{i,j,k\}$ de $\{0,1,2,3\}$.

Or en utilisant l'égalité (\ref{pim}) p. \pageref{pim}, on peut exprimer l'équation (\ref{primo}) en utilisant la dérivée covariante du $\mathcal{F}^2$ relativement à la connexion de Levi-Civita de $(M,g)$. On obtient : $\frac{\partial F_{ij}}{\partial x^k}={F_{ij}}_{/k}+
\Gamma^s_{ki}F_{sj}+\Gamma^s_{kj}F_{is}$. Ainsi l'équation (\ref{primo}) s'écrit 
\[{F_{ij}}_{/k}+{F_{ki}}_{/j}+{F_{jk}}_{/i}+A_{ijk}=0\]
où
\[A_{ijk}=\Gamma^s_{ki}F_{sj}+\Gamma^s_{kj}F_{is}+
\Gamma^s_{ij}F_{sk}+\Gamma^s_{ik}F_{js}+
\Gamma^s_{jk}F_{si}+\Gamma^s_{ji}F_{ks}
\]
Un calcul direct montre que le terme $A_{ijk}$ est nul compte tenu de l'antisymétrie de $\mathcal{F}^2$ et de la symétrie des coefficients de Christoffel  exprimée par  $\Gamma^s_{\alpha\beta}=\Gamma^s_{\beta\alpha}$.
D'où le premier groupe d'équations de Maxwell en coordonnées locales :
\begin{equation}\label{promi}
{F_{ij}}_{/k}+{F_{jk}}_{/i}+{F_{ki}}_{/j}=0
\end{equation}
La deuxième équation (tensorielle) de Maxwell dit qu'il existe un champ de vecteurs $J$ telle que $d*\mathcal{F}^2=\frac{4\pi}{c} i(J) \tau$ où $\tau $ est la forme volume riemanienne de l'espace de Minkowski.

En coordonnées locales le deuxième groupe s'écrit en notant $(\alpha,\beta)$ le couple orienté complémentaire du couple orienté $(i,j)$ dans $\{0,1,2,3\}$ :

\begin{align}
&\sum_{\underset{i<j}{i,j,k}}\varepsilon(\alpha,\beta,i,j)\frac{\partial \sqrt{\abs{\det g}}F^{\alpha\beta}}{\partial x^k}\;dx^k\wedge dx^{i}\wedge dx^j
=\frac{4\pi}{c}\sqrt{\abs{\det g}}\left(J^0 dx^1\wedge dx^2\wedge dx^3\right.\nonumber\\
&\left.-J^1 dx^0\wedge dx^2\wedge dx^3+
J^2 dx^0\wedge dx^1\wedge dx^3
-J^3dx^0\wedge dx^1\wedge dx^2\right)\label{sapo}
\end{align}

\noindent Le terme de gauche est constitué de la somme de douze termes qu'on peut décrire de la manière suivante. Le coefficient $\al$ parcourt les entiers $0,1,2$. Pour $\al=0$, le triplet $(\be,i,j)$ décrit les triplets 
$(1,2,3),(2,1,3),(3,1,2)$, pour $\al=1$ les triplets $(2,0,3), (3,1,2)$, pour $\al=2$ le triplet $(3,1,2)$. Pour chacun des $6$ quadruplets $(\al,\be,i,j)$ avec $\al<\be$ et $i<j$ l'indice $k$ prend  les deux valeurs complémentaires de $\{i,j\}$, d'où les douze termes de la somme de gauche. En développant cette somme avec la décomposition qui vient d'être  décrite par exemple on montre directement que l'équation (\ref{sapo}) équivaut aux équations suivantes :


\[\frac{\partial  \sqrt{\abs{\det g}}\;F^{ii_1}}{\partial x^{i_1}}+\frac{\partial  \sqrt{\abs{\det g}}\;F^{ii_2}}{\partial x^{i_2}}+\frac{\partial  \sqrt{\abs{\det g}}\;F^{ii_3}}{\partial x^{i_3}}=\frac{4\pi}{c}\sqrt{\abs{\det g}}J^{i}\]
où $(i_1,i_2,i_3)$ est le triplet orienté complémentaire de $i$ dans l'ensemble$\{0,1,2,3\}$, égalités qu'on écrira synthétiquement 

\begin{equation}\label{deuzio}
\forall i\in\{0,1,2,3\}\; \frac{1}{\sqrt{\abs{\det g}}}\frac{\partial \left( \sqrt{\abs{\det g}}\;F^{ij}\right)}{\partial x^{j}}=\frac{4\pi}{c} J^{i}
\end{equation} où$j$ décrit l'ensemble $\{1,2,3\}$.
\begin{proposition}\label{divinmax}
Le système d'équations (\ref{deuzio}) équivaut à 
	\begin{equation}\label{deuzcolo}{F^{ij}}_{/ j }=\frac{4\pi}{c} J^{i}.
	\end{equation}
Que l'on écrit :
\[\Div\mathcal{F}^2=\frac{4\pi}{c} J.\]
\end{proposition}
\begin{preuve}{}
En développant le terme de gauche de l'égalité (\ref{deuzio}), on a :

\noindent$\frac{\partial \ln \sqrt{\abs{det g}}}{\partial x^s}F^{is}+\frac{\partial F^{is}}{\partial x^s}=\frac{4\pi}{c} J^{i}$,
c'est à dire d'après l'exercice \ref{conlevcicris} p.\pageref{conlevcicris} :

\noindent $\Gamma^l_{ls}F^{is}+\frac{\partial F^{is}}{\partial x^s}=\frac{\partial F^{is}}{\partial x^s}+
\Gamma^{i}_{sl}F^{ls}+\Gamma^l_{sl}F^{is}=F^{is}_{/s}$ car $\Gamma^{i}_{sl}F^{ls}=0$,
ce qui permet de conclure si l'on se  souvient de  l'exercice \ref{divdivergence}.
\end{preuve}
\begin{remark}
Les équations (\ref{promi}) et (\ref{deuzio}) sont également les équations tensorielles de Maxwell dans l'espace temps de la relativité générale munie de sa connexion de Lévi-Civita.
\end{remark}
\subsubsection{Quadrivecteur densité force de Lorentz et tenseur impulsion-énergie électromagnétique}
En partant de l'égalité (\ref{lor}) p.\pageref{lor}, on définit la $1$-forme
 \begin{equation}\label{defolo}\mathcal{K}^1=-\sy_0i(u)\mathcal{F}^2\end{equation}\index{force de Lorentz (forme densité)}
  invariante par transformée de Lorentz appelée \emph{forme densité force de Lorentz}. Dans des coordonnées quelconques cette forme s'écrit :
\[\mathcal{K}^1=-i\left(\frac{J}{c}\right)\mathcal{F}^2=-\frac{1}{c}F_{ij}J^{i} dx^j=:\mathcal{K}^1_i dx^{i}.\]
On en déduit les coordonnées contravariantes du \emph{quadrivecteur densité force de Lorentz} $\mathcal{K}_1$ défini par $\mathcal{K}^1=\la\mathcal{K}_1,.\ra_{\mathcal{L}}$, c'est à dire en coordonnées locales :
\begin{equation}\label{denfololo}
\mathcal{K}_1^{i}=-\frac{1}{c}F^{i}_sJ^s
\end{equation}
\begin{exo}\label{denfolola}
Démontrer la relation (\ref{denfololo}).
\end{exo}
Il résulte des équations de Maxwell que le quadrivecteur de Lorentz est la divergence du tenseur impulsion-énergie que nous définissons ci-après.
\begin{definition}\label{tenenim}\index{tenseur impulsion-énergie électromagnétique}
On note $F^{ij}$ les cordonnées  contravariantes  du tenseur électromagnétique dans une carte
locale de l'espace de Minkowski.
On appelle tenseur impulsion-énergie électromagnétique le tenseur contravariant d'ordre $2$, $\ma{M}$, ayant dans cette carte les coordonnées :
\begin{equation}\label{toneuneu}
\ma{M}^{ij}=\frac{1}{4\pi c^2}\left(\frac{1}{4}g^{ij}F_{\al\be}F^{\al\be}-F^{ik}{F^j}_k\right)
\end{equation}
\end{definition}
\noindent La dénomination ''impulsion-énergie'' se justifie par le fait que dans des coordonnées galiléennes, on a : \begin{equation}\label{ninin}\ma{M}^{00}=-\frac{1}{8\pi c^2}(\Vert E\Vert^2+\Vert B\Vert^2)\end{equation}
qui représente  (au signe près) une densité d'énergie électromagnétique (voir [\ref{lali}] ).

\begin{exo}\label{enimp}
Démontrer l'égalité (\ref{ninin}).
\end{exo}
\begin{theorem}\label{divlornini}
Le quadrivecteur densité force de Lorentz est  la divergence du tenseur impulsion-énergie, au coefficient $-c^2$ près, ce qui se traduit par les égalités 
\[\Div( \ma{M})=-\frac{1}{c^2}\ma{K}_1.\] ou en coordonnées par 
\[\frac{1}{4\pi}\left(F^{ik}{F^j}_k-\frac{1}{4}g^{ij}F_{\al\be}F^{\al\be}\right)_{/j}=-\frac{1}{c}{F^{i}}_s J^s\]
\end{theorem}
La preuve du théorème \ref{divlornini} passe par le minuscule lemme technique suivant.
\begin{lemma}\label{ptilem}
On a les trois égalités :
\[
{F_{ks}}_{/j}F^{sj}=-\frac{1}{2}{F_{sj}}_{/k}F^{sj},
\]
\[
{F^{i}}_sF^{sj}=-F^{is}{F^j}_s,
\]
\[
\frac{1}{2}\left(F_{sj}F^{sj}\right)_{/k}={F_{sj}}_{/k}F^{sj}.
\]
\end{lemma}
\begin{preuve}{du lemme  \ref{ptilem}}
\begin{enumerate}
 \item${F_{ks}}_{/j}F^{sj}=\frac{1}{2}{F_{ks}}_{/j}F^{sj}+\frac{1}{2} {F_{kj}}_{/s}F^{js}
=\frac{1}{2}\left({F_{ks}}_{/j}-{F_{kj}}_{/s}\right)F^{sj}=..$

$..=\frac{1}{2}\left({F_{ks}}_{/j}+{F_{jk}}_{/s}\right)F^{sj}=
-\frac{1}{2}{F_{sj}}_{/k}F^{sj}$ \;\;grâce à l'équation de Maxwell (\ref{promi})
\item On a ${F^{i}}_sF^{sj}=-{F^{i}}_sF^{js}=-g_{ks}F^{ik}g^{st}{F^{j}}_t=-\delta ^t_kF^{ik} {F^{j}}_t=-F^{is}{F^j}_s$.
\item $\frac{1}{2}\left(F_{sj}F^{sj}\right)_{/k}=\frac{1}{2}{F_{sj}}_{/k}F^{sj}+\frac{1}{2}F_{sj}{F^{sj}}_{/k}$. Mais 
$\frac{1}{2}F_{sj}{F^{sj}}_{/k}=\frac{1}{2}F_{sj}{(g^{sl}g^{jh}F_{lh})}_{/k}=$

\noindent $\frac{1}{2}\left(F_{sj}g^{sl}g^{jh}\right){F_{lh})}_{/k}=\frac{1}{2}F^{lh}{F_{lh}}_{/k}=\frac{1}{2}F^{sj}{F_{sj}}_{/k}$
et on conclut.
\end{enumerate}
\end{preuve}
\begin{preuve}{du théorème \ref{divlornini}}
D'après l'équation (\ref{deuzcolo}), $\frac{1}{c}{F^{i}}_s J^s=\frac{1}{4\pi}{F^{i}}_sF^{sj}_{/j}$. 
Ainsi :

\noindent $\frac{1}{c}{F^{i}}_s J^s=\frac{1}{4\pi}\left(F^{i}_s F^{sj}\right)_{/j}-\frac{1}{4\pi} {F^{i}_s}_{/j}F^{sj}
=\frac{1}{4\pi}\left(F^{i}_s F^{sj}\right)_{/j}-\frac{1}{4\pi}g^{ik}({F_{ks}}_{/j} F^{sj}).$
Donc par le lemme \ref{ptilem}, on peut écrire :

\noindent$\frac{1}{c}{F^{i}}_s J^s=\frac{1}{4\pi}\left(F^{i}_s F^{sj}\right)_{/j}+\frac{1}{8\pi}g^{ik}{F_{sj}}_{/k}F^{sj}=
\frac{1}{4\pi}\left(F^{i}_s F^{sj}\right)_{/j}+\frac{1}{16\pi}g^{ik}(F_{sj}F^{sj})_{/k}=$

\noindent$=-\frac{1}{4\pi}\left(F^{is}{F^k}_s+\frac{1}{4}g^{ik}F_{sj}F^{sj}\right)_{/k}.$
\end{preuve}

\section{Appendice $6$ : dynamiques des milieux continus}\label{ap6}
\subsection{\'{E}léments d'intégration sur les ensembles paramétrés et sur les variétés orientables. Formulaire.}\label{crocs}
\subsubsection{un exemple  élémentaire}
Dans cet exemples on fait deux observations qui vont inspirer les généralisations exposées dans les sections qui suivent.
Soit $f$ une fonction continue sur $[a,b]$ à valeurs réelles.

\noindent \textbf{Première observation}

Nous savons que la construction de l'intégrale d'une fonction réelle $f$ disons continue, définie sur $I=[a,b]$, passe par la construction de fonctions étagées construites à partir de la subdivision de $I$ ou de l'intervalle $f(I)$ suivant qu'il s'agisse des constructions de Riemann-Darboux ou de Lebesgue, fonctions étagées dont on calcule l'intégrale. L'intégrale de $f$ sur $I$ s'obtient alors par un passage à la limite des intégrales de ces fonctions en escalier. Dans cette construction l'orientation de $I$ n'intervient pas.
 
 Cette construction effectuée on note généralement $\int_a^bf$ l'intégrale de $f$ sur $I$ ou encore $\int_I f$. Puis on rajoute la convention $\int_b^{a}f=-\int_a^bf$. Ce faisant on fait intervenir l'orientation de $I$. On introduit alors sans le mentionner le concept d'intégrale d'une forme différentielle sur $I$. C'est ce concept que l'on développe   précisément dans cette section, en commençant par l'exemples des  $1$-formes sur $\rr$.

\noindent Définissons l'intégrale sur l'intervalle fermé $I$  muni d'une orientation $o(I)$, de  la $1$-forme sur $\rr$, $\al=f dt$  où $f$ est une fonction continue, par \begin{equation}\label{outre}\int_I \al =\varepsilon (I)\int_a^bf(t)\;dt,\end{equation}
$\int_a^bf(t)\;dt$ désignant lintégrale de $f$ au sens de Riemann sur l'inervalle $I=[a,b]$.
\noindent avec $\varepsilon(I) =1$ si $o(I)$ est l'orientation de $a$ vers $b$, c'est à dire $o(I)=o(\frac{\partial}{\partial t}$) et $\varepsilon =-1$ sinon. On peut considérer dans l'intégrale de droite de l'équation (\ref{outre}), $dt$ comme la mesure de Lebesgue sur $[a,b]$.

\noindent Comme cette intégrale dépend du choix d'une orientation sur $I$ on la note également $\int_{( I,o(I))}\al$.
Considérons la paramétrisation de $I$, $\varphi : J=[c,d]\longrightarrow I$ (c'est à dire une bijection de $J$ sur $I$ réalisant un difféomorphisme de $]c,d[$ sur $]a,b[$. Supposons que :
\begin{enumerate}
\item $J$ possède une orientation $o(J)$,
\item $\varphi(c)=a,\varphi(d)=b$ et $ \varphi$ préserve l'orientation. 
\ 
\end{enumerate}
Alors \begin{equation}\label{outro}
\int_{(I,o(I))}\al=\int_{(J,o(J))}\varphi^*\al.
\end{equation}
En effet, si on note $t=\varphi(u)$, alors $\varepsilon(J)=\varepsilon(\varphi^{-1}(I))=\varepsilon(I)$ ( car $\varphi$ préserve l'orientation) et
 $\int_{(J,o(J))}\varphi^*\al=\varepsilon(J)\int_c^df\circ\varphi(u)\varphi'(u) \;du=\varepsilon(I)\int_a^bf(t)\;dt=\int_{(I,o(I))}\al$.

\noindent Notons de plus que \begin{equation}\label{biden}
\int_{(I,-o(I)}\al=-\int_{(I,o(I)}\al.
\end{equation}
qui évoque la convention concernant l'intégrale d'une fonction $f$ continue sur $[a,b]$ : $\int_a^bf(t)dt=-\int_b^af(t)dt$.

\noindent \textbf{Seconde observation}

\noindent
Notons que si $[a,b]$ est orienté de $a$ vers $b$, on peut orienter le bord $\partial [a,b]$ de $[a,b]$ en affectant à $a$ le signe $-$ et à $b$ le signe $+$, de sorte conformément à l'idée de la définition (\ref{outre}) on ait  pour toute $0$-forme $g$
\begin{equation}\label{beloutre}
\int_ {\partial [a,b]} g d\delta= g(b)-g(a),
\end{equation}
de sorte que la relation fondamentale de l'intégration s'écrit si $f$ est une fonction dérivable sur $]a,b[$, continue sur $[a,b]$ $f'(t) dt $ une $1$-forme sur $[a,b]$ orienté de $a$ vers $b$ :
\begin{equation}\label{ptistockes}
\int_{[a,b]} f'(t)dt=\int_{\partial [a,b]}f\;d\delta.
\end{equation}
Si on note $\alpha$ la $0$-forme $f$, 
l'intégrale de gauche est l'intégrale de $d\al$ sur $[a,b]$ orienté de $a$ vers $b$, et l'intégrale de droite est l'intégrale de $\al$ sur le bord de $[a,b]$ muni de l'orientation induite par l'orientation de $[a,b]$.
\subsubsection{Intégrale d'une forme différentielle sur un domaine paramétré orienté.}
Le premier concept à définir est celui de domaine paramétré orienté tel l'intervalle $I$ de la section précédente. Typiquement dans une variété différentiable  les domaines des cartes locales $(U,\psi)$ sont des domaines paramétrés, le paramétrage étant réalisé par $\psi^{-1}$. Il faut élargir légèrement cet exemple.
\begin{definition}\label{parana}\index{domaine paramétré}
Un domaine paramétré de dimension $p$ est un sous-ensemble $P$ d'une variété différentiable $M$ de dimension supérieure ou égale à $p$, pour lequel il existe un borélien $U$ de $\rr^p$ et une application $\varphi$ définie sur un voisinage ouvert de $U$, vers $M$, tels que :
\begin{enumerate}
\item $P=\varphi(U)$
\item $\varphi $ est de rang $p$ presque partout sur $U$.
\end{enumerate}
On dit que le couple $(U,\varphi)$ est une paramétrisation de $P$.

\noindent Si $U$ est un ouvert muni d'une orientation $\mathcal{O}(U)$, on dit que $P$ est un domaine paramétré orienté. Il est entièrement  défini par le triplet $(U,\mathcal{O}(U),\varphi)$.
\end{definition}
\begin{exemples}
$i)$ La sphère unité $S^2$ incluse dans $\rr^3$ est un domaine paramétré de dimension $2$. En effet,

 \noindent$S^2=\varphi(U)$ avec 
$U=[0,\pi]\times [0,2\pi[ $ et l'application $\varphi$ définie sur $\rr^2$ par 
$\varphi(\psi,\theta)=(\sin\psi\cos\theta,\sin\psi\sin\theta,\cos\psi)$. 
L'application $\varphi$ est de rang deux sur $U$ sauf aux points de $\{0\}\times [0,2\pi[\cup\{\pi\}\times [0,2\pi[.$

\noindent $ii)$ Toute courbe paramétrée différentiable $\varphi $ définie sur un intervalle fermé $I$ de $\rr$ à valeurs dans une variété différentiable $M$ définit l'ensemble paramétré $\varphi(I)$ de dimension $1$ si l'ensemble des points où la vitesse est nulle est de mesure nulle sur $I$.
\end{exemples}
\begin{exo}\label{paravari}
Un domaine paramétré de dimension $p$ a une structure de variété différentiable de dimension $p$.
\end{exo}
\begin{remark}\label{dopaetvaror}
Considérons le cas d'une variété orientée. Elle possède un atlas $\{(U_i,\varphi_i)\}$ pour lequel existe sur chaque domaine de carte $U_i$ munies des coordonnées locales $({x_i}^1,\ldots,{x_i}^n)$ une orientation $\mathcal{O}(U_i)$ déterminée par la base de champs de vecteurs $(\frac{\partial}{\partial {x_i}^1},\ldots,\frac{\partial}{\partial {x_i}^n} )$ sur $TU_i$ . Cette orientation définit l'orientation de l'ouvert
$\varphi(U_i)$ de $\rr^n$, par les coordonnées $({x_i}^1,\ldots,{x_i}^n)$. Ainsi,

$i)$ chaque carte $(U_i,\varphi_i)$ correspond au domaine paramétré $U_i$ défini par le triplet $(\varphi_i(U_i),\mathcal{O}\left(\varphi_i^{-1}(U_i)\right),\varphi_i^{})$ (voir définition \ref{parana}),

$ii)$ Les orientations coïncident sur les intersections non vides $U_i\cap U_j$
\end{remark}
\noindent On peut maintenant définir l'intégrale d'une $p$-forme différentielle sur un domaine paramétré orienté de dimension $p$.

\begin{definition}\label{intfodipara}\index{intégrale sur un domaine paramétré}
\hspace{1cm}\\
On reprend les notations de la définition \ref{parana}. Soit  $(U,\mathcal{O}(U),\varphi)$ un triplet définissant un domaine paramétré orienté $P$ de dimension $p$ d'une variété différentiable $M$.
 Soit $\al$ une $p$-forme différentielle sur $P$. Notons $(u^1,u^2,\ldots,u^p)$ des coordonnées sur $U\subset \rr^p$.
Soit $\beta=b\;du^{1}\wedge\ldots\wedge du^{p}$ une $p$- forme différentielle sur $\rr^p$.
Posons \[\varepsilon(U)=\begin{cases}1\;\mathrm{si} \;(\frac{\partial}{\partial u^1},\frac{\partial}{\partial u^2},\ldots,\frac{\partial}{\partial u^p})\in \mathcal O(U)\\
-1\;\mathrm{sinon}
\end{cases}
\]
Alors \begin{enumerate}
\item L'intégrale $\int_U\beta$ se définit par l'égalité :\[\int_U\beta=\varepsilon(U)\int_U b\;du^{1}\ldots du^{p},\]
l'intégrale de droite étant l'intégrale ordinaire de la fonction $b$ sur $U$.
\item L'intégrale $\int_{(U,\mathcal{O}(U),\varphi)}\al$ se définit par l'égalité 

\begin{equation}\label{bellaint}\int_{(U,\mathcal{O}(U),\varphi)}\al=:\int_U\varphi^*\al,\end{equation}
l'intégrale de droite étant définie dans l'item précédent.
L'intégrale $\int_{(U,\mathcal{O}(U),\varphi)}\al$ se note également $\int_{\varphi(U)}\al$.
\end{enumerate}
\end{definition}
L'intérêt immédiat de la définition  (\ref{bellaint}) est l'indépendance de l'intégrale de la paramétrisation considérée, au sens de la proposition suivante.
\begin{proposition}\label{indesgalantes}
Supposons que $P=\varphi(U)=\psi(V)$ où $(U,\varphi)$ et $(V,\psi)$ soient deux paramétrisations de  $P$ pour lesquelles
on note
$u=(u^1,\ldots,u^p)$ et $v=(v^1,\ldots,v^p)$  les coordonnées respectivement sur $U$ et $V$. On pose $\varepsilon(U,V)=1$ si le  changement de paramètres $\varphi^{-1}\psi$ respecte les orientations de $U$ et $V$ et $-1$ dans le cas contraire. Alors 
\begin{equation}\label{intdompa}\int_{(U,\mathcal{O}(U),\varphi)}\al=\varepsilon(U,V)\;\int_{(V,\mathcal{O}(V),\psi)}\al.\end{equation}
\end{proposition}

\begin{preuve}{de la proposition \ref{indesgalantes}}
 Supposons que le changement de variables $\varphi^{-1}\psi$ respecte les orientations de $U$ et de $V$ et  posons $\theta=\varphi^{-1}\psi : v\mapsto u=\theta(v)=(\theta^1(v),\ldots,\theta^p(v))$.  Cette hypothèse entraine l'égalité $\varepsilon (U)=\varepsilon(V)$.  Notons $\varphi^*\al
 =a \;du^1\wedge \ldots\wedge du^p$. Alors, si $J(\theta)$ désigne la matrice jacobienne de $\theta$, on a : $$\psi^*\al=\theta^*(\varphi^*\al)=a\circ\theta \;\det(J(\theta)) dv^1\wedge\ldots\wedge dv^p.$$ On sait également que 
 \[\int_Va\;\circ\theta \det(J(\theta))dv^1\ldots dv^p=\int_Ua\; du^1\ldots du^p,\]
 qui s'écrit, compte tenu de l'égalité $\varepsilon(U)=\varepsilon(V)$ : $\int_U\varphi^*\al=\int_V\psi^*\al$ ou encore 
 \[\int_{(U,\mathcal{U},\varphi)}\al=\int_{(V,\mathcal{V},\psi)}\al.\]
 Si maintenant $\theta$ inverse les orientations de $U$ et $V$, on obtient par cette même démonstration
 \[\int_{(U,\mathcal{U},\varphi)}\al=-\int_{(V,\mathcal{V},\psi)}\al,\] ce qui achève la preuve.
 \end{preuve}
 \begin{remark}
L'intégrale (\ref{bellaint}) dépend de $U$, de $\varphi$ et de $\mathcal{O}(U)$. Si $P=\varphi(U)=\psi(V)$, on aurait
envie d'écrire l'égalité (\ref{bellaint}) :  $\int_{(U,\mathcal{O}(U),\varphi)}\al=\int_{\varphi(U)}\al=\int_{P}\al$. 
On adoptera cette écriture :

$i)$ si on considère l'intégrale au signe près, car alors pour une autre paramétrisation $(V,\psi)$, on aurait 
$\int_{\varphi(U)}\al=\varepsilon(U,V)\int_{\psi(V)}\al.$

$ii)$ si l'orientation $\mathcal{O}(U)$ est fixée

$ii)$ et pour finir dans la situation suivante. Supposons que $I$ soit un réel qui dépend de $(U,\mathcal{O}(U),\varphi)$ et que $I=\int_{(U,\mathcal{O}(U),\varphi)}\al$ quelque soit l'orientation $\mathcal {O}(U)$. On écrira encore dans ce cas la dernière égalité $I=\int_{\varphi(U)}\al$.
\end{remark}
 Notons encore un résultat qui ne résulte que des propriétés de l'image réciproque d'une forme par la composée de deux applications différentiables.
 \begin{proposition}\label{soft}
 Soit $U$ un borélien de $\rr^p$, $M^m, N^n$ deux variétés différentiables de dimensions respectives  $m$ et $n$ supérieures ou égales à $ p$,  $\varphi : U\longrightarrow M^m$ une paramétrisation de $\varphi(U)$, $F: M^m\longrightarrow N^n$ une application différentiable telle que le rang de $F\circ\varphi$ soit égal à $p$ presque partout et enfin $\al$ une $p$-forme différentielle sur $N^n$. Alors :
 \begin{enumerate}
 \item $F\circ \varphi$ est une paramétrisation de $F\circ \varphi(U)$.
 \item $\int_{F\circ\varphi(U)}\al=\int_{\varphi (U)}F^*\al.$
 \end{enumerate}
 \end{proposition}
 \begin{exo}\label{softy}
 Démontrer la proposition\ref{soft}.
 \end{exo}
 \subsubsection{Intégrale sur une variété orientée de dimension $n$ d'une $n$-forme différentielle à support compact. Théorème de Stockes}
 Les variétés concernées par la définition de l'intégrale sont les variétés différentielles à bord dans la catégorie desquelles on inclut les variétés différentielles (à bord vide). On commence par étendre la définition de variété orientée. 
 \begin{definition}\index{variété à bord orientée}
 On dit qu'une variété différentielle à bord $M$ est orientée si elle possède un atlas tel que les orientations locales sur les domaines de carte coïncident sur leur intersection.
 \end{definition}
 \begin{proposition}\label{oribo}
 Soit $M$ est une variété de dimension $n$ de bord non vide, orientée. Alors $\partial M$ est une variété de dimension $n-1$, orientable.
 \end{proposition}
 \begin{preuve}{}
 Considérons un atlas orienté $\{(U_i,\varphi_i)\}_i$ de $M$ : sur $U_i$, on dispose de coordonnées locales $({x_i}^1,\ldots,{x_i}^n)$ de sorte que sur $U_i$ on ait la base directe de champs de vecteurs $(\frac{\partial}{\partial {x_i}^1},\ldots,\frac{\partial}{\partial {x_i}^n})$. Si $U_i$ est une carte contenant un point $x$ de $\partial M$, alors $U_i\cap M$ détermine sur $\partial M$
 une carte locale dotée des coordonnées locales $(x_i^1,\ldots,x_i^{n-1})$ (voir la définition \ref{bordeau}). On considère sur cet ouvert un champ $T_i $ de vecteurs tangeant à $M$ et transversal à $\partial M$ : en chaque point $y$ de $U_i\cap\partial M$ il est le vecteur vitesse d'un  chemin $\gamma : [0,\epsilon[\longrightarrow U_i$ différentiable sur $U_i$ de sorte que $T_i(y)=\gamma'(0^+)$. Alors si $\sigma$ est une permutation de $(1,\ldots,n-1)$ la base $\left(\frac{\partial}{\partial {x_i}^{\sigma(1)}},\ldots,\frac{\partial}{\partial {x_i}^{\sigma(n-1)}}\right)$ est directe si $ (T_i,(\frac{\partial}{\partial {x_i}^{\sigma(1)}},\ldots,\frac{\partial}{\partial {x_i}^{\sigma(n-1)}})$ est directe dans $U_i$. Il est facile de voir alors que sur les intersections $\left(\partial M\cap U_i\right)\cap\left(\partial M\cap U_j\right)$ les orientations ainsi définies coïncident, ce qui définit une orientation de la variété $\partial M$ héritée de l'orientation de $M$.  \end{preuve}
 Le théorème suivant donne la clef du passage de l'intégrale sur un domaine paramétré à l'intégrale sur une variété.

 \begin{theorem}\label{exuninte}
  Soit $M$ une variété orientée de dimension $n$. On note $\Lambda_c^n(M)$ l'espace vectoriel réel des $n$-formes différentielles à support compact sur $M$.
  Il existe sur cet espace une unique forme linéaire $\mathcal{I}$ telle que pour tout atlas orienté $\{(U_i,\varphi_i)\}_{i\in \nn}$ sur $M$, et toute $n$-forme différentielle $\al$ à support compact inclus dans $U_i$, on ait :
  \[\mathcal{I}(\al)=\int_{\varphi_i(U_i)}\al.\]
 \end{theorem}
 \begin{preuve}{}
 Soit $\mathcal{A}=\{(U_i,\varphi_i)\}_{i\in I}$ un atlas orienté sur $M$. Le support de $\al$ ne rencontre qu'un nombre fini d'ouverts de  $\mathcal{A}$ disons ceux de la famille $\{U_i\}_{i\in J} $ où $J\subset I$ est fini. Considérons une partition de l'unité 
 $\{\varepsilon_i\}_{i\in\nn}$ subordonnée au recouvrement ouvert $\{U_i\}_{i\in\nn}$. Rappelons que cela signifie que pour tout $i\in\nn$,
 on a :\index{partition de l'unité}
 
 $i)$ $\supp \varepsilon_i \subset U_i$, $ii)$ chaque fonction $\varepsilon_i$ est à valeur dans $[0,1]$, son support ne rencontre qu'un nombre fini d'ouverts de $\mathcal{A}$ et la somme (finie en chaque point de $M$), $\sum_{i\in\nn}\epsilon_i$ est égale à $1$. On pourra consulter [\ref{BG}] ou [\ref{god}] pour la preuve de l'existence d'une partition de l'unité.
 On a donc $\al =\sum_I\varepsilon_i\al=\sum_J\varepsilon_i\al$ et par suite nécessairement,
 \begin{equation}\label{labeldef}
 \mathcal{I}(\al)=\sum_I\int_{U_i}\varepsilon_i\al,
 \end{equation} 
 puisque $\supp \varepsilon_i\al\subset U_i$. Cette dernière égalité  prouve l'existence de $\mathcal{I}$  si on montre qu'elle est indépendante de l'atlas orienté choisi et de la partition de l'unité qui lui est subordonnée.
 
 Soit donc un autre atlas orienté $\mathcal{A'}=\{(U'_j,\psi_j)\}_{j\in \nn}$ et $\{\eta_j\}_{j\in\nn}$ une partition de l'unité subordonnée au recouvrement $\{U'_j\}_{j\in \nn}$. Pour tout $i\in \nn$ on peut écrire : $\varepsilon_i\al=\sum_j\eta_j\varepsilon_i\al$ avec $\supp (\eta_j\varepsilon_i\al)\subset U_i\cap U'_j\subset U'_j$. 
 D'où : 
 
 \noindent $\mathcal{I}(\al)=\sum_i\int_{U_i}\varepsilon_i\al=\sum_i\int_{U_i}\sum_j \varepsilon_i\eta_j\al=
 \sum_i\sum_j\int_{U'_j}\varepsilon_i\eta_j\al=\sum_j\sum_i\int_{U'_j}\varepsilon_i\eta_j\al=$
 
 \noindent$=\sum_j\int_{U'_j}\sum_i\varepsilon_i\eta_j\al=\sum_j\int_{U'_j}\eta_j\al$.
 Ainsi l'égalité (\ref{labeldef}) définit une application de ${\Lambda_c}^n(M)$ à valeurs réelles, dont on vérifie immédiatement la linéarité. Cela montre à la fois l'existence et l'unicité de la forme linéaire annoncéee.
   \end{preuve}
   On notera naturellement désormais $$\mathcal{I}(\al)=:\int_M\al.$$
   $\mathcal{I}(\al)$ est \emph{l'intégrale de la forme volume $\al$ à support compact sur la variété orientée $M$}.\index{intégrale d'un forme volume à support compact}
   \begin{proposition}\label{pulbakint}
   Soit $M,N$ deux variétés différentiables difféomorphes de dimension $n$. Si $F$ est un difféomorphisme de $M$ sur $N$ et $\al$ une $n$-forme sur $N$, alors $\int_M F^*\al=\int_N\al.$ 
   \end{proposition}
   \begin{preuve}{}
   Soit $\{U_i,\varphi_i)\}_I$ un atlas de $M$. Alors $(F(U_i), \varphi_i\circ F^{-1})$ est un atlas de $N$. Si $\{\varepsilon_i\}$ est une partition de l'unité subordonnée au recouvrement $\{F(U_i)\}$ alors la famille $\{\eta_i=\varepsilon_i\circ F\}_I$ est une partition de l'unité subordonnée au recouvrement $\{U_i\}_I$. D'après le deuxième item de la proposition \ref{soft}, on a : $\int_{F(U_i)}\varepsilon_i\al=\int_{U_i}F^*(\varepsilon_i\al)$. Ainsi :
   
  \noindent $\int_N\al=\sum_I\int_{F(U_i)}\varepsilon_i\al=\sum_I\int_{U_i}F^*(\varepsilon_i\al)=\sum_I\int_{U_i}\eta_iF^*\al=\int_MF^*\al.$
   \end{preuve}
   L'égalité (\ref{ptistockes}) laisse espérer qu'il y a un lien entre l'intégrale d'une $n$-forme exacte $d\al $ sur une variété à bord $M$ de dimension $n$ et l'intégrale de $\al$ sur son bord $\partial M$. Le théorème précisant ce lien est connu sous le nom de théorème de Stockes. En voici l'énoncé. \index{Stockes (théorème de)}
   \begin{theorem}[Théorème de Stockes]\label{bonstoc}
   \hspace{1cm}\\
   \noindent Soit $M$ une variété différentiable  orientée à bord de dimension $n$, $\partial M$ étant munie de l'orientation induite décrite par la proposition \ref{oribo}. Le demi-espace  $H^n$  de $\rr^n$ (voir section  \ref{varabobo}) est muni de la mesure de Lebesgue ainsi que $\partial H^n=\rr^{n-1}$ si $n>1$. Si $n=1$, on considèrera sur $\partial H^n$ la mesure de Dirac.
   On considère la $(n-1)$-forme différentielle $\al $ sur $M$. Soit  $i : \partial M\to M$   l'injection canonique. La $(n-1)$-forme  $i^*\al$ est donc une forme volume sur la variété différentiable (sans bord) $\partial M$ et on a l'égalité  suivante :
   \[\int_M d\al=\int_{\partial M}i^*\al.\]
    \end{theorem}
  \begin{preuve}{}
  On considère sur $M$ un atlas orienté $\{U_k,\varphi_k\}$, on pose $\psi_k=:\varphi_k^{-1}$ qui fait de $U_k$ un domaine paramétré. Soit également $\{\varepsilon_k\}$ une partition de l'unité subordonnée au recouvrement $\{U_k\}$.
  On a :
  $\int_Md\al=\int_M d\sum_k\varepsilon_k\al=\sum_k\int_{U_k}d(\varepsilon_k\al)=\sum_k\int_{V_k} \psi_k^*d(\varepsilon_k \al)$ si on pose $V_k=\varphi_k(U_k)$.
  Le theorème est la conséquence des deux lemmes suivants.
  \begin{lemma}\label{oulele}
  On note $(x^1,\ldots,x^n)$ les coordonnées sur $\rr^n$.
  
  \noindent Soit $\beta=a_{1,\ldots,\widehat{p},\ldots,n}\;dx^{1}\wedge\ldots\wedge dx^{{p-1}}\wedge\widehat{dx^{p}}\wedge dx^{{p+1}}\wedge\ldots\wedge dx^{{n}}$ une $(n-1)$-forme sur $H^n$ ( $\widehat{a}$ signifie que $a$  est omis dans l'expression). On suppose que le support de $\beta $ est un compact inclus dans un ouvert $V$ de $H^n$. Rappelons que $\partial H^n=:\rr^{n-1}=:\{(x^1\ldots,x^{n-1},0)\;; x^{i}\in \rr, i=1,\ldots,n-1\}$.
  Alors \begin{enumerate}
  \item Si $V\cap \rr^{n-1}=\emptyset$, on a $\int_Vd\beta=0$
  \item Si $V\cap \rr^{n-1}\not=\emptyset$, on a $\int_Vd\beta=\int _{V\cap\rr^{n-1}}{j^*}\beta$ où $j: V\cap\rr^{n-1}\to H^n$ est l'injection canonique.
  \end{enumerate}
  \end{lemma}
  \begin{preuve}{du lemme \ref{oulele}}
  On a $d\beta=\frac{\partial a_{1,\ldots,\widehat{p},\ldots,n}}{\partial x^{p}} (-1)^{(p-1)}dx^{1}\wedge\ldots\wedge dx^{{p-1}}\wedge{dx^{p}}\wedge dx^{{p+1}}\wedge\ldots\wedge dx^{{n}}$.
  Ainsi, en utilisant le théorème de Fubini et la formule fondamentale de l'intégration on a immédiatement le premier item.
  Pour le deuxième item, notons que 
  
  $\int_V d\beta=\sum_{p=1}^n(-1)^{p-1}\int_V\sum_{p=1}^n\int_V\frac{\partial a_{1,\ldots,\widehat{p},\ldots,n}}{\partial x^{p}}  dx^1\ldots dx^n=$
  
 $ =\sum_{p=1}^n(-1)^{p-1}\int_{V\cap\{x^p=0\}}\left(\int_{D_{p}\cap V}\frac{\partial a_{1,\ldots,\widehat{p},\ldots,n}}{\partial x^{p}} dx^p\right)dx^1\ldots\widehat{dx^p}\ldots dx^n$ où $D_p$ est une droite sur laquelle $x^1,\ldots \widehat{x^p},\ldots x^n$ sont fixés (application du théorème de Fubini)
 
\noindent $=(-1)^{n-1}\int_{V\cap \rr^{n-1}}-a_{1,\ldots,n-1,\widehat{n}}(x^1,\ldots,x^{n-1},0)dx^1\ldots dx^{n-1}=$

$=(-1)^{n}\int_{V\cap \rr^{n-1}}a_{1,\ldots,n-1,\widehat{n}}(x^1,\ldots,x^{n-1},0)dx^1\ldots dx^{n-1}$.

\noindent Par ailleurs $\int_V j^*\beta=(-1)^n\int_V a_{1,\ldots,n-1,\widehat{n}}(x^1,\ldots,x^{n-1},0)dx^1\ldots dx^{n-1}$ le terme $(-1)^n$ provenant de la convention sur l'orientation de $\partial H^n$ telle qu'elle est décrite dans la démonstration de la proposition \ref{oribo}.
  \end{preuve} 
 \begin{lemma}\label{oulala}
 On suppose que $V_k\cap \partial H^n\not=\emptyset$. On note $\theta_k$ la restriction de $\psi_k$ à $V_k\cap\partial H^n$,
 $j$ l'injection canonique de $\partial H^n$ dans $\rr^n$, $i$ l'injection canonique de $\partial M$ dans $M$. Alors on a le diagramme commutatif

\[
\begin{array}{ccc}
V_k\cap\partial H^n&\stackrel{j}{\longrightarrow}&V^k\\
\vcenter{\llap{$\theta_k$}}\Big\downarrow&&\Big\downarrow\vcenter{\rlap{$\psi_k$}}\\
U_k\cap\partial M&\xrightarrow[\enspace i\enspace]&U_k
\end{array}
\] 
\end{lemma}
\noindent Ce lemme est une remarque qui ne nécessite pas de démonstration.

\noindent   D'après le lemme \ref{oulele}, on a : $\int_{V_i}{\psi_i}^*({\varepsilon_i}d\al)=0 $ si $V_i\cap\rr^{n-1}=\emptyset$. 

\noindent  Si 
 $V_i\cap\rr^{n-1}\not=\emptyset$, alors $ \int_{V_k}{\psi_k}^*d(\varepsilon_k\al) =\int_{V_k}d{\psi_k}^*\varepsilon_k\al)=
 \int_{V_k\cap\partial H^n}j^*{\psi_k}^*(\varepsilon_k\al)=$
 
 \noindent $=\int_{V_k\cap H^n}{\theta_k}^*i^*(\varepsilon_k\al)$ \hspace{1cm} (d'après le lemme \ref{oulala})
 
 \noindent $=\int_{U_k\cap \partial M}i^*(\varepsilon_k\al)$,
 
 \noindent de sorte que $\int_Md\al=\sum_k\int_{U_k\cap\partial M}i^*(\varepsilon_k\al)=\int_{U_k\cap\partial M}\sum_k(\varepsilon_ki^*\al)=..$
 
 \noindent$..=\int_{\partial M}i^*\al$, conformément au théorème \ref{exuninte} appliqué à la variété 
 $\partial M$.
 \end{preuve}

  \subsubsection{Dérivée de l'intégrale  intégrale d'une forme différentielle}
  \textbf{Quelques remarques préliminaires sur les équations différentielles}
  
  \noindent Rappelons qu'un système d'équations différentielles se définit géométriquement par la donnée d'un champ de vecteurs continu sur une variété différentiable $M$ de dimension $n$. Si $X$ est un tel champ, toute courbe dérivable $t\in I\to c(t)$ où $I$ est un intervalle ouvert de $\rr$ telle que $\frac{d c}{dt}=X(c(t))$ en est \emph{une courbe intégrale}. \index{courbe intégrale}Rappelons que le système d'équations différentielles associée à $X$ s'écrit sur une carte locale $U$ munie des coordonnées $(x^1,\ldots,x^n)$ si on note $X=X^1(x^1,\ldots,x^n)\frac{\partial}{\partial x^1} +\ldots +X^n(x^1,\ldots,x^n)\frac{\partial}{\partial x^n}$ :
\[\left\{
  \begin{array}{ccc}
 \frac{d x^1}{dt}&=&X^1(x^1,\ldots,x^n)\\
  &\vdots&\\
    \frac{d x^n}{dt}&=&X^n(x^1,\ldots,x^n)
 \end{array}
 \right.
 \]
  La proposition \ref{grr}, nous indique que $X$ définit un groupe local à un paramètre de difféomorphismes $\{\varphi_t\}$ tel que pour tout $x_0\in M$ il existe un voisinage ouvert $U$ de $x_0$, et un réel $\varepsilon>0$ pour lesquels
  $t\to \varphi_t(x)$ est la courbe intégrale de $X$ définie sur un intervalle $]-\varepsilon, \varepsilon[$  vérifiant $\varphi_0(x)=x$ pour tout $x$ de  $U$.
 Le système précédent est appelé \emph{un système autonome}.\index{système différentiel autonome} Il correspond à un champ de vecteurs $X$ qui n'est pas dépendant de la variable $t$. 
  Lorsque le champ de vecteurs $X$ toujours défini sur $M$ dépend de $t$, alors le système qu'il définit s'écrit :
  \[\frac{d x^{i}}{dt}=X^{i}(t,x^1,\ldots,x^n),\;i=1,2,\ldots,n.\]
  Ce système est dit \emph{non autonome}.\index{système  différentiel non autonome}
  Dans ce cas, bien qu'on ait également un théorème d'existence et d'unicité locale : pour un point $(x_0,t_0)$ d'un ouvert
 d'une carte locale, sur lequel $X$ est défini il existe un réel $\varepsilon>0$ et une solution $x(t)$ définie sur $]t_0-\varepsilon,t_0+\varepsilon[$ telle que $x(t_0)=x_0$,  deux telles solutions coïncidant sur leur intersection (Voir [\ref{poipoi}]).
 Si on note encore $\varphi_t(x_0)$ la solution qui vaut $x_0$ à pour $t=0$,  la famille $\{\varphi_t\} $ de difféomorphismes locaux ne définit plus un groupe à un paramètres de difféomorphismes.
Pour illustrer ce qui précède donnons un exemple  et un contre-exemple.
\begin{exemple}
La variété $M$ est ici $\rr$. Soit $n$ un entier $>1$ et $x_0$ un réel $>0$. Le champ $X= x^n\frac{\partial}{\partial x}$ définit l'équation différentielle 
$x'=x^n$. On vérifie que $$\gamma :t\mapsto \left(\frac{1}{\frac{1}{x_0^{n-1}}-(n-1)t}\right)^{1/n-1}$$ est la courbe intégrale de $X$ définie sur l'intervalle ouvert (maximal) $I=\left]-\infty, \frac{x_0^{1-n}}{n-1}\right[$ vérifiant $\gamma(0)=x_0$ . On a bien $\gamma(t)
=\varphi_t(x_0)$ puisque $\gamma(0)=x_0$.

\noindent Soient $t$ et $t'$ dans $I$ de sorte que $t+t'\in I$. On a : 

$\varphi_t\left(\varphi_{t'}(x_0)\right)=\left(\frac{1}{\frac{1}{\varphi_{t'}(x_0)^{n-1}}-(n-1)t}\right)^{1/n-1}=
\left(\frac{1}{\frac{1}{x_0 ^{n-1}}-(n-1)t'-(n-1)t}  \right)^{1/n-1}=\varphi_{t+t'}(x_0).$
\end{exemple}

\begin{cexemple}
La variété est encore $\rr$. On considère le champ de vecteurs sur $\rr$ défini par $X=xf(t)\frac{\partial}{\partial x}$ où $f$ est une application continue  de $\rr$ vers $\rr$. On note $F$ la primitive de $f$ qui s'annule pour $t=0$.
Alors $\varphi_t(x_0)=x_0\exp(F(t))$ est la courbe intégrale de $X$ telle que $\varphi_0(x_0)=x_0$.

\noindent Ainsi, $\varphi_{t+t'}(x_0)=x_0\exp(F(t+t'))$ alors que $\varphi_t(\varphi_{t'}(x_0))=x_0\exp(F(t)+F(t')).$
En général $\varphi_{t+t'}(x_0)\not=\varphi_t(\varphi_{t'}(x_0))$.
\end{cexemple}
\noindent Dans un milieux continu on a classiquement des particules qui se meuvent, leur trajectoires définissant des lignes de courant. 
Pour décrire ces lignes,  Euler a proposé  de considérer en chaque point géométrique $(x,y,z)$ du milieu le vecteur vitesse $V(x,y,z,t)$ à l'instant $t$ où la particule du milieu coïncide avec $(x,y,z)$. On obtient, avec ce type de description  un champ de vecteurs sur le milieu qui dépend également du temps et ainsi les lignes de courant sont localement solution d'un système d'équations différentielles, non autonome.

Il y a une façon canonique de définir un champ de vecteurs \og autonome\fg\; à partir d'un champ \og non autonome\fg.
Décrivons la.

\noindent Soit $M$ une variété différentiable de dimension $n$ sur laquelle $V$ est un champ de vecteurs qui dépend en chaque point
de $M$ d'une variable réelle $t$. Précisons cette dépendance : sur tout domaine $U$ d'une carte locale sur laquelle on a des coordonnées $(x^1,\ldots,x^n)$, le champ $V$ s'écrit en coordonnéees locales $V=V^{i}(t,x^1,\ldots,x^n)\frac{\partial}{\partial x^{i}}$ de sorte que les applications de $\rr^{n+1}$ vers $\rr$ définies par : $V^{i} :(t,x^1,\ldots,x^n)\mapsto V^{i}(t,x^1,\ldots,x^n)$ soient continues. On considère alors la variété  $\rr\times M$ de dimension $n+1$ et sur cette variété le champ de vecteurs 
\begin{equation}\label{chauchaut}X=\frac{\partial}{\partial t}+ V\end{equation} 
C'est un champ de vecteurs autonome sur $\rr\times M$, conduisant localement au système différentiel (autonome) suivant :
\[
\left\{\begin{array}{rcl}
\frac{dt}{ds}&=&1\\
&&\\
\frac{dx}{ds}&=&V(t,x)
\end{array}
\right.
\]
\begin{proposition}\label{passautnonaut}
Soit $\Phi_s(t_0,x_0)$ le groupe local à un paramètre de difféomorphismes  de \;$\rr\times M$ engendré par $X$, et $s\to \varphi_s(x_0)$ la courbe intégrale de $V$ telle que $\varphi_0(x_0)=x_0$. Notons $p_2$ la deuxième projection de \;$\rr\times  M$ sur $M$. On a $p_2\Phi_s(t_0,x_0)=\varphi_s(x_0).$
\end{proposition}
\begin{preuve}{}
L'application $\psi$  définie sur $\rr$ par $\psi(s)=(s+t_0,\varphi_s(x_0)$ est la solution du système précédent, telle que $\psi(0)=(t_0,x_0)$. Par unicité, on a donc $\Phi_s(t_0,x_0)=\psi(s)$ et on conclut. 
\end{preuve}
\noindent\textbf{Dérivation de l'intégrale}\index{dérivation d'une intégrale}
\begin{theorem}\label{derint1}
Soit $M$ une variété différentiable, $X$ un champ de vecteurs sur $M$ et $C_k(0)$ une sous variété orientée  de $M$, de dimension $k$ et $\al^k$ une $k$-forme différentielle sur $M$. Notons $\{\varphi_t\}_{t\in ]-\varepsilon,\varepsilon[}$ le groupe local à un paramètre de difféomorphismes  de $M$ engendré par le champ de vecteurs $X$.
On définit la variété orientée $C_k(t)$ par :  $C_k(t)=\varphi_t(C_k(0))$. Alors l'application définie sur $]-\varepsilon,\varepsilon[$ par $t\mapsto \int_{C_k(t)}\al^k$ est différentiable et
\begin{equation}\label{pufi}
\frac{d}{dt}\int_{C_k(t)}\al^k=\int_{C_k(t)}L_X\al^k.
\end{equation}
\end{theorem}
\begin{corollary}\label{derint2}
Soit $M$ une variété différentiable et $C_k(0)$ une sous variété orientée  de $M$, de dimension $k$ et $\al^k$ une $k$-forme différentielle sur $M$.  Soit $V(t,x)$ un champ de vecteurs non autonome sur $M$ et $t\mapsto\varphi_t(x_0)$ la ligne de courant associée à $V$ c'est à dire : $\frac{d \varphi_t(x_0)}{dt}=V\left(t,\varphi_t(X_0)\right), t\in ]-\varepsilon,\varepsilon[$ et $ \varphi_0(x_0)=x_0$. Notons encore $C_k(t)=\varphi_t\left(C_k(0)\right)$. 
Alors l'application définie sur $]-\varepsilon,\varepsilon[$ par $t\mapsto \int_{C_k(t)}\al^k$ est différentiable et
\begin{equation}\label{pufi}
\frac{d}{dt}\int_{C_k(t)}\al^k=\int_{C_k(t)}\left(\frac{\partial \al^k}{\partial t}+L_V\al^k\right).
\end{equation}
\end{corollary}
\begin{preuve}{du corollaire \ref{derint2}}
On associe à $V$ le champ $X$ autonome défini par l'égalité (\ref{chauchaut}) et on applique le théorème \ref{derint1}.
\end{preuve}
\begin{preuve} {du théorème \ref{derint1}}
Posons $F(t)=\int_{C_k(t)}\al^k.$ On a donc $F(t)=\int_{C_k(0)}\varphi^*_t\al^k.$ D'où :

\noindent$\frac{F(t+h)-F(t)}{h}=\int_{C_k(0)}\varphi_t^*\beta(h)$, avec $\beta(h)=\frac{\varphi_h^*\al^k-\al^k}{h}$. Comme
$\lim_{h\to 0}\beta(h)=L_X\al^k$, on a : $\frac{d}{dt}F(t)=\lim_{h\to 0}\int_{C_k(0)}\varphi_t^*\beta(h)=\int_{C_k(0)}\varphi_t^*L_X\al^k=\int_{C_k(t)}L_X\al^k.$
\end{preuve}
 \subsubsection{Quelques outils pour une mécanique des milieux continus}
 Citons le principe de conservation de la masse en mécanique newtonienne : \emph{si on suit un domaine contenant des particules  de fluide au cours de leur déplacement, la masse de ce domaine est une constante du temps}. 
 
 \noindent Formalisons cet énoncé.
 \begin{theorem}{de conservation de la masse}\label{tcs}\index{conservation de la masse}
 
 \noindent Soit $C_n(t)$ une sous-variété différentiable de dimension $3$ compacte à bord de $\rr^3$ issue du déplacement de $C_n(0)$ sous l'action du flot d'un champ de vecteurs non autonome $V(t,x)$. Soit $\rho$ une fonction contin\^{u}ment différentiable et $\omega=dx\wedge dy\wedge dz$ la forme volume standard de $\rr^3$. Alors il y a équivalence entre le fait que
  $\int_{C_{n}(t)}\rho\omega$ ne dépend pas de $t$ et la relation  suivante :
\begin{equation}\label{comas}
\frac{\partial \rho}{\partial t}+\Div (\rho V)=0\end{equation}
 \end{theorem}
 \begin{preuve}{}
 Notons $m(t)=\int_{C_n(t)}(\rho\omega)$. Si $\frac{d m(t)}{dt}=0$, on a d'après le corollaire \ref{derint2} :
 $$\int_{C_n(t)}\frac{\partial\rho\omega}{\partial t}+L_V(\rho\omega)=0.$$
 Or $L_V(\rho\omega)=i(V)\left(d(\rho\omega)\right)+d(i(V)\rho\omega)=d(i(V)\rho\omega)=d(i(\rho V)\omega=\Div(\rho V)\omega$ par définition de la divergence. On en déduit que pour tout volume $C_n(t)$
 \[\int_{C_n(t)}\left(\frac{\partial \rho}{\partial t}+\Div(\rho V)\right)\omega=0,\]
 ce qui implique la relation (\ref{comas}). Pour la réciproque on remonte la démonstration précédente.
 \end{preuve}
 \begin{exo}\label{consconma}
 On se place dans les hypothèses du théorème \ref{tcs} et on suppose que  $\int_{C_{n}(t)}\rho\omega$ ne dépend pas de $t$.
 Soit $f$ une fonction à valeurs réelles  définie sur $\rr\times \rr^3$. Sa restriction à\; $\rr\times C_n(t)$ définit  une fonction dérivable de la variable $t$, notée encore $f$. Montrer que 
 \[\frac{d}{dt}\int_{C_n(t)}f\rho\omega=\int_{c_n(t)}\frac{df}{dt}\rho\omega.\]
 \end{exo}
 Une conséquence directe du théorème de Stockes est le \emph{théorème de la divergence} dont voici l'énoncé.\index{divergence (théorème de la)}
 \begin{theorem}\label{thdladiv}
 On considère une sous-variété orientée compacte à bords, $C_n$, de dimension $n$ d'une variété pseudo-riemannienne $(M,\langle ,\ra)$ de dimension $n$. On note $j$ l'injection canonique de $\partial C_n(t)$ dans $C_n(t)$. Soit $X$ un champ de vecteurs sur
  $M$ dont on note $\varphi_t$ le flot et on pose $C_n(t)=\varphi_t\left(C_n\right)$. On note $\omega^n$ la forme volume riemannienne de $M$. On considère $N$ un champ de vecteurs unitaire de $M$, défini sur un voisinage ouvert de $\partial C_n(t) $  tel que $N(m)$ soit la normale à $T_{m}\partial C_n(t)$, dirigée vers l'extérieur à $T_m\partial C_n(t)$, en tout point $m$ de $  \partial C_n(t)$. On pose 
  $\omega^{n-1}=j^*\left(i(N)\omega\right)$. 
 Alors
  \begin{equation}\label{divinenfan}
 \int_{C_n(t)}\Div X\;\omega^n=\int_{\partial C_n(t)}\la X,N\ra \omega^{n-1}.
 \end{equation}
 \end{theorem}
 \noindent Précisons que $\Div X$ est la $\omega$-divergence de $X$, définie par l'égalité $L_X\omega^3=\Div X\;\omega^3$. Elle coïncide avec la divergence d'un tenseur contravariant définie dans l'exercice \ref{divdivergence}.
 \begin{preuve}{} En utilisant successivement les théorèmes \ref{derint1} et \ref{bonstoc}, on a :
 
 $$\frac{d}{dt}\int_{C_n(t)}\omega^n=\int_{C_n(t)}L_X\omega^n=\int_{C_n(t)}di(X)\omega^n=\int_{\partial C_(t)}j^*\left(i(X)\omega^n\right).$$
 Le champ $X$ peut s'écrire dans un voisinage ouvert de $\partial C_n(t)$ sous la forme  $X=X=\la X,N\ra N+ T$, où $T$ est défini dans ce voisinage par $T=X-\la X,N\ra N$. En tout point de  $\partial C_n(t)$, le champ $T$ est tangent à $\partial C_n(t)$.
  On a (voir exercice \ref{cucu} p. \pageref{cucu}) : $j^*\left(i(T)\omega^n\right)=0$. De plus $i(\la X,N\ra N)\omega=\la X,N\ra \omega^{n-1}$, ce qui achève la démonstration.
 \end{preuve}
\subsection{Dynamique newtonnienne d'un milieu continu} \label{dynemicon}
On se place dans $\rr^3$ muni de sa structure euclidienne standard. C'est l'espace par rapport auquel on décrit le mouvement du milieu continu qui peut être un liquide ou une poutre qui se déforme. Il est considéré comme un référentiel euclidien avec ses coordonnées canoniques $(x,y,z)$. On considère un volume $\ma{V}(t)$ de ce milieu, modélisé par une sous-variété à bord de dimension $3$ de $\rr^{3}$. Les particules de ce volume sont soumises à deux types de forces :
\begin{enumerate}
 \item Des forces liées à leur masse (forces de gravitation) ou à leur charge ( si on est par exemple en présence d'un champ électromagnétique). 
On les appellera \emph{forces massiques} et on admet qu'elle se définissent à partir  d'une densité (vectorielle) de force $f$.
\item Des forces de  pression du milieu extérieur à ce volume. Les forces de pression s'exercent sur le bord $\partial \ma{V}(t)$.
Expérimentalement, ces forces en chaque point dépendent de l'orientation dans l'espace du plan tangent en ce point, ou encore de la normale 
unitaire extérieure $N$. Notons \emph{$P(N)$ la densité des forces de pression}. On veut considérer $P$ comme une fonction (vectorielle)
définie sur fibré  normal à $\partial \ma{V}(t)$. Dans ce cas si $\lambda$ est  un réel positif,  $P(\lambda N)$ n'a pas de sens physique et on doit pouvoir l'exprimer en fonction de $P(N)$. On adopte le principe physique suivant : $P(\lambda N)=\lambda P(N)$ pour tout $\lambda >0$. Ce principe est physique car il va conduire à l'existence d'un tenseur (tenseur des contraintes), concept fondamental, en particulier de la théorie de l'élasticité qui décrit conformément à la réalité expérimentale les déformations élastiques des matériaux. \emph{Ce principe suppose que l'on a défini $P$ sur le fibré en  sphères unité trivial  $\rr^3\times S^{1}$ et qu'alors pour un vecteur unitaire $N$ en un point $m$, $P(N)$ est la pression  s'exerçant en $m$ sur un domaine dont le plan tangent en ce point est orthogonal à $N$. }

Par exemple si on est dans un fluide parfait, par définition de celui-ci, on a : $P(N)=-p N$.
\end{enumerate}
\subsubsection{Les lois fondamentales de la mécanique}
Les notations concernant les formes volumes qui apparaissent dans les sections suivantes sont celles du théorème \ref{thdladiv}.

\noindent La première loi fondamentale en mécanique classique des milieux continus est le principe de conservation de la masse. : 

\noindent\emph{ Lorsqu'on suit un volume de particules de matières au cours du temps, la masse de ce volume est constante.}

\noindent Ainsi on considèrera dans tout ce qui suit que l'équation (\ref{comas}) est vérifiée.

\noindent Rappelons maintenant les lois fondamentales de la dynamique appliquées à un volume de particules de matière d'un milieu continu.

\noindent La résultante des forces extérieures s'exerçant sur le milieux enfermé dans $\ma{V}(t)$ est d'après ce qui précède : 
\begin{equation}\label{ffi}F_e=\int_{\ma{V}(t)}f\omega^3+\int_{\partial\ma{V}(t)}P(N)\omega^2.\end{equation} Pour chaque point $M$ de $\ma{V}(t)$ notons $r$ le vecteur $\overset{\longrightarrow}{OM}$ et $V$ le champ des vitesses des particules de $\ma{V}(t)$.
Les deux équations fondamentales de la dynamique s'écrivent alors :
\begin{align*}
\frac{d}{dt}\int_{\ma{V}(t)}\rho V\omega^3&=F_e\\
\frac{d}{dt}\int_{\ma{V}(t)}r\times \rho V\omega^3&=\int_{\ma{V}(t)}r\times f\omega^3+\int_{\partial\ma{V}(t)}r\times P(N)\omega^2
\end{align*}
Les membres de gauche de ces deux équations sont respectivement les dérivées par rapport au temps de la quantité de mouvement des particules de $\ma{V}(t)$
et de leur moment cinétique par rapport au point fixe $O$.
Ces deux équations s'écrivent d'après l'exercice \ref{consconma} :
\begin{align}
\int_{\ma{V}(t)}\rho \frac{dV}{dt}\omega^3&=F_e\label{zer}\\
\int_{\ma{V}(t)}r\times \rho \frac{dV}{dt}\omega^3&=\int_{\ma{V}(t)}r\times f\omega^3+\int_{\partial\ma{V}(t)}r\times P(N)\omega^2\label{rez}
\end{align}
Remarquons que les égalités (\ref{zer}) et (\ref{rez}) sont les relations fondamentales de la dynamique dans lesquelles on a pris en compte le principe de conservation de la masse.
En examinant  les relations (\ref{zer}) et (\ref{rez}), nous allons démontrer le théorème fondamental suivant.
Notons préalablement $(e_1,e_2,e_3)$ la base canonique de $\rr^3$.
\begin{theorem}\label{extencon}
Il existe un champ de $(1,1)$-tenseurs $T$, symétrique, tel que pour tout point $m$ du milieu et tout vecteur unitaire $N=N^{i}e_i$ en ce point $(T^{i}_jN^j)e_i$ soit la force de pression dûe au milieu qui s'exerce en $M$ dans la direction $N$. La symétrie se traduisant par exemple par l'égalité 
$T^{ij}=T^{ji}$ si $T^{ij}$ est la version contravariante de $T$. Soit $V$ le champ des vitesses des particules coïncidant avec chaque point du milieu. 
Ce champ de tenseurs $T$ est lié au champ des accélérations correspondant par la relation :
\begin{equation}\label{eqmomo}
\rho\frac{dV}{dt}=f+\Div T.
\end{equation}
\end{theorem}
Le champ de  tenseurs $T$ est appelé en mécanique des milieux continus le \emph{tenseur des contraintes}.\index{tenseur des contraintes}
\begin{preuve}{}
La  démonstration du théorème \ref{extencon} nécessite trois étapes.

\noindent \emph{Première étape}.  \emph{ On montre que la fonction vectorielle $P(N)$ définie sur le fibré normal à $\partial\ma{V}(t)$ est homogène :
pour tout $\lambda\in \rr$, on a : $P(\lambda N)=\lambda P(N)$. }

\noindent L'égalité précédente est vraie, par principe pour tout $\lambda >0$.
En coupant le domaine $\ma{V}(t)$ par un hyperplan $\Sigma$ transverse à $\partial \ma{V}(t)$, on obtient deux domaines $\ma{V^+}(t)$ et $\ma{V^-}(t)$  qui sont des sous-variétés de dimension $3$ de bords respectifs $\partial \ma{V}(t)^+ \cup \Sigma(t)$ et ${\partial \ma{V}(t)^-} \cup {\Sigma(t)}$ où $\partial \ma{V}(t)^+$ est la partie du bord de $\ma{V}(t)$ située dans $\ma{V^+}(t)$ et $\partial \ma{V}(t)^-$ est la partie du bord de $\ma{V}(t)$ située dans $\ma{V^-}(t)$. 
En écrivant la relation (\ref{zer}), sur  $\ma{V^-}(t)$ et $\ma{V^+}(t)$, on obtient directement :
\[\int_{\Sigma(t)}(P(N)+P(-N))\omega^2=0.\] 
Ceci étant vrai sur toute hypersurface $\Sigma$ on a immédiatement $P(-N)=-P(N)$, ce qui prouve l'homogénéité.

\noindent  \emph{Deuxième  étape}. \emph{On montre que pour tout point de $\rr^3$ et tout vecteur $N=N^{i}e_i$ unitaire en ce point, $P(N)=N^{i}P(e_i)$.}

\noindent Faisons la démonstration pour le point $O$.
L'idée classique est de considérer le domaine particulier constitué par le $3$-simplexe affine $\ma{V}$ constitué par  la pyramide de 
sommets $O, A=(a,0,0),B=(0,b,0), C=(0,0,c)$. Le bord de $\ma{V}$ est la réunions de ses $4$ faces $\sigma_1=OBC,\sigma_2=OAC,\sigma_3=OAB, \sigma_4=ABC$. On note $N$ le vecteur unitaire normal à $ABC$. En jouant sur $a,b$ et $c$, on peut former tous les vecteurs de l'ensemble $\{(a,b,c)\in \rr^3\; a\ge 0, b\ge 0, c\ge 0, a^2+b^2+c^2=1\}$  des vecteurs unitaires d'un quart d'espace, c'est à dire du cône positif d'origine $O$. Pour chacun des trois autres quarts on procèderait de manière analogue pour en construire les vecteurs unitaires.
Un petit exercice de géométrie des lycées montre que les aires des faces sont données par 
\[\ma{A}(\sigma_1)=\frac{1}{2}bc,\ma{A}(\sigma_2)=\frac{1}{2}ac,\ma{A}(\sigma_3)=\frac{1}{2}ab,\ma{A}(\sigma_4)=\frac{1}{2}\sqrt{a^2b^2+a^2c^2+b^2c^2}\]
et de plus : \[N=\frac{1}{\ma{A}(\sigma_4)}\left(\sum_{i=1}^3\ma{A}(\sigma_i)e_i\right)=:N^{i}e_i.\]
 Pour une fonction vectorielle $g=(g_1,g_2,g_3)$ continue définie sur un compact de $\rr^3$ et à valeurs dans $\rr^3$, nous noterons $\mu(g)=(\mu(g_1),\mu(g_2),\mu(g_3))$ sa valeur moyenne sur ce compact.
 On considère ici $g=P(N)$ définie sur $\partial\ma{V}$. On a :
$\int_{\partial\ma{V}}P(N)\omega^2=\sum_{i=1}^3\int_{\sigma_i}P(-e_i)\omega^2+\int_{\sigma_4}P(N)\omega^2$. On peut donc écrire en faisant apparaître les valeurs moyennes :

$\int_{\partial\ma{V}}P(N)\omega^2=\sum_{i=1}^3 \ma{A}(\sigma_i)\mu(P(-e_i))+\ma{A}(\sigma_4)\mu(P(N))$ et 

$\int_\ma{V}(\rho\frac{dV}{dt}-f)\omega^3=Vol(\ma{V})\mu\left(\rho\frac{dV}{dt}-f\right)$. D'où la réécriture de l'égalité (\ref{zer}) :
\[\mu\left(P(N)-\sum_{i=1}^3N^{i}P(e_i)\right)=\frac{\vol \ma{V}}{\ma{A}(\sigma_4)}\mu\left(\rho\frac{dV}{dt}-f\right).\]
Si on pose $a=s\al,b=s\be,c=s\gamma$ alors en faisant tendre $s$ vers $0$ on voit que $\frac{\vol \ma{V}}{\ma{A}(\sigma_4)}$ tend vers $0$. Comme $\mu\left(P(N)-\sum_{i=1}^3N^{i}P(e_i)\right)$ tend à la fois vers la valeur de $P(N)-\sum_{i=1}^3N^{i}P(e_i)$ au point $O$ et vers $0$, on a en ce point  $P(N)=\sum_{i=1}^3N^{i}P(e_i)$. 
Ce qui a été fait pour $O$ se fait pour n'importe quel point $M$ du milieu.
Ainsi en avons-nous terminé avec la deuxième étape.
La première étape associée à la deuxième montre que $P$ définit un endomorphisme  en tout point $M$
de l'espace tangent $T_M\rr^3=\rr^3$.
Notons $ T^{i}_j$ la matrice associée à $P$ dans la base canonique de $\rr^3$ où $i$ est l'indice de ligne et $j$ l'indice de colonne. $ T^{i}_j$ sont les composantes d'un tenseur mixte $(1,1)$ qu'on notera encore $T$ et pour tout vecteur $X$ de $\rr^3$, $T^{i}_j X^j$ représente le vecteur $P(X)$ ou encore la contraction du tenseur $T$ et du tenseur $1$ fois  contravariant $X$. Notons $T^{ij}=\delta^{aj} T^{i}_a$ où $\delta^{ij}$ est la version covariante du tenseur métrique standard de  $\rr^3$ : 
$\delta^{ii}=1$ et  si $i\not= jj$, $\delta^{ij}=0$. 

\noindent \emph{Troisième étape}. 
\emph{Montrons que $T^{ij}=T^{ji}$ pour tous $i,j\in\{1,2,3\}$. }

\noindent  Notons $\Div T$ le champ de vecteurs $\Div T=T^{\al\be}_{\;,\be}e_\al$ où conformément aux notations déjà utilisées
$T^{\al\be}_{\;,\be}=\sum_{\be=1}^3\frac{\partial T^{\al\be}}{\partial x^\be}$.  Cette notation sera totalement justifiée relativement
à la définition de l'exercice \ref{divdivergence}, quand on aura montré la symétrie de $T$. Si l'on considère le champ de vecteurs 
$T(N)$, on remarque qu'il peut également s'écrire $T(N)=T(N)^\al e_\al$  où $T(N)^\al =\sum_\be T^{\al\be}N^\be$ ou encore  $T(N)=\la Y^\al,N\ra e_\al$ si on pose $Y^\al=(T^{\al 1},T^{\al 2},T^{\al 3})$.
 Notons par ailleurs que l'intégrale $ \ma{I}=\int_{\partial \ma{V}(t)}T(N)\omega^2$ qui apparaît dans les forces extérieures $F_e$ définies par l'égalité (\ref{ffi}) peut s'écrire alors 
\[\ma{I}=\left(\int_{\partial \ma{V}(t)}\langle  Y^\al,N\rangle \omega^2\right) e_\al.\]
Alors le théorème \ref{thdladiv}, montre que 
\[\ma{I}=\int_{\ma{V}(t)}\Div T\;\omega^3.\]
Tenant compte de cette égalité, la relation fondamentale (\ref{zer}) entraine l'égalité 
\[
\rho\frac{dV}{dt}=f+\Div T.
\]
En tenant compte de l'égalité (\ref{eqmomo}), la relation fondamentale (\ref{rez}) se réécrit :

\[\int_{\ma{V}(t)}r\times \Div T\;\omega^3=\int_{\partial\ma{V}(t)}r\times T(N)\;\omega^2,\]
qui équivaut aux trois équations suivantes :
\begin{equation}\label{merlus}
\int_{\ma{V}(t)}\left(x^\al\frac{\partial T^{\be\gamma}}{\partial x^\gamma}-x^\be\frac{\partial T^{\al\gamma}}{\partial x^\gamma}\right)\;\omega^3=\int_{\partial\ma{V}(t)}\left(x^\al T(N)^\be-x^\be T(N)^\al\right)\;\omega^2,\;\al,\be\in\{1,2,3\}
\end{equation}
En écrivant $T(N)^\be=T^{\be\gamma}N_{\gamma}$ on peut donner à l'intégrale de droite de l'égalité précédente la forme :

\noindent$\int_{\partial\ma{V}(t)}\left(x^\al T(N)^\be-x^\be T(N)^\al\right)\;\omega^2=\int_{\partial\ma{V}(t)}(x^\al T^{\be\gamma}x^\be -T^{\al\gamma})N_\gamma\; \omega^2= $

$=\int_{\ma{V}(t)}\frac{\partial}{\partial x^\gamma}\left(x^\al T^{\be\gamma}-x^\be T^{\al\gamma}\right)\;\omega^3$, d'après le théorème de la divergence,

\noindent \hspace{0.6cm}$=\int_{\ma{V}(t)}(T^{\be\al}-T^{\al\be})\omega^3+\int_{\ma{V}(t)}\left(x^\al\frac{\partial T^{\be\gamma}}{\partial x^\be}-x^\be\frac{\partial T^{\al\gamma}}{\partial x^\gamma}\right)\omega^3$.

Ainsi, les égalités (\ref{merlus}) montrent que quelque soit le volume $\ma{V}(t)$, on a $\int_{\ma{V}(t)}(T^{\be\al}-T^{\al\be})\omega^3=0$, ce permet de conclure aux égalités
$$T^{\al\be}=T^{\be\al}.$$
\end{preuve}
\subsubsection{Le tenseur impulsion-énergie de l'espace temps newtonien et les équations du mouvement.}
En nous plaçant dans l'espace-temps de la mécanique newtonienne $\rr\times \rr^3$, nous pouvons exprimer par une seule relation vectorielle le principe de conservation de la masse et l'équation (\ref{eqmomo}), en introduisant le \emph{tenseur énergie-impulsion}.
\begin{definition}\label{teenim}\index{tenseur impulsion-énergie de l'espace temps newtonnien}
On pose $U=\frac{\partial}{\partial t}+V(x,t)$, et $M=\rho U\otimes U$.

\noindent Le tenseur  $S=M-T$, où $T$ est le tenseur des contraintes, est appelé tenseur impulsion-énergie.
Notons que dans le cas d'un fluide parfait on a $S=\rho U\otimes U+p g$ où $g$ est le tenseur métrique.
\end{definition}
Alors \begin{proposition}\label{eqmodi4}
Le tenseur $S$ vérifie l'équation du mouvement 
\begin{equation}\label{eqmodidi}
\sum_{j=0}^3\frac{\partial S^{ij}}{\partial x^j}=f^{i}
\end{equation}
Cette équation équivaut au principe de conservation de la masse et à l'équation (\ref{eqmomo}).
\end{proposition}
\begin{preuve}{}
Pour simplifier l'écriture, désignons les indices de $1$ à $3$ par des lettres grecques.

\noindent Vérifions la relation (\ref{eqmodidi}) pour $i=0$.
On a : $\frac{\partial S^{0j}}{\partial x^j}=\frac{\partial M^{0j}}{\partial x^j}$ (car $T^{0j}=0$)

\hspace{1.75cm}$=\frac{\partial \rho}{\partial t}+\frac{\partial (\rho V^\al)}{\partial x^\al}= \frac{\partial \rho}{\partial t}+\Div (\rho V)=0$ (selon l'équation (\ref{comas})).

\noindent Vérifions la relation (\ref{eqmodidi}) pour $i\not=0$. On a :

\noindent$\frac{\partial T^{\al j}}{\partial x^j}=\frac{\partial }{\partial x^j}(\rho U^\al U^j)-\frac{\partial S^{\al j}}{\partial x^j}=$

\hspace{1.8cm}\noindent$=\frac{\partial( \rho V^\al)}{\partial t}+\frac{\partial }{\partial x^\be}(\rho V^\al V^\be)-\frac{\partial S^{\al j}}{\partial x^j}$\;\;\;\;( grâce à l'équation (\ref{comas}))

\hspace{1.8cm}\noindent$=\rho\frac{\partial V^\al}{\partial t}+V^\al (\frac{\partial \rho}{\partial t}+\frac{\partial(\rho V^\be)}{\partial x^\be})
+\rho V^\be \frac{\partial V^\al}{\partial x^\be}-\frac{\partial S^{\al j}}{\partial x^j}$

\hspace{1.8cm}\noindent$= \rho(\frac{\partial V^\al}{\partial t}+V^\be \frac{\partial V^\al}{\partial x^\be})-(\Div S)^\al=\rho\frac{d V^\al}{dt}-(\Div S)^\al$.

\noindent Cette égalité s'écrit encore compte tenu de l'équation (\ref{eqmomo}):

$$(\Div T)^\al=(\Div T)^\al +f^\al-(\Div S)^\al.$$ Et on conclut.





Inversement en écrivant l'équation (\ref{eqmodidi}) pour $i=0$, on obtient le principe de conservation de la masse puis en l'écrivant pour $i\not=0$, on obtient l'équation (\ref{eqmomo}).
\end{preuve}
Considérons sur $\rr^3$ des coordonnées curvilignes quelconques $(x^1,x^2,x^3)$. On peut leur associer la connexion $\nabla$ de Levi-Civita dont la matrice locale de connexion s'écrit $\omega^k_i=\Gamma^k_{ji}dx^j$ où les coefficients $\Gamma^k_{ji}$
sont les coefficients de Christoffel définis par l'égalité (\ref{cristo}). On note $V(x,t)=V^{i}\frac{\partial}{\partial x^{i}}$ le vecteur vitesse à l'instant $t$ d'une particule de coordonnées $(x^1,x^2,x^3)$.
\begin{proposition}\label{moucocu}
On suppose le fibré tangent à  $\rr^3$  muni de la connexion de Levi-Civita associée à la métrique de tenseur $(g_{ij})$ où $g_{ij}=\la\frac{\partial}{\partial x^{i}},\frac{\partial}{\partial x^{j}}\ra$.

\noindent Compte tenu du principe de conservation de la masse, l'équation du mouvement (\ref{eqmodidi}) s'écrit dans des coordonnées curvilignes $(x_1,x^2,x^3)$ :
\begin{equation}\label{eqmoucocu}
\frac{\partial(\rho V^{i})}{\partial t}+ \frac{\nabla S^{ik}}{\partial x^k}=f^{i}
\end{equation}
\end{proposition}
\begin{preuve}{}
Dans l'équation (\ref{eqmomo}), le champ de vecteurs $\frac{dV}{dt}$ est l'accélération de la particule. En coordonnées curvilignes, le vecteur accélération a pour composantes $\Gamma^{i}=\frac{\nabla V^{i}}{dt}=\frac{d V^{i}}{dt}+\Gamma^{i}_{kj}V^kV^j$. On a donc :

\noindent$\rho \Gamma^{i}=\rho(\frac{d V^{i}}{dt}+\Gamma^{i}_{kj}V^kV^j)= \rho\left(\frac{\partial  V^{i}}{\partial t}  
+\Gamma^{i}_{kj}V^kV^j+ V^k\frac{\partial V^{i}}{\partial x^k}\right)=$

\noindent$=\rho\left(\frac{\partial V^{i}}{\partial t}+V^k\frac{\nabla V^{i}}{\partial x^k}\right) $ (voir équation (\ref{nonotes}) p. \pageref{nonotes})

\noindent$=\frac{\partial(\rho V^{i})}{\partial t}-V^{i}\frac{\partial \rho}{\partial t}+\left(\frac{\nabla(\rho V^kV^{i})}{\partial x^k}-V^{i}\frac{\nabla(\rho V^k)}{\partial x^k}\right)$. Ayant supposé que $\frac{\partial \rho}{\partial t}-\frac{\nabla(\rho V^k)}{\partial x^k}=0$, on a finalement :
\[\rho \Gamma^{i}=\frac{\partial(\rho V^{i})}{\partial t}+\frac{\nabla(\rho V^kV^{i})}{\partial x^k}=f^{i}+\frac{\nabla T^{ki}}{\partial x^k}.\]
La dernière égalité permet de conclure.
\end{preuve}
\subsection{Dynamique des milieux continus en relativité restreinte.}\label{dimicorr}
En suivant la démarche de A. Lichnérowicz dans [\ref{lichné}], montrons comment déduire les équations relativistes du mouvement dans un milieu continu en partant des deux équations de la mécanique classique qui  détermine ce mouvement, à savoir 
l'équation de conservation de la masse (\ref{comas}) et l'équation (\ref{eqmoucocu}).
Introduisons pour commencer deux définitions.
\begin{definition}\label{qutens}\index{quadritenseur}
Un tenseur symétrique  deux fois contravariant $\tau$ sur l'espace de Minkowski est un quadritenseur si ses composantes $\tau^{ij}$ et ${\tau'} ^{ij}$ dans deux référentiels galiléens $\ma{R}$ et $\ma{R}'$ vérifient  ${\tau'} ^{ij}=L^{i}_kL^j_h\tau^{kh}$ 
où $(L^{i}_j)$ est la matrice de Lorentz qui transforme les coordonnées d'un quadrivecteur dans $\ma{R}$ en ses coordonnées dans $\ma{R}'$.
\end{definition}
\begin{exo}\label{qutenvec}
Les notations sont celles de la définition\ref{qutens}.
Si $\tau $ est un quadritenseur et $X$ un quadrivecteur. Soit le vecteur $Y$  de coordonnées $\tau^{i}_j X^j$ où $\tau^{i}_j$ sont les composantes du $(1,1)$-tenseur associé à $\tau$ dans l'espace $(\rr^4,\la,\ra_L)$. Montrer que $Y$ est un quadrivecteur.
\end{exo}
Par analogie avec le tenseur des contraintes introduit dans le théorème \ref{extencon} et le quadrivecteur force de Minkowski-Lorentz, on définit un tenseur des con\-train\-tes de l'espace de Minkowski en reprenant les notations de l'exercice précédent.
\begin{exo}\label{coool}
Remarquer que $\tau^{\al\be}u_\be=0$
\end{exo}

Considérons un point quelconque  $P$ du milieu continu dans l'espace de Minkowski, c'est à dire un point de l'espace à un instant donné. On a en ce point une densité de quantité de mouvement et il existe pour $P$ un référentiel galiléen $\mathcal{R} $ par rapport auquel la vitesse est nulle. On peut alors écrire les  équations de conservation et du mouvement  au point $P$ de façon identique aux équations (\ref{comas}) et (\ref{eqmoucocu}), en notant $p$ la densité de quantité de mouvement :
\begin{align}
\frac{\partial \rho}{\partial t}+\Div (p)&=0\label{comasse}\\
\frac{\partial(p^{i})}{\partial t}+ \frac{\nabla S^{ik}}{\partial x^k}&=f^{i}\label{eqmoucoucou}
\end{align}
La différence avec le cas classique, est dans l'évaluation de la densité de masse qui participe à la quantité de mouvement : la densité relativiste $\rho=\rho_0\gamma$ est aussi égale à $E/c^2$ où $E$ est la densité d'énergie. Il faut donc tenir compte de toutes les formes d'énergie. Ici la densité d'énergie déterminée la densité de masse au repos de la particule mais également par la densité de travail fourni par les forces de pression. En notant $(e_1,e_2,e_3)$ la base canonique de $\rr^3$,
la force de pression dans la direction $e_i$ est d'après le théorème \ref{extencon} égale à $T_i^k e_k= (T^{i}_1,T^{i}_2,T^{i}_3)$ puisque $T^{i}_j=T^j_i$ et l'énergie associée dans un déplacement de vitesse $V$ est égale au produit scalaire 
$\la(T^{i}_1,T^{i}_2,T^{i}_3), (V^1,V^2,V3)\ra=T^{i}_k V^k.$ 
  Notons que $\frac{T^{i}_k }{c^2}$ a la dimension d'une densité de masse.
  \begin{exo}\label{diphy}
 Démontrer l'assertion précédente.
 \end{exo}
 Ainsi  la contribution de l'énergie dûe au travail des forces de pression à la $i$-ème composante de $p$ de l'énergie de des force de pression est $\frac{T^{i}_k}{c^2}V^k.$
   On peut donc écrire 
 \begin{equation}\label{krela}
 p^{i}=\rho_0\gamma V^{i}+\frac{T^{i}_kV^k}{c^2}.\end{equation}
 En remplaçant dans les équations (\ref{comasse}) et (\ref{eqmoucoucou}) $p^{i}$ par son expression donnée dans l'équation (\ref{krela}) on obtient les équations du mouvement relativiste en $P$ dans les coordonnées galiléennes de $\mathcal{R}$.
 On notera $u$ le quadrivecteur vitesse de la particule de vitesse $V$.
 
 Nous allons faire les hypothèses ($\mathcal{H}$) suivantes :
 \begin{enumerate}
\item Il n'y a pas de champ de gravitation créé par les particules du fluide ayant une action sur chacune d'entre elles. 
 \item Le seul champ extérieur ayant une action sur le fluide est un champ électromagnétique qui peut éventuellement être nul.
 \item Il existe sur l'espace de Minkowski un champ de $(1,1)$-quadritenseur  $\mathcal{T}$ des contraintes, tel que pour tout point $P$ et référentiel inertiel $\mathcal{R}$ vérifiant les deux points suivants.
  \begin{enumerate}
 \item  On a dans les coordonnées de $\mathcal{R}$ :
  \[ \mathcal{T}^{0}_j=\mathcal{T}^{j}_0=0, \mathcal{T}^{i}_j={T}^{j}_i \;\mathrm{pour \;tout}\; i,j\in \{1,2,3\}.\]
  \item Dans tout référentiel galiléen les composantes $\tau^\al_\be$ de $\mathcal{T}$ vérifient 
  \begin{equation}\label{ortogo}\tau^\al_\be u^\be=0.\end{equation}
  \end{enumerate}
 \end{enumerate}
La troisième hypothèse est à rapprocher de l'hypothèse qui exprime le prolongement en un quadrivecteur sur  l'espace de Minkowski de la force de Lorentz.
La deuxième hypothèse implique l'égalité $f^{i}=(\ma{K}_1)^{i}$ avec les notations de l'équation (\ref{denfololo}), p.\pageref{denfololo}.
\begin{theorem}\label{eqmvmcrr}
Désignons toujours par $u$ le quadrivecteur vitesse des particules du fluide, les équations du mouvement dans un système de coordonnées locales quelconque $(x^0,x^1,x^2,x^3)$ s'écrivent :
\begin{equation}\label{super!}\frac{\nabla}{\partial x^{\mu}}\left(\rho_0 c^2 u^{\lambda} u^{\mu}-\ma{T}^{\lambda\mu}\right)=(\ma{K}_1)^\lambda,\end{equation}
où $\ma{K}_1$ est le quadrivecteur densité de force de Lorentz.
Elles sont équivalentes aux équations (\ref{comasse}) et (\ref{eqmoucoucou}).
\end{theorem}
\begin{preuve}{}
En tenant compte du fait que $V=0$ au point $P$, les équations (\ref{comasse}) et (\ref{eqmoucoucou}) se réécrivent alors  :
 \[\left\{
 \begin{array}{rcl}
\frac{\partial \rho_0}{\partial t}+\frac{\partial}{\partial x^k}\left(\rho_0 V^k+\frac{T^{k}_jV^j}{c^2}\right)&=&0\\
\frac{\partial}{\partial t}\left(\rho_0  V^{i}+\frac{T^{i}_kV^k}{c^2}\right)- \frac{\partial T^{i}_k}{\partial x^k}&=&(\ma{K}_1)^{i}
\end{array}
 \right.
 \]
 \emph{Au vue de ces deux équations, les indices latins désigneront dans cette démonstration des entiers de $\{1,2,3\}$ et les indices grecs des entiers de $\{0,1,2,3\}$.}
 
 Ces deux équations s'écrivent dans les coordonnées $(x^0=ct,x^1,x^2,x^3)$ du référentiel $\ma{R}$ :
 \[\left\{
 \begin{array}{rcl}
c\frac{\partial \rho_0}{\partial x^0}+\rho_0\frac{\partial V^k}{\partial x^k}+\frac{1}{c^2}\frac{\partial V^j }{\partial x^k} T^k_j&=&0\\
c\rho_0 \frac{\partial V^{i}}{\partial x^0} +\frac{1}{c} \frac{ \partial V^k}{\partial x^0}T^{i}_k-\frac{\partial T^{i}_k}{\partial x^k}&=&(\ma{K}_1)^{i}
\end{array}
 \right.
 \]
 Dans les deux équations précédentes les composantes de la vitesse en fonction de celle du quadrivecteur vitesse.
 On a en effet au point $P$ les égalités $$\frac{\partial u^{i}}{\partial x^k}=\frac{1}{c}\frac{\partial V^{i}}{\partial x^k}\;\mathrm{et} \;
 \frac{\partial u^0}{\partial x^\lambda}=0.$$ D'où :
  \begin{align}
c^2\frac{\partial \rho_0}{\partial x^0}+\rho_0c^2\frac{\partial u^k}{\partial x^k}+\frac{\partial u^j }{\partial x^k} T^k_j&=0\label{bleue}\\
c^2\rho_0 \frac{\partial u^{i}}{\partial x^0} + \frac{ \partial u^k}{\partial x^0}T^{i}_k-\frac{\partial T^{i}_k}{\partial x^k}&=(\ma{K}_1)^{i}\label{rouge}
\end{align}

La troisième hypothèse permet d'écrire dans tout système de coordonnées galiléennes, selon l'équation (\ref{ortogo}), l'égalité : $\ma{T}^\mu_0\gamma=-\ma{T}^\mu_ku^k$ et en dérivant par rapport à $x^\lambda$, on obtient au point $P$ l'égalité :
\[\frac{\partial \ma{T}^\mu_0}{\partial x^\lambda}=-\frac{\partial u^k}{\partial x^\lambda}\ma{T}^\mu_k.\]
On a en particulier :
$\frac{\partial \ma{T}^0_0}{\partial x^0}=0$ et $\frac{\partial \ma{T}^k_0}{\partial x^k}=-\frac{\partial u^l}{\partial x^k}T^k_l.$

\noindent On peut réécrire le système d'équations (\ref{bleue}) et (\ref{rouge}) de façon équivalente en coordonnées galiléennes sous la forme suivante :
\begin{equation}\label{susuper!}
\frac{\partial}{\partial x^{\mu}}\left(\rho_0 c^2 u_{\lambda} u^{\mu}-\ma{T}^\mu_{\lambda}\right)={(\ma{K}_1)}_{\lambda}.\end{equation}
\noindent Montrons cette équivalence.

\noindent Notons que $\mathcal{K}_1$ étant le quadrivecteur densité de forces de Lorentz, il vérifie la relation d'orthogonalité décrite dans l'exercice \ref{secret} qui entraine qu'au point $P$ on a ${(\mathcal{K}_1)}_0=0$.
\'{E}crivons l'équation (\ref{susuper!}) pour $\lambda=0$ :
\[\frac{\partial}{\partial x^{k}}\left(\rho_0 c^2 u_{0} u^{k}-\ma{T}^k_0\right)+\frac{\partial}{\partial x^{0}}\left(\rho_0 c^2 u_{0} u^{0}-\ma{T}^0_0\right)=0\]

\noindent qui s'écrit encore :

$$c^2\frac{\partial \rho_0}{\partial x^0}+c^2\rho_0\frac{\partial u^k}{\partial x^k}-\frac{\partial \ma{T}^k_0}{\partial x^k}=
c^2\frac{\partial \rho_0}{\partial x^0}+c^2\rho_0\frac{\partial u^k}{\partial x^k}+\frac{\partial u^l}{\partial x^k}T^k_l=0.$$

\noindent Ceci montre l'équivalence des équations (\ref{rouge}) et (\ref{super!}) pour $\lambda=0$.

\noindent Réécrivons l'équation (\ref{susuper!}) pour $\lambda =i\in\{1,2,3\}$. On obtient directement :

\noindent$(\ma{K}_1)_i=\rho_0c^2\frac{\partial u_i}{\partial x^0}-\frac{\partial \ma{T}^\mu_i}{\partial x^\mu}
=\rho_0c^2\frac{\partial u_i}{\partial x^0}-\frac{\partial \ma{T}^0_i}{\partial x^0}-\frac{\partial \ma{T}^k_i}{\partial x^k}=
\rho_0c^2\frac{\partial u_i}{\partial x^0}-\frac{\partial \ma{T}^{i}_0}{\partial x^0}-\frac{\partial \ma{T}^k_i}{\partial x^k}=$

$=\rho_0c^2\frac{\partial u_i}{\partial x^0}+\frac{\partial u^k}{\partial x^0}T^{i}_k-\frac{\partial \ma{T}^k_i}{\partial x^k}$, ce qui montre l'équivalence des équations 	(\ref{rouge}) et (\ref{susuper!}) pour $\lambda\in\{1,2,3\}$.

\noindent L'équation (\ref{susuper!}) s'écrit en relevant l'indice $\lambda$ :
 \[\left(\frac{\nabla}{\partial x^\mu}(\rho_0c^2 u^\lambda u^\mu-\ma{T}^{\lambda\mu})-(\ma{K}^1)^\lambda\right)\frac{\partial}{\partial x^\lambda}=0.\]  
 Si on se réfère à l'exercice \ref{divdivergence}, p. \pageref{divdivergence}, cette équation s'écrit encore que les composante du tenseur 
$\Div \left(\rho_0c^2U\otimes U-\ma{T}\right)-\ma{K}^1$   dans un référentiel galiléen sont nulles en \emph{tout} point $P$. Ce tenseur est donc le tenseur nul, ce qui démontre le théorème.
 \end{preuve}
 \begin{definition}\label{teimen}\index{tenseur impulsion-énergie}
 Le tenseur $\ma{P}=\rho_0U\otimes U-\frac{1}{c^2}\ma{T}$ est appelé tenseur im\-pul\-sion-énergie.
 \end{definition}
 Ainsi les équations du mouvement s'écrivent 
 \[\Div \ma{P}=\frac{1}{c^2}\ma{K}_1.\]
 Les coefficients du tenseur $\ma{P}$ ont la dimension d'une densité de masse.
Dans le cas d'un \emph{fluide parfait} \index{fluide parfait}le tenseur des pressions s'écrit (par définition d'un fluide parfait)
\begin{equation}\label{tepreparf}
\ma{T}=-p(g+U\otimes U)\end{equation} où $g$ est le tenseur métrique de l'espace de Minkowski.
Dans ce cas 
\begin{equation}\label{impenparf}
\ma{P}=(\rho_0+\frac{p}{c^2})U\otimes U+\frac{p}{c^2} g.
\end{equation}
\begin{remark}\label{impclarela}
\hspace{2cm}

\noindent L'expression  (\ref{impenparf}) est la correction relativiste du tenseur d'impulsion-énergie classique dans laquelle
on a replacé le coefficient de $U\otimes U$ classique, qui est une densité de masse par une densité de masse-énergie ( voir définition \ref{teenim}).
\end{remark}
\begin{corollary}[du théorème \ref{eqmvmcrr}]\label{eqmeumeu}
Les équations du mouvement relativiste d'un fluide, en l'absence d'un champ gravitationnel s'écrivent 
\begin{equation}\label{eqmvrr}
\Div\left(\ma{P}+\ma{M}\right)=0
\end{equation}
où $\ma{P}$ est le tenseur impulsion-énergie du milieu considéré en l'absence de champ électromagnétique et 
$\ma{M} $ le tenseur impulsion-énergie électromagnétique.
\end{corollary}
\begin{preuve}{}
C'est une conséquence immédiate du théorème \ref{divlornini}.
\end{preuve}
\begin{definition}\label{poussiere}
On dit que le fluide est une poussière si $p=0$, ce qui implique  $\mathcal{T}=0$. Autrement dit quand l'énergie des forces de pression est négligeable devant $\rho_0 U\otimes U$ dans le tenseur impulsion-énergie.
\end{definition}
\begin{corollary}
Si le fluide est une poussière, les lignes de courrant sont des droites en l'absence de champ électromagnétique.
\end{corollary}
\begin{preuve}{}
On a en effet $\Div (\mathcal{P}+\mathcal{M})=\Div (\mathcal{P})=\Div(\rho_0 U\otimes U)=0$, ce que l'on peut écrire dans des coordonnées galiléennes 
\begin{equation}\label{gueut}(\rho_0 u^{i}u^j)_{/i}=(\rho_0 u^{i})_{/i}u^j+\rho_0 u^{i})u^j_{/i}=0.\end{equation} Mais le quadrivecteur $u$ étant unitaire, 
on a pour tout champ de vecteurs $X$ : $ X.\la u,u\ra=0$. Ce que l'on peut écrire en particulier pour $X=\frac{\partial}{\partial x^{i}}$ : $u_j {u^j}_{/i}=0$. En multipliant l'équation (\ref{gueut}) par $u_j$; on obtient alors $(\rho_0 u^{i})_{/i}=0$.
Comme $P=\rho_0 c u$ est le vecteur \og vitesse\fg\; de la ligne de courrant (voir équation (\ref{ainsifond})),  celle-ci a une dérivée covariante nulle, ou encore c'est une géodésique de l'espace de Minkowski, c'est à dire une droite.
\end{preuve}

\section{Solutions des exercices}\label{thesol}
\begin{sol}{de l'exercice \ref{matcom}}

\emph{Première solution}

Par hypothèse la matrice $M$ est de la forme $=\left(\begin{matrix}0&\mu&\alpha&\beta\\
-\mu&0&a&b\\
-\alpha&-a&0&\nu\\
-\beta&-b&-\nu&0\end{matrix}\right).$ 

On pose $z_{1}=x_{1}+ix_{2}$ et $z_{2}=x_{3}+ix_{4}$ le vecteur $\varphi(z_{1},z_{2})$ est représenté dans la base canonique de $\rr^{2}$ par la matrice colonne 
$^{{t}}(x_{1},x_{2},x_{3},x_{4})$

et $f\varphi(z_{1},z_{2})$ par la matrice $\left(\begin{matrix}\mu x_{2}+\alpha x_{3}+\beta x_{4}\\
-\mu x_{1}+ax_{3}  +bx_{4}\\
-\alpha x_{1}-a x_{2}+\nu x_{4}\\
-\beta x_{1}-bx_{2}-\nu x_{3}        \end{matrix}\right)$

Ainsi $\varphi^{-1}f\varphi(z_{1},z_{2})$ est représenté dans $\cc^{2}$ par la matrice
\begin{equation}\label{mama}
M_{\cc}=
\left(\begin{matrix}(\mu x_{2}+\alpha x_{3}+\beta x_{4})+i(
(-\mu x_{1}+ax_{3}  +bx_{4})\\
(-\alpha x_{1}-a x_{2}+\nu x_{4})+i(
-\beta x_{1}-bx_{2}+\nu x_{3})       
\end{matrix}\right)
\end{equation}

Si $\left(\begin{matrix}A&B\\
C&D
\end{matrix}\right)$ 
est la matrice de $f_{\cc}$ dans la base canonique de $\cc^{2}$, la matrice de l'équation (\ref{mama}) doit être égal à 
$\left(\begin{matrix}A&B\\
C&D
\end{matrix}\right)\left(\begin{matrix}z_{1}\\
z_{2}\end{matrix}\right)
$
L'identification aboutit aux égalités suivantes $ A=-i\mu, B=\alpha+ia=b-i\beta, C=-\alpha-i\beta=-b+ia,D=-i\nu$, ce qui donne $\beta=-a$ et $\alpha=b$ : on obtient bien les matrices annoncées dans dénoncé de l'exercice.

\emph{Deuxième solution}

On écrit que $f$ est complexe si et seulement si $M\ma J=\ma J M$ et on obtient directement  $\beta=-a$ et $\alpha=b$.
\end{sol}
\begin{sol}{de l'exercice \ref{exotens}}
\noindent Soit $f\in\mathcal{L}(E,F)$. On considère l'élément $\psi(f)\in \mathcal{L}^2(E\times F^*)$ défini pour tout $x\in E$ et tout $\alpha\in F^*$ par :
$\psi(f)(x,\alpha)=\alpha\left(f(x)\right)$ 
On vérifie directement que $\psi(f)$ est bien bilinéaire que $\psi $ est linéaire, injective. Les deux espaces $\mathcal{L}(E,F)$ et $\mathcal{L}^2(E\times F^*)$ ayant la même dimension, on déduit que $\psi$ est un isomorphisme. Ainsi $\mathcal{L}(E,F)$ est canoniquement isomorphe à $ \mathcal{L}^2(E\times F^*)$ lui-même canoniquement isomorphe à 
$ \mathcal{L}^2((E^*)^*\times F^*)$ qui est la définition de $E^*\otimes F$. En effet en dimension finie 
un espace vectoriel $E$ est canoniquement isomorphe à son bidual l'isomorphisme en question étant défini par \[\begin{array}{rcl}
E&\longrightarrow&(E^*)^*\\
x&\mapsto&(\alpha\mapsto\alpha(x))
\end{array}
\]
\end{sol}
\begin{sol}{de l'exercice \ref{topinembour}}
Notons tout d'abord que $U\cap V$ est un ouvert de $M$ inclus dans $U$ donc que $\varphi(U\cap V)$ est un ouvert de $\varphi(U)$, et par conséquent de $\rr^n$ puisque $\varphi(U)$ est lui-même un ouvert de $\rr^n$.

Il s'agit  ensuite de montrer que la restriction de $\varphi$ à l'ouvert $U\cap V$ de $M$ est un homéomorphisme de $U\cap V$ sur $\varphi(U\cap V)$. On remarque tout d'abord que cette restriction est une bijection de $U\cap V$ sur 
$\varphi(U\cap V)$.

 Montrons que la restriction de $\varphi$ à $U\cap V$ est une application ouverte  et continue de $U\cap V$ sur $\varphi(U\cap V)$.  
\emph{Elle est ouverte} : 

 \noindent Un ouvert de $U\cap V$ s'écrit $A\cap U\cap V=(A\cap U)\cap (U\cap V)$ où $A$ est un ouvert de $M$. On a 
 $\varphi(A\cap U\cap V)=\varphi(A\cap U)\cap \varphi(U\cap V)$. Or $\varphi(A\cap U)$ étant un ouvert de $\varphi(U)$, il existe un ouvert $B$ de $\rr^n$ tel que $\varphi(A\cap U)=B\cap \varphi(U).$ Par conséquent
  $\varphi(A\cap U\cap V)=B\cap\varphi(U)\cap\varphi(U\cap V)=B\cap\varphi(U\cap V)$ qui est un ouvert de $\varphi(U\cap V)$. Et on conclut.
  
  \emph{Elle est continue} : 
  
 \noindent Soit $W$ un ouvert de $\varphi(U\cap V)$. Il existe un ouvert $B$ de $\rr^n$ pour lequel on a : $W=B\cap \varphi(U\cap V)$ . On a : $\varphi_{\vert U\cap V}^{-1}(W)=\varphi_{\vert U\cap V}^{-1}(B)\cap(U\cap V)$.
 Or
 
 \noindent $\varphi_{\vert U\cap V}^{-1}(B)=\varphi^{-1}(B)\cap (U\cap V)$, d'où
 $\varphi_{\vert U\cap V}^{-1}(W)=\varphi^{-1}(B)\cap (U\cap V)$. Mais 
  $\varphi^{-1}(B)$ est un ouvert de $U$ qui est un ouvert de $M$. Ainsi $\varphi^{-1}(B)$ est un ouvert de $M$ et en fin de compte $\varphi_{\vert U\cap V}^{-1}(W)=\varphi^{-1}(B)\cap (U\cap V)$
  est un ouvert de $U\cap V$. D'où la continuité.
 
\end{sol}
\begin{sol}{de l'exercice \ref{vaca}}
L'application $\varphi$ est bijective de $U$ vers $\rr$. En effet en posant $x=\cos \alpha, y=\sin\alpha$
on remarque que l'application $\gamma : ]-3\pi/2,\pi/2[\longrightarrow U$ définie par  $\gamma(\alpha)= (\cos\alpha,\sin\alpha)$ est une homéomorphisme. De plus $\Gamma : \alpha\in ]-\frac{3\pi}{2},\frac{\pi}{2}[\longrightarrow \rr$ 
définie par $\Gamma(\alpha)=\frac{\cos\alpha}{1-\sin\alpha}$ est continue et strictement croissante et réalise un homéomorphisme de $]-\frac{3\pi}{2},\frac{\pi}{2}[$ sur $]-\infty,+\infty[$.
Comme $\varphi=\Gamma\circ\gamma^{-1}$ on en déduit que $\varphi$ est un homéomorphisme.
Ainsi $U$ est homéomorphe à $\rr$. Il en est de même pour $V$.
On a $\varphi(x,y)\psi(x,y)=\frac{x^2}{1-y^2}=1$ puisque $(x,y)\in S^1$, ce qui donne le changement de cartes. Il est infiniment différentiable sur $\varphi(U\cap V)$.

\end{sol}
\begin{sol}{de l'exercice \ref{didi}}
Supposons que $y>0$ alors $x=X(1-y)=X(1-\sqrt{1-x^2})$. D'où l'on déduit directement que $x=\frac{2X}{1+X^2}$ et par suite $y=\frac{X^2-1}{X^2+1}$. Même procédure si on suppose $y<0$. Si 
$y=0$ alors $X=x=1$. 
\end{sol}
\begin{sol}{de l'exercice \ref{sousous}}
On note $n_1,n_2,n_3$ respectivement les dimensions de $M_1,M_2,M_3$. Ainsi $n_1\le n_2\le n_3$.
Soit $x$ un point de $M_1$. Supposons pour commencer que $x$ soit un point de $\Int M_2$. Il existe une carte locale de 
$M_2$ notée $(W,\eta)$ telle que $\eta(W\cap M_1)=\eta(W)\cap \rr^{n_1}$. Considérant  pour $M_2$ la srtucture de sous-variété de $M_3$, on peut prendre $W$ de sorte que $W=U\cap M_2$ et $\eta=\varphi_{\vert W}$ où $(U,\varphi)$ est une carte locale de $M_3$ en $x$. Ainsi,

\noindent $\varphi(U\cap M_1)=\eta(U\cap M_1)$ (car $U\cap M_1\subset U\cap M_2$),

\noindent\hspace{1.8cm} $=\eta(W\cap M_1)$ (car $W\cap M_1=U\cap M_1$),

\noindent\hspace{1.8cm} $=\eta(W)\cap \rr^{n_1}$.

\noindent Or, $\eta(W)=\varphi(U\cap M_2)=\varphi(U)\cap \rr^{n_2}$.
D'où

\noindent $\varphi(U\cap M_1)=(\varphi(U)\cap\rr^{n_2})\cap \rr^{n_1}=\varphi(U)\cap \rr^{n_1}$ et on conclut.

Si on suppose maintenant que $x\in \partial M_2$. On reprend exactement la démonstration précédente dans laquelle on remplace $\rr^{n_2}$ par $H^{n_2}$ et on obtient finalement l'égalité : $\varphi(U\cap M_1)=\varphi(U)\cap \rr^{n_1}$
car $\rr^{n_1}\subset H^{n_2}$.
\end{sol}
\begin{sol}{de l'exercice \ref{fivar}}
Soit $e\in E$. Notons $x=p(e)$. Soit $(U,\Phi_U)$ une carte locale du fibré en $e$. On considère une carte locale $(V,\varphi)$ de la variété différentiable $M$, telle que $x\in V$ et  $V\subset U$.
L'égalité (\ref{resto}) entraine que la restriction de $\Phi$ à $p^{-1}(V)$ est un homéomorphisme de 
$p^{-1}(V)$ sur $V \times F$. Si on note $\psi$ un isomorphisme de $F$ vers $\rr^p$, alors $(\phi\times \psi)\circ\Phi_{\vert V}$ est un homéomorphisme du voisinage ouvert $p^{-1}(V)$ sur $\rr^n\times \rr^p$.
Et on conclut.
\end{sol}
\begin{sol}{de l'exercice \ref{ladeuz}}
$1$)Soit $(U'_i,\Phi'_i),(U'_j,\Phi'_j)$ deux cartes locales de $\tau'$ de domaines non disjoints, soit $(U_k,\Phi_k)$ une carte locale de $\tau$ telle que $W_{ijk}=h^{-1}(U'_i\cap U'_j)\cap U_k\not=\emptyset$. 
Alors pour tout $x\in W_{ijk}$ on a $c'_{ij}(h(x))\circ h_{jk}(x)=h_{ik}(x)$.

$2$) Vérification de la première relation (\ref{concon}) :

Avec les notations déjà utilisées, $\Phi'_iH\Phi_k^{-1}(x,f)=(h(x),h_{ik}(x)f)$

$=\Phi'_iH\Phi_j^{-1}(\Phi_j\Phi_k^{-1})(x,f)=\Phi'_iH\Phi_j^{-1}(x,c_{jk}(x)f)=(h(x),h_{ij}(x)c_{jk}(x)f)$.

Et on conclut.

\end{sol}
\begin{sol}{de l'exercice \ref{job}}
$1)$ Soit $\tau=(E,p,M)$un fibré vectoriel. Alors $p$ est ouverte. En effet soit $\cal{O}$ un ouvert de $E$. On peut écrire $\cal{O}$ sous la forme $\mathcal{O}=\bigcup_{i}\big(\mathcal{O}\cap p^{-1}(U_i)\big)$ où les ouverts $U_i$ sont homéomorphes à $U_i\times F$, où l'homéomorphisme correspondant  $\Phi_i$ vérifie $p_1\circ\Phi_i=p$. Ainsi $p(\mathcal{O})=p\left(\bigcup_{i}\big(\mathcal{O}\cap p^{-1}(U_i)\big)\right)=\bigcup_{i}p\left(\mathcal{O}\cap p^{-1}(U_i)\right)=\bigcup_{i}p_1\Big(\Phi_i\big(\mathcal{O}\cap p^{-1}(U_i)\big)\Big)$. Comme $\Phi_i\big(\mathcal{O}\cap p^{-1}(U_i)\big)$ est ouvert, que $p_1$ est ouverte est que la réunion d'ouverts est un ouvert, on conclut que 
$p(\mathcal{O})$ est un ouvert.

$2)$Le diagramme commutatif 
\[
\begin{array}{ccc}
E&\stackrel{H}{\longrightarrow}&E'\\
\vcenter{\llap{p}}\Big\downarrow&&\Big\downarrow\vcenter{\rlap{$p'$}}\\
M&\xrightarrow[\enspace h\enspace]&M'
\end{array}
\]

montre avec ce qui précède que $h$ est ouverte : si $U$ est un ouvert de $M$, on a $U=p\circ p^{-1}(U)$ car $p$ est surjective et $h(U)=h(p\circ p^{-1}(U))=(p'\circ H)(p^{-1}(U))$.

 Comme $h$ est aussi continue il suffit de montrer qu'elle est bijective. Le diagramme commutatif précédent montre directement la surjectivité. Si $y\in M'$, on a $H^{-1}(p'^{-1}(y))$ est une fibre $p^{-1}(x)$ de $\tau$. Le diagramme nous dit que $x$ est le seul antécédent possible de $y$.
 \end{sol}
 \begin{sol}{de l'exercice \ref{jobi}}
 $1)$ On sait déjà que $H$ applique $p^{-1}(b)$ sur $p'^{-1}(h(b))$. Soit deux éléments $e$ et $e'$ de $p^{-1}(b)$ et $\lambda\in \rr$. Avec les notations habituelles, on a dans la carte $(U_{i},\Phi_{i})$ de $E$ en exprimant conformément à l'équation (\ref{morfer}) $H$ dans les cartes $(U_{i},\Phi_{i})$ et $(U'_{j},\Phi'_{j})$ en $h(b)$:
 \begin{align}
 &e=\Phi_{i^{-1}}(b,f),e'=\Phi_{i}^{-1}(b ,f')\label{roc}\\
  &e+\lambda e'=\Phi_{i}^{-1}(b, f+\lambda f')=:\Phi_i^{-1}(b,f)+\Phi_i^{-1}(b,f')\label{rac}\\
&  \Phi'_j H\Phi_i^{-1}(b,f)= \Phi'_j H(e)=(h(b),h_{ji}(b)f)\label{rec}\\
& \Phi'_j H\Phi_i^{-1}(b,f')= \Phi'_j H(e)=(h(b),h_{ji}(b)f')\label{reuc}
  \end{align}
  En utilisant les équations (\ref{rec}) et (\ref{reuc}), on a:
  
  $H(e)+\lambda H(e')= {\Phi'_j}^{-1}(h(b),h_{ji}(b)f)+\lambda {\Phi'_j}^{-1}(h(b),h_{ji}(b)f').$ La structure d'espace vectoriel sur le fibres représentée par les équations (\ref{rac}) donnent alors :
  
  $H(e)+\lambda H(e')={\Phi'_j}^{-1}(h(b),h_{ji}(b)f+\lambda h_{ji}(b)f')$. En tenant compte de la linéarité des
  $h_{ji}(b)$, on a :
  
    $H(e)+\lambda H(e')={\Phi'_j}^{-1}(h(b),h_{ji}(b)(f+\lambda f'))=H\Phi_i^{-1}(h(b),f+\lambda f')=:H(e+\lambda e')$.
  
  $2)$ Si $H$ est un isomorphisme de fibré de fibre $\rr^n$ dans un fibré de fibre $\rr^p$, alors $h$ est un homéomorphisme, et on a avec les notations de l'équation (\ref{rec}):
  \begin{align*}
 &  \Phi'_j H\Phi_i^{-1}(b,f)= (h(b),h_{ji}(b)f)\\
& \Phi_i H^{-1}{\Phi'}_j^{-1}((h(b),f')=(b,\tilde{h}_{ij}(h(b))f')
  \end{align*}
  D'où $\Phi_i H^{-1}{\Phi'}_j^{-1}((h(b),h_{ji}(b)f)=(b,\tilde h_{ij}(h(b))h_{ji}(b)f)$
  
  $=\Phi_i H^{-1}{\Phi'}_j^{-1} \Phi'_j H\Phi_i^{-1}(b,f)=(bf)$.
  
  Ainsi $h_{ji}(b)$ est inversible ce qui montre que $n=p$ et $h_{ji}(b)$ est un automorphisme de $\rr^n$.
  Inversement si les $h_{ji}(b)$ sont des isomorphismes d'espace vectoriel et $h$ homéomorphisme, on démontre immédiatement que $H$ est une application inversible de $E$ vers $E'$. On sait que $H$ est continue et le diagramme commutatif de la solution de l'exercice \ref{job} montre qur $H$ est ouverte.
  C'est donc un homéomorphisme de $E$ sur $E'$.
  \end{sol}
  \begin{sol}{de l'exercice \ref{bofbof}}
  On a une relation du type $\epsilon'=\epsilon B$. De plus $^t\epsilon e=I$. D'où :
  
  $I=^t\epsilon'\;.e'=^t\epsilon' \;eA=^tB\;^t\epsilon eA=^tB.A.$ D'où $B=^tA^{-1}$. Et on conclut.
   \end{sol}
    \begin{sol}{de l'exercice \ref{nok}}
Plaçons dans des coordonnées locales où $X$ et $Y$ s'écrivent :\[X=X^{i}\frac{\partial}{\partial x^{i}},
Y=Y^j \frac{\partial}{\partial x^{j}}.\]
Alors pour toute fonction différentiable $f$, on a d'après l'équation (\ref{crobar}) :
\[\left[fX,Y\right]^{i}=fX^j\frac{\partial Y^{i}}{\partial x^{j}}-Y^j\frac{\partial(fX^{i})}{\partial x^j} .\]
On a donc :

\noindent $\left[fX,Y\right]^{i}=f\left(X^j\frac{\partial Y^{i}}{\partial x^{j}}-Y^j\frac{\partial X^{i}}{\partial x^j} \right)-Y^jX^{i}\frac{\partial f}{\partial x^j}= f[X,Y]^{i}-(L_Yf)X^{i}$ et on conclut.

\noindent Le deuxième item est évident.
\end{sol}
\begin{sol}{de l'exercice \ref{kon}}
Les trois premiers items se démontrent par un calcul direct à partir de la définition initiale.
Montrons l'item $4$. Si $f$ est une fonction au moins deux fois continument différentiable, on a :
$[X,Y].f=X.(Y.f)-Y.(X.f)=X.\left(Y^{i}\frac{\partial f}{\partial x^{i}}\right)-Y.\left(X^{i}\frac{\partial f}{\partial x^{i}}\right)
=(X.Y^{i}-Y.X^{i})\frac{\partial f}{\partial x^{i}}+Y^{i}X^j\frac{\partial^2 f}{\partial x^{i}\partial x^j}-X^{i}Y^j\frac{\partial^2 f}{\partial x^{j}\partial x^i}=..$

\noindent$..=(L_XY-L_YX).f$

\noindent Passons au cinquième item.

\noindent Pour $(a)$, si $X,Y,Z$ sont trois champs tangents à $S$, en utilisant l'item $1.$, pour l'additivité par rapport au premier argument, on a par définition de $D$  :
\[D_{X+Y}Z-D_XZ-D_YZ=(w(X+Y,Z)-w(X,Z)-w(Y,Z))N=0\] puisque ce sont des champs de vecteurs orthogonaux égaux. D'où $D_{X+Y}Z=D_XZ+D_YZ$ et $w(X+Y,Z)=w(X,Z)+w(Y,Z)$. Idem pour l'additivité par rapport au deuxième argument.

\noindent Pour $(b)$, en utilisant la première propriété de l'item $2.$, on a :
$L_{fX}Y-fl_XY=0=\left(D_{fX}Y-fD_XY\right)+\left(w(fX,Y)-f(w(X,Y)\right)N$ d'où $D_{fX}Y-fD_XY=(w(fX,Y)-f(w(X,Y))N=0$ puisque ce sont des champs de vecteurs orthogonaux égaux. Voilà pour la première égalité.

Pour la seconde, on utilise la deuxième égalité de l'item $2.$, ce qui donne :

$D_X(fY)+w(X,fY)N=(X.f)Y+f(D_XY+w(X,Y)N)$, d'où $(D_X(fY)-fD_XY-(X.f)Y=(fw(X,Y)-w(X,fY))N=0$  puisque ce sont des champs de vecteurs orthogonaux égaux. 

\noindent $(c)$ est une conséquence immédiate de l'item $3.$. puisque $\la Z,N\ra=\la X,N\ra=0$.

\noindent Pour $(d)$, on utilise l'item $4.$ : 

\noindent$L_XY=L_YX=[X,Y]=D_XY-D_YX+\left(w(X,Y)-w(Y,X)\right)N$.
D'où

\noindent $D_XY-D_YX-[X,Y]=\left(w(X,Y)-w(Y,X)\right)N=0$ puisque ce sont des champs de vecteurs orthogonaux égaux, ce qui démontre en même temps $(d)$ et $(e)$.
\end{sol}
   \begin{sol}{de l'exercice \ref{cotan}}
   
   $1$) Pour tout $j$, $dy^j=\dfrac{\partial y^j}{\partial x^{i}}dx^{i}$. D'où la relation matricielle annoncée.
   
    $2$) Ceci montre que la famille $\{dx^{i}\}$ est une base locale du fibré cotangent et l'exercice \ref{bofbof} montre que cette base est la base duale d'une base de $\rr^n$ (fibre du fibré tangent) donc une base de $1$-formes linéaire sur $\rr^n$ ou encore une base de $\Lambda^1(\rr^n)$( fibre du fibré cotangent). Ainsi localement toute $1$-forme $\al^1$ sur une variété $M$ de dimension $n$ s'écrira 
    \[\al^1=\sum_{i=1}^na_i dx^{i}.\] De même l'écriture locale de toute $p$-forme $\al^p$ sur $M$ peut s'écrire :
    \[\al^p=\sum_{i_1<i_2<\ldots<i_p} a_{i_1\ldots i_p}dx^{i_1}\wedge\ldots\wedge dx^{i_p}.\]
    \end{sol}
   
    \begin{sol}{de l'exercice \ref{exau2}}
 Soit $H_1$ et $H_2$ deux morphisme de $\tau$ vers $\tau'$ au-dessus de $h$ et $\lambda$ un réel.
 Soit $e\in E$ tel que $p(e)=b$. Alors $H_1(e)$ et $H_2(e)$ appartiennent à la fibre ${p'}^{-1}(h(b))$ qui a une structure d'espace vectoriel. Donc $H_1(e)+\lambda H_2(e)$ est défini et appartient à ${p'}^{-1}(h(b))$. Ceci définit $H_1+\lambda H_2$. Les axiomes d'espace vectoriel sur $\rr$ sont élémentaires à vérifier.
  \end{sol}
 
  \begin{sol}{de l'exercice \ref{secsec}}
  Rappelons que si $E_1$ est un espace vectoriel dont une base est $ \{e_{1i}\}$,  $E_2$ un espace vectoriel dont une base est $\{e_{2j}\}$, si $f_1\in \opop(E_1)$ et $f_2\in \opop(E_2)$, alors l'application $f_1\otimes f_2$  se définit de la façon suivante :  elle est définie sur la base la base $\{e_{1i}\otimes  e_{2j}\}$ de $E_1\otimes E_2$ par 
  \[(f_1\otimes f_2)(e_{1i}\otimes  e_{2j})=f_1(e_{1i})\otimes f_2(e_{2j})\]
  et  est étendue par linéarité sur  $E_1\otimes E_2$.
  Cette définition montre directement que $f_1\otimes f_2$ est inversible si $f_1$ et $f_2$ le sont,
  et $(f_1\otimes f_2)^{-1}=f_1^{-1}\otimes f_2^{-1}$.
  
  De ceci découle que $\{c_{ij}\otimes c'_{i'j'}$ est à valeurs dans $\mathcal{G}(F\otimes F')$.
  Par ailleurs  la relation évidente $(c_{ij}\otimes c'_{i'j'})\circ (c_{jk}\otimes c'_{j'k'})=c_{ik}\otimes c'_{i'k'}$
  montre que $\{c_{ij}\otimes c'_{i'j'}\}$ est un cocycle.
 \end{sol}
   \begin{sol}{de l'exercice \ref{pupulette}}
 Si $\al=a_{i_1\ldots i_p}dy^{i_1}\wedge\ldots\wedge dy^{i_p}$, alors $a_{i_1\ldots i_p}=\al(\frac{\partial}{\partial y^{i_1}},\ldots,\frac{\partial}{\partial y^{i_p}})$. 
 \'{E}valuons le réel $f^*\al(\frac{\partial}{\partial x^{i_1}},\ldots,\frac{\partial}{\partial x^{i_p}}).$ L'algèbre linéaire nous enseigne que $df\left(\frac{\partial}{\partial x^{i_s}}\right)=\frac{\partial f^{i}}{\partial x^{i_s}}\frac{\partial}{\partial y^{i}}$
 (interprétation des colonnes d'une matrice associée à une application linéaire). On a donc :
 
 $f^*\al(\frac{\partial}{\partial x^{i_1}},\ldots,\frac{\partial}{\partial x^{i_p}})=\al\left(df(\frac{\partial}{\partial   x^{i_1}}),\ldots,df(\frac{\partial}{\partial   x^{i_p}})\right)=
 \al\left(\frac{\partial f^{i}}{\partial x^{i_1}}\frac{\partial}{\partial y^{i}},\ldots,\frac{\partial f^{i}}{\partial x^{i_p}}\frac{\partial}{\partial y^{i}}\right)=$
 
 \noindent $=\left\vert\frac{\partial(f^{k_1}\ldots f^{k_p})}{\partial(x^{i_1}\ldots x^{i_p})}\right\vert\al\left(\frac{\partial}{\partial y^{k_1}},\ldots,\frac{\partial}{\partial y^{k_p}}\right)=\left\vert\frac{\partial(f^{k_1}\ldots f^{k_p})}{\partial(x^{i_1}\ldots x^{i_p})}\right\vert a_{k_1\ldots k_p}.$
 
 \noindent D'où $f^*\al =\left\vert\frac{\partial(f^{k_1}\ldots f^{k_p})}{\partial(x^{i_1}\ldots x^{i_p})}\right\vert a_{k_1\ldots k_p} dx^{i_1}\wedge\ldots\wedge dx^{i_p}$.

 \end{sol}
 \begin{sol}{de l'exercice \ref{exercice}}
 \noindent Avec les notations de la définition \ref{assocpul}, on a :
   $\left(f^*\left(i(Y)\alpha\right)\right)(x)\left(X_1(x),\ldots,X_{p-1}(x)\right)$
   
   \noindent$=i(Y)\alpha(f(x))\left(df(x)\big(X_1(x)\big),\ldots,df(x)\big(X_{p-1}(x)\right)=$
   
   \noindent $=\alpha(f(x))\left(df(x)\big(X(x)\big),df(x)\big(X_1(x)\big),\ldots,df(x)\big(X_{p-1}(x)\big)\right)$ et par ailleurs
   
   \noindent $\left(i(X)f^*\alpha \right)(x)\left(X_1(x),\ldots,X_{p-1}(x)\right)=(f^*\alpha)\left(X(x),
 X_1(x),\ldots,X_{p-1}(x)\right)=$
 
 \noindent$=\alpha(f(x))\left(df(x)\big(X(x)\big),df(x)\big(X_1(x)\big),\ldots,df(x)\big(X_{p-1}(x)\big)\right)$.
 
 Ce qui permet de  conclure.
 \end{sol}
 \begin{sol}{de l'exercice \ref{cucu}}
 On considère $x$ un point de $\partial N$, et $U$ un domaine ouvert d'une carte locale de $M$ de sorte que 
 $(x^1,\ldots,x^n)$ soient des coordonnées locales sur $U$ et $U\cap N$ et que $\partial N\cap U$ soit les points de $N\cap U$ tels que $x^n=0$. On peut écrire sur ce voisinage :
 $T=\sum_{i=1}^nT^{i}\frac{\partial}{\partial x^{i}}$ et $\tilde T=T+T^n\frac{\partial}{\partial x^{n}}$ où $T^n(y)=0$ si et seulement si $y\in \partial N$. Alors $i(\tilde T)\omega=\sum_{i=1}^n(-1)^{i-1}T^{i}dx^1\wedge\ldots\wedge \widehat{dx^{i}}\wedge\ldots dx^n$ et $j^*\left(i(\tilde T)\omega\right)=(-1)^{n-1}(T^n\circ j )dx^1\wedge\ldots\wedge dx^{n-1}$ car $x^n\circ j=0$. Comme $T^n\circ j=0$ on conclut.
  \end{sol}
\begin{sol}{de l'exercice \ref{konar}}

\end{sol}
On a : $L_XY=:(L_XY^{i})e_i$ si $(e_1,\ldots,e_n)$ est la base canonique de $\rr^n$.
De plus $L_XY^{i}(x)=dY^{i}(x)(X(x))=dY^{i}(x)\left(\frac{d\varphi_t(x)}{dt}\right)=\frac{d}{dt}(Y^{i}\circ\varphi_t(x))
=\lim_{t\to 0}\frac{Y^{i}\circ\varphi_t(x)-Y^{i}(x)}{t}$, ce qui entraine directement
 $$L_XY(x)=L_XY^{i}(x)e_i=\lim_{t\to 0}\frac{Y^{i}\circ\varphi_t(x)-Y^{i}(x)}{t}e_i=\lim_{t\to 0}\frac{Y\circ\varphi_t(x)-Y(x)}{t}.$$
 \begin{sol}{de l'exercice \ref{abdel}}
On a par définition des coordonnées sphériques : pour tout $(x,y,z)\in S^2$, \[\left\{\begin{array}{rcl}
x&=&\sin\theta\cos\varphi\\y&=&\sin\theta\sin\varphi\\z&=&\cos\theta
\end{array}
\right.\]

D'où les deux égalités :

 \noindent$\frac{\partial}{\partial\theta}=\cos\theta\cos\varphi\frac{\partial}{\partial x}+\cos\theta\sin\varphi\frac{\partial}{\partial y}-\sin\theta\frac{\partial}{\partial z}$, 
 $\frac{\partial}{\partial\varphi}=-\sin\theta\sin\varphi\frac{\partial}{\partial x}+\cos\varphi \sin\theta\frac{\partial}{\partial y}$.
Rappelons que si un champ de vecteurs de $\rr^3$ est défini par $(x,y,z)\mapsto X(x,y,z)=(X^1,X^2,X^3)\in\rr^3$, il réalise la dérivation suivante sur toute fonction dérivable $f$ : $X.f=X^1\frac{\partial f}{\partial x}+X^2\frac{\partial f}{\partial y}+X^3\frac{\partial f}{\partial z}.$
Ainsi $X=(1,0,0)$ réalise la dérivation $\frac{\partial}{\partial x}$. Avec cette vision, il est clair que
la base $\left(\frac{\partial}{\partial x},\frac{\partial}{\partial y},\frac{\partial}{\partial z})\right)$ est la base canonique de $\rr^3$ et qu'elle est donc orthogonale dans $\rr^3$ avec la structure euclidienne standard. On en déduit que $g_{\theta\theta}=:\la\frac{\partial}{\partial \theta},\frac{\partial}{\partial \theta}\ra=1,$
$g_{\varphi\varphi}=\sin^2\theta$, $g_{\theta\varphi}=g_{\varphi\theta}=0$. D'où $$g=\left(\begin{matrix}1&0\\0&\sin^2\theta\end{matrix}\right).
$$
$g$ est bien une structure riemannienne sur $U$.
\end{sol}
\begin{sol}{ de l'exercice \ref{sectense}}
Tout élément du produit tensoriel $\mathcal{S}_{\tau}\otimes \mathcal{S}_{\tau^*(M)}$ des modules $\mathcal{S}_{\tau}$ et $\mathcal{S}_{\tau^*(M)}$ est une combinaison linéaire de produits $s\otimes \al$ où $s\in \mathcal{S}_{\tau}$ et $\al\in \mathcal{S}_{\tau^*(M)}$. De plus pour tout $x\in M$, 
$(s\otimes \al)(x)=s(x)\otimes \al(x)$ appartient à $p^{-1}(x)\otimes T^*_x(M)$. Ainsi $s\otimes\al$ apparaît comme une section du produit tensoriel de fibrés $\tau\otimes \tau^*(M)$. En tant que section de $\tau\otimes \tau^*(M)$, notons le $\sigma(s\otimes \al)$. On vérifie alors directement la linéarité et la bijectivité  de l'application $\sigma$ de $\mathcal{S}_{\tau}\otimes \mathcal{S}_{\tau^*(M)}$ vers $\mathcal{S}_{\tau\otimes \tau^*(M)}$ qui est donc un isomorphisme de module sur les applications différentiables.
\end{sol}
\begin{sol}{de l'exercice \ref{susucre}}
1) Pour l'égalité (\ref{sucre}), on a : $\nabla^{{U}}fs=e_{U}\otimes \left(d(f\Lambda_{U})+\omega_{U}(f\Lambda_{U})\right)=$

$f\left(e_{U}\otimes (d\Lambda_{U}+\omega_{U}\Lambda_{U})\right)+(e_{U}\otimes \Lambda_{U})\otimes df=f\nabla^{{U}}s+s\otimes df$.

2)$e_{V}\otimes (d\Lambda_{V}+\omega_{V}\Lambda_{V})=$

$e_{U} C_{UV}\otimes \left(dC_{VU}\Lambda_{U}+C_{VU} d\Lambda_{U}+(C_{VU}\omega_{U}C_{UV}+C_{VU}dC_{UV})C_{VU}\Lambda_{U}\right)=$

$e_{U}\otimes (d\Lambda_{U}+\omega_{U}C_{UV})$ car $C_{UV} dC_{VU}+dC_{UV}C_{VU}=0$.
\end{sol}
\begin{sol}{de exercice \ref{pimpim}}
L'égalité  (\ref{chacal}) entraine l'égalité (\ref{chacaux}) qui implique si on tient compte de $e_{V}=e_{U}C_{UV} $ :

\noindent$d\Lambda_{U}+\omega_{U}\Lambda_{U}=C_{UV}(d\Lambda_{V}+\omega_{V}\Lambda_{V})$. D'où 
$C_{UV}\omega_{V}\Lambda_{V}=d\Lambda_{U}+\omega_{U}\Lambda_{U}-C_{UV}d\Lambda_{V}$. Ainsi,

$\omega_{V}\Lambda_{V}=C_{VU}(dC_{UV}\Lambda_{V}+C_{UV}d\Lambda_{V})+C_{VU}\omega_{U}(c_{UV}\Lambda_{V})-d\Lambda_{V}=$

 $(C_{VU}\omega_{U}C_{UV}+C_{VU}dC_{UV})\Lambda_{V}$ et on conclut.
\end{sol}

\begin{sol}{de l'exercice \ref{suichac}}
Comme le fibré en droite est trivial on prend dans toute carte locale $U$ la base de section $x\mapsto 1$. Alors $C_{UV}=1$ et $dC_{UV}=0$.
Ce qui  équivaut au caractère intrinsèque de l'expression de la différentielle $df =\frac{\partial f}{\partial x^{i}}dx^{i}$.
\end{sol}
 \begin{sol}{de l'exercice \ref{conlevcicris}}
 \begin{enumerate}
 \item Puisque $A$ est semblable à une matrice diagonale $D=(a_{ii})$, on a $\tr(A'A^{-1})=\tr(D'D^{-1})$
 et $\ln\abs{\det A}=\ln\abs{\det D}$. Par ailleurs :
 
  $\tr (D'D^{-1})=\sum_i a'_{ii}a_{ii}^{-1}=\sum_i(\ln \abs{a_{ii}})'=
 (\ln\prod_i \abs{a_{ii}})'=(\ln\abs{\det D})'$. Et on conclut. 
 \item On sait que la matrice $G$ est diagonalisables dans une base orthonormée et que $\det G\not= 0$. On peut donc appliquer ce qui précède.
 \item D'après l'égalité (\ref{cristo}), on a : 
 
 $\Gamma^{i}_{ij}=\frac{1}{2}g^{i\mu}\left(\frac{\partial g_{i\mu}}{\partial x^j}+\frac{\partial g_{\mu j}}{\partial x^{i}}-\frac{\partial g_{ij}}{\partial x^{\mu}}\right)=\frac{1}{2}g^{i\mu}\frac{\partial g_{i\mu}}{\partial x^j}$\hspace{1cm}(car $g^{i\mu}\frac{\partial g_{\mu j}}{\partial x^{i}}=g^{i\mu}\frac{\partial g_{ij}}{\partial x^{\mu}}$)
 
 $ =\frac{1}{2}\tr\left(G^{-1}\frac{\partial G}{\partial x^j}\right)=\frac{1}{2}\tr\left(\frac{\partial G}{\partial x^j}G^{-1}\right).$
D'où l'égalité (\ref{goodform}).
 \end{enumerate}
 \end{sol}
 \begin{sol}{de l'exercice \ref{trescool}}
 Le transporté parallèle de $X(t)$ le long de $\gamma^{-1}$ en $\gamma^{-1}(1)=\gamma (0)$ est un champ $Y$ le long de $\gamma^{-1}$ dont les coordonnées dans un système de coordonnées locales vérifient les équations (\ref{paral}), c.à d. 
\[\frac{dY^{i}}{dt}+\Gamma^{i}_{jk}\frac{d(\gamma^{-1})^{i}}{dt}Y^k=0.\]
On vérifie immédiatement que $Y(\gamma^{-1}(t))=:Y(t)=X(1-t)$ vérifie l'équation précédente. Le transporté parallèle de $Y\left((\gamma^{-1}(t)\right)=X(1-t)$ en $\gamma^{-1}(1)$ est $Y(\gamma^{-1}(1))=Y(1)=X(0).$ 
 \end{sol}
\begin{sol}{de l'exercice \ref{poum}}
De l'équation (\ref{pim}), on a ${g_{ij}}_{/k}=\mfrac{\partial g_{ij}}{\partial x^{k}}-\Gamma^{s}_{ki}g_{sj}-\Gamma^{s}_{kj}g_{is}$. On en déduit grâce à l'équation (\ref{christo}) : ${g_{ij}}_{/k}=\Gamma ^{s}_{ki}g_{sj}+\Gamma^{s}_{kj}g_{si}-\Gamma^{s}_{ki}g_{sj}-\Gamma^{s}_{kj}g_{is}=0$
\end{sol}
\begin{sol}{de l'exercice \ref{divdivergence}}
\begin{enumerate}
\item
La symétrie résulte de la remarque \ref{contderi}.
Puisque $T$ est un tenseur on a les transformations 
\[ T^{ii_2\ldots i_n}_{j_1\ldots j_p}=g^{k_1}_{j_1}\ldots A^{k_p}_{j_p} A^{i}_{l_1}A^{i_2}_{l_2}\ldots A^{i_n}_{l_n} \;T^{l_1\ldots l_n}_{k_1\ldots k_p}\] où $(A^{i}_j)$ est la jacobienne de l'application changement de carte. Voir par exemple [\ref{franki}] p. $62$.

\noindent Cette relation exprime que la famille de réels $\{T^{ii_2\ldots i_n}_{j_1\ldots j_p}\}$ sont les composantes dans une carte locale d'un tenseur et donne le liens avec les composantes dans une autre carte locale.

\noindent Pour montrer que les scalaires $\{(\Div T)^{i_2\ldots i_n}_{j_1\ldots j_p}\}$ définissent un tenseur il suffit de montrer que 
\[(\Div T)^{i_2\ldots i_n}_{j_1\ldots j_p}=A^{k_1}_{j_1}\ldots A^{k_p}_{j_p} A^{i_2}_{l_2}\ldots A^{i_n}_{l_n} \;(\Div T)^{l_2\ldots l_n}_{k_1\ldots k_p}.\]

En dérivant et en tenant comte de la relation $\nabla g=0$, on a :

${T^{ii_2\ldots i_n}_{j_1\ldots j_p}}_{/i}=A^{k_1}_{j_1}\ldots A^{k_p}_{j_p} A^{i}_{l_1}A^{i_2}_{l_2}\ldots A^{i_n}_{l_n} \;{T^{l_1\ldots l_n}_{k_1\ldots k_p}}_{/i}=A^{k_1}_{j_1}\ldots A^{k_p}_{j_p} A^{i_2}_{l_2}\ldots A^{i_n}_{l_n} \;\left(g^{i}_{l_1}{T^{l_1\ldots l_n}_{k_1\ldots k_p}}\right)_{/i}$

$=A^{k_1}_{j_1}\ldots A^{k_p}_{j_p} A^{i_2}_{l_2}\ldots A^{i_n}_{l_n} \;{T^{il_2\ldots l_n}_{k_1\ldots k_p}}_{/i}$, ce qui achève la démonstration.
\item Avec les notations de l'énoncé on a directement 

\noindent $i(T)\omega=T^0 dx^1\wedge dx^21\wedge dx^3-T^1 dx^2\wedge dx^3\wedge dx^0+T^2 dx^3\wedge dx^0\wedge dx^1-T^3 dx^0\wedge dx^2\wedge dx^3$. 
D'où $di(T)\omega=\left(\frac{\partial T^0}{\partial x^0}+\frac{\partial T^1}{\partial x^1}+\frac{\partial T^2}{\partial x^2}+\frac{\partial T^3}{\partial x^3}\right)\omega$ et on conclut.
\end{enumerate}
\end{sol}

\begin{sol}{de l'exercice \ref{clebard}}
\begin{enumerate}
\item  On prend $c=1$. Une transformation orthogonale est représentée dans la base canonique de $\rr^2$ par une matrice $\left(\begin{matrix}\al&\be\\\gamma&\delta\end{matrix}\right)$ telle que $\al^2-\gamma^2=-1,\be^2-\delta^2=1,-\al\be+\gamma\delta=0$. On en déduit qu'il existe deux réels $\theta$ et $\theta'$ et $4$ réels $\varepsilon_1,
\varepsilon_2,\varepsilon_3,\varepsilon_4$ appartenant à $\{-1,1\}$ tels que :
$\al=\varepsilon_1 \sinh\theta,\gamma=\varepsilon_2 \cosh\theta,\delta=\varepsilon_3 \sinh\theta',
\be=\varepsilon_4\cosh\theta'$ et $\varepsilon_1\varepsilon_4\cosh\theta'\sinh\theta-\varepsilon_2\varepsilon_3\sinh\theta'\cosh\theta=0$.
On peut distinguer deux cas : ($i$) $\al=\sinh \theta$, ($ii$) $\al=-\sinh\theta$.
En discutant toutes les possibilités dans chacun de ces deux cas, discussion laborieuse mais élémentaire, on finit par mettre en évidence les deux types annoncés.
Considérons le  \textbf{premier cas } : $\varepsilon _1=1$.

\textbf{Premier sous-cas} :  $\varepsilon_2=1$

Si $(\varepsilon _3,\varepsilon _4)=(1,1)$, alors $\theta=\theta'$ et on a une symétrie.\\
Si $(-\varepsilon _3,\varepsilon _4)=(1,1)$, alors $\theta=-\theta'$ et on a une symétrie.\\
Si $(\varepsilon _3,-\varepsilon _4)=(1,1)$, alors $\theta=-\theta'$ et on a une rotation.\\
Si $(-\varepsilon _3,-\varepsilon _4)=(1,1)$, alors $\theta=\theta'$ et on a une  rotation.

\textbf{Deuxième sous-cas} :  $\varepsilon _2=-1$

Si $(\varepsilon _3,\varepsilon _4)=(1,1)$, alors $\theta=-\theta'$ et on a une rotation.\\
Si $(-\varepsilon _3,\varepsilon _4)=(1,1)$, alors $\theta=\theta'$ et on a une rotation.\\
Si $(\varepsilon _3,-\varepsilon _4)=(1,1)$, alors $\theta=\theta'$ et on a une symétrie\\
Si $(-\varepsilon _3,-\varepsilon _4)=(1,1)$, alors $\theta=-\theta'$ et on a une  symétrie

Le \textbf{deuxième cas}, $a=-\cosh\theta$, i.e. $\varepsilon _1=-1$ se traite de manière équivalente.
\item 
$(a)$ La transformation faisant passer des coordonnées galiléennes $(ct',x')$ aux coordonnées galiléennes $(ct,x)$ est une transformation de Lorentz, donc d'après la question précédente une rotation hyperbolique ou une symétrie hyperbolique. Quand les deux référentiels coïncident cette transformation est l'identité donc il s'agit d'une rotation.

$(b)$ Il existe donc un réel $\theta$ tel que $\left\{\begin{matrix}ct'=ct\cosh \theta +x\sinh \theta\\
x'=ct\sinh\theta+x\cosh \theta
\end{matrix}\right.$

\noindent Comme $x(O)=0$, on obtient les égalités $\left\{\begin{matrix}t'(O)=t(O)\cosh \theta \\
x'(O)=ct(O)\sinh\theta
\end{matrix}\right.$

Un observateur lié à $\mathcal{R}_0$ voit s'éloigner $O$ à la vitesse $-\frac{v}{c}=\frac{x'(O)}{t'(O)}=\tanh\theta.$ On en déduit que $\cosh\theta=\frac{1}{\sqrt{1-\frac{v^2}{c^2}}}=:\gamma$ et que $\sinh\theta =-\gamma \frac{v}{c}$, ce qui permet de conclure.
\item Un calcul direct montre avec les notations de la question $2$.b) que $cdt'\wedge dx'=c dt \wedge dx$.
\end{enumerate}
\end{sol}

\begin{sol}{de l'exercice \ref{brut}}
\begin{enumerate}
\item On reprend le système de la question 2)c) de l'exercice \ref{clebard} et on multiplie la première équation par $c$ et on remplace dans la deuxième équation $t$ par $c{x^0}$
\item L'équation de l'axe $(x^0)'$ est $(x^1)'=o$ qui s'écrit d'après la deuxième équation du système (\ref{rafu}) : $x^0=\frac{c}{v}x^1$. Idem pour l'axe $(x^1)'$.
\item Pour un observateur lié à $\mathcal{R}$, l'équation de l'ensemble des événements simultanés à $E$ s'écrit $(x^0)'=a'$
que l'on traduit avec la première équation du système (\ref{rafu}) par $x^0-\frac{v}{c}x^1=a-\frac{v}{c}b$ et on conclut.
\item Les événements $E_1$ et $E_2$ ont pour coordonnées réduites dans $\mathcal{R}$ respectivement $((\tau^1)',\al')$ et $((\tau^2)',\be')$. D'après (\ref{rafu}), on a : $(\tau^1)'=\gamma(\tau^1-\frac{v}{c}a)$ et $(\tau^2)'=\gamma(\tau^2-\frac{v}{c}a)$.
D'où $T=(\tau2)'-(\tau^1)'=\gamma T_0$.
\item Un observateur lié à $\mathcal{R}$ décide de mesurer la longueur de la règle à un instant $t'$. En inversant le système (\ref{rafu}), on a les deux relations : $a=\gamma(a'+vt')$ et $b=\gamma(b'+vt')$, d'où $l_0=\abs{b-a}=\gamma \abs{b'-a'}=\gamma l$.
\item De la question précédente on a directement $V=\gamma V_0$. La mesure de Lebesgue définie sur les boréliens de $\rr^3$ est caractérisée par ses valeurs prises sur les parallélépipèdes. On en déduit $dV=\gamma dV_0$.
\end{enumerate}
\end {sol}

\begin{sol}{de l'exercice \ref{secret}}
On a $\la u,u\ra_L=-\gamma^2+\frac{\gamma^2}{c^2}v^2=\gamma^2(\frac{v^2}{c^2}-1)=-1$.

\noindent D'où $\la u ,F\ra_L=
\frac{1}{m_0c}\la u,\frac{du}{d\tau}\ra_L=\frac{1}{2m_0c}\frac{d\la u,u\ra_L}{d\tau}=0$.
\end{sol}
\begin{sol}{de l'exercice \ref{falco}}
Supposons que $X$ soit invariant par $h$. On a pour tout $x\in M$ et tout champ de vecteur $Y$, $ h^*\al(x)(Y(x))=
g\left(X(h(x)), dh(x)\big(Y(x)\big)\right)$ car par définition $$h^*\al(x)(Y(x))=\al(h(x))\left(dh(x)(Y(x))\right).$$
Ainsi 

\noindent$ h^*\al(x)(Y(x))=g(dh(x)(X(x)), dh(x)(Y(x)))=g(X(x),Y(x))$ (car $h$ est une isométrie)

$=\al(Y)(x)$. Ce qui montre que $h^*\al=\al$. Réciproquement si $h^*\al=\al$, alors pour tout champ $Y$ et tout $x\in M$, on a :

$g\left(X(h(x)),dh(x)(Y(x))\right)=g(X(x),Y(x))=$

$=g(dh(x)X(x),dh(x)Y(x))$ (Car $h$ est une isométrie).

\noindent On a donc $ g(X(h(x)-dh(x)(X(x)),dh(x)Y(x))=0$ pour tout $Y$. On en déduit $(dh(X)-X\circ h)(x)=0$ pour tout $x$ ce qui montre l'invariance de $X$ par $h$.

 \end{sol}

\begin{sol}{de l'exercice \ref{gral}}
Notons tout d'abord que par définition du produit vectoriel on a dans $\rr^{3}$ : $\la v\times B,\;\ra=vol^{3}(v,B,\;)=-i(v)\ma{B}^{2}.$ 

On a  donc $f^{{1}}=:\la F_{\ma{ML}},\;\ra_{L}=\gamma e\big(-\frac{1}{c}\la E,v\ra dx^0+\ma{E}^{{1}}-\frac{1}{c}i(v)\ma{B}^{2}\big)$.
Et par ailleurs $ei(u)\ma{F}^{2}
=e\gamma i(\frac{\partial}{\partial x^0}+\frac{v}{c})(\ma{E}^1\wedge dx^0+\ma{B}^2)=
\gamma e\big(-\ma{E}^{{1}}+\frac{1}{c}i(v)\ma{B}^{2}+\frac{1}{c}\la v,E\ra dt\big)=-f^1$. 
\end{sol}

\begin{sol}{de l'exercice \ref{quadteconmag}}
Partant du fait que $\ma{F}_{\ma{LM}}$ est un quadrivecteur,
pour toute transformation de Lorentz $\varphi$, on a, conformément à la remarque \ref{vecoufor?}  $\varphi^* f_1=f_1$. Considérant l'égalité 
(\ref{paliezno}) et le fait que $u$ soit invariant par $\varphi$, on a : $\varphi^*\left(i(u)\mathcal{F}^2\right)=i(u)(\varphi^*\mathcal{F}^2)=i(u)\mathcal{F}^2$. Comme ceci est vrai pour tout quadrivecteur vitesse $u$, on en déduit que $\mathcal{F}^2=\varphi^{*}\mathcal{F}^2$.
Inversement si $\ma{F}^2$ est invariant par transformée de Lorentz, les égalités précédentes montrent qu'il en est de même pour $f^1$ ou encore que $\ma{F}_\ma{LM}$ est un quadrivecteur.
\end{sol}
\begin{sol}{de l'exercice \ref{ortint}}
\begin{enumerate}
\item Notons $X=X^0 \frac{\partial}{\partial t}+X^{i}\frac{\partial}{\partial x^{i}}$. Alors $\al=-X^0dt+\sum_{i=1}^3X^{i}dx^{i}$.
Posant $Y=Y^0 \frac{\partial}{\partial t}+Y^{i}\frac{\partial}{\partial x^{i}}$, on a :
$i(Y)\al=-X^0Y^0+\sum_{i=1}^3X^{i}Y^{i}=\la X,Y\ra_L$ et on conclut.
\item D'après l'égalité (\ref{lor}), $i(u)f^1=-e\mathcal{F}^2(u,u)=0$, ce qui permet de conclure d'après la question précédente.
\end{enumerate}
\end{sol}
\begin{sol}{de l'exercice \ref{gloups}}
Posons $\ma{L}(t',x',y',z')=(t,x,y,z)$. Alors $\ma{L}^*\ma{F}^2=\ma{F}^2$ s'écrit :

\noindent $\ma{L}^*\left(c(E^1dx+E^2dy+E^3dz)\wedge dt)+B^1dy\wedge dz+B^2 dz\wedge dx+B^3 dx\wedge dy\right)
=$

\noindent $=c(E'^1dx'+E'^2dy'+E'^3dz')\wedge dt')+B'^1dy'\wedge dz'+B'^2 dz'\wedge dx'+B'^3 dx'\wedge dy'$, ce qui équivaut, compte tenu de l'expression de $\ma{L}$ (voir exercice \ref{clebard}) à l'égalité :

\noindent$cE^1(\gamma^2(vdt'+dx')\wedge (dt'+\frac{v}{c^2}dx')+c\gamma E^2 dy'\wedge(dt'+\frac{v}{c^2}dx')+c\gamma E^3dz'\wedge(dt'+\frac{v}{c^2}dx')+B^1 dy'\wedge dz'+\gamma B^2 dz'\wedge(vdt'+dx')+\gamma B^3(vdt'+dx')\wedge dy'=c(E'^1dx'+E'^2dy'+E'^3dz')\wedge dt')+B'^1dy'\wedge dz'+B'^2 dz'\wedge dx'+B'^3 dx'\wedge dy'$. 

En identifiant dans les deux termes précédents les coefficients de la base $\{dt'\wedge dx',dt'\wedge dy',dt'\wedge dz',dx'\wedge dy',dx'\wedge dz',dy'\wedge dz'\}$, on obtient les relations annoncées.

\end{sol}
\begin{sol}{de l'exercice \ref{rougi}}
\noindent $d\al=\left(\frac{\partial a_{ij}}{\partial x^k}dx^k\wedge dx^{i}\wedge dx^j+\frac{\partial a_{0i}}{\partial x^k}dx^k\wedge dt\wedge dx^{i}\right)+\frac{\partial a_{ij}}{\partial t}dt\wedge dx^{i}\wedge dx^j.$

\noindent$=\left(\frac{\partial a_{ij}}{\partial x^k}dx^k\wedge dx^{i}\wedge dx^j+\frac{\partial a_{oi}}{\partial x^k}dx^k\wedge dt\wedge dx^{i}\right)+\left(\frac{\partial a_{ij}}{\partial t} dx^{i}\wedge dx^j\right)\wedge dt.$

\noindent$=\bold{d}\al +\frac{\partial \al}{\partial t}\wedge dt=\bold{d}\al +dt\wedge\frac{\partial \al}{\partial t}$
\end{sol}

\begin{sol}{de l'exercice \ref{klaciko}}
\begin{enumerate}
\item En écrivant $X=X^1 \frac{\partial}{\partial x}+X^2\frac{\partial}{\partial y}+X^3\frac{\partial}{\partial z}$ et $Y=Y^1 \frac{\partial}{\partial x}+Y^2\frac{\partial}{\partial y}+Y^3\frac{\partial}{\partial z}$, on a
$\zeta=X^1dx+X^2dy+X^3dz$ et l'égalité $d\zeta=i(Y)\omega $ s'écrit :
\[
\left\{\begin{array}{ ccc}
Y^1&=&\frac{\partial X^3}{\partial x}-\frac{\partial X^2}{\partial z}\\
Y^2&=&\frac{\partial X^1}{\partial y}-\frac{\partial X^3}{\partial x}\\
Y^3&=&\frac{\partial X^2}{\partial z}-\frac{\partial X^1}{\partial y}
\end{array}
\right.\]

ce qui permet de conclure.
\item $d\xi$ est une $3$-forme ainsi que $\omega$. Il en résulte que ces formes sont proportionnelles.
\item D'après la première équation de Maxwell, $d{\mathcal{E}}^1=i(\rot E)\omega=-\frac{1}{c}i(\frac{\partial B}{\partial t})\omega$. D'où l'égalité $\rot E=-\frac{1}{c}\frac{\partial B}{\partial t}$. De même selon la deuxième équation de Maxwell, $\bold{d}\mathcal{B}^2=0$, ce qui  équivaut à $\Div B=0$
\end{enumerate}
\end{sol}
\begin{sol}{de l'exercice \ref{contrac}}
\begin{enumerate}
\item Cette première question est un prétexte pour réviser quelques notions de base du calcul tensoriel.
Plaçons nous sur un ouvert $U$ de la variété possédant une base locale $e=(e_1,\ldots,e_n)$ de base duale 
$\epsilon=(\epsilon^1,\ldots,\epsilon^n)$. Donnons-nous sur cet ouvert une deuxième base $e=(e'_1,\ldots,e'_n)$
de base duale 
$\epsilon'=(\epsilon'^1,\ldots,\epsilon'^n)$. Si $X$ est un champ de vecteurs  et $\al$ une $1$-forme sur la variété leurs restrictions à $U$ s'écrit $X=\lambda^{i}e_i=
\lambda'^{a}e'_a$ et $\al=\xi_i\epsilon^{i}=\xi'_{b}{\epsilon'}^{b}$. Si on note $A=(A^{i}_a)$ la matrice de passage de $e$ à $e'$, $\Lambda,\Lambda'$ les matrices colonnes des composantes de $X$ dans les bases $e$ et $e'$, et de même $\Xi,\Xi'$ les matrices colonnes des composantes de $\al$ dans les bases $\epsilon $ et $\epsilon'$,  on sait que $\Lambda=A\Lambda'$ et $\Xi=A^{-1}\Xi'$. Cette dernière égalité traduit les équations $\xi_{i}=A_i^b\xi_b$ si l' pose $A^{i}_aA^{a}_j=\delta^{i}_j$.
Avec cette écriture les changements de coordonnées d'un tenseur $T=T^{i}_{jkl}e_i\otimes \epsilon^j\otimes\epsilon^k\otimes \epsilon ^l=T'^{a}_{bcd}e'_a\otimes \epsilon'^b\otimes\epsilon'c\otimes \epsilon '^d$ s'écrivent $T^{i}_{jkl}=A^{i}_aA^{b}_jA^{c}_kA^{d}_lT'^{a}_{bcd}$. De même pour un tenseur $\tau=\tau_{ij}\epsilon^{i}\otimes \epsilon^j$ on a : $\tau_{ij}=A_i^{a}A_j^{b}\tau'_{ab}$. Un critère de tensorialité classique (voir [\ref{rinrin}] ou [\ref{lichné}]) dit qu'une famille de nombres $\tau_{ij}$ définit bien un tenseur deux fois covariant  exprimé par l'égalité $\tau=\tau_{ij}\epsilon^{i}\otimes \epsilon^j$ dans la base $e$ si 
$\tau_{ij}\lambda_1^{i}\lambda_2^j$ est un scalaire pour tous champs de vecteurs $\lambda_1^{i}e_i,\lambda_2^{i}e_i$. Utilisons ce critère et montrons que 
$T_{ij}\lambda_1^{i}\lambda_2^j=T'_{ab}{\lambda'}_1^{a}{\lambda'_2}^b$. On a :

 \hspace{1cm}$T_{ij}\lambda_1^{i}\lambda_2^j=T^k_{ikj}\lambda_1^{i}\lambda_2^j=(A^k_aA_i^{b}A_k^cA_j^d{T'}^{a}_{bcd})(A^{i}_eA^j_f\lambda^{e}\lambda^f)=...$

\noindent $...={T'}^{a}_{bcd}\lambda^{e}\lambda^f\delta^b_e\delta^d_fA\delta^{a}_c={T'}^{a}_{bad}\lambda^{b}\lambda^c={T'}_{bd}\lambda^{b}\lambda^c$.
\item Notons $c$ la contraction des indices $i$ et $k$ dans $T^{i}_{jkl}$ et montrons que $\left( c(T^{i}_{jkl})\right)_{/s}=c\left({T^{i}_{jkl}}_{/s}\right)$.
On a : $\left( c(T^{i}_{jkl})\right)_{/s}={T_{ij}}_{/s}=\frac{\partial T_{ij}}{\partial x^s}-\omega^\al_{si}T_{\al j}-\omega^\al_{sj}T_{i\al }=\frac{\partial T^k_{ikj}}{\partial x^s}-\omega^\al_{si}T^k_{\al k j}-\omega^\al_{sj}T^k_{i k\al }$.
Par ailleurs, 

${T^{k}_{ilj}}_{/s}=\frac{\partial T^k_{ilj}}{\partial x^s}+\omega^k_{s\al}T^\al_{ilj}-\omega^\al_{si}T^k_{\al lj}
-\omega^\al_{sl}T^k_{i\al j}-\omega^\al_{sj}T^k_{il\al }$. D'où :

$c\left({T^{i}_{jkl}}_{/s}\right)=\frac{\partial T_{ij}}{\partial x^s}+\underbrace{\omega^k_{s\al}T^\al_{ikj}-\omega^\al_{sk}T^k_{i\al j}}_{=0}-\omega^\al_{si}T^k_{\al kj}-\omega^\al_{sj}T^k_{ik\al }$. Et on conclut.
\end{enumerate}
\end{sol}
\begin{sol}{de l'exercice \ref{cuscacon}}
$R^{i}_i=g^{ik}R_{ik}=:R$
\end{sol}
\begin{sol}{de l'exercice \ref{versrelati}}
On a :

\noindent$G(e_i,e_i)=..$

$=\epsilon_i\sum_{\{j;j\not=i\}}K(e_i,e_j)-\frac{\epsilon_i}{2}\left(\sum_{\{j;j\not=i\}}K(e_i,e_j)-\sum_{\{k;k\not=i\}}\sum_{\{j;j\not=k\}}K(e_k,e_j)\right)$

\vspace {0.25cm}

$=\frac{\epsilon_i}{2}\sum_{\{j;j\not=i\}}K(e_i,e_j)-\frac{\epsilon_i}{2}\left(\sum_{\{k;k\not=i\}}K(e_k,e_i)+\sum_{\{(k,j);k\not=i,k\not=i,j\not=i\}}K(e_k,e_j)\right)$

\vspace {0.25cm}

$=-\frac{\epsilon_i}{2}\sum_{\{(k,j);k\not=i,k\not=i,j\not=i\}}K(e_k,e_j)=
-\epsilon_i\sum_{\{(k,j);k<j,k\not=i,j\not=i\}}K(e_k,e_j).$
\end{sol}


\begin{sol}{de l'exercice \ref{topfi}}
Cet exercice est un prétexte de révision de topologie générale. On dispose d'une topologie sur un ensemble $E$ lorsque qu'on a défini sur celui-ci une famille d'\emph{ouverts}. Une famille $\mathcal{O}=\{\mathcal{O}\}_{i\in I}$ 
de sous-ensembles de $E$ est une famille d'ouverts si $i)$ $E\in \mathcal{O}, \emptyset\in \mathcal{O}$, $ii)$ toute intersection d'un nombre fini de parties de $\mathcal{O}$ appartient à $\mathcal{O}$, $iii)$ une réunion quelconque de parties de $\mathcal{O}$ appartient à $\mathcal{O}$. 
\noindent 
Le couple $(E,\mathcal{O})$ est un espace topologique.

 \noindent C'est la structure minimale pour définir la continuité : une application de $E$ vers $F$ où $E$ et $F$ sont dotés d'une topologie est \emph{continue} si l'image réciproque par $f$ de tout ouvert de $F$ est un ouvert de $E$.
De même on dit que $f$ est \emph{ouverte} si l'image directe de tout ouvert de $E$ est un ouvert de $E$.

\noindent Rajoutons à ce rappel que si $f$ est une application de l'ensemble $E$ vers l'ensemble $F$ et si $\{A_i\}_i$ est une famille de parties de $F$, alors :
\[f^{-1}\left(\bigcap_{j\in J}A_j\right)=
\bigcap_{j\in J}\left(f^{-1}(A_j)\right)\;\mathrm{ et }\;f^{-1}\left(\bigcup_{j\in J}A_j\right)=
\bigcup_{j\in J}\left(f^{-1}(A_j)\right).\]
Une fois ces rappels faits l'exercice est immédiat.
\end{sol}
\begin{sol}{de l'exercice \ref{toqo}}
Cette assertion est une conséquence directe de l'exercice \ref{topfi}.
\end{sol}
\begin{sol}{de l'exercice \ref{sature}}
Supposons que $q$ soit ouverte. Si $A$ est un ouvert de $E$, alors $q(A)$ est un ouvert de $E/\sim$ et donc par définition de la topologie quotient $\mathrm{Sat}(A)=q^{-1}\big(q(A)\big)$ est un ouvert de $E$.

Supposons que pour tout ouvert $A$ de $E$, l'ensemble $q^{-1}\big(q(A)\big)$ soit  un ouvert de $E$, alors $q(A)$ est un ouvert de $E/\sim$ donc $q$ est ouverte.

\end{sol}
\begin{sol}{de l'exercice \ref{efisoso}}
On considère l'application $f$ de $G$ vers $\mathcal{H}(E)$ définie par $f(g) : x\in E\mapsto g.x$.
Cette application est un morphisme de groupe. En effet, pour tout $x\in E$, on a : $f(g_1g_2)(x)= (g_1g_2).(x)=g_1.(g_2.x)=
\big(f(g_1)\circ f(g_2)\big)(x)$ d'où l'égalité : $\forall g_1,g_2 \in G,\; f(g_1g_2)=f(g_1)\circ f(g_2)$.
\'{E}valuons $\ker f$. Un élément $g$ de $G$ appartient à $\ker f$ si et seulement si  $f(g)=Id(E)$ c'est à dire : pour tout $x$ de $E$ on a $g.x=x$, ou encore $g\in G_x$ pour tout $x$. Puisque l'opération est effective, on a $g=e$ et $f$ est injective. C'est donc un isomorphisme de groupe de $G$ sur $f(G)$.
Et on conclut.
\end{sol}

\begin{sol}{de l'exercice \ref{gibus}}
On note dans la définition \ref{morficus} respectivement 
$\{c_{ij}\}$ et $\{c'_{ij}\}$ le cocycle associé aux recouvrements $\{(U_i,\Phi_i)\}$ et $\{(U'_j ,\Phi'_j)\}$.
On a les égalités :
\begin{eqnarray}
\forall x\in h^{-1}(U'_i\cap U'_j)\cap U_k,\; c'_{ij}(h(x))\circ h_{jk}(x)&=&h_{ik}(x)\label{sot}\\
\forall x\in h^{-1}(U'_i)\cap U_j\cap U_k,\; h_{ij}(x)\circ c_{jk}(x)&=&h_{ik}(x)\label{sotte}
\end{eqnarray}
Comme les $h_{ij}(x)$ sont inversibles, posons $h_{ij}(x)^{-1}=\tilde{h}_{ji}(h(x))$. Alors les équations (\ref{sot}) et (\ref{sotte}) se réécrivent
\begin{eqnarray*}
\forall x\in h^{-1}(U'_i\cap U'_j)\cap U_k,\;  \tilde{h}_{ki}(h(x))\circ c'_{ij}(h(x))&=&\tilde{h}_{kj}(h(x))\\
\forall x\in h^{-1}(U'_i)\cap U_j\cap U_k,\; c_{jk}(x)\circ \tilde{h}_{ki}(h(x))&=&\tilde{h}_{ji}(h(x))
\end{eqnarray*}
Qui du fait de la bijectivité de $h$ s'écrivent de façon équivalente :
\begin{eqnarray}
\forall x'\in U'_i\cap U'_j)\cap h(U_k),\;  \tilde{h}_{ki}(x')\circ c'_{ij}(x')&=&\tilde{h}_{kj}(x')\label{sots}\\
\forall x'\in U'_i\cap h( U_j\cap U_k),\; c_{jk}(h^{-1}(x'))\circ \tilde{h}_{ki}(x')&=&\tilde{h}_{ji}(x')\label{sottes}
\end{eqnarray}
Les égalités (\ref{sots}) et (\ref{sottes}) montrent, d'après la proposition \ref{drac}, que la famille $\{\tilde{h}_{ij}\}$ définit un morphisme $K$ du fibré $\tau'$ vers le fibré $\tau$ au-dessus de $h^{-1}$. 
On vérifie alors directement à partir du deuxième item de la définition \ref{morficus} que $H\circ K$ est le morphisme identité de $\tau'$ et $K\circ H$ le morphisme identité de $\tau$.
\end{sol}

\begin{sol}{de l'exercice \ref{hopfi}} 
On observe que $\varphi_{{V}}^{-1}$ est une bijection de $\cc\setminus\{0\}$ sur $S^2\setminus \{(0,0,1),(0,0,-1)\}$. Soit $w\in\cc\setminus\{0\}$. On a $\varphi_{U}^{-1}\left(\frac{1}{w}\right)=
\frac{\left(\frac{1}{\vert w\vert^2}-1,\frac{2}{w}\right)}{1+\frac{1}{\vert w\vert^2}}=\frac{(1-\vert w\vert^2,2\bar{w})}{1+\vert w\vert^2}=\varphi_{V}^{-1}(w)$ et on conclut.
\end{sol}
\begin{sol}{de l'exercice \ref{smutch}}
Plaçons-nous sur une carte locale  $(U,x)$. On note $\mathcal{S}_n$ le groupe des permutations de $\{1,2,\ldots,n\}$.

\noindent Par la définition \ref{canfori}, on a $\sigma (\frac{\partial}{\partial x^{1}},\ldots,\frac{\partial}{\partial x^{n}})=\epsilon
\sqrt{\abs{{\det g}}}$.
Puisque $X_i={X_i}^k\frac{\partial}{\partial x^k}$, on a :

\noindent $\sigma(X_1,\ldots,X_n)=X_1^{i_1}\ldots X_n^{i_n}
\;\sigma\left(\frac{\partial}{\partial x^{i_1}},\ldots,\frac{\partial}{\partial x^{i_n}}\right)=\sum_{s\in \mathcal{S}_n}
\epsilon(s)X_1^{s(1)}\ldots X_n^{s(n)} \sigma\left(\frac{\partial}{\partial x^{1}},\ldots,\frac{\partial}{\partial x^{n}}\right)$

\noindent $=\sqrt{\abs{\det g}}\;\sum_{s\in \mathcal{S}_n}
\epsilon(s)X_1^{s(1)}\ldots X_n^{s(n)}=\sqrt{\abs{\det g}}\det (X_i^j).$
On peut donc écrire 
\[\sigma(X_1,\ldots,X_n)\sigma(Y_1,\ldots,Y_n)=\abs{\det g} \;\det (X_i^j)\;\det (Y_i^j).\]
Or $g(X_i,Y_j)=X_i^\alpha Y_j^\beta g_{\alpha\beta}$. On reconnait là le terme général du produit de trois matrices et par conséquent $\det\left(g(X_i,Y_j)\right)=\det g\;\det \left(X_i^\alpha\right)\;\det \left(Y_j^\beta\right).$ Ainsi la forme volume riemanienne vérifie l'équation (\ref{smol}).

En prenant $(X^1,\ldots,X^n)=(Y^1,\ldots,Y^n)=\left(\frac{\partial}{\partial x^{1}},\ldots,\frac{\partial}{\partial x^{n}}\right)$, on obtient :
 \[\omega\left(\frac{\partial}{\partial x^{1}},\ldots,\frac{\partial}{\partial x^{n}}\right)=\sqrt{\abs{\det g}},\] ce qui montre l'unicité.
\end{sol}
\begin{sol}{de l'exercice \ref{proj}}
$i$) Au $(n+1)$-uplet  $\tilde{u}= (u^1,\ldots,u^{i-1},1,u^{i+1},\ldots,u^{n+1})$ on associe le $n$-uplet $u=(u^1,\ldots,u^{i-1},u^{i+1},\ldots,u^{n+1})$.
On a avec les notations du texte, $\varphi_i^{-1}(u)=q(\tilde{u})$ et de même :
$\varphi_j^{-1}\left(\frac{u^1}{u^j},\ldots,\frac{u^{j-1}}{u^j},\frac{u^{j+1}}{u^j},\ldots,\frac{u^{n+1}}{u^j}\right)=q\left(\frac{u^1}{u^j},\ldots,\frac{u^{j-1}}{u^j},1,\frac{u^{j+1}}{u^j},\ldots,\frac{u^{n+1}}{u^j}\right)
=q(\tilde{u}),$ ce qui permet de conclure pour la première question.

$ii$) Regardons deux cas particuliers pour se convaincre. 

Premier cas : Examinons le changement de cartes $ \varphi_2\circ\varphi_1^{-1}$ sur $\rr P^2$.

Alors  $u=(u^2,u^3)$. On a donc $\varphi_2\varphi_1^{-1}(u)=(\frac{1}{u^2},\frac{u^3}{u^2})$.

D'où $D(\varphi_2\varphi_1)^{-1}(u)=\left(\begin{matrix} -\frac{1}{(u^2)^2}&0\\-\frac{u^3}{(u^2)^2}&\frac{1}{u^2}
\end{matrix}\right)$, dont le déterminant vaut $-\frac{1}{(u^2)^3}$.

Deuxième cas : Examinons le changement de cartes $\varphi_3\circ\varphi_1^{-1}$ sur $\rr P^3$.

Alors $u^1=1$ et $u=(u^2,u^3,u^4)$. On a donc $\varphi_3\varphi_1^{-1}(u)=(\frac{1}{u^3},\frac{u^2}{u^3},\frac{u^4}{u^3})$.

D'où $D(\varphi_3\varphi_1^{-1})(u)=\left(\begin{matrix}0& -\frac{1}{(u^3)^2}&0\\\frac{1}{u^3}&-\frac{u^2}{(u^3)^2}&0\\0&-\frac{u^4}{(u^3)^2}&\frac{1}{u^3}
\end{matrix}\right)$, dont le déterminant vaut $-\frac{1}{(u^3)^4}$.

Le lecteur voudra bien écrire le cas général.

\end{sol}

\begin{sol}{de l'exercice \ref{leder}}
\noindent On reprend les notations de la démonstration. 

Soit $u=(x,y)$ et $ U_u=p\left(]x-\frac{1}{2},x+\frac{1}{2}[\times \rr\right)$.
Alors l'application

 $\Phi_U : \pi^{-1}(U)\longrightarrow U\times \rr^2$ définie par 
$\Phi_U\left([(s,t,a,b)]\right)=\left([(s,t)],(a,b)\right)$ est un homéomorphisme.
Si $U$ et $V$ sont deux cartes non disjointes de $\mathcal{M}$, on a :

\noindent $\Phi_U^{-1}\big([(s,t)],(a,b)\big)=[(s,t,a,b)]$ et $\Phi_V^{-1}\big([(s+1,-t)],(a,-b\big)=[s+1,-t,a,-b)]=[(s,t,a,b)]$. Ainsi,
$\Phi_V\Phi_U^{-1}\big([(s,t)],(a,b)\big)=\big([(s,t)],(a,-b)\big)$, ce qui montre que les cocycles sont ceux qui déterminent le fibré tangent à $\mathcal{M}$. Ils sont décrits dans la première étape de la démonstration de la proposition \ref{mimo}. Et on conclut.
\end{sol}
\begin{sol}{de l'exercice \ref{isogigi}}
On note toujours $\sigma $ la forme volume riemannienne sur $M$.
Soit $\alpha\in\Lambda^p(M)$. Plaçons nous sur un domaine ouvert d'une carte locale sur lequel existe une base orthonormée de champs de vecteurs. Notons $(\omega^1,\ldots,\omega^n)$ la base duale.
Alors sur $U$ on a l'écriture $\alpha=a_{i_1\ldots i_p} \omega^{i_1}\wedge\ldots\wedge\omega^{i_p}$
où $i_1<i_1<\ldots<i_p$. Soit $\mathcal{I}=(i_1,\ldots,i_p)$. Notons,en adoptant les notations de la page \pageref{oh!G}, $\omega^{\mathcal{I}}$ la $p$-forme $\omega^{i_1}\wedge\ldots\wedge\omega^{i_p}$ et $e_{\bar{\mathcal{I}}}$ le $(n-p)$-uplet
$(e_{j_1},\ldots,e_{j_{n-p}})$ .
En remarquant que l'on a pour toute fonction $f$ et toute forme $\beta$,$*(f\beta)=f(*\beta)$,
il suffit de montrer que $$**(\omega^{\mathcal{I}})=(-1)^{p(n-p)}\omega^{\mathcal{I}}.$$
 Or, $*(\omega^{i_1}\wedge\ldots\wedge\omega^{i_p})(e_{j_1},\ldots,e_{j_{n-p}})\sigma=\omega^{i_1}\wedge\ldots\wedge\omega^{i_p}\wedge\gamma(e_{j_1})\wedge\ldots\wedge \gamma(e_{j_{n-p}})=\omega^{i_1}\wedge\ldots\wedge\omega^{i_p})\wedge(\omega^{j_1}\wedge\ldots\wedge\omega^{j_{n}})=\epsilon(i_1,\ldots,i_p, j_1,\ldots,j_{n-p})\sigma$, d'après la remarque \ref{cokette} et en utilisant le fait que l'on est dans une base orthonormée ce qui implique $\gamma(e_j)=\omega^j$. Avec les notations introduites, nous venons de montrer que  $*(\omega^{\mathcal{I}})=\epsilon(\mathcal{I},\bar{\mathcal{I}})\omega^{\bar{\mathcal{I}}}$ (écriture non sommatoire ici!)
 On en déduit que 
$$*(* \omega^{\mathcal{I}})  =  \epsilon(\mathcal{I},\bar{\mathcal{I}})\epsilon(\bar{\mathcal{I}},{\mathcal{I}})   \omega^{\mathcal{I}}=(-1)^{p(n-p)}    \omega^{\mathcal{I}}$$ et on conclut.                      
\end{sol}
\begin{sol}{de l'exercice \ref{maxou}}
Démonstration écrite avec $c=1$.
\begin{enumerate}
\item En utilisant les définitions exprimées par les égalités (\ref{houa}) et (\ref{houaoua}),
on obtient :

\noindent $\star\mathcal{E}^1=\sqrt{\abs{\det g}}\left(\varepsilon(3,1,2)E^3dx\wedge dy+\varepsilon(2,1,3)E^2dx\wedge dz+\varepsilon(1,2,3)E^1dy\wedge dz\right)$. Avec nos hypothèses

\noindent$\abs{\det g}=1$ d'où
$\star\mathcal{E}^1=E^3 dx\wedge dy+E^2dz\wedge dx+E^1 dy\wedge dz=\mathcal{E}^2.$
\item De même, $\star\mathcal{B}^2=\varepsilon(2,3,1){(\mathcal{B}^2)}^{23}dx+
\varepsilon(1,3,2){(\mathcal{B}^2)}^{13}dy+\varepsilon(1,2,3){(\mathcal{B}^2)}^{12}dz=$
$=B^1dx-(-B^2)dy+B^3 dz=\mathcal{B}^1.$
\item On a $*\mathcal{B}^2=*\left(B^1 dy\wedge dz+B^2 dz\wedge dx+B^3 dx\wedge dy\right)$
Un calcul direct utilisant les égalités (\ref{houa}) et (\ref{houaoua}) montre que 
$*(dy\wedge dz)=dt\wedge dx, *(dz\wedge dx)=dt\wedge dy,*(dx\wedge dy)=dt\wedge dz$.
On en déduit que $*\mathcal{B}^2=-(B^1dx+B^2 dy+B^3 dz)\wedge dt=-\mathcal{B}^1\wedge dt.$

\item Par définition, $\mathcal{E}^1\wedge dt=-E_1 dt\wedge dx-E_2 dt\wedge dy-E_3 dt\wedge dz$, d'où :

\noindent $*(\mathcal{E}^1\wedge dt)=\varepsilon(2,3,0,1)(\mathcal{E}^1\wedge dt)^{23}dt\wedge dx+\varepsilon(1,3,0,2)(\mathcal{E}^1\wedge dt)^{13}dt\wedge dy+\varepsilon(1,2,0,3)(\mathcal{E}^1\wedge dt)^{12}dt\wedge dz
+\varepsilon(0,3,1,2)(\mathcal{E}^1\wedge dt)^{03}dx\wedge dy+\varepsilon(0,2,1,3)(\mathcal{E}^1\wedge dt)^{02}dx\wedge dz+\varepsilon(0,1,2,3)(\mathcal{E}^1\wedge dt)^{01}dy\wedge dz$.

Or $(\mathcal{E}^1\wedge dt)^{ij}=0$ si $0\notin \{i,j\}$,
$(\mathcal{E}^1\wedge dt)^{03}=-E^3,(\mathcal{E}^1\wedge dt)^{02}=-E^2$ et enfin $(\mathcal{E}^1\wedge dt)^{01}=-E^1$. Ainsi 

$*(\mathcal{E}^1\wedge dt)=E^3 dx\wedge dy+E^2 dz\wedge dx+E^1 dy\wedge dz=\mathcal{E}^2$.
\item Le dernier item résulte des deux précédents.
\end{enumerate}
\end{sol}
\begin{sol}{de l'exercice \ref{max2etc}}
\begin{enumerate}
\item On a $\mathcal{J}^3=i(\sy c\frac{\partial}{\partial x^0}+j)\tau=(\sy c )dx\wedge dy\wedge dz+i(j)\tau=\sigma^3-\left(i(j)dx\wedge dy\wedge dz\right)\wedge dx^0$ car $j=j^k\frac{\partial}{\partial x^k}$. Et on conclut.
\item D'après l'exercice \ref{rougi}, on a $d(\mathcal{B}^1\wedge dx^0)=\bold{d}(\mathcal{B}^1\wedge dx^0)=\bold{d}\mathcal{B}^1\wedge dx^0$ car $dx^0\wedge \frac{\partial(\mathcal{B}^1\wedge dx^0)}{\partial x^0}=0$ et
$d(\star\mathcal{E}^1)=\bold{d}(\star\mathcal{E}^1)+dx^0\wedge\frac{\partial(\star\mathcal{E}^1)}{\partial x^0}.$

Ainsi de l'égalité (\ref{max2}), on obtient :

\noindent $\bold{d}(\star\mathcal{E}^1)+dx^0\wedge\frac{\partial(\star\mathcal{E}^1)}{\partial x^0}-\bold{d}\mathcal{B}^1\wedge dx^0=\bold{d}(\star\mathcal{E}^1)+dx^0\wedge\frac{\partial(\star\mathcal{E}^1)}{\partial x^0}-\bold{d}(\star\mathcal{B}^2)\wedge dx^0
=\frac{4\pi}{c}(\sigma^3-j^2\wedge dx^0)$, ce qui permet de conclure le deuxième item.
\item On a $\star\mathcal{E}^1=\mathcal{E}^2 $. Ainsi la première équation de l'item $ii$)
s'écrit $\bold{d}\mathcal{E}^2=\frac{4\pi}{c} \sigma^3=\bold{d}\left( i(E) dx\wedge dy\wedge dz\right)=(\Div E)\; dx\wedge dy\wedge dz$, d'où $\Div E=4\pi\sy$.

L'équation $ii$) de l'item précédent s'écrit :

\noindent  $i(\rot B) dx\wedge dy\wedge dz=i\left(\frac{4\pi}{c} j+\frac{1}{c}\frac{\partial E}{\partial t}\right)dx\wedge dy\wedge dz$, d'où l'équation :
$$\rot B=\frac{4\pi}{c} j+\frac{1}{c}\frac{\partial E}{\partial t}.$$

\end{enumerate}
\end{sol}
\begin{sol}{de l'exercice \ref{bebete}}
\begin{enumerate}
\item Résulte directement de la définition de $\mathcal{F}^2$ par l'égalité (\ref{tenconmag}).
\item Résulte directement de l'item $5$ de la proposition \ref{champetre}.
\end{enumerate}
\end{sol}
\begin{sol}{de l'exercice \ref{denfolola}}

\noindent Par définition, on a : $(\ma{K}^1)_s=-\frac{1}{c}F_{is}J^{i}$. D'où $g^{\al s}(\ma{K}^1)_s=: (\ma{K}_1)^\al=
-\frac{1}{c}g^{\al s}F_{is}J^{i}=-\frac{1}{c}F_i^\al J^{i}$
\end{sol}
\begin{sol}{de l'exercice \ref{enimp}}
\'{E}criture avec $c=1$.
On a : $F^{oj}{F^0}_j=F^{0j}g^{0k}F_{kj}=-F^{0j}F_{0j}$. Or $F^{0j}=g^{os}g^{tj}F_{st}=-F_{0j}$ car en coordonnées galiléennes $g^{ij}=\delta^{ij}$ si $i>0$ et $j>0$ et $g^{00}=-1$. D'où
$F^{0j}F^0_j=\sum_j(F_{0j})^2=\Vert E\Vert^2$ d'après l'exercice \ref{bebete}.
Ainsi

 $-\ma{M}^{00}=\frac{1}{4\pi}\Vert E\Vert^2+2\times\frac{1}{16\pi} \left(F_{0j}F^{0j}+F_{1j}F^{1j}+F_{2j}F^{2j}+F_{3j}F^{3j}\right.)=$
 
 $=\frac{1}{4\pi}\Vert E\Vert^2+\frac{1}{8\pi}\left(-\Vert E\Vert^2+\Vert B\Vert^2)=\frac{1}{8\pi}(\Vert E\Vert^2+\Vert B\Vert^2\right).$
\end{sol}
\begin{sol}{de l'exercice \ref{paravari}}
Soit $P$ un ensemble paramétré par $\varphi$ défini sur un borélien $U$ de $\rr^p$.
On considère un point $x$ de $P$. Il existe $t\in U$ tel que $x=\varphi(t)$. Il existe un voisinage ouvert $V_t$ de $t$ dans $\rr^p$ tel que $\varphi_{\left\vert V_t\right.}$ soit un difféomorphisme de $V_t$ sur $\varphi(V_t)$. Si on considère un autre voisinage de $W_t$ de $t$ sur lequel $\varphi $ définit également un difféomorphisme sur son image, alors le changement de carte est l'identité. On a donc défini sur $\varphi(I)=P$ un atlas $C^\infty$ qui en fait une variété différentiable.
\end{sol}
\begin{sol}{de l'exercice \ref{softy}}
Pour le premier item on applique directement la définition \ref{parana}. Pour le deuxième item,
on a : $(F\circ\varphi)^*\al=\varphi^*(F^*\al)$. Ainsi, $\int_{F\circ\varphi(U)}\al=:\int_U(F\circ\varphi)^*\al=\int_U\varphi^*(F^*\al)=:
\int_{\varphi(U)}F^*\al$.
\end{sol}
\begin{sol}{de l'exercice \ref{consconma}}
En posant $X=\frac{\partial}{\partial t}+V(t,x)$, on a : $\frac{d}{dt}\int_{c_n(t)}f\rho\omega=\int_{c_n(t)}L_X(f\rho\omega)$.
Comme $L_X$ est une dérivation, on a : $L_X(f\rho\omega)=(L_Xf)\rho\omega+fL_X(\rho\omega).$

\noindent Mais 
$L_Xf=\frac{\partial f}{\partial t}+L_Vf=\frac{\partial f}{\partial t}+\frac{\partial f}{\partial x}\frac{\partial x}{\partial t}
+\frac{\partial f}{\partial y}\frac{\partial y}{\partial t}+\frac{\partial f}{\partial z}\frac{\partial z}{\partial t}=\frac{df}{dt}$.
Par ailleurs d'après un calcul déjà vu dans la démonstration du théorème \ref{tcs}, $L_X(\rho\omega)=\frac{\partial\rho}{\partial t}+\Div(\rho V)=0$ par hypothèse (en utilisant  l'équivalence démontrée dans le  théorème \ref{tcs}).
Et on conclut.
\end{sol}

\begin{sol}{de l'exercice \ref{qutenvec}}
On reprend les notations de la définition \ref{qutens}. Le symbole $\delta^{i}_j$ est le symbole de Kroncker : il vaut $0$ sauf quand $i=j$, auquel cas il vaut $1$.

On a : ${\tau'}^{i}_jX'^{j}=L^{i}_kL^h_j\tau^k_lL^j_sX^s=L^{i}_k(L^h_jL^j_s)(\tau^k_hX^s)=L^{i}_k\delta^h_s(\tau^k_hX^s)
=L^{i}_k(\tau^k_sX^s)$, soit encore $Y'^{i}=L^{i}_kY^k$.
\end {sol}
\begin{sol}{de l'exercice \ref{coool}}
On a : $ \tau^{i}_ju^j=\tau^{ik}g_{kj}u^j=\tau^{ik}u_k.$
\end {sol}

\begin{sol}{de l'exercice \ref{diphy}}
Notons respectivement $[m], [l], [t]$ les dimensions de masse, longueur, temps. Si $x$ est une grandeur $[x]$ désigne sa dimension. Soit $T$ une composante du tenseur contrainte. Alors $[T][ l^2]=[m] [l]/[t^2]$ qui est la dimension d'une masse par une accélération. Ainsi $[T]=[m]/[t^2][l]$ et $[T/c^2]=[m]/[l^3]$ qui est la dimension d'une densité de masse.

\end {sol}

 \printindex
\end{document}